\pdfoutput=1
\documentclass[fontsize=11pt,
paper=a4,
twoside=true,
numbers=noendperiod,
headings=big,
parskip=half,
captions=bottombeside,
bibliography=totoc]{scrbook}
\usepackage[a4paper,twoside,left=2.75cm,right=3.5cm,top=3cm,bottom=3cm]{geometry}

\usepackage[headsepline]{scrpage2}
\clearscrheadfoot
\ohead{\pagemark}
\ihead{\headmark}
\automark[section]{chapter}

\usepackage[english]{babel}
\usepackage{graphicx}
\usepackage{amsmath,amsfonts,amssymb,amsbsy,dsfont,amsthm,amscd,amstext}
\usepackage[percent]{overpic}
\usepackage[perpage,para,symbol*]{footmisc}
\DefineFNsymbols*{lamportnostar}[math]{\dagger\ddagger\S\P\|{\dagger\dagger}{\ddagger\ddagger}}
\setfnsymbol{lamportnostar}
\usepackage[utf8]{inputenc}
\usepackage[T1]{fontenc}

\usepackage{mathtools}
\usepackage{float}
\usepackage{rotating,standalone}
\usepackage{tikz}
\usepackage{stmaryrd}
\usepackage{dsfont}
\usepackage{hyphenat}

\usepackage[titletoc,toc,title]{appendix}
\usepackage{braket}

\usepackage[super,sort&compress,comma]{natbib}
\usepackage{hyperref}

\usepackage{makeidx}

\usepackage{color}

\usepackage{booktabs}
\usepackage{url}

\usepackage{psfrag}

\usepackage{calc}
\usepackage{tabularx}
\usepackage{array}
\newcolumntype{C}[1]{>{\centering\arraybackslash}m{#1}}
\newcolumntype{R}[1]{>{\raggedleft\arraybackslash}m{#1}}

\usepackage{ulem}  
\usepackage{setspace}  

\usepackage{nomencl}
\makenomenclature
 
\usepackage{etoolbox}
\renewcommand\nomgroup[1]{%
  \item[\bfseries
  \ifstrequal{#1}{A}{Physics Constants}{%
  \ifstrequal{#1}{B}{Symbols and abbreviations}{%
  \ifstrequal{#1}{C}{Other Symbols}{}}}%
]}
\newcommand{\nomunit}[1]{\renewcommand{\nomentryend}{\hspace*{\fill}#1}}

\makeatletter
\newcommand*{\@advisor}{}
\newcommand{\advisor}[1]{\gdef\@advisor{#1}}
\newcommand*{\@firstreader}{}
\newcommand{\firstreader}[1]{\gdef\@firstreader{#1}}
\newcommand*{\@secondreader}{}
\newcommand{\secondreader}[1]{\gdef\@secondreader{#1}}
\newcommand*{\@accdate}{\parbox[b]{5cm}{\hrulefill}}
\newcommand{\acceptancedate}[1]{\gdef\@accdate{#1}}
\newcommand*{\@authorinfo}{\parbox[b]{5cm}{\hrulefill}}
\newcommand{\moreauthor}[1]{\gdef\@authorinfo{#1}}
\makeatother

\newcommand{\vc}[1]{{\mathbf{#1}}}
\newcommand{\ssection}[1]{{\noi  \it #1:}}

\newcommand{\id}{\mathds{1}}
\newcommand{\sub}[2]{{#1}_{\mbox{\!\! \scriptsize #2}}}

\newcommand{\bv}[1]{\mathbf{ #1 }}

\def\noi{\noindent}
\def\beq{\begin{equation}}
\def\eeq{\end{equation}}

\newcommand{\rref}[1]{Ref.~\citenum{#1}}
\newcommand{\fref}[1]{Fig.~\ref{#1}}
\newcommand{\frefplural}[4]{Figs.~\ref{#1}(#2)#3\ref{#1}(#4)}

\newcommand{\frefp}[2]{Fig.~\ref{#1}#2}

\newcommand{\eref}[1]{Eq.~(\ref{#1})}

\newcommand{\sref}[1]{section~\ref{#1}}

\newcommand{\Sref}[1]{Section~\ref{#1}}
\newcommand{\Chref}[1]{Chapter~\ref{#1}}

\newcommand{\Tref}[1]{Table~\ref{#1}}
\newcommand{\aref}[1]{appendix~\ref{#1}}
\newcommand{\Aref}[1]{Appendix~\ref{#1}}

\newcommand{\op}[1]{\ensuremath{\hat{#1}}}
\newcommand{\im}{\ensuremath{\mathrm{i}}}

\newcommand{\echarge}{\ensuremath{\mathrm{e}}}

\newcommand{\dvol}{\ensuremath{\mathrm{d}^3}}

\newcommand{\dscalar}{\ensuremath{\mathrm{d}}}
\newcommand{\bohrradius}{\ensuremath{a_\mathrm{B}}}
\newcommand{\rydbergunit}{\ensuremath{\mathrm{R_{y}}}}
\newcommand{\finestructurec}{\ensuremath{\alpha_{\mathrm{fs}}}}
\newcommand{\drad}{\ensuremath{\mathfrak{d}}}
\newcommand{\dsphere}{\ensuremath{\mathfrak{S}}}
\newcommand{\lstate}[1]{\ensuremath{\mathrm{#1}}}
\newcommand{\spinorbitsymbol}{\ensuremath{\mathrm{so}}}
\newcommand{\posrep}{\ensuremath{_{\mathcal{PR}}}}
\newcommand{\ClebchGordanSymb}{\ensuremath{\mathcal{C}}}
\newcommand{\dreduced}{\ensuremath{\mathfrak{D}}}
\newcommand{\kettext}{ket }

\newcommand{\ketstext}{kets }
\newcommand{\brastext}{bras }
\newcommand{\nullop}{\ensuremath{\mathbb{O}}}
\newcommand{\quantax}{\ensuremath{\vc{q}_{\mathrm{a}}}}

\newcommand{\symbvecspace}{\ensuremath{V}}
\newcommand{\symbspinspace}{\ensuremath{S}}
\newcommand{\stylecs}[1]{\ensuremath{\mathfrak{#1}}}

\newcommand{\rep}{\ensuremath{\mathrm{rep}}}
\newcommand{\adj}{\ensuremath{\mathrm{adj}}}

\newcommand{\spaces}{\ensuremath{\symbvecspace}}
\newcommand{\spacel}{\ensuremath{\mathcal{\symbvecspace}}}

\newcommand{\spaceallspins}{\ensuremath{\mathcal{\symbspinspace}}}
\newcommand{\spacepinsmaller}[1]{\ensuremath{\mathcal{\symbspinspace}^{(#1)}}}

\newcommand{\spacefull}{\ensuremath{\stylecs{\symbvecspace}}}
\newcommand{\spacefullj}[1]{\ensuremath{\stylecs{\symbvecspace}_{#1}}}

\newcommand{\textinfig}[2]{\tikz [baseline]{ \node[#1] {#2};}}

\newcommand{\basiss}{\ensuremath{B[\spaces]}}
\newcommand{\basisl}{\ensuremath{B[\spacel]}}

\newcommand{\basissallspins}{\ensuremath{B[\spaceallspins]}}
\newcommand{\basisspinsmaller}[1]{\ensuremath{B[\mathcal{\symbspinspace}^{(#1)}]}}

\newcommand{\basissfullj}[1]{\ensuremath{B[\stylecs{\symbvecspace}_{#1}]}}

\newcommand{\Eddmax}{\ensuremath{E_{\mathrm{dd}}^{\mathrm{max}}}}
\newcommand{\Rmin}{\ensuremath{R_{\mathrm{min}}}}

\makeindex

\begin{document}

\title{Interplay of excitation transport and atomic motion in flexible Rydberg aggregates}
\subject{Dissertation}
\author{Karsten Leonhardt}
\moreauthor{geboren am 30.12.1986 in Rodewisch}
\date{14.04.2016}
\advisor{Prof. Dr. Jan Michael Rost}
\firstreader{Prof. Dr. Peter Schmelcher}
\secondreader{ }

\frontmatter
\pagenumbering{gobble}

\pagestyle{empty}
{
\usefont{\encodingdefault}{\familydefault}{\seriesdefault}{\shapedefault}
\newcommand{\tpdeflen}[2]{%
    \expandafter\newlength\csname #1\endcsname
    \expandafter\setlength\csname #1\endcsname{#2}%
}

\tpdeflen{tpvspace}{1.0cm}


{
\makeatletter

\begin{titlepage}
\bfseries
	\centering
	{ \Huge \sffamily \@title\par}
	\vspace{\tpvspace}
	{\textsc{\huge  \mdseries \@subject} \par}
	\vspace{\tpvspace}
	{\Large  zur Erlangung des akademischen Grades\par}
	\vspace{\tpvspace}
	{\Large  Doctor rerum naturalium\\ (Dr. rer. nat.)\par}
	\vspace{\tpvspace}
	{\Large  vorgelegt\par}
	\vspace{\tpvspace}	
	{\Large  der Fakult\"{a}t Mathematik und Naturwissenschaften\\
         der Technischen Universit\"{a}t Dresden\par}
	\vspace{\tpvspace}	
	{\Large  von\par}
	\vspace{\tpvspace}	
	{\Large  \mdseries \@author\par}
	\vspace{\tpvspace}	
	{\Large   \mdseries \@authorinfo\par}
	\vspace{\tpvspace}	
	{\Large  Eingereicht am \@date \\Verteidigt am 18.10.2016\par}
\vfill\null
\vfill\null
\quad\\
\vspace*{\fill}
{\mdseries
\begin{minipage}[b]{\textwidth}
     1. Gutachter: \@advisor\\
     2. Gutachter: \@firstreader\\
\end{minipage}}
\end{titlepage}
}
\makeatother
}

\clearpage
\thispagestyle{empty}
\par\vspace*{.35\textheight}{\centering\Large \textit{To my grandpa~(1930-2016).}\par}

\pagestyle{empty}
\addchap*{Abstract}\label{abstract} 
Strong resonant dipole-dipole interactions in flexible Rydberg aggregates enable the formation of excitons, many-body states which collectively share excitation between atoms. Exciting the most energetic exciton of a linear Rydberg chain whose outer two atoms on one end are closely spaced causes the initiation of an exciton pulse for which electronic excitation and diatomic proximity propagate directed through the chain. The emerging transport of excitation is  largely adiabatic and is enabled by the interplay between atomic motion and dynamical variation of the exciton.  

Here, we demonstrate the coherent splitting of such pulses into two modes, which induce strongly different atomic motion, leading to clear signatures of nonadiabatic effects in atomic density profiles. The mechanism exploits local nonadiabatic effects at a conical intersection, turning them from a decoherence source into an asset. The conical intersection is a consequence of the exciton pulses moving along a linear Rydberg chain and approaching an additional linear, perpendicularly aligned Rydberg chain.
The intersection provides a sensitive knob controlling the propagation direction and coherence properties of exciton pulses.

We demonstrate that this scenario can be exploited as an exciton switch, controlling direction and coherence properties of the joint pulse on the second of the chains.
Initially, we demonstrate the pulse splitting on planar aggregates with atomic motion one-dimensionally constrained and employing isotropic interactions. Subsequently, we confirm the splitting mechanism for a fully realistic scenario in which all spatial restrictions are removed and the full anisotropy of the dipole-dipole interactions is taken into account. Our results enable the experimental observation of non-adiabatic electronic dynamics and
entanglement transport with Rydberg atoms. The conical intersection crossings are clearly evident, both in atomic mean position information and excited state spectra of the Rydberg system. This suggests flexible Rydberg aggregates as a test-bench for quantum chemical effects in experiments on much inflated length scales. The fundamental ideas discussed here have general implications for excitons on a dynamic network.

{
\hypersetup{hidelinks}
\pagestyle{empty}
\tableofcontents
\cleardoublepage
}
\pagestyle{scrheadings}

\mainmatter
\setcapindent{0em}
\setkomafont{caption}{\small}
\setkomafont{captionlabel}{\bfseries}

\addchap{Introduction}\label{chap:intro}
At all times humans were seeking for a deeper understanding of nature. Already Greek philosophers during the classical age had a big debate about the nature of reality, with two opposing theories. Leucippus and his disciple Democritus proposed that all matter consists of smallest, indivisible particles --- atoms --- and established eventually the natural philosophy of atomism\cite{democritus:the_atomist_ancient_greece}. In contrast, many intellectuals believed in matter as being a continuum and hence unendingly divisible. The question about what ``[b]inds the world’s innermost core together''\footnote{from Goethe's Faust: The First Part of the Tragedy. Dr. Faust studied his whole life long, anyhow he is dissatisfied and expresses that mankind can know nothing. Despite this, he feels the urge to continue questioning about nature and eventually emphasizes his impulse with the phrase cited above.} concerned philosophers and later on physicists for centuries. It was not before the nineteenth and early twentieth century that theoretical models could satisfy empirical data from spectroscopic measurements. In particular, for a theoretical understanding of observed spectral lines in Hydrogen\cite{angstroem_recherches_solaire,balmer:spectral_lines} and subsequently other chemical elements, the development of quantum mechanics\cite{bohr1913:bohr_model1,bohr1913:bohr_model2} was essential. The atomism of ancient times appeared to be right after all. Consequently, atomic physics was established with the theoretical calculation of atomic energy spectra and spectroscopic measurements. A breakthrough for the further progress of atomic physics was the development of the laser\cite{maiman1960:laser} in 1960, following a proposal of Schawlow and Townes\cite{schawlow1958:laser_theory}. It opened the way of high precision spectroscopy\cite{demtroder1973:high_res_spec_lasers}, notably by the development of narrow-linewidth tunable lasers\cite{haensch1972:tunable_dye_laser}. This, in turn, led to the development driving laser cooling\cite{haensch1975:laser_cooling_gases,phillips1985:laser_cooling,aspect1988:laser_cooling_below_recoil,kasevich1992:laser_cooling_nK} of atomic ensembles to the nK regime. Since then, experiments with atoms in the ultracold temperature regime had massive impact on quantum optics, condensed matter and quantum statistical physics. Particularly worth mentioning is the experimental observation of Bose-Einstein condensation\cite{anderson1995:bec_observ,davis1995:bec_observ2} and the renaissance of Rydberg atoms\cite{book:gallagher} in the early noughties. Finally, the development of optical lattices\cite{jessen1992:observ_quant_motion_optical_lattice,verkerk1992:spatial_order_cesium_optical_lattice,hemmerich1993:2datomic_crystal_optical_lattice,raithel1997:cooling_optical_lattice,guidoni1997:quasiperiodic_optical_lattices,bloch2005:ultracold_quantum_gases_optical_lattices} allowed the realization of condensed matter many-body Hamiltonians\cite{jaksch1998:bosonic_atoms_optical_lattices,lewenstein2007:mimicking_condensed_matter_physics_ultracold_atomic_gases,bloch2008:many_body_physics_ultracold_gases} with ultracold atoms by spatially ordering and trapping them with counterpropagating laser beams. Although already intensively studied in the early days of spectroscopy, a new era began for highly excited atoms --- Rydberg atoms --- with the possibility to cool\cite{haensch1975:laser_cooling_gases,phillips1985:laser_cooling,aspect1988:laser_cooling_below_recoil,kasevich1992:laser_cooling_nK,klots1985:evaporative_cooling,hess1986:evaporative_cooling_magnetic_trapped_spin_polarized_hydrogen,masuhara1988:evaporative_cooling_spin_polarized_hydrogen,adams1995:evaporative_cooling_crossed_dipole_trap,ketterle1996:evaporative_cooling_trapped_atoms}, trap\cite{raab1987:trapping_of_neutral_sodium_atoms,monroe1990:ultracold_atoms_trapped_vapor_cell,chu1992:laser_trapping,jessen1992:observ_quant_motion_optical_lattice,verkerk1992:spatial_order_cesium_optical_lattice,hemmerich1993:2datomic_crystal_optical_lattice,raithel1997:cooling_optical_lattice,guidoni1997:quasiperiodic_optical_lattices,bloch2005:ultracold_quantum_gases_optical_lattices} and coherently excite them, both as mesoscopic ensembles\cite{reetz2008:rabi_oscillations_mesoscopic_frozen_gas,reetzlam:rabiblockade} and individually\cite{johnson2008:rabi_oscilltation_ground_rydberg}, within the ultracold temperature regime. 
Although Rydberg atoms were intensively studied since the very beginning of atomic physics, the discovery of the \emph{dipole blockade}\cite{lukin2001:dipole_blockade,singer2004:dipole_blockade_experimental_evidence,urban:twoatomblock}, which inhibits the excitation of more than one atom to a Rydberg state within a small volume, drew the attention of many ultracold atom physicists towards them. The dipole blockade is a consequence of extremely large, long-range interactions, such that Rydberg atoms affect each other on distances of several micrometers.
Based on this blockade, there were proposals for quantum information processing\cite{jaksch:dipoleblockade,lukin2001:dipole_blockade,urban:twoatomblock,gaetan:twoatomblock,saffman2010:quantuminformation_ryd_atoms}, the simulation of spin systems\cite{Weimer-Buchler-Rydbergquantumsimulator-2010,lesanovsky:kinetic,rvb:quantmagdressed,glaetzle:frustquantmag} and quantum optical nonlinear effects, such as electromagnetically induced transparency~(EIT)\cite{pritchard2010:eit,petrosyan2011:eit,gorshkov2011:photon-photon_interaction_via_blockade,cenap2011:eit,sevincli2011:nonloc_nonlin_optics_rydberg_gas,peyronel2012:quantum_nonlinear_optics_single_photons}. The latter effectively leads to a strong photon-photon interaction and remarkably, saturation of probe field transmission with few photons. The strong interactions yield spatially ordered structures of Rydberg excitations\cite{pohl2010:dynamical_crystallisation_dipole_blockade,schauss2012:observation_spatially_ordered_structures_Rydberg_gas} and other exotic phases of matter in Rydberg dressed ground-state atoms, such as supersolidity\cite{cinti2010:supersolid_rydberg,nils2010:supersolids,henkel2012:vortex_crystals,henkel2013:phd_thesis}.
Besides these features concerning the Rydberg excitation, new forms of matter, particularly bound states of ground-state and Rydberg atoms forming Rydberg molecules were theoretically predicted\cite{Greene:LongRangeMols,boisseau2002:rydberg_molecules_dimer,liu:prediction_trylobite_ryd_mol,rittenhouse2010:polyatomic_ryd_atoms} and experimentally observed\cite {bendkowsky2009:rydberg_molecules_observ,bendkowsky2010:rydberg_trimer_quantum_refl,butscher2010:rydberg_molecule,butscher2011:phd_thesis,nipper2012:phd_thesis,li2011:homonuc_ryd_mol_perm_electr_dip}.

Although Rydberg atoms mutually perpetrate strong forces, most of the aforementioned processes can occur on a much smaller time-scale than the initiation of atomic motion, allowing to consider the atomic ensemble as a frozen Rydberg gas\cite{mourachko1998:frozen_rydberg_gas}. The unavoidable (thermal) motion of the atoms constitutes then a limiting source of noise and decoherence \cite{wilk:entangletwo,mueller:browaeys:gateoptimise}.
 This situation changes by doping the Rydberg gas with an additional angular momentum excitation, which results in dominant resonant dipole-dipole interactions, in contrast to dominant van-der-Waals~(vdW)\nomenclature[B]{vdW}{van-der-Waals}\index{dipole-dipole interaction!van-der-Waals} interactions for non-doped Rydberg gases. Such doped Rydberg systems are called \emph{flexible Rydberg aggregates}\cite{moebius:cradle}, where aggregation refers to the formation of excitons, many-body states which collectively share excitation between atoms due to strong resonant interactions, and flexibility refers to the significance of atomic motion for the transfer of excitation. In fact, spatial migration of the doping occurs on the same time-scale as atomic motion. In contrast to other systems featuring resonant energy transfer~(RET)\nomenclature[B]{RET}{resonant energy transfer}\index{resonant energy transfer}, where spatial motion of constituents perturbs migration of electronic excitation, the strong interactions of Rydberg atoms link electronic and spatial degrees of freedom to allow for directed transport and ultimately generate combined pulses of doping migration and diatomic proximity, called \emph{exciton pulses}\cite{wuester:cradle,moebius:cradle}.
These pulses provide an efficient way to transfer entanglement, which is a purely quantum mechanical quantity, with high fidelity. This result is remarkable, since the spatial degrees of freedom can be thought of as a bath to the electronic system, and typically, couplings to a bath are a source of decoherence and destroy entanglement. 

Previous studies of flexible Rydberg aggregates revealed 
combined transport of electronic coherence along with atomic mechanical momentum in reduced dimensional geometries\cite{cenap:motion,wuester:cradle,moebius:cradle,wuester:CI,wuester:dressing,zoubi:VdWagg,moebius:bobbels,wuester:cannon,genkin:dressedbobbles,moebius:cat,leonhardt:switch}, in particular one-dimensional linear\cite{cenap:motion,wuester:cradle,moebius:cradle,wuester:dressing,zoubi:VdWagg,moebius:bobbels,wuester:cannon,genkin:dressedbobbles,moebius:cat,leonhardt:switch} and ring configurations\cite{wuester:CI}. The exciton pulse is initiated by preparing the aggregate in a localized exciton state for which excitation resides on a single atom-pair by chosing their interatomic distance to be small compared to all other spatial atomic spacings. Directing the pulse requires to position the diatomic proximity at one end of an otherwise equidistant linear Rydberg chain.

Flexible Rydberg aggregates are also useful to study controlled non\-adiabatic dynamics, which allow transitions to other excitons. The rich structure of the dipole-dipole interactions together with the possibility to tune them with the atomic configuration of the aggregates allows to engineer conical intersections~(CI)s\nomenclature[B]{CI}{conical intersection}\index{conical intersection}\cite{neumann1929:ci_first_paper,teller1937:ci_crossing_of_pes,yarkony1996diabolical,yarkony2001conical,domcke2004:book_cis,spiridoula2007:ci_molec_systems_review,wuester:CI},
 which are genuine energy surface crossings of two or more electronic states and provide radiationless transitions allowing for instance isomerization of molecules or the transformation of excitons.
A previous study of a flexible Rydberg trimer with atoms confined on a ring showed nonadiabatic dynamics due to a CI\cite{wuester:CI}.

Both RET and CI-dynamics are important features of chemical and biological processes. RET is essential for molecular aggregates\cite{ritschel2011:absence_quantum_oscillations_FMO,olbrich2011:atomistic_modeling_to_excitation_transfer_FMO,lambert2013:quantum_biology,saikin:excitonreview}, assemblies of molecules with strong near-field interactions between electronic excitations in the individual subunits and thereby featuring coherent energy transport.
Prominent examples of a molecular aggregates in nature are light-harvesting complexes~(LHC)s\cite{kuhlbrandt1991:3d_structure_lhc,kuehlbrandt1994:atomic_model_lhc,mcdermott1995:crystal_structure_lhc}\nomenclature[B]{LHC}{light-harvesting complex}\index{light-harvesting complex}. They are functional units in photosynthesis with an optimized self-assembled structure to maximally absorb solar photons in order to subsequently transport photoinduced energy to a certain reaction center, enabled by resonant dipole-dipole interactions between chlorophylls, which are important biomolecules. The transport mechanism was first described by Förster\cite{foerster1948:RET} and is called \emph{Förster resonance energy transfer}~(FRET)\nomenclature[B]{FRET}{Förster resonance energy transfer}\index{Förster resonance energy transfer}.
Many theoretical and experimental investigations were and still are carried out with LHCs.
Under debate is the root of high efficiency of transport, which is remarkably performed at ambient temperatures. Also the role of quantum effects for the transport\cite{ritschel2011:absence_quantum_oscillations_FMO,olbrich2011:atomistic_modeling_to_excitation_transfer_FMO,lambert2013:quantum_biology} is ambiguous. Distinguishing between quantum and classical transport is challenging for experiments, since a controlled decoupling of the aggregate from the environment would be necessary or even isolation of aggregate subunits, which is very difficult.

The role of CIs was under-estimated for a long time, when they were considered to be rather mathematical artifacts than features with physical impact. During the last two decades studies revealed that they appear more frequently than initially thought. Moreover, there is experimental evidence for their significance in chemical and biological processes. For example, CIs appear as functional junctions in vision, where they provide ultrafast relaxation and thereby enable photoisomerization of rhodopsin\cite{hahn2002:ultrafast_cis_trans_photoswitching,ben-nun2002:cis_trans_photoisomerization,levine2007:isomerization_through_CI,polli2010:ci_photoisomerization,martinez2010:seaming_believing}. In general it is today widely accepted that CIs serve as radiationless decay channel\cite{domcke2012:review_ci,robb1995:CI_as_mechanistic_feature_organic_photochemistry} in organic photochemistry.

Rydberg aggregates differ in many aspects from molecular aggregates and the usage of the term aggregation in connection with Rydberg systems is limited to the ability of the atoms to collectively share excitation due to resonant dipole-dipole interactions. In the following we discuss the differences between both types of aggregates. Molecular aggregates have spatial dimensions $\sim$~nm. The internal structure of them, even of individual monomers\footnote{A molecule or a compound of molecules which can undergo polymerization, thereby contributing constitutional units to the essential structure of a macromolecule\cite{IUPAC1997}.}, is very complex and they are often embedded in an structured environment, which stabilizes a particular configuration of the aggregate and also makes them resistant against heat. This complexity demands many approximations and efficient methods\cite{strunz1996:linear_quantum_state_diffusion,strunz1998:non_markovian_quantum_state_diffusion,strunz1999:non_markov_qsd_prl,ritschel2011:FMO_efficient_method,ritschel2015:diss,valleau2012:alternatives_bath_correlators_quantum_classical_simualtions,suess2014:hierarchy_stochastic_states_oqs} to theoretically describe them.
On the other hand, Rydberg aggregates show RET on much inflated length, $\sim\ \mu$m,  and time scales, $\sim\ \mu$s\cite{white:coherence:moljunction,wuester:cradle,moebius:cradle}, due to the strong interactions. However, they do not appear naturally and are therefore artificial aggregates which have to be synthetically prepared in an ultracold experimental setup.
The experimental progress in ultracold atomic physics provides highly tunable model systems regarding the adjustability of interaction strengths, spatial arrangements and also the degree of coupling with environmental degrees of freedom. This allows the study of coherent transport without decoherence sources, in contrast to molecular aggregates. However, influences from an environment can naturally be added by embedding Rydberg aggregates in a background gas of ground-state atoms\cite{schoenleber:immag}, which is experimentally automatically realized. The huge spatial dimensions facilitate a direct experimental observation of nonadiabatic and exciton dynamics, with several novel observation techniques, based on optical monitoring\cite{olmos:amplification,guenter:EIT,guenter:EITexpt,schoenleber:immag,schempp:spintransport}, microwave spectroscopy\cite{celistrino_teixeira:microwavespec_motion} and position sensitive field ionization\cite{thaicharoen:trajectory_imaging}. This indicates their eligibility as quantum simulators for biological and chemical processes, e.g. to identify key elements of energy transfer in LHCs. Specifically, our aim is to clarify the impact of constituent motion on excitation transfer in higher-dimensional systems. Exciton pulses in flexible Rydberg aggregates represent a transport mechanism that essentially relies on atomic motion for adiabatic transport. The studied aggregates were one-dimensional for which combined pulse propagation could be be theoretically confirmed. However, the spatial configurations of molecular aggregates are higher-dimensional and excitation transfer can be affected by sources of nonadiabaticity, for instance CIs. It is a priori unclear how exciton pulses are affected by CIs and if they at all can be sustained after traversing such an intersection region.

In this thesis we study flexible Rydberg aggregates with higher-dimensional configurations to investigate exciton pulses under the influence of CIs. We aim for demonstrating the suitability of flexible Rydberg aggregates as test bench for quantum transport.
The geometry of the aggregates is T-shaped, which is realized by perpendicularly aligning linear Rydberg chains. The exciton pulse is initiated on one of the two linear Rydberg chains and hits a CI when approaching the adjacent Rydberg chain. We demonstrate the coherent splitting of a single exciton mediated by a CI which results in the creation of a coherent superposition of two states of the excitons. This process is enabled through strong couplings between spatial and electronic degrees of freedom and ultimately results in entanglement between both. 
The CI prevents exciton pulse propagation on the adjacent chain, however, choosing a slightly asymmetric T-shape configuration allows for redirecting the exciton pulse on the adjacent chain. Moreover, we demonstrate the control of directionality for pulse propagation on the adjacent chain by tuning a single internal dimension of the aggregate.

\section*{Organization of the thesis}
A review of Rydberg atoms is given in \Chref{part:fd::chap:ryd_a}, where we first recall the Hydrogen energy levels and quantum states to get an intuition of energy spacings and properties of the wave function. Furthermore it sets the basic notation, which we use henceforward. This chapter introduces the essential interactions between alkali Rydberg atoms --- dipole-dipole interactions --- and contains formulae for transition matrix elements describing the resonant swap of a Rydberg \lstate{p} and Rydberg \lstate{s} state of two atoms. We also discuss the treatment of vdW interactions and compare them to their resonant counterpart.

In \Chref{part:fd::chap:theoretical_framework} we introduce the process of RET in both frozen and unfrozen systems. In frozen systems excitation is transferred via Rabi-Oscillations. However, excitons are time-independent since they are eigenstates of the electronic Hamiltonian therefore transport is only possible with localized excited states. In contrast to this, in unfrozen systems, excitons can be tuned from localizing to delocalizing excitation since the flexibility of the spatial arrangement varies the interaction between the constituents. This opens the way to transport excitation with excitons, which is called adiabatic transport.
Using the example of a Rydberg trimer, we illustrate this dependency of the excitons on the spatial configuration and show that an equilateral triangle configuration features a CI.
%
%
Finally, we motivate the use of T-shape aggregates to study combined exciton pulse propagation and CI dynamics.
The T-shape configuration arises as a combination of two perpendicularly aligned linear Rydberg chains, referred to as horizontal and vertical chain, respectively, whereas the exciton pulse is always initiated on the horizontal chain.

\Chref{part:rs::chap:planar_aggregates} is dedicated to the investigation of planar flexible Rydberg aggregates and employs a simple model of interactions, which we assume to be isotropic. The aggregate atoms are all excited to the same Rydberg \lstate{s} state and additionally doped with a Rydberg \lstate{p} excitation of the same principal quantum number manifold, which due to the interactions is collectively shared between atoms, dependent on the atomic configuration.
At the beginning we present the theoretical framework, including the treatment of resonant and off-resonant interactions, preparation of the initial state and a method to obtain the quantum dynamics.

To understand how a CI affects the excitation transfer, we first study a minimal T-shape aggregate with four atoms, two atoms on each chain. The configuration is symmetric with the horizontal dimer centered with respect to the vertical one.
In order to investigate whether excitation transfer can be sustained after the exciton pulse traversed the vicinity of the CI, we then study an extended T-shape aggregate with four atoms on the vertical and three atoms on the horizontal chain, again with the horizontal chain centered relative to the vertical one. 
Finally, we demonstrate how high fidelity exciton pulse propagation can be obtained after redirection on the vertical chain by shifting the horizontal chain in vertical direction and hence introducing an asymmetry.
The main result is that the pulse propagation direction can be controlled by varying a single interatomic distance. 
%

To reduce the experimental challenges in observing these features, in particular the CI mediated exciton splitting, we remove in \Chref{part:rs::chap:unconstrained_aggregate} all restrictions and simplifications but keep the setup of transferring a single \lstate{p} excitation. Different to \Chref{part:rs::chap:planar_aggregates}, the atomic motion is here completely unconstrained and the anisotropy of the interactions is taken into account. The studied system is a four atom T-shape aggregate, similar to the planar four atom aggregate.
We will demonstrate that if initiated in a low dimensional space, entangled atomic motion in the continuum will remain confined to this space despite the possibility for all particles (ions and electrons) to move in full space and that the CI retains the function as an exciton splitter.
Together with advances in the newest generation experiments on Rydberg gases beyond the frozen gas regime, involving microwave spectroscopy\cite{celistrino_teixeira:microwavespec_motion} or position sensitive field ionization\cite{thaicharoen:trajectory_imaging}, the results enable quantum simulation of chemical processes in flexible Rydberg aggregates as an experimental science and render now the rich dynamics of Rydberg aggregates fully observable\cite{leonhardt:3dswitch}.

In the final \Chref{part:app::chap:tun_int_mag_field} we illustrate how dipole-dipole interactions can be modified by applying an external magnetic field. The resonant interactions of the planar aggregates in \Chref{part:rs::chap:planar_aggregates} are assumed to be isotropic with negative amplitudes. However, both these properties can only be easily realized for Rydberg aggregates. In \Chref{part:app::chap:tun_int_mag_field}, we demonstrate how to approximately realize the simple interaction model by applying an external magnetic field. We use block-diagonalization techniques to derive a formula describing effective interactions. Finally we compare isotropic and effective interactions with the complete interaction model, which also takes spin-orbit coupling into account.

\section*{Publications}
In the context of this thesis, the following articles have been submitted and published:

\begin{itemize}
 \item K. Leonhardt, S. Wüster, and J. M. Rost, ``Switching Exciton Pulses Through
Conical Intersections,'' Phys. Rev. Lett. \textbf{113}, 223001 (2013).
\item  K. Leonhardt, S. Wüster, and J. M. Rost, ``Orthogonal flexible Rydberg aggregates,'' Phys. Rev. A \textbf{93}, 022708 (2016).
\item  K. Leonhardt, S. Wüster, and J. M. Rost, ``Exciton induced directed motion of unconstrained atoms in an ultracold gas,'' arXiv:1602.01032, submitted (2016).
\end{itemize}

\chapter{Rydberg atoms}\label{part:fd::chap:ryd_a}
The name Rydberg is closely connected to the early days of spectroscopy.
The swedish physicist \AA{}ngström was able to identify hydrogen by performing a spectral analysis of sunlight\cite{angstroem_recherches_solaire}. The swiss physicist Balmer found a formula\cite{balmer:spectral_lines} to compute the wavelengths of a series of hydrogen spectral lines, matching the experimentally found values of \AA{}ngström. The series describes the transitions of the hydrogen electron from higher excited states to the first excited state. Quantum mechanics in its early days gained a lot of attention with the successful theoretical explanation of the experimentally found discrete spectral lines. It was the \emph{Bohr model}\cite{bohr1913:bohr_model1,bohr1913:bohr_model2}, introduced by Niels Bohr in 1913, which reduced the classical options for the dynamics of the electron surrounding the proton by allowing only specific orbits, such that the spectrum of orbital energy levels is discrete. It was furthermore postulated that the atom is stable when the electron remains on an orbit implying that no emission of radiation occurs. The spectral lines are the result of a quantum jump, a transition of the electron between two orbitals with different energy.
Besides Hydrogen, highly excited atoms of other species with a single valence electron have spectral lines similar to the spectral lines of Hydrogen, when the formula is readjusted with the individual atomic mass and atomic number. Since Rydberg was the first to observe the spectral lines for Hydrogen, highly excited atoms are called \emph{Rydberg atoms}. 
Their inner structure is thus very similar to Hydrogen and can be well described by the Bohr or the extended \emph{Bohr-Sommerfeld model}\cite{sommerfeld1916:quant_theory_spec_lines}. Since these old quantum mechanical models do not account for all quantum features with their quantization scheme, they are \emph{semiclassical} models and were later corrected by the new quantum mechanics --- wave\cite{deBroglie1923:waves_and_quanta,schroedinger1926:quantisierung_eigenwertproblem,schroedinger1926:uebergang_mikro_makromechanik,schroedinger1926:undulatory_theory_phys_rev} and matrix\cite{heisenberg1925:qm_reinterpretations,born1925:zur_quantenmechanik,born1926:zur_quantenmechanik_2} mechanics.
Based on this consideration, Rydberg atoms can be thought of being on the edge between classical and quantum mechanics. In fact, Rydberg atoms where the orbits are almost circular, behave virtually as classical objects.
Although Rydberg atoms were studied in detail already one hundred years ago, they underwent a renaissance in the last twenty years, which is closely connected to the physics of quantum computation and information. The elementary units performing logical operation on a quantum level, quantum gates, require highly coherent systems with large interactions between the constituents. This brought the attention to Rydberg atoms which have both properties, allowing for fast gate operations\cite{jaksch:dipoleblockade,saffman2010:quantuminformation_ryd_atoms}. Intimately connected with the large interactions between Rydberg atoms is the dipole-blockade, which inhibits transitions into all but singly excited states\cite{lukin2001:dipole_blockade}. Our studies utilize both properties of Rydberg atoms as well but to create entanglement between atomic motion and dynamics of electronic excitation. A detailed analysis of Rydberg atoms can be found in Ref.~\citenum{book:gallagher,gallagher2008:dip_dip_interactions_ryd_atoms,choi2007:cold_rydberg_atoms,saffman2010:quantuminformation_ryd_atoms}.

This chapter starts with the Hydrogen problem in \Sref{part:fd::chap:ryd_a::sec:hyd_ryd_a}, to review the solution for the energy levels and stationary wave functions for Hydrogen-like atoms and to set basic notations. Since the Rydberg aggregates are based on lithium atoms, we present the treatment of alkali Rydberg atoms in \Sref{part:fd::chap:ryd_a::sec:alk_ryd_a} with the definition of dipole matrix elements and a discussion of how to calculate them. Based on this, \Sref{part:fd::chap:ryd_a::sec:dip_dip_int} finally contains the derivation of the dipole-dipole interaction formula, the evaluation of their transition matrix elements and vdW~interactions\index{dipole-dipole interaction!van-der-Waals} as a result of couplings to off-resonant states.

%
%

%
%
%
\section{Hydrogen}\label{part:fd::chap:ryd_a::sec:hyd_ryd_a} 
Hydrogen\index{Hydrogen|(} is the most important reference element in Rydberg physics, since its wave function and spectra can be calculated analytically and a lot of properties only slightly change for alkali Rydberg atoms. 
We introduce the vector $\vc{r}$, pointing from the core to the position of the electron and denote with $r=|\vc{r}|$ the distance between both charges.
The relevant properties of Hydrogen are described through wave functions, $\varphi(\vc{r})$, which depend only on the relative coordinate vector $\vc{r}$ and are solutions of the time-independent Schrödinger equation\index{Schrödinger equation!time-independent} 
\begin{equation}
 \hat{H}(\vc{r})\varphi(\vc{r}) = E\varphi(\vc{r}).
\label{eq:ryd_atoms:time_indep_schroedinger_equ}
\end{equation}
\nomenclature[B]{$\hat{H}$}{notation for an Hamiltonian}
According to the statistical interpretation of quantum mechanics, the absolute square value of the wave function is interpreted as spatial probability density. This requires for bound states where $E<0$ the normalization condition
\begin{equation}
 \int_{\mathbb{R}^3}|\varphi(\vc{r})|^2\,\dvol r = 1.
\label{eq:ryd_atoms:wave_function_normalisation}
\end{equation}

Due to boundary conditions and the normalization condition in \eref{eq:ryd_atoms:wave_function_normalisation}, each wave function of the bound states solutions of \eref{eq:ryd_atoms:time_indep_schroedinger_equ} is uniquely defined by a set of quantum numbers, $\Gamma = \{\gamma_1,\dots,\gamma_N\},\, N\geq 1$. We denote each of these functions with $\ket{\Gamma}$, which is a so called \kettext - the function not expanded in a basis. For each \kettext $\ket{\Gamma}$, there is a ``bra'' $\bra{\Gamma}$ - the corresponding function from the dual space, such that $\bra{\Gamma}\cdot \ket{\Gamma} = \braket{\Gamma|\Gamma} = \|\Gamma\|^2$.
Projecting the \kettext to an eigenstate of the position operator, we get the wave function in position representation, $\varphi_{\Gamma}(\vc{r}) :=\braket{\vc{r}|\Gamma}$. 
Furthermore we denote for an operator $\hat{A}$ with
\begin{equation}
 \braket{\hat{A}}_{\Gamma}^{\Gamma'}:=\bra{\Gamma}\hat{A}\ket{\Gamma'},
\label{eq:ryd_atoms:def_transition_matrix_element_operator_abstract}
\end{equation}
the matrix element for a transition $\ket{\Gamma} \to \ket{\Gamma'}$ and abbreviate $\braket{\hat{A}}_{\Gamma}:=\braket{\hat{A}}_{\Gamma}^{\Gamma}$ for expectation values. If a position-representation $\hat{A}\posrep$ for the operator $\hat{A}$ exists, the calculation of the matrix elements can explicitly be perfomed with
\begin{equation}
 \braket{\hat{A}}_{\Gamma}^{\Gamma'} = \int_{\mathbb{R}^3}\varphi_{\Gamma}^{*}(\vc{r})\hat{A}\posrep\varphi_{\Gamma'}(\vc{r})\dvol r.
\label{eq:ryd_atoms:def_transition_matrix_element_operator_pos_rep}
\end{equation}
We continue to denote the operators without specification of the representation, unless it is needed. If an operator acts on pure \kettext or ``bra'' vectors, the operator is also not expanded in a basis. Otherwise, if the operator is followed by a state in a certain representation, the operator has to be in the same representation.

The wave functions in \eref{eq:ryd_atoms:time_indep_schroedinger_equ} are eigenstates of the Hamiltonian $\hat{H}(\vc{r})$, which for Hydrogen is given by
\begin{equation}
 \hat{H}(\vc{r}) = \dfrac{1}{2\mu}\Bigl(-\hbar^2\Delta_r + r^{-2}\hat{\vc{L}}^2\Bigr) + V(r).
\label{eq:ryd_atoms:hamiltonian_spherical_symmetric_potential}
\end{equation}
The Hamiltonian in \eref{eq:ryd_atoms:hamiltonian_spherical_symmetric_potential} is the result of reducing the two-body problem of a proton and an electron to an effective one-body problem, with the reduced mass $\mu := m_{\mathrm{c}} m_{\echarge}/ ({m_{\mathrm{c}} + m_{\echarge}})$, where $m_\echarge$\nomenclature[A]{$m_{\echarge}$}{electron mass \nomunit{$9.10938291(40)\cdot10^{-31}$~kg}} is the electron mass and $m_{\mathrm{c}}$ is the mass of the core, which is for Hydrogen the proton mass $m_{\mathrm{p}}$\nomenclature[A]{$m_{\mathrm{p}}$}{proton mass \nomunit{$1.672621777(74)\cdot10^{-27}$~kg $= 1836.2\,m_{\echarge}$}}.
Since the eigenvalues of the Hamiltonian are the possible total energies, it has to consist of an operator for the kinetic energy,
$\hat{T} := ({2\mu})^{-1}\hat{\vc{p}}^2$
and a potential $V(r)$.
For Hydrogen the potential is spherically-symmetric and thus the orbital angular momentum is a conserved quantity. This makes it convenient to use for the squared momentum operator the position-representation identity $\hat{\vc{p}}^2:= -\hbar^2\Delta_r + r^{-2}\hat{\vc{L}}^2$, where $\Delta_r$ is the purely radially dependent part of the Laplace~operator and $\hat{\vc{L}}^2$ the squared orbital angular momentum operator, which is related to the angularly dependent part of the Laplace~operator, $\Delta_{\theta,\phi}=-\hat{\vc{L}}^2/(\hbar r)^2$. Both parts of the Laplace~operator have the specific form
\begin{align}
 \Delta_r&:=\dfrac{1}{r^2}\dfrac{\partial}{\partial r}\left(r^2\dfrac{\partial}{\partial r}\right),
\label{eq:ryd_atoms:laplace_op_radial}
\\
\Delta_{\theta,\phi} &:= \dfrac{1}{r^2\sin^2 \theta}\left[
\sin \theta \dfrac{\partial}{\partial \theta}\left( \sin \theta \dfrac{\partial}{\partial \theta}\right) + \dfrac{\partial^2}{\partial \phi^2}
\right].
\label{eq:ryd_atoms:laplace_op_spherical}
\end{align}
The angles $(\theta,\phi)$ are the azimuthal and polar angle of the vector $\vc{r}$, whose azimuthal axis is chosen to be the quantization axis, which we denote with \quantax\nomenclature[B]{\quantax}{quantization axis}.
Since Hydrogen is a system of two elementary charges $\echarge$\nomenclature[A]{$\echarge$}{elementary electric charge \nomunit{$1.6021766208(98)\cdot 10^{-19}$~C}}, the core positive and the electron negative, its potential is purely coulombic,
\begin{equation}
 V(r) = \dfrac{-Z\echarge^2}{4\pi \epsilon_0} \dfrac{1}{r},
\label{eq:ryd_atoms:potential_hydrogen}
\end{equation}
with the atomic number\index{atomic number} $Z=1$, the number of protons in the core. By chosing this number higher, Hydrogen-like atoms can be described. This is helpful to get rough analytical approximations for the wave functions of alkali Rydberg states.
The product ansatz of the wave function
\begin{equation}
 \varphi(\vc{r}) = r^{-1}R(r)Y(\theta, \phi),
\label{eq:ryd_atoms:separation_wave_function}
\end{equation}
with a radial part $R(r)$ and an angular part $Y(\theta, \phi)$ leads to 
a successful separation of variables in \eref{eq:ryd_atoms:time_indep_schroedinger_equ}, by choosing for the angular part the spherical harmonics $Y_{\ell, m}(\vartheta, \varphi)$, which are eigenstates of $\hat{\vc{L}}^2$ and $\hat{L}_z$, the $z$-component of the angular momentum operator:
\begin{align}
 \hat{\vc{L}}^2Y_{\ell, m}(\theta, \phi) &= \hbar^2 \ell (\ell+1)Y_{\ell, m}(\theta, \phi),\label{eq:ryd_atoms:eig_functions_Lsquared}\\
\hat{L}_z Y_{\ell, m}(\theta, \phi) &= \hbar m Y_{\ell, m}(\theta, \phi).
\label{eq:ryd_atoms:eig_functions_Lz}
\end{align}
The two indices of the spherical harmonics are the azimuthal quantum number $\ell = 0,1,2,3,\dots$, which we also denote with \lstate{s}, \lstate{p}, \lstate{d}, \lstate{f} in ascending order, and the magnetic quantum number $m \in [-\ell,\ell]\cap \mathbb{Z}$. Each azimuthal quantum number gives the wave function a characteristic angular dependency, known as orbital\index{atomic orbital}. The magnetic quantum number specifies the orientation of the orbital relativ to the chosen quantization axis.
The \kettext of each spherical harmonic, $\ket{\ell,m}$, has the relation $Y_{\ell,m}(\theta, \phi)=\braket{\vc{e}_{r}| \ell,m}$, where $\vc{e}_{r} = \vc{r}/r$ denotes the unit vector along $\vc{r}$ in spherical coordinates.
\begin{figure}[!t]
\centering
\includegraphics{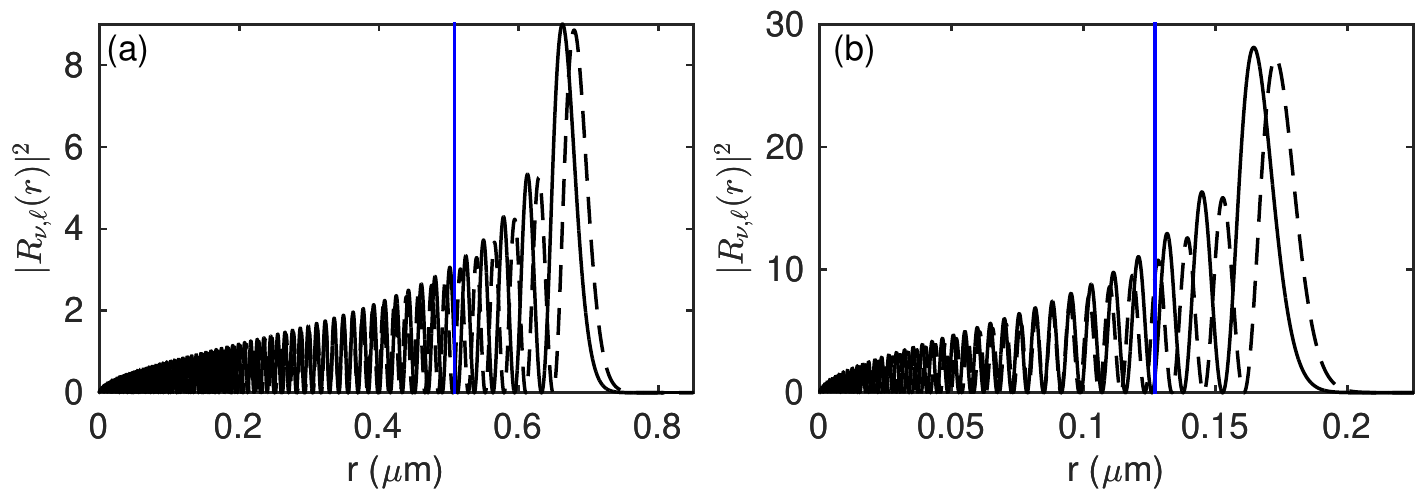} 
\caption{Absolute square of selected Hydrogen radial wave functions with $\ell=0$~(solid lines) and $\ell=1$~(dashed lines) according to \eref{eq:ryd_atoms:radial_wavefunctions_Hydrogen} for (a) $\nu =80$ and (b) $\nu=40$. The blue lines mark the mean value $\braket{\hat{r}}_{\nu,\ell}$.
\label{fig:radial_wavefunctions_hydr}
}
\end{figure}

With the product ansatz and the use of spherical harmonics for the angular dependency of the wave function, \eref{eq:ryd_atoms:time_indep_schroedinger_equ} reduces to the following one-dimensional time-independent Schrödinger equation:
\begin{equation}
\hat{H}_{\mathrm{rad}}(r)R(r) = ER(r),\label{eq:ryd_atoms:time_indep_schroedinger_equ_pure_radial}
\end{equation}
where the Hamiltonian here is the purely radially dependent part, defined by
\begin{equation}
 \hat{H}_{\mathrm{rad}}(r) := \dfrac{-\hbar^2}{2\mu}\dfrac{\partial^2}{\partial r^2} + \dfrac{\hbar^2 \ell(\ell+1)}{2\mu r^2} + V(r).
\label{eq:ryd_atoms:radial_part_Hamiltonian}
\end{equation}
For Hydrogen and Hydrogen-like atoms, with the potential given in \eref{eq:ryd_atoms:potential_hydrogen}, \eref{eq:ryd_atoms:time_indep_schroedinger_equ_pure_radial} has the following analytical solutions for bound states:
\begin{align}
 R_{\nu, \ell}(r) &= D_{\nu, \ell}\, r \mathrm{e}^{-Z r/(\nu\cdot\bohrradius)}\left(\dfrac{2 Z r}{\nu\cdot\bohrradius}\right)^\ell L_{\nu+\ell}^{2\ell+1}\left(\dfrac{2 Z r}{\nu\cdot\bohrradius}\right),
\label{eq:ryd_atoms:radial_wavefunctions_Hydrogen}
\\
E_{\nu}&=-Z^2\dfrac{\mu}{m_\echarge}\dfrac{\rydbergunit}{\nu^2}\approx -Z^2\left(1-\dfrac{m_\echarge}{m_\mathrm{c}}\right)\dfrac{\rydbergunit}{\nu^2}.
\label{eq:ryd_atoms:energy_levels_Hydrogen}
\end{align}
Both, the radial wave functions and the energy levels are additionally quantized by the principal quantum number $\nu \in \mathbb{N}_{+}$\nomenclature[B]{$\mathbb{N}_{+}$}{set of natural numbers excluding 0}. The bound states in \eref{eq:ryd_atoms:radial_wavefunctions_Hydrogen} require an additional limitation by the azimuthal quantum number to $\ell \in [0,\nu-1] \cap \mathbb{Z}$\nomenclature[B]{$\mathbb{Z}$}{set of all integers}.
The radial wave functions scale intrinsically with the Bohr radius,\index{Bohr radius} $\bohrradius=0.529$~\AA{},\nomenclature[A]{$\bohrradius$}{Bohr radius \nomunit{$0.529$~\AA{}}} and contain the associated Laguerre polynomials\index{associated Laguerre polynomials} $L_{k}^{\alpha}(x)$ and a normalization constant, given by
\begin{equation}
 D_{\nu, \ell} = \left(\dfrac{Z}{\bohrradius}\right)^{3/2}\dfrac{2}{\nu^2(\nu+\ell)!}\sqrt{\dfrac{(\nu-\ell-1)!}{(\nu+l)!}}.
\label{eq:ryd_atoms:nomalization_constant_radial_wavefunctions}
\end{equation} 
We denote the corresponding \kettext with $\ket{\nu,\ell}$, with the property $R_{\nu,\ell}(r)=\braket{r|\nu,\ell}$. In \fref{fig:radial_wavefunctions_hydr} the absolute square of Hydrogen radial wave functions are shown for highly excited states. For these Rydberg states the electron is far separated from the nucleus, which is apparent from the micrometer scaling.

The energy levels in \eref{eq:ryd_atoms:energy_levels_Hydrogen} scale with the Rydberg constant, $1~\rydbergunit$ $= m_\echarge \echarge^4/(8 \epsilon_0^2 h^2)$ $\approx 13.606$~eV\index{Rydberg constant}\nomenclature[A]{$\rydbergunit$}{Rydberg constant\nomunit{$13.60569253(30)$~eV}} and are not dependent on the quantum numbers $(\ell,m)$.
If an atom is excited to a state with large principal quantum number, it is called a Rydberg atom. For them, the relative spacing between neighbouring states is $\Delta E/E_\nu \approx 2 \nu^{-1}$.

\subsection{Spin-orbit coupling}
The treatment of the Hydrogen problem is not complete so far, since relativistic effects $(\mathcal{RE})$ were not taken into account.
A relativistic treatment results in the so called spin-orbit coupling,
which couples the orbital angular momentum, already introduced and the spin of the electron. The spin is an additional property of the electron, which can be described by an operator, which we denote with $\hat{\vc{S}}$. This operator fulfills all conditions to be an angular momentum operator with azimuthal quantum number $s=1/2$ and magnetic quantum number $m_{S} \in \{-1/2,1/2\}$. However, in contrast to the orbital angular momentum, the spin has no classical analog and thus has to be treated as an additonal independent quantity.
We abbreviate the \kettext of the spin-eigenstates with $\ket{m_s}:=\ket{s=1/2,m_s}$. A complete description of a quantum system including spin is given by
\begin{equation}
 \ket{\Psi(\vc{r})} = \sum_{m_s = \pm 1/2}\psi_{m_s}(\vc{r})\ket{m_s},
\label{eq:ryd_atoms:def_spinor}
\end{equation}
which is called a spinor\index{spinor}, a two-component entity. The functions $\psi_{m_s}(\vc{r})$ describe the system conditioned to be in the specific spin-eigenstate.

The spin-orbit coupling leads to an additional contribution to the Hamiltonian, given by
\begin{equation}
 \hat{H}_{\spinorbitsymbol}(r) = V_{\spinorbitsymbol}(r)\,\hat{\vc{L}}\cdot\hat{\vc{S}},
\label{eq:ryd_atoms:spin_orbit_Hamiltonian}
\end{equation}
with the potential
\begin{equation}
 V_{\spinorbitsymbol}(r) = \dfrac{1}{2\mu^2 c^2} \dfrac{1}{r}\dfrac{\dscalar V(r)}{\dscalar r},
\label{eq:ryd_atoms:spin_orbit_potential}
\end{equation}
where $V(r)$ is the non-relativistic potential, used in \eqref{eq:ryd_atoms:time_indep_schroedinger_equ_pure_radial}. Finding the bound state solutions of Hydrogen including spin-orbit coupling, we have to replace the Hamiltonian in \eref{eq:ryd_atoms:hamiltonian_spherical_symmetric_potential}, $\hat{H}(r) \to \hat{H}(r) + \hat{H}_{\spinorbitsymbol}(r)$, and solve the eigenproblem. The Hamiltonian including spin-orbit coupling is no longer diagonal in eigenstates of $\hat{\vc{L}}$, instead it is in eigenstates of the total angular momentum, defined by
\begin{equation}
 \hat{\vc{J}} := \hat{\vc{L}} + \hat{\vc{S}}.
\label{eq:ryd_atoms:def_total_angular_momentum}
\end{equation}
We denote the corresponding \kettext eigenstates with $\ket{j,m_j}$, where $j$ is the azimuthal and $m_j$ the magnetic quantum number of $\hat{\vc{J}}$. For electrons, the value of the azimuthal quantum number is restricted to be $j = l\pm 1/2$.
The spin-orbit coupling operator can be expressed as
\begin{equation}
 \hat{\vc{L}}\cdot\hat{\vc{S}} = \dfrac{1}{2}\left(\hat{\vc{J}}^2 - \hat{\vc{L}}^2 - \hat{\vc{S}}^2\right),
\end{equation}
and its eigenstates are then given by the generalized spherical harmonics\index{generalized spherical harmonics}, which are spinors defined by
\begin{equation}
 \ket{\mathcal{Y}_{\ell,j,m_j}(\theta,\phi)} := \sum_{m_s = \pm 1/2}\ClebchGordanSymb_{\ell,m_j-m_s;s,m_s}^{j,m_j}Y_{\ell,m_j-m_s}(\theta,\phi)\ket{m_s}.
\label{eq:ryd_atoms:def_generalized_spherical_harmonics}
\end{equation}
The prefactor is given by\cite{newton2013:scatter_theory_of_waves_and_particles}
\begin{equation}
\ClebchGordanSymb_{\ell,m_j-m_s;s,m_s}^{j,m_j}=
\dfrac{1}{\sqrt{2\ell+1}}\cdot
\begin{cases}
\sqrt{\ell+\dfrac{1}{2}\pm m_j}, & j-\ell = \frac{1}{2}\\
\mp\sqrt{\ell+\dfrac{1}{2}\mp m_j}, & j-\ell = -\frac{1}{2}
\end{cases},\qquad m_s = \pm \frac{1}{2},\ |m_j-m_s| \leq \ell,
\label{eq:ryd_atoms:clebsch_gordan_coeffs_for_generalized_sph_harmonics}
\end{equation}
which is a so called Clebsch-Gordan coefficient, which arises in a transformation from an uncoupled product basis of two angular momenta to a basis of a coupled angular momentum. The definition of the Clebsch-Gordan coefficient\index{Clebsch-Gordan coefficient} is given via
\begin{equation}
 \ClebchGordanSymb_{\ell,m;s,m_s}^{j,m_j}:=\braket{\ell,m;s,m_s|j,m_j}:=\left(\bra{\ell,m}\otimes\bra{s,m_s}\right)\ket{j,m_j}.
\end{equation}
The symbol $\otimes$ denotes the tensor product. If we string together \ketstext or \brastext without this symbol, we actually mean the tensor product between these states.

The eigenequation of the spin-orbit coupled operator is then given by
\begin{equation}
 \hat{\vc{L}}\cdot\hat{\vc{S}}\ket{\mathcal{Y}_{\ell,j,m_j}(\theta,\phi)} = \dfrac{\hbar^2}{2}\left[j(j+1)-\ell(\ell+1)-3/4\right]\ket{\mathcal{Y}_{\ell,j,m_j}(\theta,\phi)}.
\end{equation}
A separation of the wave functions into radial and angular part to solve the full eigenproblem is successful by modifying the product ansatz in \eref{eq:ryd_atoms:separation_wave_function} and choosing for the angular dependent  the generalized spherical harmonics, defined in \eref{eq:ryd_atoms:def_generalized_spherical_harmonics}. The bound state radial functions are again solutions of \eref{eq:ryd_atoms:time_indep_schroedinger_equ_pure_radial}, but with a modified radial Hamiltonian, $\hat{H}_{\mathrm{rad}} \to \hat{H}_{\mathrm{rad}} + \bra{\mathcal{Y}_{\ell,j,m_j}(\theta,\phi)}\hat{\vc{L}}\cdot\hat{\vc{S}}\ket{\mathcal{Y}_{\ell,j,m_j}(\theta,\phi)}V_{\spinorbitsymbol}(r)$. This makes them additionally dependent on $j$, $R_{\nu,\ell}(r) \to R_{\nu,j,\ell}(r)$.

Next to the spin-orbit coupling, there is an additional correction from the relativistic consideration, called the Darwin term. Combining these two relativistiv corrections, we get small energy shifts compared to the non-relativistic energy levels. Up  to first order, they are given by
\begin{equation}
 \dfrac{\Delta E^{(\mathcal{RE})}_{\nu,j}}{E_\nu} = \dfrac{(Z\cdot\finestructurec)^2}{\nu}\left(\dfrac{1}{j+1/2}-\dfrac{3}{4\nu}\right).
\label{eq:ryd_atoms:energy_correction}
\end{equation}
These relativistic corrections scale with the second power of the dimensionless finestructure constant\index{finestructure!constant}\nomenclature[A]{$\finestructurec$}{finestructure constant\nomunit{$0.007297352533(27)$}}, $\finestructurec^2 \approx 137^{-2}$, which makes them a very small effect for Hydrogen. 

%

%
%
%
%
%
%
\index{Hydrogen|)}

\section{Alkali atoms}\label{part:fd::chap:ryd_a::sec:alk_ryd_a}
Alkali atoms consist of more than two elementary particles, but have a single valence electron only, similar to Hydrogen. Their Rydberg level structure and as well the shape of their Rydberg wave functions is not much different from Hydrogen. This is due to the fact, that only a small part of the Rydberg electron orbit is close to the other electrons and the nucleus, which form an effectiv singly charged positive-ion core. Alkali Rydberg atoms can thus approximately be treated as an effective two-body problem, but with a potential which is not purely coulombic, since the charge of the ion core is radially dependently screened. The difficulty in the two-body treatment of Alkali Rydberg atoms is that there is no rigorous derivation of the effective potential for the Rydberg electron. However, a model potential was found, which describes observed phenomena well
\cite{marinescu1994:model_potential_alkalis}.
The form of this potenial is
\begin{equation}
 V_{\ell}(r) = -\dfrac{Z_{\ell}(r)}{r} - \dfrac{\alpha_{\mathrm{c}}}{2r^4}[1-e^{-(r/r_{\mathrm{c}})^6}],
\label{eq:ryd_atoms:alkali_potential}
\end{equation}
where $\alpha_{\mathrm{c}}$ is the static dipole polarizability and $Z_{\ell}(r)$ the radial charge, given by
\begin{equation}
 Z_{\ell}(r) = 1+(Z-1)\mathrm{e}^{-a_{1}r}-r(a_3+a_4 r)\mathrm{e}^{-a_2 r}.
\label{eq:ryd_atoms:alkali_radial_charge}
\end{equation}
The values of the five parameters $(a_1,a_2,a_3,a_4,r_{\mathrm{c}})$ are the result from a nonlinear fit, such that the eigenenergies of \eref{eq:ryd_atoms:time_indep_schroedinger_equ_pure_radial} with the model potential in \eref{eq:ryd_atoms:alkali_potential} match to the experimental values of the Rydberg energies, which were successfully measured extremely precisely\cite{haroche:li_finesplitting,fabre1978:spectroscopy_ryd_sodium_quant_defects,martin1981:spectroscopy_ryd_sodium_quant_defects,han2006:spectroscopy_ryd_rb85_quant_defects,li2003:spectroscopy_ryd_rb_quant_defects,lorenzen1983:spectroscopy_ryd_potassium_rb_quant_defects,goy1982:spectroscopy_ryd_cesium_quant_defects}. The parameters have to be fitted for each specific alkali atom and for each azimuthal quantum number $\ell$.

\subsection{Quantum defects}
\label{part:fd::chap:ryd_a::sec:alk_ryd_a_quantum_defects}
The determination of the alkali eigenenergies from \eref{eq:ryd_atoms:time_indep_schroedinger_equ_pure_radial} is
 only possible numerically.
This makes it useful to write \eref{eq:ryd_atoms:time_indep_schroedinger_equ_pure_radial} in atomic units, where the values of $\hbar, \echarge, m_\echarge$ and furthermore $4\pi \epsilon_0$ are set to unity. Distances are then measured in units of the Bohr radius, $\bohrradius$, and energies in units of the Hartree energy\index{Hartree energy}, $E_{h} = \finestructurec^2 m_\echarge c^2$\nomenclature[A]{$E_{h}$}{Hartree energy\nomunit{$27.21138505(60)$~eV}}. Note that \eref{eq:ryd_atoms:alkali_potential} and \eref{eq:ryd_atoms:alkali_radial_charge} are already formulated in atomic units.
The energy levels for the alkali Rydberg states can be written in the form
\begin{equation}
 E_{\nu,\ell,j} = -\dfrac{\rydbergunit^{(\mathrm{alk})}}{(\nu -\delta_{\nu,\ell,j})^2},
\label{eq:ryd_atoms_alkali_energy_levels}
\end{equation}
where $\delta_{\nu,\ell,j}$ are called the quantum defects and $\rydbergunit^{(\mathrm{alk})}$ is the Rydberg constant for the specific alkali atom.
The quantum defects can be expanded in the series
\begin{equation}
 \delta_{\nu,\ell,j} = \delta^{(0)}_{\ell,j} + \dfrac{\delta^{(2)}_{\ell,j}}{(\nu-\delta^{(0)}_{\ell,j})^2} + \dfrac{\delta^{(4)}_{\ell,j}}{(\nu-\delta^{(0)}_{\ell,j})^4} + \dots \qquad.
\end{equation}
Since we use lithum atoms in this work, we present the values of the $\lstate{s},\lstate{p}$-states quantum defects in \Tref{tab:ryd_atoms_lithium_quantum_defect_values} for this element. Furthermore, the Rydberg constant for $^{6}$Li is $\rydbergunit{(^6\mathrm{Li})}=3.289541926(2)\times10^9$~MHz and for for $^{7}$Li is $\rydbergunit{(^7\mathrm{Li})}=3.289584728(2)\times10^9$~MHz.
\begin{table}[!ht]
\centering
\begin{tabular}{C{\widthof{Xmass number of isotope (A)X}}C{\widthof{Xstate ($\ell_{j}$)X}}C{\widthof{X0.3995101(10)X}}C{\widthof{X0.0290(5)X}}C{0.01cm}}
\hline
\hline
mass number of isotope (A) & state ($\ell_{j}$)& $\delta^{(0)}_{\ell,j}$ & $\delta^{(2)}_{\ell,j}$&{\phantom{g}}\vspace{2cm}\\
\hline
6,7	& $\lstate{s}_{1/2}$ & $0.3995101(10)$ & $0.0290(5)$&\\
6  	& $\lstate{p}_{1/2}$ & $0.0471835(20)$ & $-0.024(1)$&\\
6 	& $\lstate{p}_{3/2}$ & $0.0471720(20)$ & $-0.024(1)$&\\
7 	& $\lstate{p}_{1/2}$ & $0.0471780(20)$ & $-0.024(1)$&\\
7 	& $\lstate{p}_{3/2}$ & $0.0471665(20)$ & $-0.024(1)$&\\
\hline
\hline
\end{tabular}
\caption{
\label{tab:ryd_atoms_lithium_quantum_defect_values}
Some quantum defect parameters for $^{6}$Li and $^{7}$Li, see~\rref{haroche:li_finesplitting}.
}
\end{table}
Having the eigenenergies at hand, the wave functions of the Rydberg states can be determined numerically from \eref{eq:ryd_atoms:time_indep_schroedinger_equ_pure_radial} via Numerov's method\cite{noumerov1924,numerov1927}. Analytic expressions are provided by quantum defect theory\cite{seaton1958:quantum_defect_theory,seaton1983:review_quantum_defect_theory}. Besides the principal quantum number, the energy level structure of alkali atoms is essentially dependent on the azimuthal quantum number of the orbital angular momentum, $\ell$. This is the main difference to Hydrogen. There also is a dependency on the azimuthal quantum number of the total angular momentum, $j$. Since this dependency is weak, we neglect it in the description of the bound state radial wave functions.

\subsection{Calculation of dipole matrix elements}\label{part:fd::chap:ryd_a::sec:alk_ryd_a:calc_dip_matrix_elements}
The interactions between Rydberg atoms used later are dipole-dipole interactions, where every atom is treated as a dipole in a simplified picture. Therefore, it is useful to evaluate the dipole matrix elements beforehand. Classically, in the simplest case of two point charges are a dipole, defined as
\begin{equation}
 \vc{d} := |q|\vc{r}
\label{eq:ryd_atoms_def_dipole}
\end{equation}
where one particle carries a charge $+q$ and the other one $-q$. The vector $\vc{r}$ is the separating distance between both charges. In quantum mechanics, the distance has to be replaced by the corresponding operator, $\vc{r} \to \hat{\vc{r}}$. If the evaluation of matrix elements is performed in the position-representation, than the operator consists of the classical vector, $\hat{\vc{r}} = \vc{r}$. Using the definition of a dipole in \eref{eq:ryd_atoms_def_dipole} for a Rydberg atom, we have to use the distance vector $\vc{r}$ between Rydberg electron and nucleus as the charge separating distance and set $|q|=\echarge$. Since the distance between both charges is large, treating the Rydberg atom as a dipole is a good approximation. The evaluation of matrix elements becomes simple by expanding the dipole in a spherical basis:
\begin{equation}
 \hat{\vc{d}} = \sum_{\mu \in \{\pm 1,0\}}\hat{d}_{\mu}\vc{b}_{\mu}.
\label{eq:alk_ryd_atoms:r_spherical_decomp}
\end{equation}
The spherical components and basis vectors are given by
\begin{align}
 \hat{d}_{\mu}&:=
\begin{cases}
 -\left(\mu \hat{d}_x+\im \hat{d}_y\right), & \mu = \pm 1\\
 \hat{d}_z, & \mu = 0
\end{cases},\\
\vc{b}_{\mu}&:=
\begin{cases}
 \left(-\mu\vc{e}_{x} + \im \vc{e}_{y}\right)/2, & \mu = \pm 1\\
\vc{e}_{z}, & \mu = 0
\end{cases},
\label{eq:alk_ryd_atoms:def_r_zeta}
\end{align}
where $\hat{d}_{\xi} = \echarge \cdot \hat{\xi}$ is the cartesian dipole component and $\vc{e}_\xi$ the cartesian unit vector in $\xi$-direction, where for both quantities $\xi \in \{x,y,z\}$.
%
%
%
We evaluate first the matrix elements of the single spherical components, neglecting the finestructure. For an evaluation of the dipole matrix elements accounting for finestructure, see \Aref{part:app::calcs_ryd_atoms__spherical_matrix_element}.
We calculate the matrix elements $\braket{\hat{d}_{\mu}}^{\Gamma'}_{\Gamma} = \bra{\Gamma}\hat{d}_{\mu}\ket{\Gamma'}$, between two bound states, $\ket{\Gamma} = \ket{\nu,\ell,m}$ and $\ket{\Gamma'} = \ket{\nu',\ell',m'}$.
In position-representation, the kets become $\ket{\Gamma} \to \varphi_{\Gamma}(\vc{r}) = r^{-1}R_{\nu,\ell}(r)Y_{\ell,m}(\theta,\phi)$ and the spherical components of the dipole $d_\mu = \echarge\, \sqrt{\frac{4\pi}{3}}r Y_{1,\mu}(\theta,\phi)$.
The calculation of the dipole matrix elements results in
\begin{equation}
 \braket{\hat{d}_{\mu}}^{\Gamma'}_{\Gamma} = \drad_{\nu,\ell;\nu',\ell'}\,\dsphere_{\ell,m;\ell',m'}^{(\mu)},
\label{eq:alk_ryd_atoms:calc_matrix_element_r_component}
\end{equation}
with $\drad_{\nu,\ell;\nu',\ell'}$ the radial dipole matrix element, defined by
\begin{equation}
 \drad_{\nu,\ell;\nu',\ell'}:= \echarge \int_{0}^{\infty}R_{\nu,\ell}(r) r R_{\nu',\ell'}(r) \dscalar r = \echarge\braket{\hat{r}}_{\nu,\ell}^{\nu',\ell'}
\label{eq:alk_ryd_atoms:def_radial_matrix_element}
\end{equation}
and $\dsphere_{\ell,m;\ell',m'}^{(\mu)}$ the spherical dipole matrix,
\begin{equation}
 \dsphere_{\ell,m;\ell',m'}^{(\mu)}:=\sqrt{\dfrac{4\pi}{3}}\int Y^{*}_{\ell,m}(\theta,\phi) Y_{1,\mu}(\theta,\phi) Y_{\ell',m'}(\theta,\phi) \dscalar \Omega.
\label{eq:alk_ryd_atoms:def_spherical_matrix_element}
\end{equation}
Note that the integration is over the entire unit sphere, $\int \dscalar \Omega = \int_{0}^{\pi}\int_{0}^{2\pi}\sin \theta\, \dscalar \theta\, \dscalar\phi$.
The spherical dipole matrix elements can be related to Clebsch-Gordan coefficients by applying the Wigner-Eckart Theorem\cite{eckart1930:wigner_eckart_theorem,wigner1927:wigner_eckart_theorem}\index{Wigner-Eckart theorem} to it:
\begin{equation}
\dsphere_{\ell,m;\ell',m'}^{(\mu)}=\sqrt{\dfrac{2\ell'+1}{2\ell+1}} \ClebchGordanSymb_{\ell',0;1,0}^{\ell,0} \ClebchGordanSymb_{\ell',m';1,\mu}^{\ell,m}.
\label{eq:alk_ryd_atoms:spherical_matrix_element_eval}
\end{equation}
The matrix elements of the entire dipole are then given by
\begin{equation}
 \braket{\hat{\vc{d}}}_{\Gamma}^{\Gamma'}=\dreduced_{\nu,\ell}^{\nu',\ell'}\vc{D}_{\ell,m}^{\ell',m'},
\label{eq:alk_ryd_atoms:radial_matrix_element}
\end{equation}
with $\dreduced_{\nu,\ell}^{\nu',\ell'}$ the reduced matrix element\index{reduced matrix element},
\begin{equation}
 \dreduced_{\nu,\ell}^{\nu',\ell'}:=\sqrt{\dfrac{2\ell'+1}{2\ell+1}}\,\ClebchGordanSymb_{\ell',0;1,0}^{\ell,0}\,\drad_{\nu,\ell;\nu',\ell'},
\label{eq:alk_ryd_atoms:reduced_matrix_element}
\end{equation}
which is not dependent on the orientation of the dipole, and $\vc{D}_{\ell,m}^{\ell',m'}$ the vector containing the Clebsch-Gordan coefficients in each basis direction:
\begin{equation}
 \vc{D}_{\ell,m}^{\ell',m'} := \sum_{\mu=\pm 1, 0}\ClebchGordanSymb_{\ell',m';1,\mu}^{\ell,m}\vc{b}_{\mu}.
\label{eq:alk_ryd_atoms:anisotropy_dipole_matrix_element}
\end{equation}
The Clebsch-Gordan coefficients enforce the selection rules 
$\Delta m \in \{0,\pm 1\}$ with $\Delta m:=m'-m$ and $\Delta \ell \equiv \ell'-\ell = \pm 1$. The only element specific quantity is the radial dipole matrix element, which depend on the radial wave functions. The relevant transitions in this work are $\lstate{s} \to \lstate{p}$ of $^7$Li, which we list in \Tref{tab:ryd_atoms_lithium_radial_dipole_matrix_elements}. These transition matrix elements scale in good approximation quadratically with the principal quantum number.
\begin{table}[!ht]
\centering
\begin{tabular}{C{\widthof{X$\nu$X}}C{\widthof{X$d_{\nu,s;\nu,p}$[a.u.]X}}C{0.01em}}
\hline
\hline
$\nu$ & $\drad_{\nu,1;\nu,0}$[a.u.] &{\phantom{g}}\vspace{2cm} \\
\hline
$35$	& $1579$ &\\
$44$  	& $2498$ &\\
$60$ 	& $4649$ &\\
$80$	& $8265$ &\\
\hline
\hline
\end{tabular}
\caption{
\label{tab:ryd_atoms_lithium_radial_dipole_matrix_elements}
Radial dipole matrix elements of $\lstate{s} \to \lstate{p}$ transitions for $^7$Li\cite{wuester:radial_matrix_elements_li}
}
\end{table}

\section{Dipole-dipole interactions}\label{part:fd::chap:ryd_a::sec:dip_dip_int}
Atoms can interact in many ways, giving rise to possibly very complex systems such as large biomolecules. Rydberg atoms open the way to more clearly study fundamental properties of interacting systems, since their long-range interactions allow many body systems with large interatomic separations. This allows a description of interactions in the lowest non-vanishing order, the dipole-dipole form\index{dipole-dipole interaction}.
We present here the derivation of the dipole-dipole Hamiltonian and start with a classical picture of how dipole-dipole interactions arise.
Detailed considerations can be found in Ref.~\citenum{jackson:ed_book,nolting:ed_3__book}.

In general, there is a charge distribution\index{charge distribution} $\rho_{\mathrm{el}}(\vc{r})$, which yields an electric potential\index{electric potential}
\begin{equation}
 \Phi({\vc{r}})=\dfrac{1}{4\pi \epsilon_0}\int \dvol r' \dfrac{\rho_{\mathrm{el}}(\vc{r'})}{|\vc{r}-\vc{r'}|}.
\label{eq:alk_ryd_atoms:electric_potential}
\end{equation}
We are interested in the interaction of this potential with 
another, far separated, charge distribution. This allows to use the far field approximation\index{electric potential!far field approximation} ($\mathcal{FF}$) of the potential, where $|\vc{r}-\vc{r'}|^{-1}$ is small, such that
\begin{equation}
 \dfrac{1}{|\vc{r}-\vc{r}'|} = \exp\left(-\vc{r'}\nabla\right)\dfrac{1}{r} \approx \dfrac{1}{r} + \dfrac{\vc{r}\cdot\vc{r'}}{r^3} 
\label{eq:alk_ryd_atoms:taylor_expansion_r_minus_rprime}
\end{equation}
Inserting \eref{eq:alk_ryd_atoms:taylor_expansion_r_minus_rprime} in \eref{eq:alk_ryd_atoms:electric_potential}, we get
\begin{equation}
 \Phi_{\mathcal{FF}}({\vc{r}}) \approx \dfrac{1}{4\pi \epsilon_0}\left(\dfrac{q_{\mathrm{el}}}{r} + \dfrac{\vc{d}\cdot \vc{r}}{r^3}\right),
\label{eq:alk_ryd_atoms:electric_potential_far_field_approximation}
\end{equation}
with the total charge, $q_{\mathrm{tot}}$, and the dipole moment $\vc{d}$, defined by
\begin{align}
 q_{\mathrm{tot}} &:= \int \dvol r\, \rho_{\mathrm{tot}}(\vc{r}),\label{eq:alk_ryd_atoms:def_total_charge}\\
 \vc{d} &:= \int \dvol r\, \rho_{\mathrm{el}}(\vc{r})\vc{r}.\label{eq:alk_ryd_atoms:def_dipole_moment_general}
\end{align}
The dipole moment is equivalent to the definition in \eref{eq:ryd_atoms_def_dipole} for two pure point charges.
Having a single atom as a source for the electric potential, we have $q_{\mathrm{tot}}=0$ and thus, the far field approximation of an atom leads in lowest order to a dipole potential $\Phi_{\mathrm{d}}(\vc{r})$, and field $\vc{E}_{\mathrm{d}}(\vc{r}):= -\nabla \Phi_{\mathrm{d}}(\vc{r})$:
\begin{align}
 \Phi_{\mathrm{d}}(\vc{r}) &:= \dfrac{1}{4\pi \epsilon_0}\dfrac{\vc{d}\cdot \vc{r}}{r^3},
\label{eq:alk_ryd_atoms:def_dipole_potential}\\
\vc{E}_{\mathrm{d}}(\vc{r})&:= \dfrac{1}{4\pi \epsilon_0}\left( -\dfrac{\vc{d}}{r^3} + 3 \dfrac{\left(\vc{r}\cdot\vc{d}\right)\vc{r}}{r^5}\right).
\label{eq:alk_ryd_atoms:def_dipole_field}
\end{align}
We denote with $\Phi^{(k)}(\vc{r})$ the potential and with $\rho^{(k)}_{\mathrm{el}}(\vc{r})$ the charge distribution of atom~$k=1,2$, furthermore the dipole moment with $\vc{d}^{(k)}$ and the interatomic distance with $\vc{R}_{12}$, pointing from atom~1 to atom~2. The interaction energy is then given by
\begin{equation}
 W_I = \int \dvol r \rho^{(2)}_{\mathrm{el}}(\vc{r})\Phi^{(1)}_{\mathrm{d}}({\vc{r}}),
\label{eq:alk_ryd_atoms:interaction_energy}
\end{equation}
assuming a large interatomic distance, which justifies the use of the dipole potential. The spread, where the charge distribution of the second atom is non-vanishing, is small compared to the interatomic distance. This allows for an additional approximation, namely that the variation of the dipole potential of atom~1 is small inside the relevant interaction volume. Setting the coordinate origin to the center of atom~1, this second approximation is a taylor expansion of $\Phi^{(1)}_{\mathrm{d}}({\vc{r}})$ at $\vc{R}_{12}$ up to first order, which gives
\begin{align}
 W_I &\approx - \vc{E}_{\mathrm{d}}^{(1)}(\vc{R}_{12})\cdot \vc{d}^{(2)},\\
&= \dfrac{1}{4\pi \epsilon_0}\left[ \dfrac{\vc{d}^{(1)}\cdot\vc{d}^{(2)}}{R_{12}^3} - 3 \dfrac{\left(\vc{R}_{12}\cdot\vc{d}^{(1)}\right)\left(\vc{R}_{12}\cdot\vc{d}^{(2)}\right)}{R_{12}^5}\right],
\label{eq:alk_ryd_atoms:dip_dip_interaction}
\end{align}
for two far separated charge distributions, with total charge vanishing for both.
The interaction in \eref{eq:alk_ryd_atoms:dip_dip_interaction} is called dipole-dipole interaction. For a quantum mechanical expression, we replace the dipole moments, with their corresponding operators, $\vc{d}^{(k)} \to \hat{\vc{d}}^{(k)}$. Furthermore we use \eref{eq:ryd_atoms_def_dipole} with $|q| = \echarge$, such that we have $\hat{\vc{d}}^{(k)}:=\echarge \cdot\hat{\vc{r}}^{(k)}$, with $\hat{\vc{r}}^{(k)}$ the Rydberg electron position operator of atom $k=1,2$, relative to its nucleus. The distance between the two dipoles, $\vc{R}_{12}$, remains parameterized classically. The dipole-dipole interaction Hamiltonian of two Rydberg atoms is then given by
\begin{equation}
 \hat{V}_{\mathrm{dd}}(\vc{R}_{12}):=\dfrac{1}{4\pi \epsilon_0}\left[ \dfrac{\hat{\vc{d}}^{(1)}\cdot\hat{\vc{d}}^{(2)}}{R_{12}^3} - 3 \dfrac{\left(\vc{R}_{12}\cdot\hat{\vc{d}}^{(1)}\right)\left(\vc{R}_{12}\cdot\hat{\vc{d}}^{(2)}\right)}{R_{12}^5}\right].
\label{eq:alk_ryd_atoms:dip_dip_interaction_qm}
\end{equation}
This formula can be simplified by expanding the dipoles and the distance vector in their spherical basis, which results in (see \Aref{part:app::calcs_ryd_atoms__eval_matrix_elements_dip_dip_interaction})
\begin{equation}
 \hat{V}_{\mathrm{dd}}(\vc{R}_{12})=-\dfrac{1}{4\pi \epsilon_0}\dfrac{\sqrt{24\pi}}{R_{12}^3}\sum_{\mu,\mu'=\pm 1, 0}
\begin{pmatrix}
 1 & 1 & 2\\
\mu & \mu' & -(\mu + \mu')                                                                                                                                                                                                                                                                                                                                                                                                         \end{pmatrix}
Y_{2,-(\mu + \mu')}(\vartheta_{12},\phi_{12})\hat{d}_{\mu}^{(1)}\hat{d}_{\mu'}^{(2)},
\label{eq:alk_ryd_atoms:dip_dip_interaction_qm_spherical_expansion}
\end{equation}
\index{dipole-dipole interaction!binary!operator, general}
with polar angle $\theta_{12}$ and azimuthal angle $\phi_{12}$ of the distance vector. The six numbers in parantheses denote the Wigner 3-j symbol, which appears equivalently to the Clebsch-Gordan coefficients due to coupling angular momenta. 

\subsection{Transition matrix elements}
\label{part:fd::chap:ryd_a::sec:dip_dip_int_subsec_transition_dip_moments}
\begin{figure}[!t]
\centering
\includegraphics{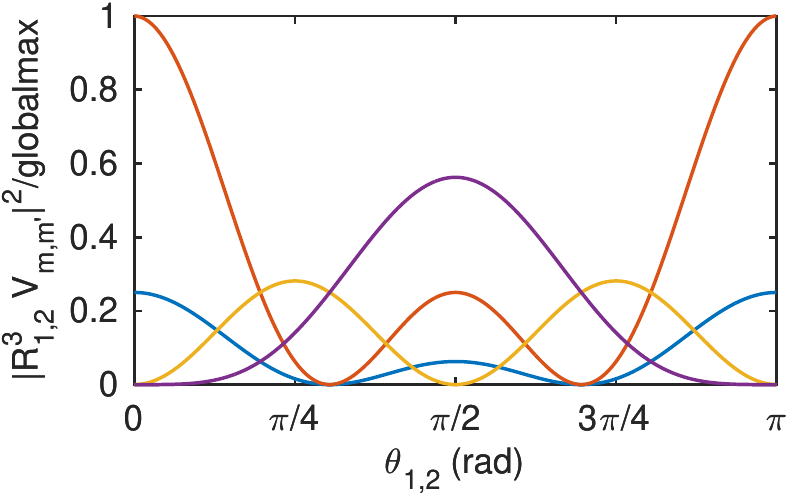} 
\caption{Angular dependence of transition dipole strengths for resonant $\ket{\lstate{ps}},\ \ket{\lstate{sp}}$ states, measured in units of the global maximum, defined by $\mathrm{globalmax} := \max_{m,m',\theta_{1,2}} |R_{1,2}^{3} V_{m,m'}(\vc{R}_{1,2})|^2$.
There are 4 different strengths: $(|\Delta m|,m) = (0,\pm 1)$~(blue line), $(|\Delta m|,m) = (0,0)$~(red line), $|\Delta m| = 1$~(yellow line) and $|\Delta m| = 2$~(violet line), with $\Delta m := m'-m$.
\label{fig:resonant_sp_interactions_dens_angular_dependence}
}
\end{figure}

We calculate now transition matrix elements of the dipole-dipole interactions between the states $\ket{\Gamma_1;\Gamma_2}$ and $\ket{\Gamma_1';\Gamma_2'}$, where $\ket{\Gamma_k}=\ket{\nu_k,\ell_k, m_k}$ and $\ket{\Gamma_k'}=\ket{\nu_k',\ell_k', m_k'}$ are alkali bound states of atom $k=1,2$. The notation $\ket{X;Y} := \ket{X}_{1}\otimes\ket{Y}_{2}$ is an abbreviation, with 
$\ket{X}_{1}$ a state of dipole~1 and $\ket{Y}_{2}$ a state of dipole~2. The matrix elements are given by\index{dipole-dipole interaction!binary!transition matrix element!general}
\begin{multline}
 \bra{\Gamma_1;\Gamma_2}\hat{V}_{\mathrm{dd}}(\vc{R}_{12})\ket{\Gamma_1';\Gamma_2'} = 
-\sqrt{24\pi}\dfrac{\braket{\hat{d}^{(1)}_{-\Delta m_1}}_{\Gamma_1}^{\Gamma_1'} \braket{\hat{d}^{(2)}_{-\Delta m_2}}_{\Gamma_2}^{\Gamma_2'}}{4\pi \epsilon_0 R_{12}^3}\\
\times
\begin{pmatrix}
 1 & 1 & 2\\
-\Delta m_1 & -\Delta m_2 & \Delta m_1 + \Delta m_2                                                                                                                                                                                                                                                                                                                                                                                                      \end{pmatrix}Y_{2,\Delta m_1 + \Delta m_2}(\theta_{12},\phi_{12}).
\label{eq:alk_ryd_atoms:dip_dip_interaction_qm_matrix_elements_general}
\end{multline}
The selection rules of the Clebsch-Gordan coefficients yield contributions only in the spherical components $-\Delta m_k$ for atom $k$, where $\Delta m_k:=m_k' - m_k$. In particular we are interested in the transition matrix elements between the resonant two atom states $\ket{\lstate{ps},m}:=\ket{\nu,\lstate{p},m;\nu,\lstate{s},0}$ and $\ket{\lstate{sp},m'}:=\ket{\nu,\lstate{s},0;\nu,\lstate{p},m'}$ of the same species.
We abbreviate $V_{m,m'}(\vc{R}_{12}):=\bra{\lstate{ps},m}\hat{V}_{\mathrm{dd}}(\vc{R}_{12})\ket{\lstate{sp},m'}$, which have the explicit expression\cite{robicheaux2004:resonant_ddint_sp_formula} (see \Aref{part:app::calcs_ryd_atoms__eval_matrix_elements_dip_dip_interaction_s_p}):\index{dipole-dipole interaction!binary!transition matrix element!resonant between $\ket{sp}\leftrightarrow\ket{ps}$}
\begin{align}
 V_{m,m'}(\vc{R}_{12})&=
 -\sqrt{\dfrac{8\pi}{3}}\dfrac{\drad_{\nu,0;\nu,1}^2}{4\pi \epsilon_0 R_{12}^3}(-1)^{m'}
\begin{pmatrix}
 1 & 1 & 2\\
m & -m' & m' - m                                                                                                                                                                                                                                                                                                                                                                                                       \end{pmatrix}Y_{2,m' - m}(\theta_{12},\phi_{12}).
\label{eq:alk_ryd_atoms:dip_dip_interaction_qm_matrix_elements_s_p}
\end{align}
Since the principal quantum number enters only in the radial matrix elements, the transition between states with different principal quantum numbers can be easily adjusted by replacing the radial matrix elements.
The derived resonant dipole-dipole interaction between the energetically resonant states $\ket{\lstate{ps},m}$ and $\ket{\lstate{sp},m'}$ in \eref{eq:alk_ryd_atoms:dip_dip_interaction_qm_matrix_elements_s_p} is the fundamental interaction in flexible Rydberg aggregates for the transport of a single \lstate{p}-excitation. We show the angular dependence of the corresponding dipole strengths in \fref{fig:resonant_sp_interactions_dens_angular_dependence} and find that for values $\theta_{1,2} \in \{0, \pi/2, \arccos(1/\sqrt{3}) , \pi \}$, individual matrix elements can vanish which simplifies the description of interactions.

\subsection{Resonant and van-der-Waals interactions}
\label{part:fd::chap:ryd_a::sec:resonant_and_vdW_interactions} 
The transition matrix elements would give the direct interactions between two coupled pair states in the absence of other states. Since the dipole-dipole interaction couples many available pair states, a diagonalization of the interaction matrix has to be performed to get the molecular/di-atomic interaction potentials. We can distinguish two types of states: The first are energetically resonant pair states, which have non-vanishing transition matrix elements. An example is the resonant manifold $\{\ket{\lstate{sp},m},\ket{\lstate{ps},m}\}_{m=\{-1,0,1\}}$, where each transition matrix element can be evaluated with \eref{eq:alk_ryd_atoms:dip_dip_interaction_qm_matrix_elements_s_p}. This type of interaction is called resonant dipole-dipole interaction and scales with the third inverse power of the interatomic distance. However, the dipole-dipole interaction couples as well to states which are off-resonant. A diagonalization of the interaction matrix yields leading terms, that are proportional to the sixth inverse power of the interatomic distance. This is the so called van-der-Waals interaction.
When between two resonant pair-states the transition dipole matrix element vanishes, the leading term of interaction is then in most cases the van-der-Waals interaction. However, if couplings to (quasi)-resonant states appear, the dominant interaction can again scale more or less as the resonant interactions with the third inverse power of the interatomic distance. The appearance of quasi-resonances is often connected to a certain choice of principal quantum number\cite{henkel2013:phd_thesis}.

To understand the different scaling between resonant and van-der-Waals interactions, we present here a minimal example consisting of two pair states which we label with $\ket{0},\ket{1}$. We set the energy level of $\ket{0}$ as the zero point energy and assume the state $\ket{1}$ is energetically detuned by $\Delta >0$. The two states are dipole coupled and we denote the dipole transition matrix element with $V$, which we assume to be real valued. The Hamiltonian describing this two level system is then given by
\begin{equation}
 \hat{H} = \begin{bmatrix}
            0 & V\\
	    V & \Delta
           \end{bmatrix}
\label{eq:alk_ryd_atoms:vdW_interactions__min_example_H}
\end{equation}

Diagonalizing this simple Hamiltonian leads to the eigenenergies
\begin{equation}
 E_{\pm} = \dfrac{1}{2}\left( \Delta \pm \sqrt{4V^2 + \Delta ^2}\right)
\label{eq:alk_ryd_atoms:vdW_interactions__min_example_eigenenergies}
\end{equation}
We can now distinguish two limiting cases: When the detuning is much larger than the resonant interaction, $\Delta \gg V$, the eigenenergies are approximately
\begin{equation}
 E_{\pm}/\Delta \approx  \left(1\pm 1\right)/2  \pm \left( V/\Delta \right)^2 \mp \mathcal{O}\left(\left( V/\Delta \right)^4\right),
\label{eq:alk_ryd_atoms:vdW_interactions__min_example_eigenenergies_large detuning}
\end{equation}
For small interactions between the states, $E_{-}$ is asymptotically connected to the state $\ket{0}$ and $E_{+}$ to the state $\ket{1}$.
The absolute value of effective interactions for both states is according to \eref{eq:alk_ryd_atoms:vdW_interactions__min_example_eigenenergies_large detuning} given by $V^2/\Delta$. Since $V$ scales with the third inverse power of the interatomic distance, $\sim R^{-3}$, the interaction to off-resonant states with large detuning is $\sim R^{-6}$. The sign of the interaction is dependent on the sign of the detuning. State $\ket{0}$ is coupled to another state with positive detuning which yields positive interactions. For state $\ket{1}$ it is the opposite case.

\begin{figure}[!t]
\centering
\includegraphics{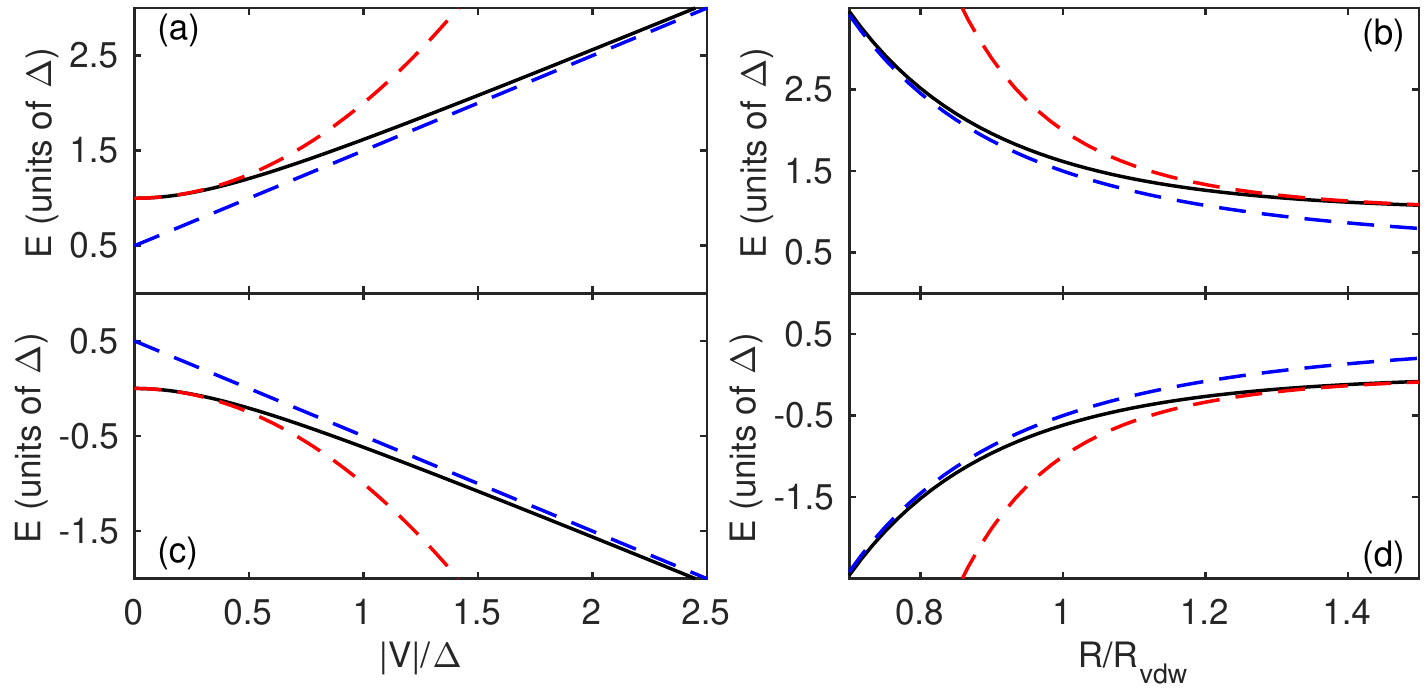} 
\caption{Comparison of eigenenergy solutions corresponding to state $\ket{1}$ (a,b) and state $\ket{0}$ (c,d).
The exact solutions~(black lines) from \eref{eq:alk_ryd_atoms:vdW_interactions__min_example_eigenenergies} are plotted together with the solutions for large detuning~(red dashed lines), given in \eref{eq:alk_ryd_atoms:vdW_interactions__min_example_eigenenergies_large detuning}, and small detuning~(blue dashed lines), given in \eref{eq:alk_ryd_atoms:vdW_interactions__min_example_eigenenergies_small_detuning}, relative to the transition dipole interaction.
(a,c) Plot over $|V|/\Delta$, without specifying the transition dipole interaction $V$.
(b,d) Plot over $R/R_{\mathrm{vdw}}$, where $V =\Delta (R/R_{\mathrm{vdw}})^{-3}$.
\label{fig:resonant_vs_van_der_waals_int}
}
\end{figure}

The other limiting case is for small detunings compared to the dipole transition element, $\Delta \ll V$. In this case, the eigenenergies are approximately
\begin{equation}
 E_{\pm}/\Delta \approx 1/2 \pm |V|/\Delta + \mathcal{O}\left(\Delta/|V|\right)
\label{eq:alk_ryd_atoms:vdW_interactions__min_example_eigenenergies_small_detuning},
\end{equation}
When we treat the transition dipole interaction with the right dependency on the interatomic distance $R$, such that explicitly we have $V(R) = \Delta (R/R_{\mathrm{vdw}})^{-3}$, we see from \fref{fig:resonant_vs_van_der_waals_int}~(b,d) that for $R \gg R_{\mathrm{vdw}}$, the interaction scales as $\Delta (R/R_{\mathrm{vdw}})^{-6}$. The other limiting case, $R \ll R_{\mathrm{vdw}}$, lets the interactions get more and more of resonant type, such that it scales as $\Delta(R/R_{\mathrm{vdw}})^{-3}$.
 
In reality, the dipole-dipole interaction couples to infinitely many off-resonant states. Calculating the van-der-Waals interactions is then the sum over all effective interactions, transition dipole strength squared divided by the detuning, from each off-resonant coupled state. Often, it is sufficient to consider the coupling to only a few more states, such that the series converges. Since the van-der-Waals interaction scales with $\sim R^{-6}$, it can be written as\index{dipole-dipole interaction!van-der-Waals!two state model}
\begin{equation}
 V_{\mathrm{vdw}}(R) = -C_{6}/  R^{6},
\label{eq:alk_ryd_atoms:vdW_interactions__vdw_general_formula}
\end{equation}
where the $C_6$ is the so called dispersion coeffient\index{dispersion coefficient}, which for each state individually characterizes the strength of the van-der-Waals interaction. A detailed analytical evaluation of dispersion coefficients for alkali Rydberg atom pairs is given in \rref{singer:VdWcoefficients} and a numerical evaluation for Rubidium in \rref{henkel2013:phd_thesis}.
We present in \Aref{part:app::calcs_ryd_atoms__vdw_ints_from_block_diag} a more formal treatment and show how van-der-Waals interactions can be calculated with a generalized formula of second order perturbation theory, which is a result of block-diagonalization.

Overall already from the minimal examples presented above we  see that for Rydberg atoms, the van-der-Waals interactions can be extremely large even on micrometer scaling. This is due to the fact that the principal quantum number scaling of the transition dipole strength is $V \sim \nu^4$ and for the detunings $\Delta \sim \nu^3$, which results for the van-der-Waals interaction in a scaling of $V^2/\Delta \sim \nu^{11}$.

\section{Properties of Rydberg atoms and their interactions---an overview}
After presenting the essential level structure of Rydberg atoms and discussing their interactions, we list here important properties of them, which are useful for the investigation of Rydberg aggregates. The large transition dipole moments due to high principal quantum numbers are reflected in extreme values for both intrinsic and interaction properties:

\begin{itemize}
 \item Strong long-range dipole-dipole interactions both resonant, $\propto \nu^4$, and vdW interactions, $\propto \nu^{11}$. Of main importance are resonant interactions in this thesis.
 \item The radial transiton matrix elements between neighbouring states scales $\propto \nu^2$.
\item Mesoscopic scaling of the wave function $\sim \mu$m.
\item Mesoscopic dipole blockade\cite{lukin2001:dipole_blockade,urban:twoatomblock} of radius
\begin{equation}
R_{\mathrm{bl}}\approx\left(\dfrac{|C_{6}|}{\hbar\Omega}\right)^{1/6},
\label{eq:alk_ryd_atoms:blockade_radius}                                                                          \end{equation}
which is the radius of a spherical volume in which one Rydberg excitation inhibits the excitation of a second one. The radius is $\sim \mu$m for Rydberg atoms and sets the minimal spatial distances for initial configurations of Rydberg aggregates. The definition of the blockade radius assumes dominant vdW interactions. The preparation of Rydberg aggregates requires as a first step the excitation from ground to equal Rydberg states, such that during the excitation process vdW interactions are in fact dominant. The blockade is dependent on the collective Rabi frequency $\Omega$, which is typically a two-photon Rabi frequency necessary to excite the Rydberg state.
\item The lifetime\cite{gounand1979:drad_and_lifetimes_alkalis_theoretical} is large and for zero temperature it can be parameterized by\footnote{The scaling is with the effective principal quantum number, which is the principal quantum number corrected by the quantum defect, $\nu_{\ast} \equiv \nu - \delta_{\ell}$}
\begin{equation}
\tau_{\nu,\ell} = \tau^{(0)}_{\ell} \nu_{\ast}^{\alpha_{\ell}},
\label{eq:alk_ryd_atoms:lifetime_ryd_atoms}                                                                          \end{equation}
with an almost constant exponent for alkalis, $\alpha_{\ell} \simeq 3$. However, the other scaling parameter is dependent on the azimuthal quantum number $ \ell$. For ${}^7$Li, which we utilize in this thesis as constituents of the aggregates, it is given by $\tau^{(0)}_{\mathrm{s}} = 0.8431$~ns for \lstate{s} states and $\tau^{(0)}_{\mathrm{p}} = 2.8807$~ns for \lstate{p} states, respectively. This gives a total lifetime of roughly $70\ \mu$s ($232\ \mu$s) using $\nu=44$, and $400\ \mu$s ($1386\ \mu$s) using $\nu=80$ for \lstate{s} states (\lstate{p} states)\cite{beterov:BBR}. For finite temperatures the lifetime is decreased by blackbody radiation~(BBR) to an effective lifetime $\tau^{\mathrm{eff}}_{\nu, \ell}$. Since the experimental preparation of Rydberg aggregates requires ultracold temperatures\cite{hofmann2014:experimental_approach_many_body_phenomena_ryd_aggrs,weber2015:superatom_1D_confinement} $\sim \mu $K and the relative lifetime decrease due to BBR is given by $\left(\tau_{\nu,\ell} - \tau^{\mathrm{eff}}_{\nu,\ell}\right)/\tau_{\nu,\ell} \approx 6.8 \times 10^{-8} \nu_{\ast}^{-2} (T/\mu\mathrm{K})/(\tau_{\nu,\ell}/\mu\mathrm{s})$\cite{beterov:BBR}, the BBR correction can be neglected. For relevant principal quantum numbers $\nu > 40$ in the ultracold temperature regime this correction is smaller than $10^{-12}$.
\end{itemize}
The features indicate that dynamics of interacting Rybderg atoms including atomic motion will occur on spatial distances of micrometers and time scales of microseconds.
\chapter{Resonant energy transfer: From Rabi oscillations to directed transport}\label{part:fd::chap:theoretical_framework}
Resonant energy transfer\index{resonant energy transfer} is well understood in frozen systems, where interacting constituents, such as biological complexes, molecules or atoms, can not move spatially, but still transport excitation through long range interactions. We review this in \sref{part:fd::chap:theoretical_framework_spatially_frozen system}. 
However, when the interactions are strong enough, this frozen gas approximation is not valid anymore since during the transport time-scale the constituents can significantly move. Instead of Rabi oscillations, which transports the excitation between the resonant states back and forth, unidirectional RET takes place, due to motion of the constituents. We introduce in \sref{part:fd::chap:theoretical_framework_spatially_unfrozen system} some basic concepts of unfrozen systems and demonstrate special features with a trimer system in \sref{part:fd::chap:theoretical_framework_spatially_unfrozen system_trimers}. The end of this chapter outlines how directed transport can be controlled, which is the basic motivation for the results in \Chref{part:rs::chap:planar_aggregates} and \Chref{part:rs::chap:unconstrained_aggregate}.
%
%
%
%
\section{Spatially frozen systems}
\label{part:fd::chap:theoretical_framework_spatially_frozen system}
\begin{figure}[!t]
\centering
\begin{overpic}[tics=2]
{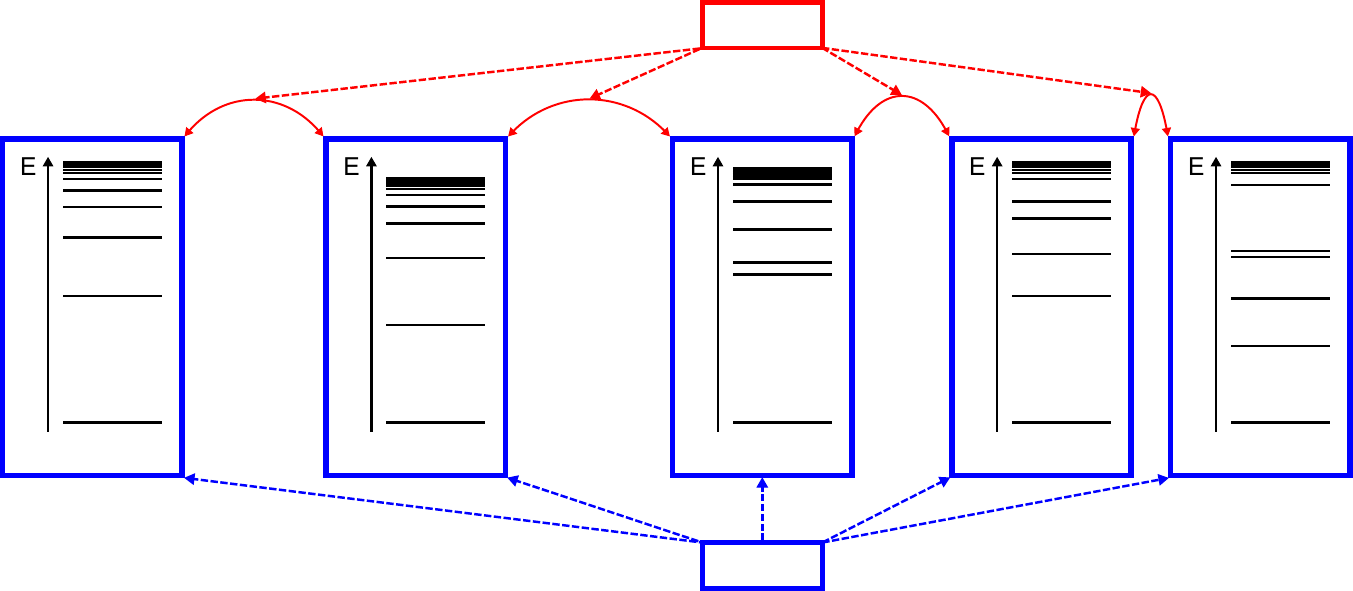}
\put (54.1,1.8) {\textinfig{text=blue,scale=.75}{$\hat{\mathcal{H}}_0$}}
\put (54.15,41.8) {\textinfig{text=red,scale=.75}{$\hat{\mathcal{V}}$}}
\end{overpic}
\caption{Sketch of a collection of individual systems~(blue boxes), each of them with their own energy level structure and eigenstates. All the couplings are collected in the interaction operator~(red), which creates a compound system with different energy levels and eigenstates.
\label{fig:sketch_general_hamiltonian_with_interactions}
}
\end{figure}

Let us consider a system of interacting, but frozen constituents.
%
%
We can always write for the Hamiltonian
\begin{equation}
 \hat{\mathcal{H}}_{\mathrm{el}} = \mathcal{\hat{H}}_{0} + \mathcal{\hat{V}},
\label{eq:theoret_framew:H_el_full}
\end{equation}
where $\mathcal{\hat{H}}_{0}$ is the collection of the individual constituent Hamiltonians and $\mathcal{\hat{V}}$ the operator describing all the interactions between them, sketched in \fref{fig:sketch_general_hamiltonian_with_interactions}. We are interested in how the interactions change eigenstates and -energies of the non-interacting system and the time evolution of quantum states. 

Resonant energy transfer can be thought of a process where two neighboring constituents excite and de-excite simultaneously, such that over time some initially excited constituents transport their excitation radiationless by other constituents. The excitation can thus be thought of hopping between the sites of the constituents. 
The distribution of excitation among the constituents is a superposition of resonant states, each of them localizes the entire excitation on an individual constituent.
The collection of these localized states is thus a basis, which we denote with $B_{\mathrm{el}}$. We can restrict the electronic Hamiltonian, $\hat{\mathcal{H}}_{\mathrm{el}}$, to this resonant subspace:
\begin{equation}
 \hat{H}_{\mathrm{el,0}}\equiv \hat{\mathcal{H}}_{\mathrm{el}}\bigr|_{B_{\mathrm{el}}} =  \hat{\id}_{\mathrm{el}}\hat{\mathcal{H}}_{\mathrm{el}}\hat{\id}_{\mathrm{el}},
\label{eq:theoret_framew:H_el_res}
\end{equation}
which includes the interactions within this manifold. The operator $\hat{\id}_{\mathrm{el}}$ is the projector onto the subspace of resonant states, which acts inside this subspace as the identity operator. Since all energy levels are degenerate, the diagonal of the Hamiltonian can be set to zero.

If we can ensure that couplings to off-resonant states are negligible, the description with the Hamiltonian in \eref{eq:theoret_framew:H_el_res}
is sufficient and we can set
\begin{equation}
 \hat{H}_{\mathrm{el}} = \hat{H}_{\mathrm{el,0}}
\label{eq:theoret_framew:H_el_neglib_oresonant_coupl}
\end{equation}
as final Hamiltonian.
For the other case, we have to add an operator containing the couplings to off-resonant states and their detunings, which we denote with $\hat{W}$
\begin{equation}
 \hat{H}_{\mathrm{el}} = \hat{H}_{\mathrm{el,0}} + \hat{W}
\label{eq:theoret_framew:H_el_with_oresonant_coupl}
\end{equation}
To say the energy transfer is resonant implies that the coupling to off-resonant states remains small during the time evolution, such that the dominant transfer is within the resonant Hamiltonian. The coupling to off-resonant states is given by the squared interaction matrix elements between the resonant and the off-resonant states divided by the detuning. It is weak when it is much smaller then the interaction between the resonant states.
Since the Hamiltonian is not time-dependent, the time evolution of an initial state $\ket{\psi_0}$ is given by $\ket{\psi(t)} = \exp(-\im \hat{H}_{\mathrm{el}} \cdot t /\hbar)\ket{\psi_0}$. Eigenstates of the Hamiltonian can only gain a phase factor, such that they are the stationary states of the time-independent Hamiltonian.

In the following we will demonstrate RET with a minimal example.

\subsection{Rabi oscillations as resonant energy transfer in a minimal model system}
\begin{figure}[!t]
\centering
\includegraphics{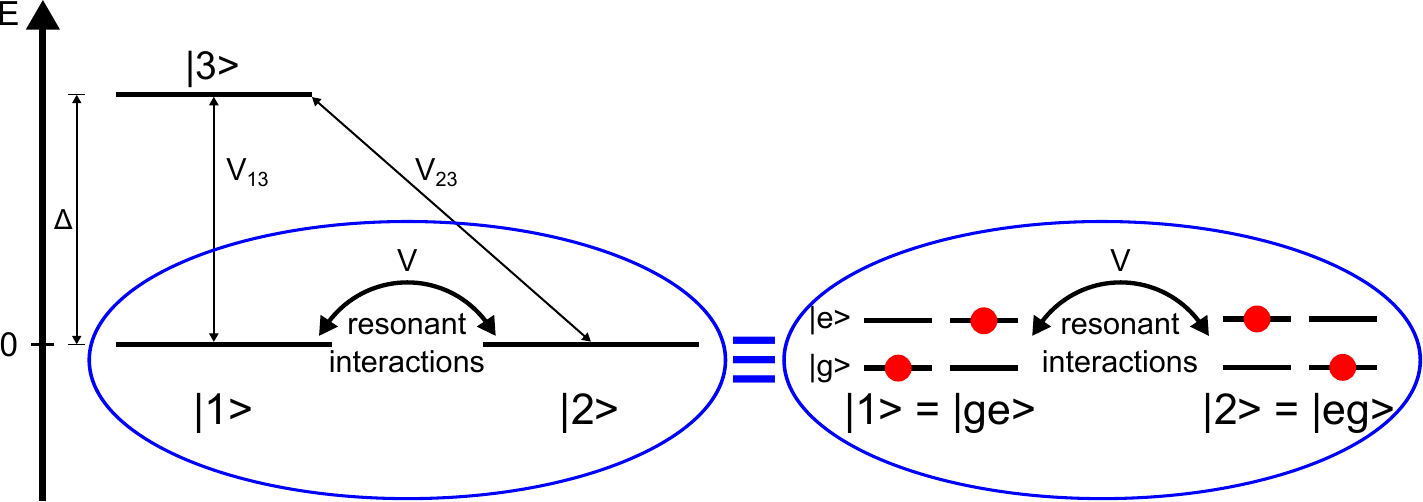} 
\caption{Energy level diagramm for a three state system. The resonant states $\ket{1}$ and $\ket{2}$ are coupled with the interaction strength $V$ and form the resonant subsystem, defined in \eref{eq:theoret_framew:min_example_H_el0}. An additional off-resonant state, $\ket{3}$, is assumed to be weakly coupled via $V_{13}$, $V_{12}$ to the resonant system. On the right side, we sketch the possibility of the resonant states to be realizations of singly excited many body states, where only one of the two constituents is in the excited state $\ket{e}$, the other in the ground state $\ket{g}$.
\label{fig:sketch_resonant_offresonant_interactions}
}
\end{figure}

Let us assume we have two resonant states, $\ket{1}$ and $\ket{2}$, whose energy is set to zero. We could think of these two states being resonant pair-states of a system with two constituents, where each constituent is described by two states, $\ket{g}$ and $\ket{e}$, where $\ket{g}$ is the ground state and $\ket{e}$ is the excited state. The resonant pair-states are then the single excited states, where only one constituent is excited. With this we could set $\ket{1} = \ket{ge} \equiv \ket{g}\otimes \ket{e}$ and $\ket{2} = \ket{eg}$, as sketched in the right half of \fref{fig:sketch_resonant_offresonant_interactions}.
In fact, the resonant states can be many-body states from a more complex systems with more constituents.

\begin{figure}[!t]
\centering
\includegraphics{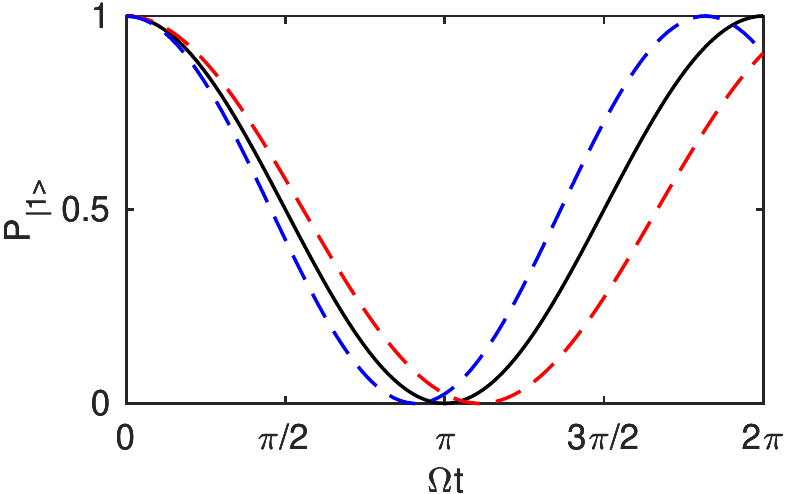} 
\caption{Probability to find the system in state $\ket{1}$ over time, when initiated with the same state, according to \eref{eq:theoret_framew:min_example_rabi_formula}. The Rabi oscillation of the system with negligible off-resonant couplings~(black line) has the Rabi frequency $\Omega$ and is the reference system for two other systems with weak, but not negligible off-resonant couplings, resulting in changed Rabi frequencies, $\tilde{\Omega} = 0.9\Omega$~(red, dashed line) and $\tilde{\Omega} = 1.1\Omega$~(blue, dashed line).
\label{fig:rabi_oscillations}
}
\end{figure}

Both states are coupled with the interaction matrix element $V$, which we assume to be real.
The resonant Hamiltonian can then be written
as\nomenclature[B]{$\mathrm{H.c.}$}{Hermitian conjugate}
\begin{align}
 \hat{H}_{\mathrm{el,0}} = V\ket{1}\bra{2} + \mathrm{H.c.} = \begin{pmatrix}
    0 & V\\
    V & 0
   \end{pmatrix},
\label{eq:theoret_framew:min_example_H_el0}
\end{align}
The eigenstates and -energies of this Hamiltonian are
\begin{align}
 E_{\pm,0} &= \pm V
\label{eq:theoret_framew:min_example_E0}
\\
 \ket{\varphi_{\pm},0} &= \left(\ket{1} \pm \ket{2}\right)/\sqrt{2}.
\label{eq:theoret_framew:min_example_states-0}
\end{align}
The resonant states are additionally coupled to the $\Delta$-detuned state $\ket{3}$, where we denote with $V_{13}$ the coupling between states $\ket{1}-\ket{3}$ and with $V_{23}$ the coupling between states $\ket{2}-\ket{3}$. For simplicity, we assume the couplings to be real valued.
With this we can write the coupling operator to the off-resont state as 
\begin{equation}
 \hat{W} = V_{13} \ket{1}\bra{3} + V_{23} \ket{2}\bra{3} + \mathrm{H.c.} + \Delta \ket{3}\bra{3} = 
\begin{pmatrix}
    0 & 0 &V_{13}\\
    0 & 0 &V_{23}\\
    V_{13} & V_{23} &\Delta
   \end{pmatrix},
\label{eq:theoret_framew:min_example_W}
\end{equation}
such that the electronic Hamiltonian can be written as
\begin{equation}
 \hat{H}_{\mathrm{el}} = 
\begin{pmatrix}
\hat{H}_{\mathrm{el,0}} & \begin{matrix}V_{13}\\ V_{23} \end{matrix} \\
\begin{matrix}V_{13}& V_{23} \end{matrix} & \Delta
\end{pmatrix}.
\label{eq:theoret_framew:min_example_rewritten_Hel}
\end{equation}
Since we assume the coupling to the off-resonant states to be weak, we can block-diagonalize the Hamiltonian according to \eref{eq:app::blockdiagonalization__sol_Hxx_HxTx} of \Chref{part:app::chap:block_diag__van_der_Waals}. This yields a Hamiltonian, which is restricted to the resonant manifold:
\begin{align}
\hat{H}'_{\mathrm{el}} &= \hat{H}_{\mathrm{el,0}} + \hat{W}',
\label{eq:theoret_framew:min_example_rewritten_Hel_prime}\\
\hat{W}' &= \Delta^{-1}\begin{pmatrix}
                                         V^2_{13} & V_{13}V_{23}\\
					 V_{13}V_{23} & V^2_{23}
                                        \end{pmatrix}.
\label{eq:theoret_framew:min_example_rewritten_W_prime}
\end{align}
We can again treat the new operator, $\hat{W}'$, as a perturbation, which allows us to use perturbation theory to get the eigenstates and -energies approximately:
\begin{align}
 E_{\pm} & \approx E_{\pm,0} + \dfrac{\left(V_{13} \pm V_{23}\right)^2}{2\Delta}
\label{eq:theoret_framew:min_example_E_approx}\\
\ket{\varphi_{\pm}} &\approx \dfrac{1}{\sqrt{1+w^2}}\left(\ket{\varphi_{\pm,0}} \pm w \ket{\varphi_{\mp,0}}\right), \qquad w:= \dfrac{V_{13}^2 - V_{23}^2}{4\Delta V}
\label{eq:theoret_framew:min_example_states_approx}
\end{align}
To get information about the RET, we set as an initial state $\ket{1}$ and calculate the probability that the system is at time $t$ in state $\ket{1}$,  $P_{\ket{1}}(t) := |\bra{1}\exp(-\im \hat{H}_{\mathrm{el}}' \cdot t /\hbar)\ket{1}|^2$. The probability that the system is in the other state is then $P_{\ket{2}}(t) = 1-P_{\ket{1}}(t)$.
We get for this probability
\begin{equation}
%
 P_{\ket{1}}(t) \approx \cos^2\bigl(\tilde{\Omega}t/2\bigr) + \mathcal{O}\left(w^2\right),\quad  \tilde{\Omega} := 2\left(V+ \Delta^{-1}V_{13}V_{23}\right)/\hbar.
\label{eq:theoret_framew:min_example_rabi_formula}
\end{equation}
The periodic behaviour of the state occupation probability is known as \emph{Rabi oscillation}\index{Rabi oscillation}. For small couplings to the detuned state, only the \emph{Rabi frequency}\index{Rabi frequency}, which is the angular frequency of the Rabi oscillation, changes from $\Omega:=2 V/\hbar$ to $\tilde{\Omega}$, displayed in \fref{fig:rabi_oscillations}. The excitation is transferred completely for $\tilde{\Omega} t = \pi$ which means the system is in state $\ket{2}$. This changes for stronger off-resonant couplings. In this case the neglected terms in \eref{eq:theoret_framew:min_example_rabi_formula}, which are $\sim \mathcal{O}(w^2)$, become important and prevent complete transfer of excitation. Then, for $\tilde{\Omega} t \approx \pi$, there is a non-negligible probability that the system remains in state $\ket{1}$.
Only fully RET can thus drive the system entirely from one resonant state to another.

To conclude, RET within spatially frozen systems causes periodic excitation exchange between the constituents. However, eigenstates of the Hamiltonian are stationary due to a time-independent Hamiltonian and can thus not transport excitation. We will see that this changes when we allow the constituents to move, which makes the interactions between them dependent on their spatial configuration and due to motion also implicitly time-dependent.

\section{Unfrozen systems}
\label{part:fd::chap:theoretical_framework_spatially_unfrozen system}
For strongly interacting systems the transport time-scale can be of the order where constituents start to move due to the interactions. The frozen gas approximation does not hold any longer then. The motion of the constituent dynamically changes the spatial configuration and thus the eigenenergies and -states as well. The coupling to the spatial degrees of freedom is often regarded as a decoherence source for the transport. However, a strong coupling allows for a combined transport of excitation and mechanical quantities, such as momentum, which we will see later. A fundamental feature of unfrozen systems is the ability to tune the excitation distribution of eigenstates from being spatially localized to delocalization. This makes it promising to transport excitation directly with eigenstates, which is not possible in frozen systems. 

Technically, we have to account for the spatial configuration of the constituents, which we do by introducing the vector $\vc{R}$, containing all positions of constituents. The interactions are dependent on the configuration, such that the electronic Hamiltonian changes dependent on the spatial configuration, $\hat{H}_{\mathrm{el}} = \hat{H}_{\mathrm{el}}(\vc{R})$. The eigenproblem of the electronic Hamiltonian is connected with the concept of excitons\index{exciton} and Born-Oppenheimer~(BO)\nomenclature[B]{BO}{Born-Oppenheimer} surfaces\index{Born-Oppenheimer surface}.

For a Hamiltonian $\hat{H}(\vc{R})$ we call the eigenstates, denoted with $\ket{\varphi(\vc{R})}$, excitons and the eigenenergies BO surfaces which we denote with $U(\vc{R})$. The eigenequation of the Hamiltonian is then
\begin{equation}
 \hat{H}(\vc{R})\ket{\varphi_{k}(\vc{R})} =U_{k}(\vc{R}) \ket{\varphi_{k}(\vc{R})},
\label{eq:theoret_framew:eigenequation_H}
\end{equation}
for each exciton state $\ket{\varphi_{k}(\vc{R})}$, with $k=1,..., \dim(\hat{H})$. We are specifically interested in the transport of single excitations. The excitons are then coherent superpositions of different localized excitation states, which is the result of interactions.

The evolution of the full quantum mechanical system requires the introduction of a wave function for the combined system of electronic Hamiltonian and spatial degrees of freedom. We denote it with $\ket{\Psi(\vc{R},t)}$.
The total Hamiltonian,
\begin{equation}
 \hat{H}_{\mathrm{T}}(\vc{R}) := -\sum_{\alpha=1}^{N}\dfrac{\hbar^2}{2M_{\alpha}}\nabla^2_{\vc{R}_{\alpha}} + \hat{H}_{\mathrm{el}}(\vc{R}),
\label{eq:theoret_framew:flex_aggr__total_Hamiltonian}
\end{equation}
describes an interacting system of $N$ constituents and contains besides the electronic Hamiltonian the operators corresponding to the kinetic energy, where $\vc{R}_{\alpha}$ denotes the position of the constituent labeled with $\alpha = 1, \dots, N$.
The evolution of the total wave function is described by the time-dependent Schrödinger equation,
\begin{equation}
 \im \hbar \dfrac{\partial}{\partial t} \ket{\Psi(\vc{R},t)}=\hat{H}_{\mathrm{T}}(\vc{R})\ket{\Psi(\vc{R},t)}.
\label{eq:theoret_framew:flex_aggr__schroedinger_equ_total_Hamiltonian}
\end{equation}
Unfortunately, the numerical effort grows rapidly with increasing number of spatial degrees of freedom, which makes it impossible to obtain the dynamics in reasonable time. 
However, many quantum-classical schemes are available to approximately get the quantum dynamics. In \sref{part:rs::chap:planar_aggregates_dynamical methods_quantum_classical_method}, we outline a method called 'Fewest switching surface hopping'\cite{tully:hopping2,tully:hopping:veloadjust}, which we use to solve the linked RET and atomic motion of the investigated Rydberg aggregates in \sref{part:rs::chap:planar_aggregates} and \sref{part:rs::chap:unconstrained_aggregate}.

In the following we outline how adiabatic transport with exciton states is possible in unfrozen systems using the example of a trimer.

\subsection{Directed transport and conical intersections in unfrozen trimers}
\label{part:fd::chap:theoretical_framework_spatially_unfrozen system_trimers}
We solve the eigenproblem of a trimer with different geometries and vary selected distances between the constituents to see how the excitons and BO surfaces depend on the geometry.
Each individual constituent is assumed to be described by a two level system with the ground state $\ket{g}$ and excited state $\ket{e}$. The combined system is excited from the ground state, $\ket{G} := \ket{ggg}$, to the single excitation manifold, which is spanned by the states $\ket{1} := \ket{egg}$, $\ket{2} := \ket{geg}$ and $\ket{3} := \ket{gge}$ and we are interested in how this single excitation gets distributed among these resonant states due to interactions.

%

\begin{figure}[!t]
\centering
\includegraphics{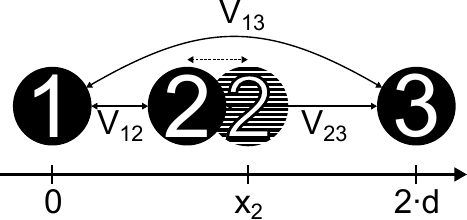} 
\caption{Sketch of a linear trimer configuration with fixed positions of constituent~1 and 3 and variable position of constituent~2.
\label{fig:sketch_linear_trimer}
}
\end{figure}

\begin{figure}[p]
\centering
\includegraphics{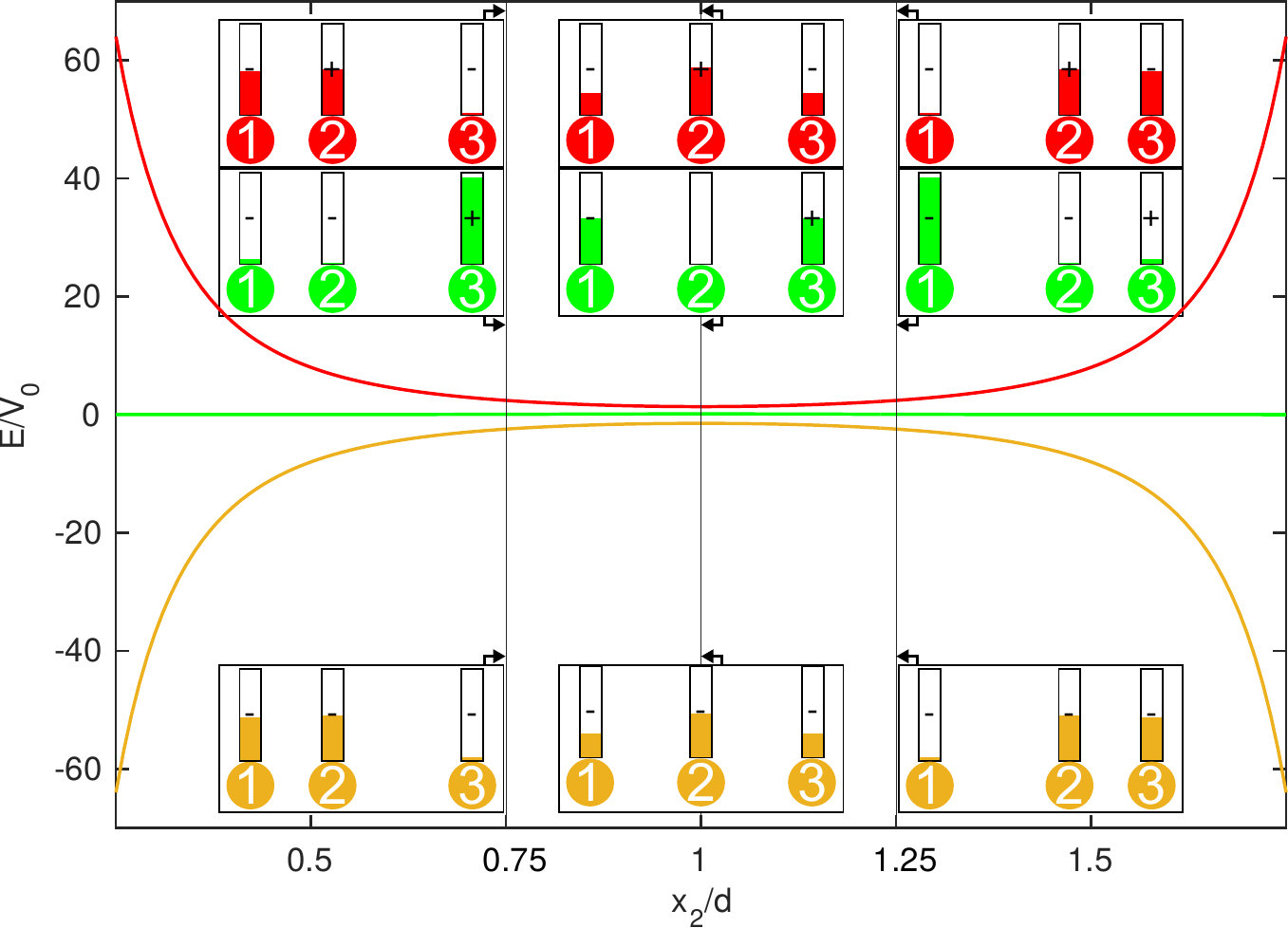} 
\caption{BO surfaces and excitons~(bars in insets) for a linear trimer with varying position of constituent~2, $x_{2}$, and resonant dipole-dipole interactions\index{dipole-dipole interaction!resonant} of the form $V_{kl} = -V_{0}(x_{kl}/d)^{-3}$. We obtain three BO surfaces, one leads to repulsive forces~(red line) and another to attractive forces~(orange line) on atom~2. The remaining BO surface is almost constant and yields effectively no interactions~(green line). The insets depict the geometry of the system for $x_{2}/d=0.75$~(left side), $x_{2}/d=1$~(center) and $x_{2}/d=1.25$~(right side), with bars visualising the excitation amplitude on each constituent, according to $\bigl|c^{\pm/0}_{k}\bigr|^2$ with  $c^{\pm/0}_{k}:=\braket{k|\varphi_{\pm/0}}$ and ``+'' for $c^{\pm,0}_{k}>0$ and ``minus'' for $c^{\pm,0}_{k}<0$. The red bars in the upper row represent the exciton corresponding to the BO surface with repulsive forces, the green bars in the middle row represent the exciton to the BO surface which yields no forces and the lower row the exciton which is connected with attractive forces due to the BO surface.
The solutions are obtained by numerical diagonalization of the Hamiltonian in \eref{eq:theoret_framew:lin_trimer_H_el} and a comparison with \eref{eq:theoret_framew:lin_trimer_Epm_phi_pm_approx} and \eref{eq:theoret_framew:lin_trimer_E0_phi_0_approx} yields perfect agreement, such that the analytical solutions can be used as well.
\label{fig:linear_trimer}
}
\end{figure}

\ssection{Linear trimer}
We set the positions of the constituents to $(x_1,x_2,x_3) = (0, x_2, 2d)$, as sketched in \fref{fig:sketch_linear_trimer}, where $x_k$ is the position of the $k$th constituent and $d$ is a unit of length. The only parameter to be varied is $x_2 \in (0,2d)$.
Furthermore we denote with $V_{kl}\equiv V(x_{kl})$ the interaction between state $\ket{k}$ and $\ket{l}$, which is equivalent for a transition of the excitation between constituent $k$ and $l$. We denote the spacings between the constituents with $x_{kl}\equiv |x_k-x_l|$.
We assume the binary interactions to be real valued and can write for the electronic Hamiltonian of the resonant manifold:
\begin{align}
 \hat{H}_{\mathrm{el}} = \sum_{\substack{k,l=1\\ k \neq l}}^{3} V_{kl} \ket{k}\bra{l} &= 
\begin{pmatrix}
 0 & V_{12} & V_{13}\\
 V_{12} & 0 & V_{23}\\
V_{13} & V_{23} & 0
\end{pmatrix}
\label{eq:theoret_framew:lin_trimer_H_el}
\end{align}
Let us assume the binary interactions to be proportional to a certain power of inverse distances, $V_{kl} \sim 1/|x_k-x_l|^{\alpha},\ \alpha > 1$. Since the distance $x_{13}$ is fixed and the largest in the system, $V_{13}$ is the smallest binary interaction. Our main interest lies in the study of resonant dipole-dipole interactions\index{dipole-dipole interaction!resonant}, which scale with $\alpha=3$. For this case, we can in good approximation set the interaction $V_{13}$ to zero. Introducing now a ratio of the remaining two binary interactions, $\tilde{V} = V_{23}/V_{12}$, we can rewrite the electronic Hamiltonian as
\begin{equation}
 \hat{H}_{\mathrm{el}} \approx V_{12}
\begin{pmatrix}
 0 & 1 & 0\\
 1 & 0 & \tilde{V}\\
 0 & \tilde{V} & 0
\end{pmatrix}.
\label{eq:theoret_framew:lin_trimer_H_el_approx}
\end{equation}
The excitons and BO surfaces for the approximate electronic Hamiltonian have the following analytic expressions
\begin{align}
 U_{\pm}(\tilde V) & \approx \pm V_{12}\sqrt{1+\tilde{V}^2} & 
\ket{\varphi_{\pm}(\tilde V} &\approx \dfrac{1}{\sqrt{2}}\left(\dfrac{\ket{1} + \tilde V \ket{3}}{\sqrt{1+\tilde{V}^2}} \pm \ket{2}\right)
\label{eq:theoret_framew:lin_trimer_Epm_phi_pm_approx}\\
U_{0} & \approx 0 & \ket{\varphi_{0}(\tilde{V})} &\approx\dfrac{1}{\sqrt{1+\tilde{V}^2}}\left(-\tilde{V}\ket{1}+\ket{3}\right),
\label{eq:theoret_framew:lin_trimer_E0_phi_0_approx}
\end{align}
We additionally perform a numerical diagonalization of the electronic Hamiltonian in \eref{eq:theoret_framew:lin_trimer_H_el} with resonant dipole-dipole interactions\index{dipole-dipole interaction!resonant} and the dependency of the excitons and the BO surfaces from the position of constituent~2 shown in \fref{fig:linear_trimer}. For a homogeneous configuration with equidistant spacings between all constituents, $x_2=d$, the excitons corresponding to the BO surfaces with highest~($U_{+}$, red line) and lowest energies~($U_{-}$, green line) have delocalized the single excitation over all constituents. When the spatial symmetry is broken, the two BO surfaces mentioned before, increase their absolute energy value, the more asymmetric the system gets. This happens for $x_2 \ll d$ and $(2d-x_2) \ll d$, respectively. Furthermore the excitation on the corresponding two excitons gets localized between the two atoms with smaller spacing, significant already for $x_2 \approx 0.75d$. A strong asymmetry is equivalent to values for the interaction ratio $\tilde V \ll 1$ or $\tilde V \gg 1$, respectively, and we see that then the excitons $\ket{\varphi_{\pm}}$ converge to the dimer eigenstates, $\ket{\varphi_{\pm}} = \left(\ket{1} \pm \ket{2}\right)/\sqrt{2}$ for $\tilde V \ll 1$ and $\ket{\varphi_{\pm}} = \left(\ket{2} \pm \ket{3}\right)/\sqrt{2}$ for $\tilde V \gg 1$.

If we vary the position of constituent~2 such that at the beginning it is close to constituent~1 and ends near constituent~3, we can transport the excitation spatially in a directed way. Since the energetically highest BO surface yields repulsive forces on constituent~2~($F_{2} = -\partial U_{+}/\partial x_{2}$), it can bring the trimer from one to the other asymmetric configuration and thus seems a promising tool for this purpose. In fact, the full quantum dynamics is goverened by the time-dependent Schrödinger equation with a total Hamiltonian consisting besides the electronic Hamiltonian of an operator for the kinetic energy. Studies of this dynamics were performed with linear chains of dipole-dipole interacting Rydberg atoms\cite{cenap:motion,wuester:cradle,moebius:cradle}, where an equidistantly spaced chain of Rydberg atoms has an additional dislocated Rydberg atom on one side, with a smaller distance $a$ to its neighbor than other neighbor distances $d$. For a dislocation ratio of $a/d = 0.4$, excitation gets almost perfectly localized on the two dislocated atoms for the exciton connected with the BO surface yielding repulsive motion.
Directed transport of excitation, linked with a transport of diatomic proximity was confirmed. The dynamics in these linear chains remains perfectly adiabatic, which means mixing with other excitons is insignificant.
\newpage
\ssection{Isosceles trimer}
Systems with linear geometries and long range interactions can in many cases be well approximated by two-body interactions of nearest neighbors. This situtation changes already for two-dimensional geometries, where three bodies can simultaneously interact with the same strength. For isotropic interactions, this is realized for an equilateral triangle configuration.

\begin{figure}[!t]
\centering
\includegraphics{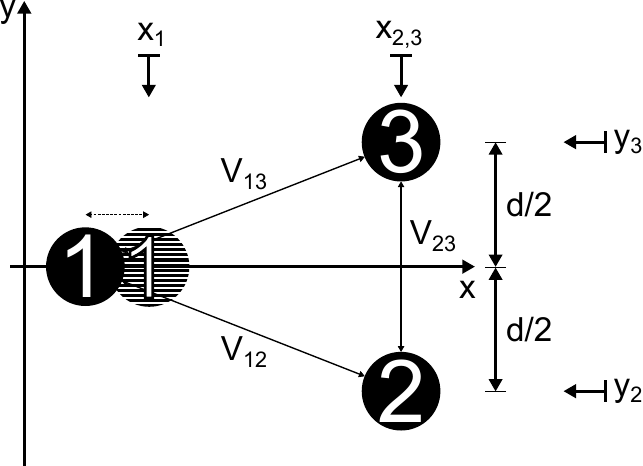} 
\caption{Sketch of an isosceles trimer configuration with variable horizontal position of constituent~1, $x_{1}$.
\label{fig:sketch_isosceles_trimer}
}
\end{figure}

\begin{figure}[p]
\centering
\includegraphics{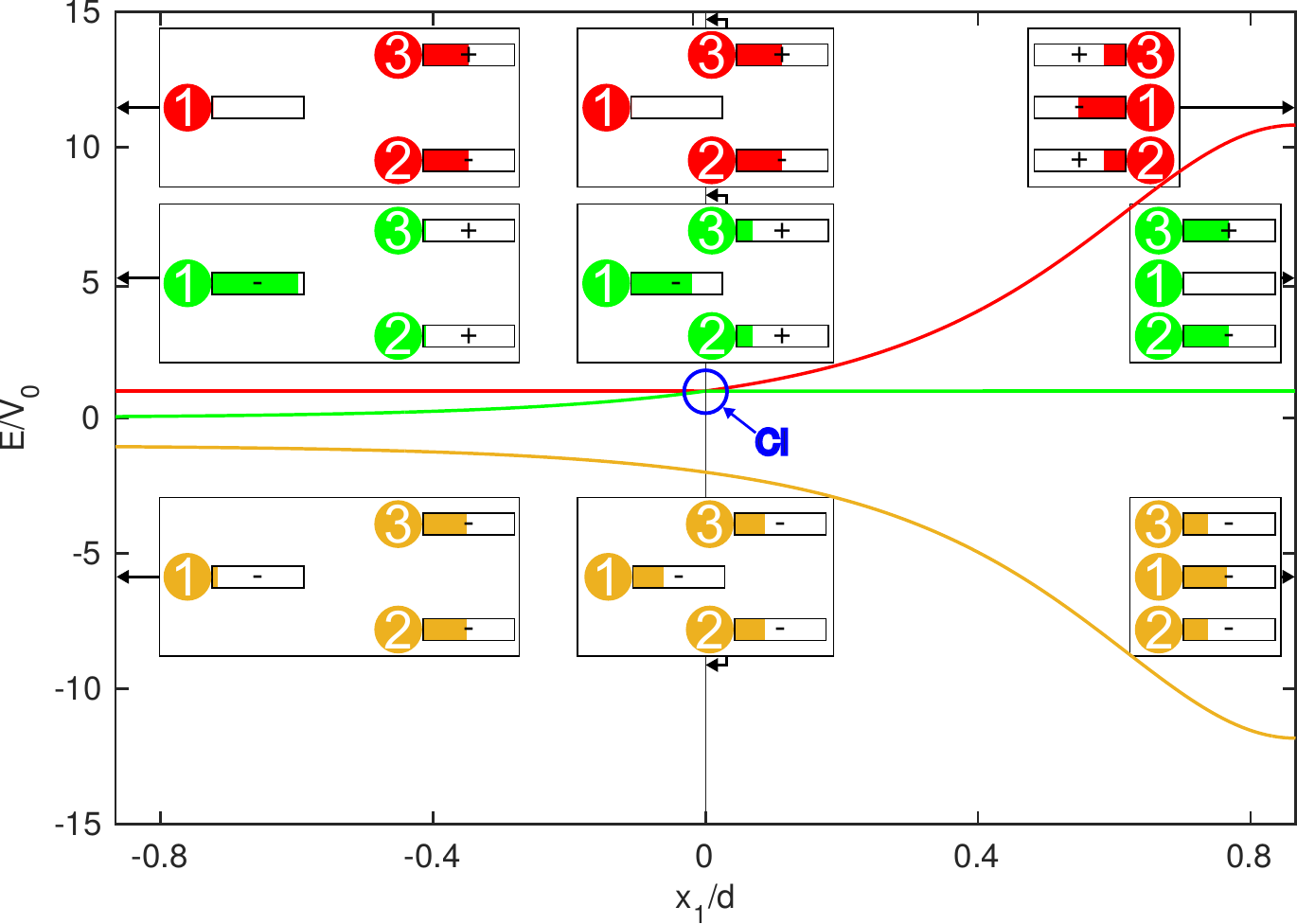} 
\caption{BO surfaces and excitons~(bars in insets) for an isosceles trimer, illustrated in \fref{fig:sketch_isosceles_trimer}, with varying position of constituent~1, $x_{1}$. The results are for resonant dipole-dipole interactions\index{dipole-dipole interaction!resonant} as in \fref{fig:linear_trimer} and follows also its illustration scheme of BO surfaces, excitons and sketches of trimer configurations. The insets on the left correspond to the trimer configuration with $x_1/d = -\sqrt{3}/2$, the middle to $x_1/d = 0$ and the right to $x_1/d = \sqrt{3}/2$. For $x_1=0$, the trimer is in an equilateral triangle configuration, where two surfaces~(red and green line) conically intersect~(marked with blue circle, CI), allowing to study highly nonadiabatic dynamics.
\label{fig:isosceles_trimer}
}
\end{figure}

Here we discuss the BO surfaces and excitons of an isosceles triangle configuration with resonant dipole-dipole interactions\index{dipole-dipole interaction!resonant}, depicted in \fref{fig:sketch_isosceles_trimer}, with $(x_{1},y_{1}) = (x_{1},0)$, $(x_{2},y_{2}) = (\sqrt{3}d/2,-d/2)$ and $(x_{3},y_{3}) = (\sqrt{3}d/2,d/2)$. The pair $(x_m,y_m)$ gives the coordinates in the horizontal x-direction and vertical y-direction. We vary only the horizontal position of constituent~1. For $x_1 \ll 0$, constituent~1 is far separated from the two others and the system is thus decomposable into a dimer, build by constituent~2 and 3, and the remaining constituent~1, which is quasi-isolated. The excitation distribution of the excitons for the case $x_1/d = -\sqrt{3}/2$ in \fref{fig:isosceles_trimer} reflects this situation:
The excitons of the energetically highest and lowest BO surfaces are dimer states, where excitation is shared between constituent~2 and 3.
The remaining exciton corresponds to the isolated constituent~1, such that the complete excitation is localized on it. For $x_1/d=\sqrt{3}/2$, a linear trimer with equispaced distances is realized, which we already discussed.
Different to these two cases, which can be explained by linear arrangements, for $x_1=0$, two BO surfaces intersect~(red and green line within the blue marked region) which is a consequence of the fact that due to the equilateral triangle configuration all binary interactions have the same strength. This lets the two-body interaction of constituent~2 and 3 be equally strong as the three-body interaction.
It is known that this type of intersections --- conical intersections\index{conical intersection} --- can change the dynamics drastically and is a cause of non-adiabaticity since it is a junction of BO surfaces. Nonadiabatic dynamics was studied in a Rydberg ring trimer\cite{wuester:CI}, where a wave packet initiated on a repulsive BO surface hits a CI and as a consequence gets split on two BO surfaces.

\subsection{Combining directed transport with conical intersections: An outline to guide exciton pulses}
\label{part:fd::chap:theoretical_framework_spatially_unfrozen system_outline_of_tasks}
The trimers as minimal systems helped us to understand how directed transport could work and to see the feature of conically intersecting BO surfaces in two-dimensional arrangements. A requirement of having directed transport are strong and long range interactions and lightweight constituents, such that they can move significantly within the Rydberg state lifetime. Highly excited Lithium atoms satisfy all these conditions and directed transport of RET linked with diatomic proximity was demonstrated for linear arrangements\cite{cenap:motion,wuester:cradle,moebius:cradle} of Lithium Rydberg aggregates in simulations. The scheme is as follows: The atoms get excited to Rydberg states with the same principal quantum number and a microwave performs a transition from the Rydberg state $\ket{\lstate{S}} = \ket{\lstate{s}\dots \lstate{s}}$, where all atoms are in the $\ket{\lstate{s}} \equiv \ket{\nu \lstate{s}}$ state, to the energetically highest exciton state of a single $\lstate{s} \equiv (\nu,\lstate{s})$-excitation. The initial configuration of the Rydberg chain has a dislocation on one end with smallest interatomic distance, such that the \lstate{p}-excitation of this exciton gets localized on the spatially dislocated atoms. The BO surface yields repulsive forces such that directed motion of the atoms is induced which leads to combined transport of diatomic proximity and electronic excitation, called an exciton pulse. The excitation is transported from one to the other end of the linear Rydberg chain, perfectly adiabatic, staying on the same BO surface during the evolution with very high probability. Directed transport of excitation is thus realized with this scheme.

This thesis is dedicated to the investigation of directed transport in higher dimensions, using flexible Rydberg aggregates. Next to their strong interactions, Rydberg atoms can be isolated and trapped very well. Furthermore a precise positioning in arbitrary geometries is possible\cite{sherson2010:single_atom_imaging_trapping,nogrette2014:single_atom_trapping_2D}.
As we saw from the isosceles trimer, conical intersections can already appear in two-dimensional arrangements and their occurrence is not rare\cite{wuester:CI}. Since directed transport in linear configurations is conditioned to dynamics taking place on BO surfaces with globally repulsive interactions, it is a priori not clear if an exciton pulse survives when changing the BO surface due to a CI. 

\begin{figure}[p]
\centering
\includegraphics{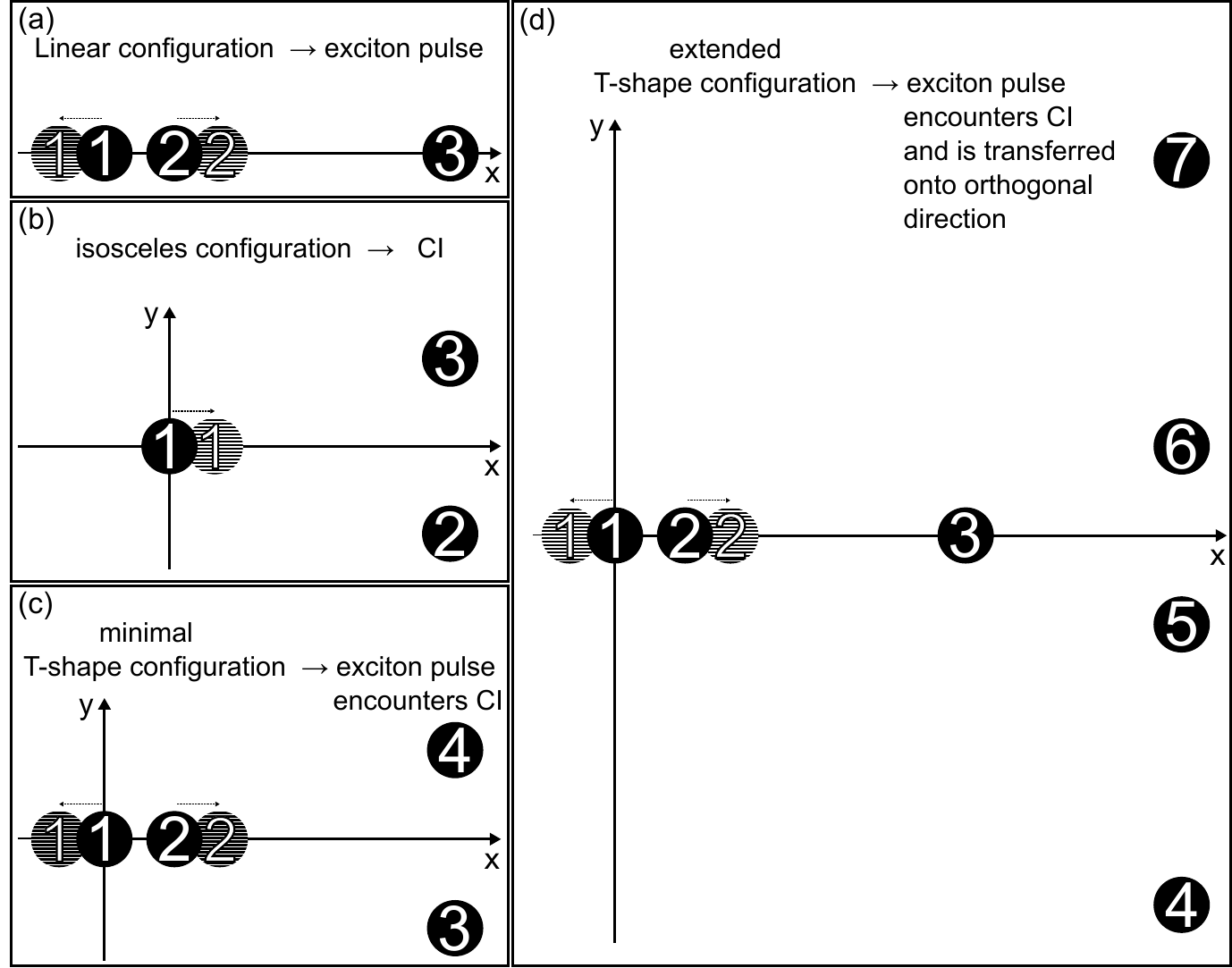} 
\caption{Different geometries of aggregates and their specific features. (a)~Exciton pulses can be created with linear arrangements. (b)~Higher dimensional configurations possess generically CIs, a cause of highly nonadiabatic dynamics. Here we show an isosceles triangle configuration where moving atom~1 towards atom~2 and 3 generates an equilateral triangle configuration such that a CI can be encountered. (c)~A minimal T-shape configuration with smallest initial spacing between atom~1 and 2, such that excitation gets localized on them. Initiating the system in the repulsive BO surface lets atom~2 approach atom~3 and 4, such that the exciton pulse hits a CI when atom~2-4 are in an equilateral triangle configuration. (d)~Extended T-shape configuration which allows for exciton pulse propagation on the horizontal chain, as in (c). The pulse hits a CI approximately when atoms~3, 5 and 6 build an equilateral trimer  and possibly continues to propagate in orthogonal direction to the initial propagation direction after the intersection region.
\label{fig:sketch_compare_configurations}
}
\end{figure}

To adress the transport properties of exciton pulses undergoing a conical intersection transition, we add to the isosceles trimer an additional atom on the horizontal axis, as sketched in \fref{fig:sketch_compare_configurations}~(c). We call this a T-shape configuration, after the ring trimer\cite{wuester:CI} the simplest system where an exciton pulse can encounter a CI. Compared to circular arrangements, T-shape confined systems are experimentally more easily realizable, since they require only linear confinement in two orthogonal directions.
Preparing the initial distance between atom~1 and 2 to be the smallest such that an exciton gets localized on them with repelling forces, atom~2 can approach the vertically placed atoms. In this way an equilateral triangle between atom~2-4 is formed where the BO surface of propagation conically intersects with another one, as seen in \fref{fig:isosceles_trimer}. The T-shape system thus reaches the CI position.
Studying exciton pulses in such a minimal T-shape system is our first objective. Extending the T-shape system as sketched in \fref{fig:sketch_compare_configurations}~(d) is intended to answer the question, whether exciton pulse propagation can be maintained after a CI transition. Furthermore we will try to control and guide the exciton pulse by modifying the arrangement with a vertical displacement of the horizontal chain and tuning the distance between atom~5 and 6. With these modifications, a transition from a CI to an avoided crossing is possbible and the size of the energy gap can be tuned, leading to different transport scenarios.

\chapter{Planar aggregates with isotropic interactions}\label{part:rs::chap:planar_aggregates}
The investigation of exciton pulses traversing a CI is our main interest. As elaborated in \sref{part:fd::chap:theoretical_framework_spatially_unfrozen system_outline_of_tasks}, aggregates in T-shape configurations satisfy the requirements for both, exciton pulse\index{exciton!pulse} propagation and the possibility to have conically intersecting\index{conical intersection} BO~surfaces. We intend to add complexity to the aggregates in several stages. This chapter restricts the spatial dynamics to a plane, supressing the third dimension. Furthermore we employ an isotropic interaction model for simplicity. 
Taking into account the full anisotropy of the dipole-dipole interactions and releasing the dynamics from spatial constrains, we aim in \Chref{part:rs::chap:unconstrained_aggregate} to simulate the dynamics of a T-shape aggregate which could soon be realized in an experiment.

The organization of the chapter is as follows:
In \sref{part:rs::chap:planar_aggregates_theoretical_framework} we present the theoretical framework, including a description of the spatial configuration of T-shape aggregates, the interaction model and methods to solve for the dynamics.
Before we answer the question, whether the presence of a CI can be utilized to guide the excitation transfer, we analyze its effect on exciton pulse propagation in \sref{part:rs::chap:planar_aggregates_nonad_dynamics}.
For a minimal aggregate consisting of two perpendicular dimers,
the essential mechanism of the CI is analyzed in \sref{sec:nges4}.
To investigate if exciton pulses can be sustained after undergoing a highly nonadiabatic transition region, we study its dynamics in an extended T-shape aggregate with more atoms after traversing the CI in \sref{sec:nges8}.
Finally we demonstrate in \sref{sec:exciton_switch} how exciton pulses can be controlled while redirecting them into an orthogonal direction.

\section{Theoretical Framework}
\label{part:rs::chap:planar_aggregates_theoretical_framework}
We study $N$ Rydberg atoms of the species $^7$Li with mass $M=11000$~a.u., all with the same principal quantum number $\nu$. We restrict $\nu$ for the sake of clarity to the two cases $\nu=44$ and $\nu=80$. With $N_x$ of the atoms  constrained on the $x$ axis, and $N_y$ on the $y$ axis, their total number is $N=N_x+N_y$, as sketched in \frefplural{fig:sketch_compare_configurations}{c}{ and }{d}.
We call the sub-system of atoms~1 to $N_x$ the horizontal chain and atoms $N_x+1$ to $N$ the vertical chain.

All atoms are constrained to move freely in only one-dimension, with their positions described by the vector $\vc{R}~=~(\vc{R}_1, \dots, \vc{R}_N)^{\mathrm{T}}$. The restriction to a T-shape configuration gives the following specific form of the individual atomic position vectors
\begin{equation}
 \vc{R}_{\alpha} = \begin{cases}
           (x_{\alpha},0), & {\alpha} \leq N_x\\
	  (\Delta x_{\mathrm{offset}},y_{\alpha}), & {\alpha} > N_x,
          \end{cases}
\label{eq:planar_aggr:set_R_n}
\end{equation}
where the notation $(x,y)$ stands for the two-dimensional vector $x\vc{e}_x + y\vc{e}_y$, with $\vc{e}_{x,y}$ the unit vector in $x,\ y$ direction. Note that $\Delta x_{\mathrm{offset}}$ is a fixed horizontal offset of the vertical chain from the co-ordinate origin. 
Since the co-ordinate of each atom is fixed in one dimension and variable in the other, the effective dimensionality of $\vc{R}$ is $N$.

The one-dimensional confinement could for example be realized by running laser fields and optical trapping of alkali-metal Rydberg atoms~\cite{Li:kuzmich:atomlighentangle}, or earth alkali-metal Rydberg atoms through their second valence electron~\cite{rick:Rydberglattice}.

\begin{figure}[!t]
\centering
\includegraphics{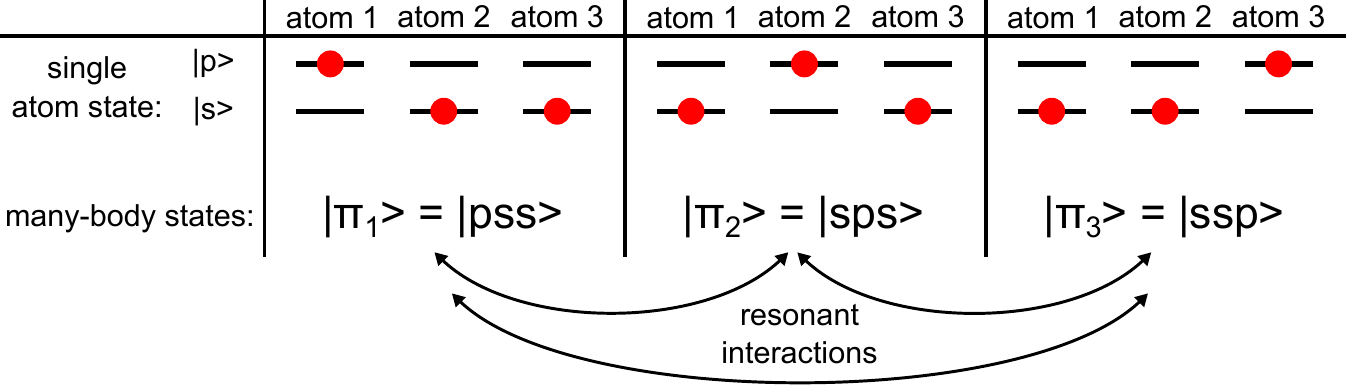} 
\caption{Illustration of localized, singly \lstate{p} excited states according to \eref{eq:planar_aggr:def_pi_states}. Here we demonstrate their construction for a three atom aggregate. They build the electronic basis and allow a diabatic representation of the electronic Hamiltonian.
\label{fig:localized_p_states_isotopic}
}
\end{figure}

For the investigation of RET in flexible Rydberg aggregates we also have to specify resonant many-body states and their interactions. 
We use the simplest model, which is the transport of a single Rydberg \lstate{p} excitation. The atoms have to be prepared such that only one atom is in an angular momentum $\lstate{p} \equiv (\nu,\lstate{p})$ state, all the other atoms are in angular momentum $\lstate{s}\equiv (\nu,\lstate{s})$ states. 
This allows us to expand the electronic wave function in the single excitation basis $B_{\mathrm{el}}:= \{\ket{\pi_\alpha}\}$, where 
\begin{equation}
 \ket{\pi_\alpha}:=\ket{\lstate{s}\dots \lstate{p}\dots \lstate{s}}
\label{eq:planar_aggr:def_pi_states}
\end{equation}
denotes\nomenclature[B]{$\ket{\pi_\alpha}$}{singly excited electronic aggregate state, for the use with isotropic interactions}\nomenclature[B]{$\ket{\pi_\alpha,m}$}{singly excited electronic aggregate state, for the use with anisotropic interactions} a state with the $\alpha$th atom in the \lstate{p} state~\cite{cenap:motion,wuester:cradle}. An illustration of these states is shown in \fref{fig:localized_p_states_isotopic}. Since we assume the interactions to be isotropic, we suppress the magnetic quantum number of the \lstate{p} states.
Using the basis $B_{\mathrm{el}}$ is only valid when no significant admixtures from off-resonant states occur, which is approximately ensured for interatomic distances larger than the dipole blockade radius for neighboring atoms in Rydberg \lstate{s} states. In particular for the initial spatial configuration of the aggregate the interatomic distances have to be larger than the blockade radius to ensure that all atoms can be excited to Rydberg \lstate{s} states before microwave\index{microwave} excitation to an exciton state spanned by the basis $B_{\mathrm{el}}$ can be performed.
Small admixtures from off-resonant states during the dynamics can be treated perturbatively, which we will show later.

\subsection{Rydberg-Rydberg interactions and the electronic Hamiltonian}
\label{part:rs::chap:planar_aggregates_electronic_Hamiltonian}
Interaction potentials between Rydberg atoms can be determined by diagonalizing a dimer Hamiltonian in a restricted electronic state space, 
using the dipole-dipole approximation \cite{book:gallagher}. 
We derive the electronic Hamiltonian\index{Hamiltonian!electronic}, which in general can be written according to \eref{eq:theoret_framew:H_el_full} as a collection of individual constituent Hamiltonians,
\begin{equation}
\hat{\mathcal{H}}_{0} := \sum_{k} \hat{H}^{(k)},
\label{eq:planar_aggr:theoretical_framework_elec_setup_H0_def}
\end{equation}
with $\hat{H}^{(k)}$ the single atom Hamiltonian of the $k$th atom, and an interaction operator, 
\begin{equation}
 \hat{\mathcal{V}}:= \dfrac{1}{2}\sum_{\substack{k,l:\\ k \neq l}}\hat{V}_{\mathrm{dd}}^{(k,l)},
\label{eq:planar_aggr:theoretical_framework_elec_setup_Vop_def}
\end{equation}
\index{dipole-dipole interaction!operator, entire system} with $\hat{V}_{\mathrm{dd}}^{(k,l)}$\index{dipole-dipole interaction!binary!operator} the dipole-dipole operator for interactions between atoms~($k$,$l$). Matrix elements of the electronic Hamiltonian, $\hat{\mathcal{H}}_{\mathrm{el}} = \hat{\mathcal{H}}_{0} + \hat{\mathcal{V}}$, within the single excitation manifold evaluate to
\begin{equation}
 \bra{\pi_{\alpha}}\hat{\mathcal{H}}_{\mathrm{el}}\ket{\pi_{\beta}} = \begin{cases}
E_{\mathrm{aggr}} := (N-1) E_{s} + E_{p}, & \alpha = \beta \\
V(R_{\alpha\beta}), & \alpha \neq \beta
\label{eq:planar_aggr:theoretical_framework_elec_setup_matrix_elements}
\end{cases}.
\end{equation}
We use for the interatomic distances the abbreviation $R_{\alpha\beta}\equiv |\vc{R}_{\alpha} - \vc{R}_{\beta}|$. The dipole-dipole interactions are assumed to be isotropic and set to 
\begin{equation}
 V(r):=-\mu^2/r^3,
\label{eq:planar_aggr:theoretical_framework_elec_setup_binary_interactions}
\end{equation}
\index{dipole-dipole interaction!binary!isotropic}
determined by the scaled radial matrix element  $\mu=d_{\nu,1;\nu,0}/\sqrt{6}$. In \Chref{part:app::chap:tun_int_mag_field} we discuss how this simplification can be realized using a magnetic field and isolating specific azimuthal angular momentum states.
Shifting the energy scale to have $E_{\mathrm{aggr}}$ as zero, the Hamiltonian containing the resonant dipole-dipole interactions\index{dipole-dipole interaction!resonant} can be written in the following way:
\begin{equation}
 \op{H}_{\mathrm{dd}}(\vc{R}):=-\mu^2\sum_{\substack{\alpha,\beta=1;\\ \alpha\neq \beta}}^{N}R_{\alpha \beta}^{-3}\ket{\pi_\alpha}\bra{\pi_\beta}.
  \label{eq:elechamiltonian-dd}
\end{equation}
\nomenclature[B]{$\op{H}_{\mathrm{dd}}$}{resonant dipole-dipole Hamiltonian}The resonant states are also coupled to off-resonant states, yielding vdW~interactions\index{dipole-dipole interaction!van-der-Waals}. Ideally the interatomic distances are large enough such that off-resonant couplings can be neglected. For instance, this can be ensured for the propagation along a globally repulsive BO~surface in one-dimensional configurations. However, since we are specifically interested in nonadiabatic dynamics, evolution can involve BO~surfaces with interactions allowing atoms in the dynamics to approach each other close enough for vdW~interactions\index{dipole-dipole interaction!van-der-Waals} to become important. This occurs when almost no \lstate{p} excitation resides on both atoms. Since any too close encounter of atoms would invalidate our simple model and possibly lead to ionisation, these are problematic. To prevent them, principal quantum numbers can be chosen for which vdW~interactions\index{dipole-dipole interaction!van-der-Waals} are repulsive such that they effectively act as a stopping mechanism at small interatomic distances. The formula for vdW~interactions\index{dipole-dipole interaction!van-der-Waals} can be derived via block-diagonalization, outlined in \sref{part:app::chap:block_diag__effective_interactions} and \sref{part:app::chap:block_diag__van_der_Waals}. For the aggregate's basis states, given in \eref{eq:planar_aggr:def_pi_states}, this procedure is equivalent to second-order Rayleigh-Schrödinger perturbation theory.
The vdW~interaction\index{dipole-dipole interaction!van-der-Waals} for the state $\ket{\pi_\alpha}$ can thus be calculated by
\begin{equation}
 h_{\mathrm{vdw}}^{\ket{\pi_\alpha}} = \sum_{\substack{\ket{Y}:\\E_{Y}\neq E_{\mathrm{aggr}}}}\dfrac{\left|\bra{\pi_\alpha}\hat{\mathcal{V}}\ket{Y}\right|^2}{E_{\mathrm{aggr}} - E_{Y}},
\label{eq:planar_aggr:formula_vdW_aggr_states_start}
\end{equation}
\index{dipole-dipole interaction!van-der-Waals!aggregate states!calculation formula}
with a summation over all $N$-body states, denoted by $\ket{Y}$, whose corresponding energy $E_{Y}$ is different to the aggregate's energy. The vdW~interaction\index{dipole-dipole interaction!van-der-Waals} formula decomposes into a sum over binary contributions~(see \aref{part:app::calcs_planar_aggr__vdw_aggr}):
\begin{equation}
  h_{\mathrm{vdw}}^{\ket{\pi_\alpha}}  = \sum_{k \neq \alpha}\sum_{\substack{\ket{y}:\\E_{y}\neq E_{\lstate{ps}}}}\dfrac{|\bra{\lstate{ps}}\hat{V}_{\mathrm{dd}}^{(\alpha,k)}\ket{y}|^2}{E_{\lstate{ps}}-E_{y}} + \dfrac{1}{2}\sum_{k \neq \alpha}\sum_{l\neq k}\sum_{\substack{\ket{y}:\\E_{y}\neq E_{\lstate{ss}}}}\dfrac{|\bra{\lstate{ss}}\hat{V}_{\mathrm{dd}}^{(k,l)}\ket{y}|^2}{E_{\lstate{ss}}-E_{y}}.
\label{eq:planar_aggr:formula_vdW_aggr_states_sum_over_binary_interactions}
\end{equation}
The states $\ket{\lstate{ps}} \equiv \ket{\nu,\lstate{p};\nu,\lstate{s}}, \ket{\lstate{sp}} \equiv \ket{\nu,\lstate{s};\nu,\lstate{p}}$ and $\ket{y}$ are pair states with energy $E_{\lstate{ps}}, E_{\lstate{sp}}$ and $E_{y}$, respectively.  Using the definition of $C_{6}$-dispersion coefficients\index{dispersion coefficient} given in \eref{eq:alk_ryd_atoms:vdW_interactions__C6_ss_states},
we finally arrive at\index{dipole-dipole interaction!van-der-Waals!aggregate states!expression!general}
\begin{equation}
  h_{\mathrm{vdw}}^{\ket{\pi_\alpha}}(\vc{R}) = -\sum_{\substack{k,l:\\ k \neq l}}\dfrac{C_{6}^{\lstate{ss}}}{2R_{kl}^6} - \sum_{k \neq \alpha}\dfrac{C_{6}^{\lstate{ps}} - C_{6}^{\lstate{ss}}}{R_{k\alpha}^6},
\end{equation}
with $C_{6}^{\lstate{ss}}$ the dispersion coefficient for the pair state $\ket{\lstate{ss}}\equiv\ket{\nu,\lstate{s};\nu,\lstate{s}}$ and $C_{6}^{\lstate{ps}}$ for $\ket{\lstate{ps}}$, respectively. 
Since the vdW~interactions serve here the practical purpose discussed above, we simplify their structure by setting the two different dispersion coefficients\index{dispersion coefficient} equal, $C_{6}\equiv C_{6}^{\lstate{ps}} = C_{6}^{\lstate{ss}}$. 
Their difference in reality can give rise to interesting effects at shorter distances\cite{zoubi:VdWagg}, which are not relevant here.
Using only a single dispersion coefficient, the vdW~interactions get independent of the position\index{dipole-dipole interaction!van-der-Waals!aggregate states!expression!simplified} of the \lstate{p} excitation,
\begin{equation}
 h_{\mathrm{vdw}}(\vc{R}) := -\sum_{\substack{k,l:\\ k \neq l}}\dfrac{C_{6}}{2R_{kl}^6},
\label{eq:planar_aggr:setup_single_p_state__vdw_scalar_function}
\end{equation}
and therefore the corresponding Hamiltonian is simply diagonal,
\begin{equation}
 \hat{H}_{\mathrm{vdw}}(\vc{R}) := h_{\mathrm{vdw}}(\vc{R})\cdot \hat{\id},
\label{eq:planar_aggr:setup_single_p_state__vdw_Hamiltonian}
\end{equation}
\nomenclature[B]{$\op{H}_{\mathrm{vdw}}$}{van-der-Waals Hamiltonian}where $\hat{\id}$ denotes the identity operator of the electronic space spanned by the single excitation states.
Adjusting $C_6 < 0$ in \eqref{eq:planar_aggr:setup_single_p_state__vdw_scalar_function} ensures vdW~interactions to be repulsive.
The final electronic Hamiltonian combines resonant and off-resonant interactions and is given by
\begin{equation}
 \hat{H}_{\mathrm{el}}(\vc{R}) := \hat{H}_{\mathrm{dd}}(\vc{R}) + \hat{H}_{\mathrm{vdw}}(\vc{R}).
\label{eq:planar_aggr:setup_single_p_state__def_Hel}
\end{equation}
\nomenclature[B]{$\op{H}_{\mathrm{el}}$}{electronic Hamiltonian}We sketch, in \Chref{part:app::chap:tun_int_mag_field}, how this
 simple model of interactions arises from the full molecular
 physics of interacting Rydberg atoms using a magnetic field
and selected total angular momentum states. Solving the eigenvalue problem of this Hamiltonian for fixed atomic positions yields eigenstates that are called Frenkel excitons\index{exciton!Frenkel}~\cite{frenkel_exciton}, denoted by $\ket{\varphi_k(\vc{R})}$. The eigenenergies define BO~surfaces evaluated at fixed atomic positions, which we denote with $U_{k}(\vc{R})$.

\begin{figure}[!t]
\centering
\includegraphics{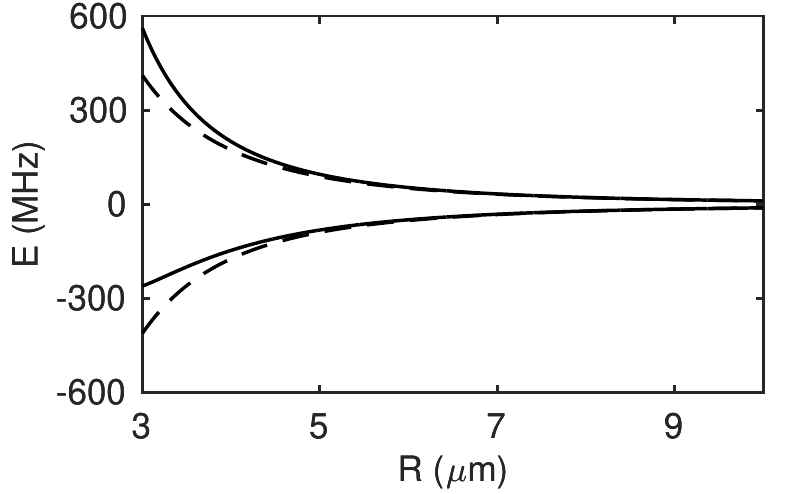} 
\caption{Dimer BO~surfaces according to \eref{eq:planar_aggr:setup_single_p_state__dimer_BO_surfaces} with~(solid lines) and without~(dashed lines) vdW interactions\index{dipole-dipole interaction!van-der-Waals} for $\nu = 80$, which corresponds to $\mu = 3374$~a.u. and $C_{6}=-7.6\times 10^{20}$~a.u.. At small interatomic distances, the vdW~interactions clearly changes the surfaces, making $U_{+}$~(>0) more repulsive and $U_{-}$~(<0) less attractive.
\label{fig:dimer_compare_resonant_vs_van_der_Waals_strength}
}
\end{figure}

\begin{figure}[!t]
\centering
\includegraphics{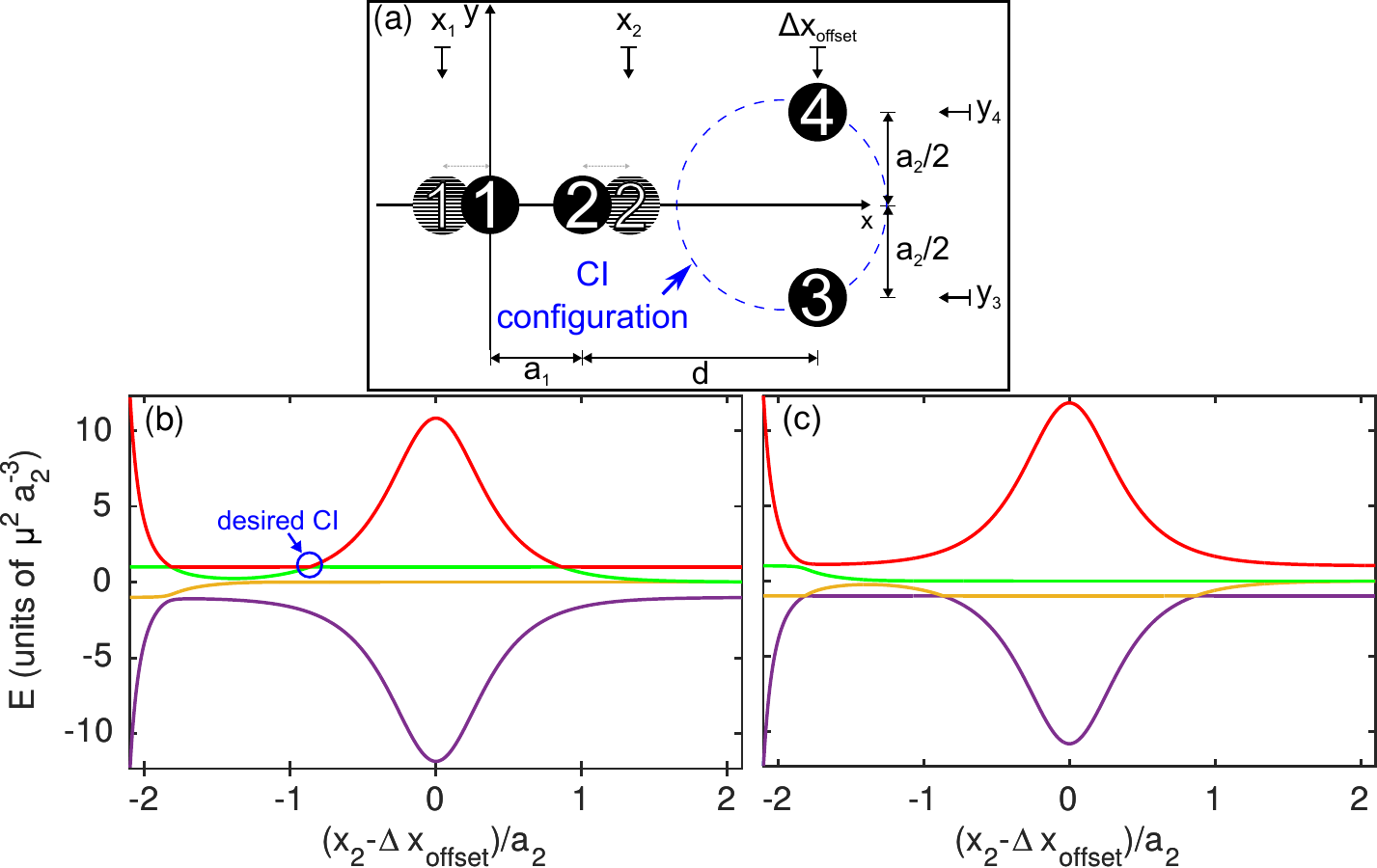} 
\caption{(a) Sketch of a four atom T-shape aggregate with configuration $(x_1,x_2,y_3,y_4) = (-x,a_1+x,-a_2/2,a_2/2)$ and vertical chain offset by $\Delta x_{\mathrm{offset}}= a_1+d$. (b), (c) Energy spectra of $\hat{H}_{\mathrm{dd}}(\vc{R})$~(b) and $-\hat{H}_{\mathrm{dd}}(\vc{R})$~(c), with a Hamiltonian according to \eref{eq:elechamiltonian-dd}. We vary the positions of atoms~(1,2) equally and plot the spectra over the distance between atom~2 and the vertical dimer axis. The T-shape aggregates of interest with closer distance between atoms~(1,2) than between atom~(3,4), then two specific states are localized on atoms~(1,2), with surfaces shown as red and violett line. Note that only for interactions with negative amplitude~[(b)] the repulsive surface~(red line) features a CI~(location marked with blue circle), different to the case of interactions with positive amplitude~[(c)]. Since the exciton pulse requires an initiation on the repulsive surface, negative interaction amplitudes are required to get access to a CI.
The parameters for the calculation of the energy spectra are set to 
$a_1 = 0.4211 a_2$ and $d = 2.1053a_2$ and correspond to the values used for the aggregate under investigation in \sref{sec:nges4}.
%
\label{fig:trimer_4atom_aggr_spectra_pl_min_int}
}
\end{figure}

The excitons for a dimer are given by
\begin{equation}
 \ket{\varphi_{\pm}} = \left(\ket{\pi_{1}}\mp \ket{\pi_{2}}\right)/\sqrt{2},
\label{eq:planar_aggr:setup_single_p_state__dimer_states}
\end{equation}
and the corresponding BO~surfaces by
\begin{equation}
 U_{\pm}(R) = \pm \dfrac{\mu^2}{R^3} - \dfrac{C_{6}}{R^6}.
\label{eq:planar_aggr:setup_single_p_state__dimer_BO_surfaces}
\end{equation}
The vdW~interactions amplify the repulsive character of the repulsive BO~surface, and, it diminish the attractive character of the attractive surface, depicted in \fref{fig:dimer_compare_resonant_vs_van_der_Waals_strength}.
Note that the dimer excitons are not dependent on the distance between the atoms. However, they are entangled states where both atoms share the excitation equally. 
It was theoretically demonstrated, that two clouds of ultracold atoms in the dipole-blockade regime can eject a single Rydberg atom per cloud after excitation from ground to Rydberg and from Rydberg to exciton state. Both ejected atoms are prepared in a dimer exciton state and thus EPR\nomenclature[B]{EPR}{Einstein-Podolsky-Rosen} correlated, violating Bell's inequalities\index{Bell inequalities}. The creation of entangled, far separated single atom pairs is possible with this setup\cite{wuester:cannon}.
As demonstrated in \sref{part:fd::chap:theoretical_framework_spatially_unfrozen system_trimers}, the excitation distribution of excitons becomes dependent on the atomic configuration for more than two atoms. Directed transport of excitation linked with diatomic proximity, a so called exciton pulse\index{exciton!pulse}, requires an initialization on a repulsive BO~surface. Only for negative resonant interaction amplitudes as in \eref{eq:planar_aggr:theoretical_framework_elec_setup_binary_interactions}, the repulsive surface in a T-shape configuration conically intersects\index{conical intersection} when a trimer subunit decouples and its atoms form an equilateral triangle configuration. 
We illustrate this for a four atom T-shape aggregate~[sketch depicted in \fref{fig:trimer_4atom_aggr_spectra_pl_min_int}(a)], and vary the interatomic distance between atoms~(1,2). The energy spectrum is shown for a resonant dipole-dipole Hamiltonian with negative binary interactions in \fref{fig:trimer_4atom_aggr_spectra_pl_min_int}(b), where the repulsive surface~(red line) clearly intersects with the second most energetic surface at the position marked with a blue circle. In contrast, the energy spectrum of the same Hamiltonian with positive binary interactions, shown in \fref{fig:trimer_4atom_aggr_spectra_pl_min_int}(c), the repulsive surface features no CI.
\newpage

\subsection{Initial state}
\label{initial_state}
We assume initially no quantum correlations between spatial and electronic degree of freedom, such that the total wave function can be written as a direct product,
%
\begin{equation}\label{initialPsi}
\ket{\Psi_{0}(\vc{R})} = \chi_{0}(\vc{R})\ket{\psi_{\mathrm{el},0}(\vc{R})}
\end{equation}
where $\ket{\psi_{\mathrm{el},0}(\vc{R})}$ is the initial electronic state and $\chi_{0}(\vc{R})$ the initial nuclear wave function.
The atoms are initially trapped in approximately harmonic potentials around the mean atomic positions. Due to the ultracold environment the nuclear wave function is in the ground state of the trapping potentials, which motivates to use for each atom a Gaussian 
nuclear distribution and finally yields a product of them for the total nuclear wave function,
\begin{equation}
\chi_{0}(\vc R)=\left(2\pi\sigma_{0}^2\right)^{-\frac{N}{4}}\prod_{n=1}^N \exp{  \left(-\frac{|\bv{R}_{n} - \bv{R}^{(0)}_n | ^2}{4 \sigma_{0}^2}  \right)}.
\label{inipos}
\end{equation}
We denote the standard deviation with $\sigma_0$, which is controlled by the trapping width. The vector $\vc{R}_{0}\equiv(\bv{R}^{(0)}_1 \dots \bv{R}^{(0)}_N)^T$ denotes the initial mean atomic configuration.
Note that each atom is constrained to move only along one direction.
The initial electronic state should approximately be a repulsive dimer state on atoms~(1,2),
\begin{equation}
 \ket{\psi_{\mathrm{el},0}}\approx \ket{\varphi_{\mathrm{rep}}} =  \left(\ket{\pi_1}-\ket{\pi_2}\right)/\sqrt{2},
\end{equation}
such that excitation is localized on them and initiates the exciton pulse due to repulsive forces.
\subsection{Dynamical methods}
\label{part:rs::chap:planar_aggregates_dynamical methods} 
So far the dependency of excitons and BO~surfaces on the atomic configuration appeared parametrically. Changing the spatial arrangement will change both quantities. On the other hand, atomic motion is induced by forces due to the surfaces, such that the spatial and electronic degrees of freedom are dynamically interlinked. 
In the following we first show the exact equations of motion in \sref{part:rs::chap:planar_aggregates_dynamical methods_exact_methods}. Since the numerical effort increases drastically for increasing number of atoms, we present a quantum-classical method in \sref{part:rs::chap:planar_aggregates_dynamical methods_quantum_classical_method}, which we use to approximately find the dynamics of the flexible Rydberg aggregates.
The methods are presented using the example of aggregates with isotropic interactions, but are straightforward to adjust for the use with general binary interactions.

\subsubsection{Exact method}
\label{part:rs::chap:planar_aggregates_dynamical methods_exact_methods} 
The full quantum dynamics is governed by the Schrödinger equation of the total system where all information is encoded in the total wave function\index{wave function}. Its evolution is determined by a Hamiltonian\index{Hamiltonian!total} which includes the electronic Hamiltonian and kinetic terms, and is given by
\begin{equation}
 \hat{H}_{\mathrm{T}}(\vc{R}) := -\dfrac{\hbar^2}{2M}\nabla^2_{\vc{R}} + \hat{H}_{\mathrm{el}}(\vc{R}),
\label{eq:quant_dyn_methods_exact:flex_aggr__total_Hamiltonian}
\end{equation}
where $\nabla^2_{\vc{R}}$ is the Laplacian of the coordinate vector $\vc{R}$, containing all atomic positions and $M$ is the mass of the atomic species. The time evolution of the total wave function is determined by the Hamiltonian in \eref{eq:quant_dyn_methods_exact:flex_aggr__total_Hamiltonian} and its equation of motion is the time-dependent Schr{\"o}dinger equation\index{Schrödinger equation!total},
\begin{equation}
 \im \hbar \dfrac{\partial}{\partial t} \ket{\Psi(\vc{R},t)}=\hat{H}_{\mathrm{T}}(\vc{R})\ket{\Psi(\vc{R},t)}.
\label{eq:quant_dyn_methods_exact:flex_aggr__schroedinger_equ_total_Hamiltonian}
\end{equation}
To numerically solve the Schr{\"o}dinger equation, it is helpful to expand the wave function in an appropriate basis of the electronic space. Using the localized \lstate{p}~states, we get the \emph{diabatic representation}\index{wave function!diabatic representation} of the wave function,
\begin{equation}
 \ket{\Psi(\vc{R},t)} = 
\sum_{\alpha=1}^{N}
\chi_{\alpha}(\vc{R},t)\ket{\pi_{\alpha}}.
\label{eq:quant_dyn_methods_exact:flex_aggr__diabatic_expansion_total_wave_function}
\end{equation}
The absolute square values of the functions $\chi_{\alpha}(\vc{R},t)$ describe the space and time\hyp{}dependent probability density for finding the \lstate{p}~excitation localized on the $\alpha$th atom. The expansion coefficients, $|\chi_{\alpha}(\vc{R},t)|^2$, are the \emph{diabatic densities}.
Another common expansion uses the excitons of the electronic Hamiltonian,
\begin{equation}
 \ket{\Psi(\vc{R},t)} = 
\sum_{k=1}^{N}
\tilde{\chi}_{k}(\vc{R},t)\ket{\varphi_{k}(\vc{R})}.
\label{eq:quant_dyn_methods_exact:flex_aggr__adiabatic_expansion_total_wave_function}
\end{equation}
which is called the \emph{adiabatic representation}\index{wave function!adiabatic representation}, also known as the Born-Oppenheimer expansion\cite{born1954:born_oppenheimer_expansion}, where $|\tilde{\chi}_{k}(\vc{R},t)|^2$ are the \emph{adiabatic densities}.
Integrating out the spatial degrees of freedom of the densities we get \emph{populations},
\begin{align}
 P_{\alpha}(t) & :=  \int \bigl|\,\chi_{\alpha}(\vc{R},t)\,\bigr|^2\, \dscalar^{N} \vc{R},
\label{eq:quant_dyn_methods_exact:flex_aggr__diabatic_pops}\\
\tilde{P}_k(t) & := \int \bigl|\,\tilde{\chi}_{k}(\vc{R},t)\,\bigr|^2\, \dscalar^{N} \vc{R},
\label{eq:quant_dyn_methods_exact:flex_aggr__adiabatic_pops}
\end{align}
where $P_{\alpha}(t)$ are the diabatic and $\tilde{P}_k(t)$ the adiabatic populations, respectively. Note that $\int \dscalar^{N} \vc{R}$ denotes integration over all atomic coordinates.
 The larger the adiabatic population of an exciton, the more prominent it is in the system and its dynamics. The change of adiabatic populations over time measures the adiabaticity. We say a process is \emph{nonadiabatic} within a time interval, when the adiabatic populations change significantly therein.

Both expansions of the total wave function make the complexity of the Schr\"odinger equation in \eref{eq:quant_dyn_methods_exact:flex_aggr__schroedinger_equ_total_Hamiltonian} fully apparent.
Using the diabatic representation leads to the following set of coupled partial differential equations:
\begin{align}
 \im \hbar \dfrac{\partial}{\partial t} \chi_{\alpha}(\vc{R},t) = \left( -\dfrac{\hbar^2}{2M}\nabla^2_{\vc{R}} + h_{\mathrm{vdw}}(\vc{R})\right)\chi_{\alpha}(\vc{R},t) + 
%
 \sum_{\substack{\beta \neq \alpha}}\,
%
V(\vc{R}_{\alpha\beta})\chi_{\beta}(\vc{R},t).
\label{eq:quant_dyn_methods_exact:flex_aggr__diabatic_expansion_schroedinger_equation}
\end{align}
Using the adiabatic representation, we get the following set of differential equations:
\begin{equation}
 \im \hbar \dfrac{\partial}{\partial t} 
\tilde{\chi}_{k}(\vc{R},t) = \left( -\dfrac{\hbar^2}{2M}\nabla^2_{\vc{R}} + U_{k}(\vc{R})\right)\tilde{\chi}_{k}(\vc{R},t) -  \sum_{\substack{l=1}}^{N}\Lambda_{k,l}(\vc{R})\tilde{\chi}_{l}(\vc{R},t).
%
%
\label{eq:quant_dyn_methods_exact:flex_aggr__adiabatic_expansion_schroedinger_equation}
\end{equation}
Each adiabatic density is coupled to the others through the \emph{nonadiabatic couplings} $\Lambda_{k,l}(\vc{R})$, which are a result of the kinetic energy operator acting on the $\vc{R}$-dependent expansion coefficients and excitons.
They by themselves consist of two parts:
\begin{equation}
 \Lambda_{k,l}(\vc{R}) := \dfrac{\hbar^2}{2M}\bigl(2\vc{F}_{k,l}(\vc{R})\nabla_{\vc{R}} + G_{k,l}(\vc{R})\bigr),
\label{eq:quant_dyn_methods_exact:flex_aggr__nonadiabatic_couplings}
\end{equation}
where the nonadiabatic
\emph{derivative couplings} are given by
\begin{equation}
 \vc{F}_{k,l}(\vc{R}):=\bra{\varphi_{k}(\vc{R})}\nabla_{\vc{R}}\ket{\varphi_{l}(\vc{R})},
\label{eq:quant_dyn_methods_exact:flex_aggr__derivative_couplings}
\end{equation}
and the nonadiabatic
\emph{scalar couplings} have the form
\begin{equation}
 G_{k,l}(\vc{R}) := \bra{\varphi_{k}(\vc{R})}\nabla_{\vc{R}}^2\ket{\varphi_{l}(\vc{R})}.
\label{eq:quant_dyn_methods_exact:flex_aggr__scalar_couplings}
\end{equation}
Note that the derivative couplings are antihermitian, $\vc{F}^\dagger_{k,l}(\vc{R}) = -\vc{F}_{l,k}(\vc{R})$.
Although the determining equations for the adiabatic expansion coefficients have a more difficult structure than those for the diabatic expansion coefficients, they have the advantage to decouple when the nonadiabatic couplings vanish. A big class of systems can be well described by neglecting the nonadiabatic couplings, which is known as \emph{adiabatic approximation}\cite{ballhausen1972:adiabatic_approximation}.
Then, the dynamics is dictated by a single Schr\"odinger equation with a single BO~surface taking over the role of the potential for the atoms. It was shown that for exciton pulses in linear Rydberg chains, the dynamics stays largely adiabatic\cite{wuester:cradle,moebius:cradle} and the adiabatic approximation would be justified. However, near CIs one can never use the adiabatic approximation.

The framework presented so far does not contain any further approximations.
The adiabatic representation of the coupled Schr{\"o}dinger equations in \eqref{eq:quant_dyn_methods_exact:flex_aggr__adiabatic_expansion_schroedinger_equation} appears equivalently in the dynamics of molecules. In either case, solving the coupled Schr{\"o}dinger equations is a hard problem. For further reading about quantum chemical methods, see \rref{domcke2004:book_cis}.


\subsubsection{Quantum-classical method}
\label{part:rs::chap:planar_aggregates_dynamical methods_quantum_classical_method} 
The systems of our interest can practically not be studied with the exact equations, since they have too many spatial degrees of freedom. Furthermore, the propagation passes nonadiabatic regions, such that an adiabatic approximation is not possible.
We therefore use a quantum-classical method, Tully's surface hopping algorithm\cite{tully:hopping2} with fewest-switches\cite{tully:hopping,barbatti:review_tully}~(FSSH)\nomenclature[b]{FSSH}{fewest-switches surface hopping}\index{fewest-switches surface hopping}. 
 The algorithm is a trajectory based approach, where classical trajectories for the atoms are propagated according to Newton's equations, with forces from individual BO~surfaces. The BO~surface of propagation can switch over time for each individual trajectory, such that nonadiabatic couplings are accounted for. Initial positions and velocities are chosen such that their statistics are according to the Wigner distribution of the initial nuclear wave function. In the following we briefly sketch the method. Let $\chi_{0}(\vc{R})$ be the initial nuclear wave function. Then, the probability to find the system at position $\vc{R}_0$ is $|{\chi_{0}(\vc{R_0})}|^2$, while velocities $\dot{\vc{R}}_0$ are distributed according to $|\mathcal{FT}[{\chi_{0}(\vc{R_0})}]|^2$, the Fourier transform of the position space wave function.
For each classical trajectory we thus randomly select a pair of initial positions and velocities, $\gamma=\{\vc R_0,\dot{\vc R}_0\}$, distributed according to the probability distributions above, derived from the wave function.
Furthermore an initial BO~surface is selected for the system.
The propagation of a classical trajectory is then fully determined by Newton's equation,
\begin{equation}
 M\ddot{\vc{R}}=-\nabla_{\vc{R}}U_{\zeta(t)}(\vc{R}),
\label{eq:quant_dyn_methods_tully:newton_equation}
\end{equation}
where $\zeta(t)$ is the index of the instantaneously propagated BO~surface. Since we wish to allow for trajectories following different BO~surfaces, the index is time-dependent with stochastic modifications, which we explain later.
Note that the position vector is also time-dependent, $\vc{R} = \vc{R}(t)$.

Simultaneously to \eref{eq:quant_dyn_methods_tully:newton_equation}, the electronic Schrödinger equation\index{Schrödinger equation!electronic} is propagated,
\begin{equation}
\im \hbar\dfrac{\partial}{\partial t}\ket{\psi_{\mathrm{el}}(\vc{R}(t))} = \hat{H}_{\mathrm{el}}(\vc{R}(t))\ket{\psi_{\mathrm{el}}(\vc{R}(t))}.
\label{eq:quant_dyn_methods_tully:el_Schroedinger_equation}
\end{equation}
These two equations are coupled, since the electronic wave function\index{wave function!electronic} and the electronic Hamiltonian parametrically depend on the atomic positions, which vary according to \eref{eq:quant_dyn_methods_tully:newton_equation}. To numerically solve both equations, Newton's equation propagates the atomic positions in each time step forward according to forces from a certain BO~surface, which was obtained by diagonalization of the electronic Hamiltonian in the previous time step.
As for the exact method, the wave function can be expanded in different ways. The adiabatic expansion is in the basis of time-dependent excitons, with expansion coefficients $\tilde{c}_k(t):=\braket{\varphi_{k}(\vc{R}(t))|\psi_{\mathrm{el}}(\vc{R}(t))}$. The diabatic expansion uses the localized \lstate{p}~states, leading to the expansion coefficients $c_\alpha(t):=\braket{\pi_{\alpha}|\psi_{\mathrm{el}}(\vc{R}(t))}$. A diabatic expansion of the electronic wave function is convenient, turning \eref{eq:quant_dyn_methods_tully:el_Schroedinger_equation} into
\begin{equation}
 \im \hbar \dot{c}_{\alpha}(t)  = h_{\mathrm{vdw}}(\vc{R}(t))c_{\alpha}(t) + \sum_{\beta\neq \alpha} V(R_{\alpha\beta}(t))c_{\beta}(t).
\label{eq:quant_dyn_methods_tully:el_Schroedinger_equation_diab_coeffs}
\end{equation}
To account for nonadiabatic dynamics, for each classical trajectory the BO~surface of propagation can change over time, which is realized by sudden transitions, also called jumps.
The jumps are not deterministic, but randomly occur according to a specific probability distribution. The presciption for possible jumps and the linked probability distribution distinguish different surface hopping algorithms. We use the most common variant, the fewest-switches method, where a check for transitions is performed in each time step.
In FSSH, the probability for a transition between two surfaces is set to the relative change of population, which for a jump from $U_m$ to $U_n$ is given by
\begin{equation}
 g_{m,n} = \max\left(0,\dfrac{b_{n,m}\Delta t}{a_{m,m}}\right),
\label{eq:quant_dyn_methods_tully:jump_probabilities}
\end{equation}
with $\Delta t$ the current numerical time step of propagation and 
\begin{align}
 a_{n,m}&:=c_{n}c^{*}_m,
\label{eq:quant_dyn_methods_tully:anm}
\\
b_{n,m}&:=-2\Re\left(a_{n,m}^{*}\braket{\dot{\vc{R}},\vc{F}_{n,m}}\right).
\label{eq:quant_dyn_methods_tully:bnm}
\end{align}
A transition is accepted, if two conditions are fulfilled. The first condition compares the probability to a uniformly selected random number, $x \in [0,1]$, and is satisfied when
\begin{equation}
 \sum_{p=1}^{n-1}g_{m,p} < x <\sum_{p=1}^{n}g_{m,p}
\label{eq:quant_dyn_methods_tully:test_acc_prob}
\end{equation}
is fulfilled. Since the jump to another surface creates a difference in the potential energy, the velocities of the atoms have to be adjusted to achieve conservation of the total energy. The velocity adjustment is chosen in the direction of the non\-adiabatic derivative coupling vector corresponding to the transition between the two surfaces, $\vc{F}_{m,n}$, such that we have the ansatz for the velocities
\begin{equation}
 \dot{\vc{R}}(t) = \dot{\vc{R}}(t-\Delta t) - \Delta_{m,n}^{v}\vc{F}_{m,n}/\|\vc{F}_{m,n}\|_{2}.
\end{equation}
The velocity adjustment is given by
\begin{equation}
 \Delta_{m,n}^{v} := A_{m,n} \pm \sqrt{B_{m,n}}, \quad A_{m,n}  \lessgtr 0
\end{equation}
with
\begin{align}
 A_{m,n} &:= \braket{\dot{\vc{R}}(t-\Delta t),\vc{F}_{m,n}}/\|\vc{F}_{m,n}\|_{2},\\
B_{m,n} &:= A_{m,n}^2 - 2(U_n-U_m)/M.
\end{align}
It ensures energy conservation, when the second condition is satisfied, that is $B_{m,n}\geq 0$. Each single trajectory has a stochastic sequence for the index of propagated BO~surfaces, $\zeta(t)$, since the jumps between the surfaces occur randomly. Therefore, typically a large number of trajectories need to be propagated to sample the atomic probability distribution, which is also called atomic density. However, the number of required propagated trajectories is specifically dependent on the observables of interest. A consistency check whether nonadiabaticity is correctly considered is to compare the average adiabatic population of each surface with the average fraction of trajectories propagating along the same surface for a large enough number of trajectories such that both quantities are statisitically converged. The fraction of the BO~surface labeled with $\ket{\varphi_{n}{\vc{R}(t)}}$ is defined by
\begin{equation}
 f_{n}(t):=\dfrac{1}{N_{\mathrm{traj}}}\sum_{i=1}^{N_{\mathrm{traj}}} \delta_{n,\zeta_{i}(t)},
\label{eq:quant_dyn_methods_tully:fractions}
\end{equation}
where $N_{\mathrm{traj}}$ denotes the number of propagated trajectories and $\zeta_{i}(t)$ is the stochastic sequence of BO~surface indices on which propagation takes place for the $i$th trajectory. Nonadiabatic transitions are accurately described when the average values of adiabatic populations and fractions for each BO~surface match. 

A comparison between FSSH and the exact propagation according to \eref{eq:quant_dyn_methods_exact:flex_aggr__schroedinger_equ_total_Hamiltonian} was performed for flexible trimer aggregates in one-dimension, finding very good agreement\cite{wuester:cradle,moebius:cradle,moebius:bobbels,leonhardt:switch}.

The surface hopping method is reviewed in detail in \rref{barbatti:review_tully}.
\section{Nonadiabatic dynamics}
\label{part:rs::chap:planar_aggregates_nonad_dynamics}
After we provided the tools to describe excitons, BO~surfaces and the dynamics of the aggregate, we discuss nonadiabatic dynamics of exciton pulses in T-shape aggregates in this section. The focus lies on understanding how a CI affects an exciton pulse, such that we can utilize it to direct and manipulate exciton pulses. A detailed study of the mechanism of a CI is presented for a minimal T-shape aggregate in \sref{sec:nges4}, consisting of two perpendicular aligned dimers, each of them constrained to move along a single direction. Subsequently, we extend the T-shape aggregate 
in \sref{sec:nges8} to seven~atoms, where the investigations are focused on the possibility of excitation transfer after the wave packet traverses a CI and redirection on an orthogonal direction.
%
\subsection{Two perpendicular dimers}
\label{sec:nges4}
\begin{figure}[!t]
\centering
\includegraphics{{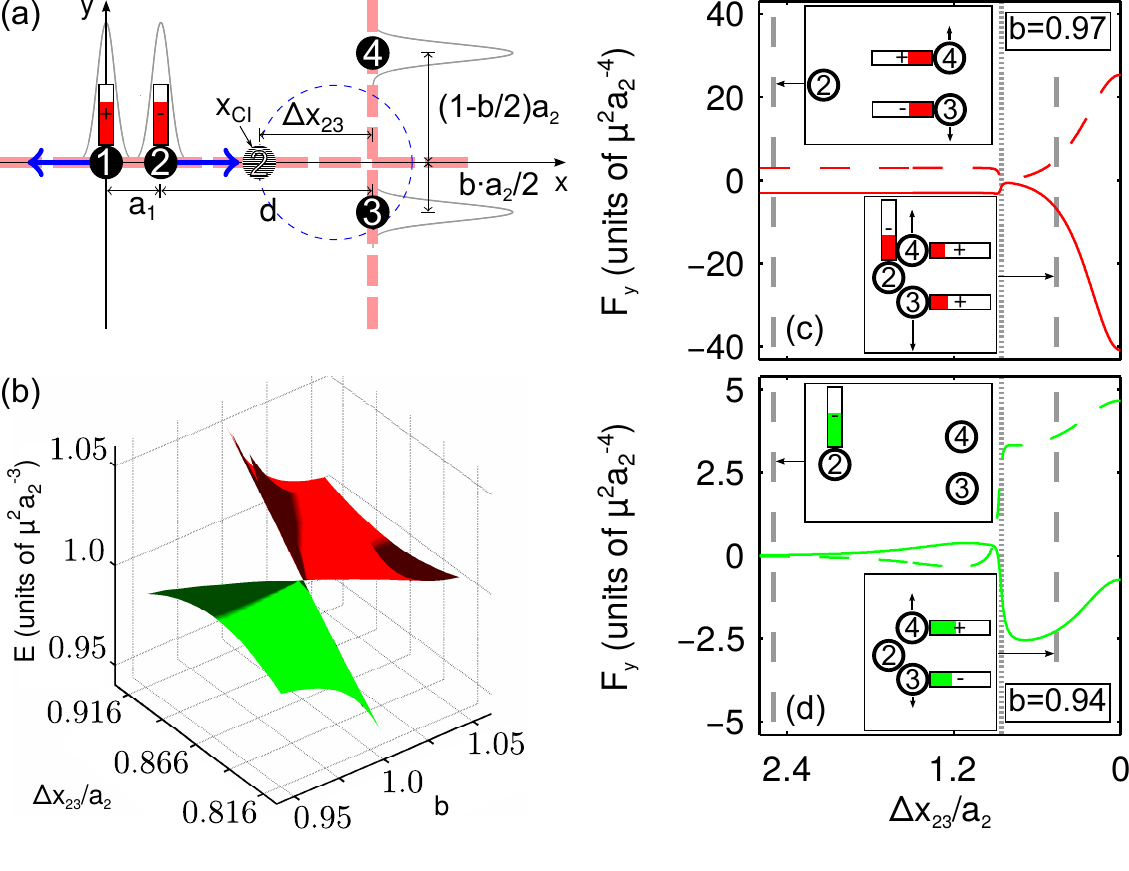}}
\caption{(a) Orthogonal atom chains with one Rydberg dimer each. Atoms~1 and 2 initially share an excitation. Due to the ensuing repulsion (blue arrows) atom~2 reaches the conical intersection at $\sub{x}{CI}$\index{conical intersection}. The colored bars visualise the excitation amplitude on each atom $c_n = \braket{\pi_n|\sub{\varphi}{rep}}$, with ``+'' for $c_n>0$ and ``-'' for $c_n<0$. The origin of the coordinate system is set to the mean initial position of atom~1. (b)~The repulsive energy surface $U_{\mathrm{rep}}$ (red) and adjacent surface $U_{\mathrm{adj}}$ (green) of the trimer subunit (atom~2, 3 and 4) near the CI. (c) and (d) Forces on atom~3 (solid lines) and atom~4 (dashed lines), for the repulsive surface [red, (c)] and adjacent surface [green, (d)]. The insets show atomic positions and the excitation distribution $c_n$ of exciton states and forces for the values $\Delta x_{23}/a_2 = 2.5, 0.46$, marked as gray, dashed vertical lines, where $\Delta x_{23}$ denotes the distance between atom~2 and the vertical chain. The parameter $b$ controls the degree of symmetry of the trimer, where $b=1$ corresponds to an isosceles trimer configuration. The gray, dotted vertical line 
marks the configuration with $\Delta x_{23}/a_2 = \sqrt{3}/2$, where for $b=1$ the CI is located.
\label{system_sketch_4}}
\end{figure}

We use Rydberg states with principal quantum number $\nu=44$, leading to a transition dipole moment of $\mu=1000$~atomic units. For simplicity we set $C_6=0$ in this section. The atomic configuration is described by the distance between the atoms on the horizontal direction, $a_1\equiv R^{(0)}_{12}$, the distance between the atoms of the vertical dimer, $a_2\equiv R^{(0)}_{34}$, and the horizontal distance between atom~2 and the $x$ position of the vertical dimer, denoted by $d$. The horizontal offset of the vertical dimer is then given by $\Delta x_{\mathrm{offset}} \equiv a_{1}+d$. A sketch of the atomic configuration is shown in \fref{system_sketch_4}(a). Specifically, we set the parameters to $(a_1, a_2,d) = (2.16,5.25,8.5)\mathrm{\ \mu m}$. The width of the Gaussian nuclear wave function is set for each atom to $\sigma_{0}=0.5\ \mathrm{\mu m}$.

Although, we position atoms~(3,4) such that their mean $y$ position is mirror symmetric to the $x$ axis, single realizations of the atomic configuration can still be asymmetric due to the Wigner distribution, such that $|y_3/y_4| \neq 1$. It is useful to introduce the asymmetry parameter\nomenclature[B]{$b$}{asymmetry parameter, quantifies the asymmetry of single trajectories relative to the $x$ axis},
\begin{equation}
 b := 2\dfrac{|y_3/y_4|}{|y_3/y_4|+1}, \quad b \in [0,2],
\label{eq:planar_aggr_nonad_dyn:def_asymmetry_param}
\end{equation}
to quantify the asymmetry for each atomic configuration and subsequently for each trajectory. We say that trajectories with $b \approx 1$ are symmetric and for $b\lessgtr 1$ asymmetric. We will later see that this 	distinctive feature of trajectories is the main reason for the occurrence of two different dynamics.
The bars in \frefp{system_sketch_4}(a) visualize the excitation amplitude of the exciton on the repulsive BO~surface initially. For each atom the length of the bar represents the amplitude of the diabatic coefficient, $c_n=\braket{\pi_n|\varphi_\mathrm{rep}(\vc{R}_0)}$, 
where the sign ``+'' is chosen for positive and ``-'' for negative values. As one can see, the single $p$ excitation is initially localized on atoms~(1,2). On the BO~surface $k$, the force on atom $n$ is given by $\bv{F}_{nk}=-\nabla_{\bv{R}_n} U_k(\vc{R})$.
Due to the initial repulsive force $F_{n,\mathrm{rep}}$, as indicated by the blue arrows, atom~2 moves and eventually reaches the position $x_\mathrm{CI}$, where atoms~(2--4) form a triangular subunit corresponding to the \emph{ring trimer} studied in \cite{wuester:CI}.
The CI of the trimer is realized for $b=1$, $d=\sqrt{3}/2 a_2$ in \fref{system_sketch_4}(a), where the three atoms form an equilateral triangle. To illustrate the CI, we show in \fref{system_sketch_4}(b) the two intersecting energy surfaces as a function of two selected atomic position variables. The upper surface~(shaded in red) will be hereafter referred to as the repulsive surface $U_{\rep}$, with corresponding exciton state $\ket{\varphi_{\rep}}$, as it always entails repulsive interactions of nearby atoms. The lower surface at the intersection~(shaded in green) will be referred to as the adjacent surface $U_{\adj}$, with corresponding exciton state $\ket{\sub{\varphi}{adj}}$. Further surfaces are not shown and play no significant role.
 
We will now systematically construct and interpret the atomic motion triggered by the initial excitation, firstly by analyzing typical trajectories and their energy spectra, then by investigating the atomic densities of the repulsive and adjacent adiabatic surface, which finally will enable us to understand the full evolution of the atomic densities. We consider its evolution in time, spatially resolved, in terms of population of adiabatic surfaces and regarding the purity of the state.

\subsubsection{Evolution and energy spectra of typical trajectories}
\label{sec:doubledimer:typicaltrajs}
The quantum-classical FSSH method is based on the propagation of classical trajectories. Studying the characteristics of single trajectories is an essential step to characterize the dynamics of the wave packet. Thus, FSSH as an approximate scheme is not only a necessary tool to determine the system's dynamics at all but moreover significantly aids our physical understanding.

\begin{figure}[!t]
\centering
\includegraphics{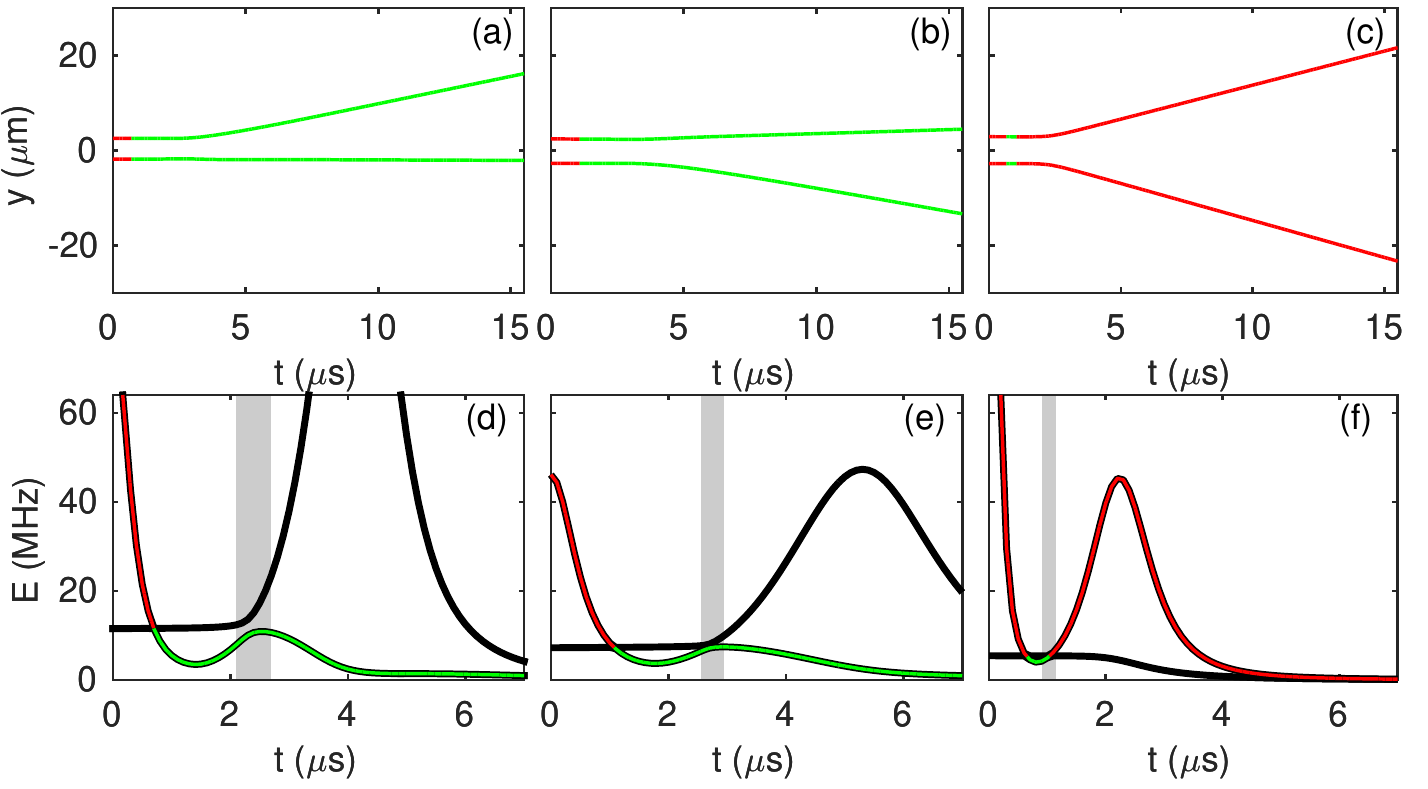}
\caption{\label{fig:Nges4_single_trajectories}
Selected single trajectories $\vc R(t)$ of atoms~(3,4) in~(a)--(c) with corresponding time resolved energy spectra~(black lines) and potential energy $U_{\zeta(t)}(\vc R(t))$~(colored line) in~(d)--(f).
For both, trajectories and potential energy, evolution on the repulsive surface is marked with red lines and evolution on the adjacent surface with green lines.
(a), (b) Two characteristic trajectories that stay on the adjacent surface. The two trajectories are almost mirror symmetric around the $x$ axis.
(c) A characteristic trajectory jumping back to the repulsive surface after passing the vicinity of the CI.
The gray area marks the vicinity of the CI between the adjacent and the repulsive surfaces, governing the dynamics. 
The earlier crossings are less relevant for our dynamics here and trivial in the sense that all trajectories undergo a transition there. We explain this in the main text and study this type of crossings in \aref{part:app::calcs_planar_aggr__trivial_crossings}.
}
\end{figure}

The initially localized excitation on atoms~(1,2) yields strong repulsive forces between them which ultimately provokes atom~2 to move towards the vertical dimer. Initially, the propagation occurs for all trajectories adiabatically on the repulsive surface, for selected single trajectories in \frefplural{fig:Nges4_single_trajectories}{a}{--}{c} and corresponding energy spectra in \frefplural{fig:Nges4_single_trajectories}{d}{--}{f} we highlight this surface as red lines. Before atoms~2, 3, and 4 form a trimer subunit, it is apparent from all energy spectra in \frefplural{fig:Nges4_single_trajectories}{d}{--}{f}, that a first transition from the repulsive to the adjacent surface~(green lines) occurs shortly before one microsecond. Physically, the aggregate continues propagation with the exciton state of the horizontal dimer, however, its BO~surface has a crossover with the repulsive BO~surface of the vertical dimer, leading to a transition in the energy spectra. At the time of the transition, both dimers are weakly coupled, which is the reason why all trajectories change the global BO~surface. For further details concerning this trivial transition, see \aref{part:app::calcs_planar_aggr__trivial_crossings}.

The physical situation remaining unchanged, atom~1 separates itself from the remaining atoms and since atom~2 gets closer to the vertical dimer, the exciton is consequently transformed from a dimer to a trimer state, thereby transferring excitation to the vertical dimer.

Already shortly before atom~2 reaches the vicinity of the CI configuration [this position of atom~2 is marked as $x_{\mathrm{CI}}$ in \fref{system_sketch_4}(a)], excitation is transferred to atoms~(3,4). Interlinked with it are forces on the vertical dimer, inducing motion of its atoms. A characteristic feature is the sudden increase of forces in the vicinity of the CI configuration, which is almost instantaneous~[apparent from the forces shortly before atom~2 reaches the position marked as gray, dotted line in \fref{system_sketch_4}(d)]. In the energy spectra of selected trajectories in \frefplural{fig:Nges4_single_trajectories}{d}{--}{f}, the vicinity of the CI is marked as gray area. After the trajectories traverse this region, they start to evolve differently. The degree of mirror asymmetry along the horizontal axis, which is quantified by the parameter $b$ defined in \eref{eq:planar_aggr_nonad_dyn:def_asymmetry_param}, distinguishes the further dynamics of trajectories.
We consider for a moment the configuration where atom~2 is at the position marked with $x_{\mathrm{CI}}$, such that for a perfectly symmetric trajectory ($b=1$), the CI configuration is realized. With increasing asymmetry (which is quantified by increasing deviations of the parameter $b$ from one), the deviation from the equilateral triangle configuration increases and with it the energy gap between repulsive and adjacent surface, resulting in an avoided crossing instead of a conical intersection. For further details about the relation of the energy gap to the asymmetry, see \aref{part:app::calcs_planar_aggr__rel_energy_gap_CI}. The two trajectories in \frefplural{fig:Nges4_single_trajectories}{a}{--}{b} with their corresponding energy spectra in \frefplural{fig:Nges4_single_trajectories}{d}{--}{e}, respectively, are representatives for asymmetric trajectories. A symmetric trajectory is shown in \fref{fig:Nges4_single_trajectories}(a), with corresponding energy spectrum in \fref{fig:Nges4_single_trajectories}(f). 
The greater energy gap makes it more likely for asymmetric trajectories to stay on the adjacent surface whereas symmetric trajectories have a higher chance to make a transition back to the repulsive BO~surface.

Staying on the adjacent surface, the excitation on atom~2 is quickly and entirely transferred to the two vertical atoms after the nonadiabatic region is traversed. Forces act only on atoms which share the excitation, due to dominant resonant interactions~(we actually neglect vdW~interactions\index{dipole-dipole interaction!van-der-Waals} in this system for simplicity). This implies that during the process of transferring the excitation from atom~2 to the vertical dimer, atoms~(3,4) are repelled from atom~2 until no excitation is left on the latter. 
Since asymmetric trajectories tend to stay on the adjacent surface, the induced forces on both vertical atoms are asymmetric as well, leading to a repulsion of only one atom. Two examples of this are the trajectories in \frefplural{fig:Nges4_single_trajectories}{a}{--}{b}. Both trajectories are almost reflections of each other by mirroring along the $x$ axis. Surprisingly, the adjacent surface provides repulsion of the vertical atom with larger distance to atom~2\footnote{For propagation along the adjacent surface, shortly after atom~3 traversed the nonadiabatic region, all excitation is transferred to the vertical dimer with most excitation residing on the atom farther apart from atom~3, as depicted in \fref{system_sketch_4}(d), which consequently is stronger repelled}. 

A transition back to the repulsive surface in the vicinity of the CI leads to a very different dynamical scenario.
Then, atom~2 does not suddenly transfer all its excitation to the vertical dimer. Instead, all three atoms share the excitation for a longer time and during this period atoms~(3,4) strongly repel from atom~2 with the magnitude of the forces a factor of ten larger compared to forces on the adjacent surface~[compare \fref{system_sketch_4}(c) with \fref{system_sketch_4}(d)]. The interactions increase until all three atoms form a linear trimer~[as evident from the peak of the red energy surface around $2.5\ \mu s$ in \fref{system_sketch_4}(f)], which is due to decreasing interatomic distances between atoms~2 and 3, and atoms~2 and 4, respectively. Since only symmetric trajectories with high probability access the repulsive surface get, the strength of repulsion is almost equal on both atoms of the vertical dimer.
In \fref{fig:Nges4_single_trajectories}(c) we show as an example a trajectory with the characteristics of the repulsive BO~surface.

The energy spectra such as shown in \frefplural{fig:Nges4_single_trajectories}{d}{--}{f} should experimentally be accessible with micro-wave spectroscopy of Rydberg aggregates, similar to \rref{park:dipdipbroadening,celistrino_teixeira:microwavespec_motion}.
\subsubsection{Exciton splitting}
\label{sec:doubledimer:exciton_splitting}
The discussion of single trajectories showed us that the CI divides classical trajectories into two classes, which indicates that
the total wave packet is also drastically affected by it.
In this section we will present and discuss the final results of the four atom T-shape aggregate, which we obtained by a numerical simulation with the FSSH method. The discussion uses typical quantum observables but resorts to the knowledge we gained from the investigation of single trajectories.
To study the spatial dynamics of the wave packet it is suitable to calculate time-resolved atomic densities, which sum the spatial probability densities of each individual atom to plot them combined over the same space coordinates. Very helpful are partial atomic densities, which are fractions of the total atomic density which evolve on single BO~surfaces only. Their density profiles are understood well with the help of single trajectories. Additionally, to investigate the dynamics, we use adiabatic populations, which indicate the participating BO~surfaces and quantify their contribution to the dynamics. Finally we also calculate the purity\index{purity} of the electronic density matrix to obtain informations about the system being in a pure or mixed state.

In the following we formally introduce atomic densities. The total atomic density is defined by
\begin{equation}
 n(\vc{r},t) :=  \dfrac{1}{N}\sum_{j=1}^{N}\int \dscalar^{N-1}\vc{R}_{\{j\}}|\Psi(\vc{R},t)|^2\bigr|_{\vc{R}_j = \vc{r}},
\label{eq:planar_aggr_nonad_dyn:def_atomic_density_quant}
\end{equation}
whereas a partial atomic density for a BO~surface corresponding to an exciton $\varphi(\vc{R})$, is given by
\begin{equation}
 n_{\varphi}(\vc{r},t) := \dfrac{1}{N}\sum_{j=1}^{N}\int \dscalar^{N-1}\vc{R}_{\{j\}}|\braket{\varphi(\vc{R})|\Psi(\vc{R},t)}|^2\bigr|_{\vc{R}_j = \vc{r}}.
\label{eq:planar_aggr_nonad_dyn:def_partial_atomic_density_quant}
\end{equation}
The integration $\int \dscalar^{N-1}\vc{R}_{\{j\}}$ is over all but the coordinates of the $j$th atom.
These definitions are based on using the exact quantum method with a full propagation of the Schrödinger equation. However, since we use the quantum-classical FSSH method, the definitions have to be readjusted. The following definitions are adapted to a trajectory based quantum-classical method. There, the spatial coordinates need to be defined on a grid, which we set in the following way:
\begin{align}
\xi_{k}^{\mathrm{grid}} := \xi_{1}^{\mathrm{grid}} + (k-1)\Delta^{\mathrm{\xi-grid}}, \quad \xi \in \{x,y,z\}, 
\label{eq:planar_aggr_nonad_dyn:def_grid_atomic_dens_semiclassical}
\end{align}
with grid spacing $\Delta^{\mathrm{\xi-grid}}$ in $\xi$ direction. The next step is to calculate the histogram of how frequently atoms visit a certain spatial volume. This procedure is based on classical trajectories, where each atom has a completely determined position at every time step.
Since the T-shape aggregates in this chapter are restricted to the $x$-$y$ plane and the atoms are free to move in a single direction only, we separately define two histograms, for the horizontal Rydberg chain
\begin{equation}
 n^{\mathrm{x-grid}}_{i}(t) := \dfrac{1}{N_{\mathrm{traj}}N_x} \sum_{k=1}^{N_{\mathrm{traj}}}\sum_{j=1}^{N_x}\Theta\left(\frac{\Delta^{\mathrm{x-grid}}}{2} - |x_{j}^{(k)}(t) -  x_{i}^{\mathrm{grid}}|\right),
\label{eq:planar_aggr_nonad_dyn:def_x_atomic_dens_hist_semiclassical}
\end{equation}
and for the vertical Rydberg chain
\begin{equation}
 n^{\mathrm{y-grid}}_{i}(t) := \dfrac{1}{N_{\mathrm{traj}}(N-N_x)} \sum_{k=1}^{N_{\mathrm{traj}}}\sum_{j=(N_x+1)}^{N}\Theta\left(\frac{\Delta^{\mathrm{y-grid}}}{2} - |y_{j}^{(k)}(t) -  y_{i}^{\mathrm{grid}}|\right),
\label{eq:planar_aggr_nonad_dyn:def_y_atomic_dens_hist_semiclassical}
\end{equation}
where $x_{j}^{(k)}(t)$ ($y_{j}^{(k)}(t)$) is the horizontal (vertical) coordinate of the $j$th atom for the $k$th trajectory, $N_{\mathrm{traj}}$ is the total number of propagated trajectories and $\Theta$ denotes the Heaviside function. Turning the histograms into densities, equivalent to the definition in \eref{eq:planar_aggr_nonad_dyn:def_partial_atomic_density_quant}, is possible by assigning a density value to all coordinate values between the grid points, which is realized by
\begin{equation}
 n(\xi,t) :=\dfrac{1}{\Delta^{\mathrm{\xi-grid}}} \sum_{i=1}^{N_{\mathrm{grid}}}\Theta\left(\frac{\Delta^{\mathrm{\xi-grid}}}{2} - |\xi -  \xi_{i}^{\mathrm{grid}}|\right)n^{\mathrm{\xi-grid}}_{i}(t), \quad \xi \in \{x,y\},
\label{eq:planar_aggr_nonad_dyn:def_atomic_dens_final_semiclassical}
\end{equation}
where $N_{\mathrm{grid}}$ is the number of grid points. Defining the partial atomic densities equivalently to \eref{eq:planar_aggr_nonad_dyn:def_partial_atomic_density_quant} for the quantum-classical method requires to condition the histograms in \eref{eq:planar_aggr_nonad_dyn:def_x_atomic_dens_hist_semiclassical} and \eref{eq:planar_aggr_nonad_dyn:def_y_atomic_dens_hist_semiclassical} such that they only measure when trajectories evolve on the BO~surface of interest. 

\begin{figure}[!t]
\centering
\includegraphics{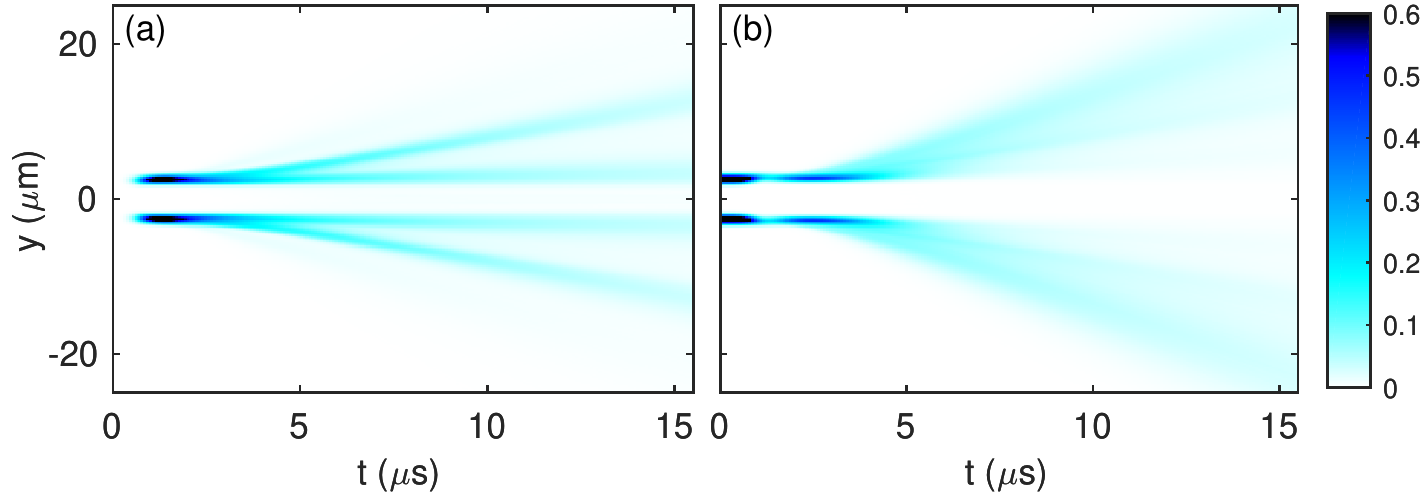}
\caption{\label{fig:Nges4_partial_densities_chain2}
Partial atomic densities of atoms~(3,4), for (a) evolution on the adjacent surface and (b) evolution on the repulsive surface. The computation is according to \eref{eq:planar_aggr_nonad_dyn:def_atomic_dens_final_semiclassical} but with frequency measures restricted to single BO~surfaces. We propagated $N_{\mathrm{traj}}=10^6$ trajectories.
Each density plot is renormalized to the global maximum value of both
densities. Furthermore we plot all density values between $0.6$ and $1$ with the same color, to highlight details at lower density values.
}
\end{figure}

The signature of the CI is very well visible in the partial atomic densities of the adjacent and repulsive BO~surface for atoms~(3,4), shown in \fref{fig:Nges4_partial_densities_chain2}. After the wave packet initially populates the repulsive surface entirely, at around one microsecond the density decreases slightly thereon due to the trivial crossing, which we discussed earlier. The entire wave packet undergoes a transition to the globally adjacent surface, apparent due to maximum density values for the adjacent surface around $1$ to $1.5$ microseconds in \fref{fig:Nges4_partial_densities_chain2}(a).
Subsequently, the CI transition follows at around $4$--$5\ \mu$s, where the transition time is broadly distributed. This leads to a splitting of the wave packet in two almost equal parts. Each part evolves on one of the two participating BO~surfaces. 
The investigation of single trajectories revealed already, that propagation along different BO~surfaces features different dynamics. 

On the repulsive surface, atoms~(3,4) experience a strong repulsion, almost equal for both atoms. This explains the strong repulsion in the partial density of the repulsive BO~surface in \fref{fig:Nges4_partial_densities_chain2}(b). Another feature is the large broadening of the atomic position distribution after the CI transition. As already mentioned, the time of a return transition from the adjacent to the repulsive surface is already broadly distributed, such that the position of atom~2 varies strongly when the propagation starts again on the repulsive surface. Since the atoms~(3,4) repel mainly from atom~2, the forces induced on the vertical atoms vary strongly, leading to a broad distribution of velocities and consequently broad position distributions.


Initial configurations with large upwards or downwards shifts of the vertical dimer away from its symmetric position relative to the horizontal axis are highly asymmetric and the resulting trajectories consequently propagate along the adjacent surface. Due to the high asymmetry of the trajectories, the adjacent surface repels effectively only the atom of the vertical dimer, which is farther apart from the horizontal axis. Since this can be either atom~3 or atom~4, two different motions of the vertical dimer can be observed which explains the branching of the atoms' position distribution into two parts, due to the fact that each atom can either rest or be repelled.
The result is a total four-fold branching in the partial density of the adjacent surface. Another characteristic of the surface are the sharp profiles of each branch, compared to the very broad atomic distributions on the repulsive surface. Evolution on the adjacent surface occurs when no return transition to the repulsive surface appears. Thus, a broadening due to different transition times can be excluded. Moreover we found for trajectories propagating along the adjacent surface, that the forces act significantly only for a short time on the vertical dimer, when atom~2 traverses the vicinity of the CI configuration. At this point, the distances between atom~2 and the other two atoms are still quite large, which besides the weaker induced forces on the vertical dimer also explains a smaller relative variation of them, compared with a propagation along the repulsive surface.
This eventually yields rather localized position distributions, as can be seen in \fref{fig:Nges4_partial_densities_chain2}(a).
%

\begin{figure}[!t]
\centering
\includegraphics{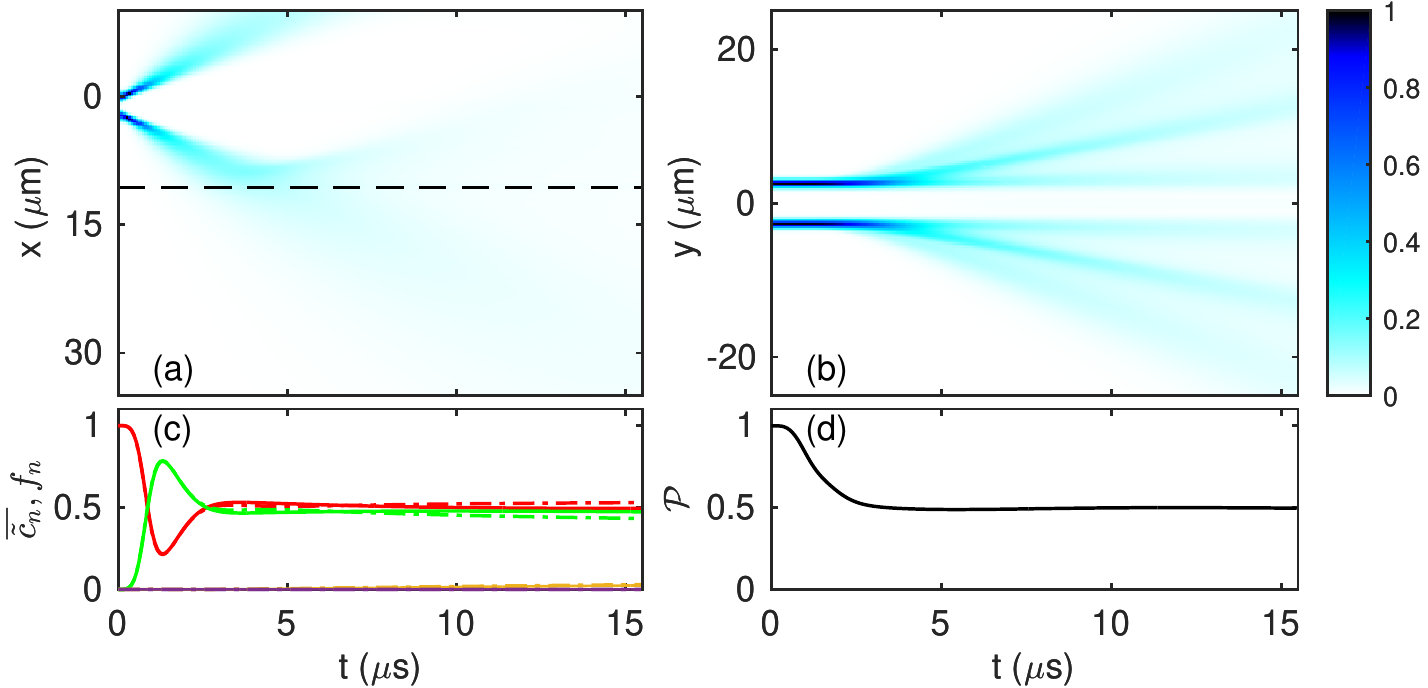}
\caption{\label{fig:Nges4_total_densities_pops_purity}
Total atomic densities of the aggregate together with measures to quantify nonadiabatic transitions and entanglement. (a) Atomic density of the horizontal dimer and (b) of the vertical dimer. The maximum value of the data in (a) and (b) is individually set to one. (c) Adiabatic populations, $\sigma_{n,n}(t)$, of the BO~surfaces, which can be calculated according to \eref{eq:planar_aggr_nonad_dyn:traj_average_binary_diab_coeffs}. The participating surfaces are the repulsive (red) and adjacent one (green). All other BO~surfaces are negligible for the dynamics. A self consistency check for the nonadiabatic transitions to be correctly treated is to compare the adiabatic populations to the trajectory fractions, defined in \eref{eq:quant_dyn_methods_tully:fractions}. The fractions~(dotted lines) are in good agreement with the populations.
(d) Purity\index{purity}, $\mathcal{P}(t)$, defined in \eref{eq:planar_aggr_nonad_dyn:def_purity}, which measures the entanglement between nuclear and electronic degrees of freedom.
To obtain the results we propagated $10^6$ trajectories with FSSH.
%
}
\end{figure}

With the interpretation of the partial atomic densities at hand, we finally can understand the total atomic densities.
We present in \fref{fig:Nges4_total_densities_pops_purity}(a) the total density of the horizontal and in \fref{fig:Nges4_total_densities_pops_purity}(b) the total density of the vertical dimer, respectively.
The density of the horizontal dimer shows that atom~2 experiences an increasing broadening of its spatial distribution, with dynamics ranging from transmission to reflection shortly before reaching the horizontal position of the vertical dimer. The reflection is due to the part of the wave packet propagating along the repulsive surface. The potential energy reaches a maximum value for the linear trimer configuration, apparent from the energy spectra in \fref{fig:Nges4_single_trajectories}(f). If the kinetic energy of atom~2 is smaller than this barrier, the atom is reflected. The transmission is due to propagation along the adjacent surface, where the atom does not experience this barrier and it thus can continue its motion without inversion of direction.
The atom's wave packet of the vertical dimer fans out after around $2.8 \mu $s due to the CI transition, such that simultaneous evolution on two BO~surfaces occurs. This is indicated also by the adiabatic populations of the repulsive and adjacent surfaces in \fref{fig:Nges4_total_densities_pops_purity}(c) which eventually are half populated. The earlier drop of the adiabatic population of the repulsive surface around $1\mu s$ is due to the trivial crossing discussed earlier.
The CI eventually splits the nuclear wave packet and the electronic state simultaneously in two equal parts, which we deliberately designed by choosing an appropriate initial configuration for the aggregate. The initial total wave function is given by 
$\ket{\Psi_{0}(\vc{R})} = \chi_{0}(\vc{R})\ket{\psi_{\mathrm{el,0}}}$, where $\chi_{0}(\vc{R})$ is the nuclear wave function and $\ket{\psi_{\mathrm{el,0}}}$ the electronic wave function, which is approximately the localized repulsive exciton, $\ket{\psi_{\mathrm{el,0}}} \approx \ket{\varphi_{\mathrm{rep}}(\vc{R}_0)}$. After the CI transition, the total wave function is transformed to $\ket{\Psi_{\mathrm{fin}}(\vc{R})} = \chi_{\mathrm{rep}}(\vc{R})\ket{\varphi_{\mathrm{rep}}(\vc{R})} + \chi_{\mathrm{adj}}(\vc{R})\ket{\varphi_{\mathrm{adj}}(\vc{R})}$, which is a coherent splitting of the wave packet where the simultaneous presence of two excitons allow for a superposition of different nuclear wave functions.
 This final state of the aggregate indicates entanglement\index{entanglement of formation} between the atomic configuration and the electronic state. To quantify this, we measure the purity of the electronic density matrix, which is defined by
\begin{equation}
\mathcal{P}(t) := \mathrm{tr}(\hat{\sigma}^2(t)),
\label{eq:planar_aggr_nonad_dyn:def_purity}
\end{equation}
\nomenclature[B]{$\mathcal{P}$}{purity}\index{purity}where 
\begin{equation}
 \hat{\sigma}(t):=\int \dscalar^{N} \vc{R} \ket{\Psi(\vc{R},t)}\bra{\Psi(\vc{R},t)}
\label{eq:planar_aggr_nonad_dyn:def_elec_density_matrix}
\end{equation}
\nomenclature[B]{$\hat{\sigma}$}{electronic density matrix}
is the electronic density matrix. Using FSSH as dynamical method, the calculation of the density matrix is best performed in its diabatic representation, 
\begin{equation}
 \hat{\sigma}(t) = \sum_{n,m=1}^{N}\sigma_{n,m}(t)\ket{\pi_n}\bra{\pi_m},
\label{eq:planar_aggr_nonad_dyn:elec_dens_matrix_diabatic_repr}
\end{equation}
where the coefficients are trajectory averages over binary products of diabatic coefficients\cite{wuester:cradle,moebius:cradle,leonhardt:switch},
\begin{equation}
 \sigma_{n,m}(t) := \overline{c_{n}(t)c_{m}^{*}(t)}.
\label{eq:planar_aggr_nonad_dyn:traj_average_binary_diab_coeffs}
\end{equation}
The purity\index{purity} drops from one to one half, which indicates a transition from a pure to a mixed state. The mixed state is the result of entanglement\index{entanglement of formation} between nuclear and electronic degrees of freedom. 

\subsection{The seven atom T-shape aggregate}
\label{sec:nges8}
\begin{figure}[!t]
\centering
\includegraphics{{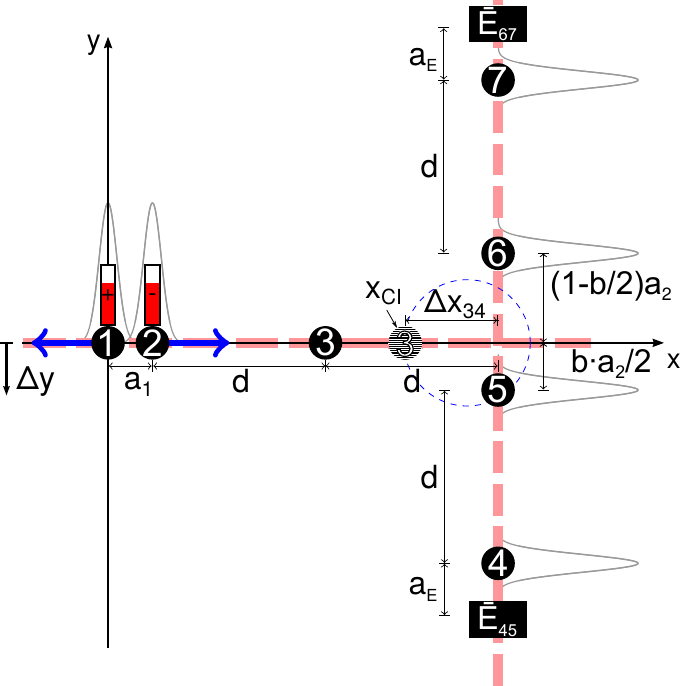}} 
\caption{\label{system_sketch_7}
Sketch of a seven atom system combining features of adiabatic entanglement transport and CI dynamics. 
Three atoms are placed on the horizontal and four on the vertical direction. Atoms~(1,2) are prepared in a repulsive exciton initially, resulting in the excitation distribution shown by red bars, as described in the caption of \fref{system_sketch_4}. Atom~3 is in the vicinity of a CI at position $x_{\mathrm{CI}}$. 
Around that configuration, the excitation can either almost exclusively reside on atoms~3, 5 and 6, when the distance between atoms~(5,6) is adjusted sufficiently small or resides on atoms~(2,3) for large distances between atoms~(5,6). Atom~(5,6) are accelerated at this time on the two different surfaces, already described in the four atom section. We analyze whether or not the atom pair~(5,4) or (6,7) finally build a combined exciton-motion pulse. This is quantified by the binary entanglement $E_{45}$ and $E_{67}$ at the moment that atom~4 and atom~7 reaches the location indicated by a black box, termed "entanglement readout", respectively. We finally use the indicated displacement ${\Delta}y$ of the horizontal chain to steer entanglement transport upwards or downwards.
}
\end{figure}

For one-dimensional Rydberg chains with a dislocation at one end, the initiation of exciton pulses was demonstrated\cite{wuester:cradle, moebius:cradle}, where diatomic proximity combined with excitation is transported. This mechanism of transporting energy and momentum can be regarded as a quantum analogue of the classical Newton's cradle. However, the behavior of exciton pulses in higher dimensional arrangements is a priori unclear due to new features such as CIs.
The four atom T-shape aggregate already indicated that the CI drastically affects the exciton dynamics. This minimal T-shape aggregate revealed the mechanism of the CI, but investigating transport features on the vertical direction was not possible due to too few atoms.
In this section we change this, with specific interest in the possibility of continued exciton pulse propagation after redirection on an orthogonal direction. T-shape aggregates are convenient, since the one-dimensional restriction of the atomic motion technically simplifies the implementation. Furthermore, two-dimensional effects such as the CI appear only in a small volume, namely when a trimer sub-unit is formed between atoms from the horizontal and vertical chain. 
To investigate transport features on the vertical chain we extend it to four atoms. They are positioned such that transport in upwards and downwards direction can occur. Moreover the horizontal chain is extend to three atoms to allow transfer of excitation and momentum already on the horizontal chain.
Thus, altogether, we investigate a seven atom T-shape aggregate as sketched in~\fref{system_sketch_7}. The interatomic distance between atoms~(1,2), denoted with $a_1$, is adjusted to be the smallest, which allows for initiating the exciton pulse on both atoms by populating the repulsive BO~surface. Propagation along this surface ensures exciton pulse propagation towards the vertical chain, transferring 
excitation and momentum first to atom~3 which eventually reaches the CI position [marked with $x_{\mathrm{CI}}$]. Atoms~3, 5, and 6 can form a trimer subunit for a sufficiently small spacing between atoms~(5,6),
so that repulsive and adjacent surface get closely spaced, which ultimately splits the exciton pulse as for the four atom aggregate.

We specifically use Rydberg states with principal quantum number $\nu =80$, which corresponds to a scaled radial dipole moment of $\mu = 3371$ a.u.. The geometric parameters of the configuration are set to $a_1=6\ \mu$m, $a_2 = 9.5\ \mu$m, $d= 22\ \mu$m. The horizontal offset of the vertical chain is $\Delta x_{\mathrm{offset}} = a_1 + 2d$.
The width of each atom's nuclear wave function is set to $\sigma_0 = 0.5\ \mu$m initially. The remaining parameter is $\Delta y$ which adjusts the vertical shift of the horizontal chain. 
In the following we explicitly consider vdW~interactions and set $C_{6} = -7.6\times 10^{20}$ atomic units. In this section we set $\Delta y = 0$, to realize a symmetric T-shape configuration. Note that we define $\Delta y$ to be positive for shifts in downwards $y$ direction. 
We will discuss this vertical shift as control parameter for the dynamics on the vertical chain in \sref{sec:exciton_switch}.
 \subsubsection{Nondirectional transport}
\label{sec:nges8_nondirectional}
When the exciton pulse encounters a CI, the total wave function is coherently split, as for the four atom aggregate, discussed in \sref{sec:nges4}. 
The investigation of single trajectories yields an intuitive understanding of the dynamics. An exciton pulse is initiated on atoms~(1,2) and transfers momentum and excitation to atom~3 which hence approach the vicinity of the CI~[marked with $x_{\mathrm{CI}}$ in \fref{system_sketch_7}].
During the time of exciton pulse propagation along the horizontal chain, the only difference to the four atom aggregate is the appearance of three instead of one trivial crossings~(see discussion in \sref{sec:doubledimer:typicaltrajs} and \aref{part:app::calcs_planar_aggr__trivial_crossings}) before the configuration of the nontrivial CI is reached. The reason for the appearance of three trivial crossings is that the potential energy reaches a local minimum twice, which is energetically below the repulsive BO~surface of the vertical chain. This takes place when atom~2 and atom~3 carry maximum momentum, respectively. At the position where atoms~(2,3) are in closest proximity, the surface of propagation is above the most energetic surface of the vertical chain. Together this implies the appearance of three (avoided) crossings of the BO surface of propagation with the BO~surface of the vertical chain, before the vicinity of the nontrivial CI crossing is reached.

The nontrivial CI branches the wave packet to evolve on two BO~surfaces as we observed it already for the four atom aggregate. We start the discussion of the dynamics with an investigation of single trajectories.

\begin{figure}[!t]
\centering
\includegraphics{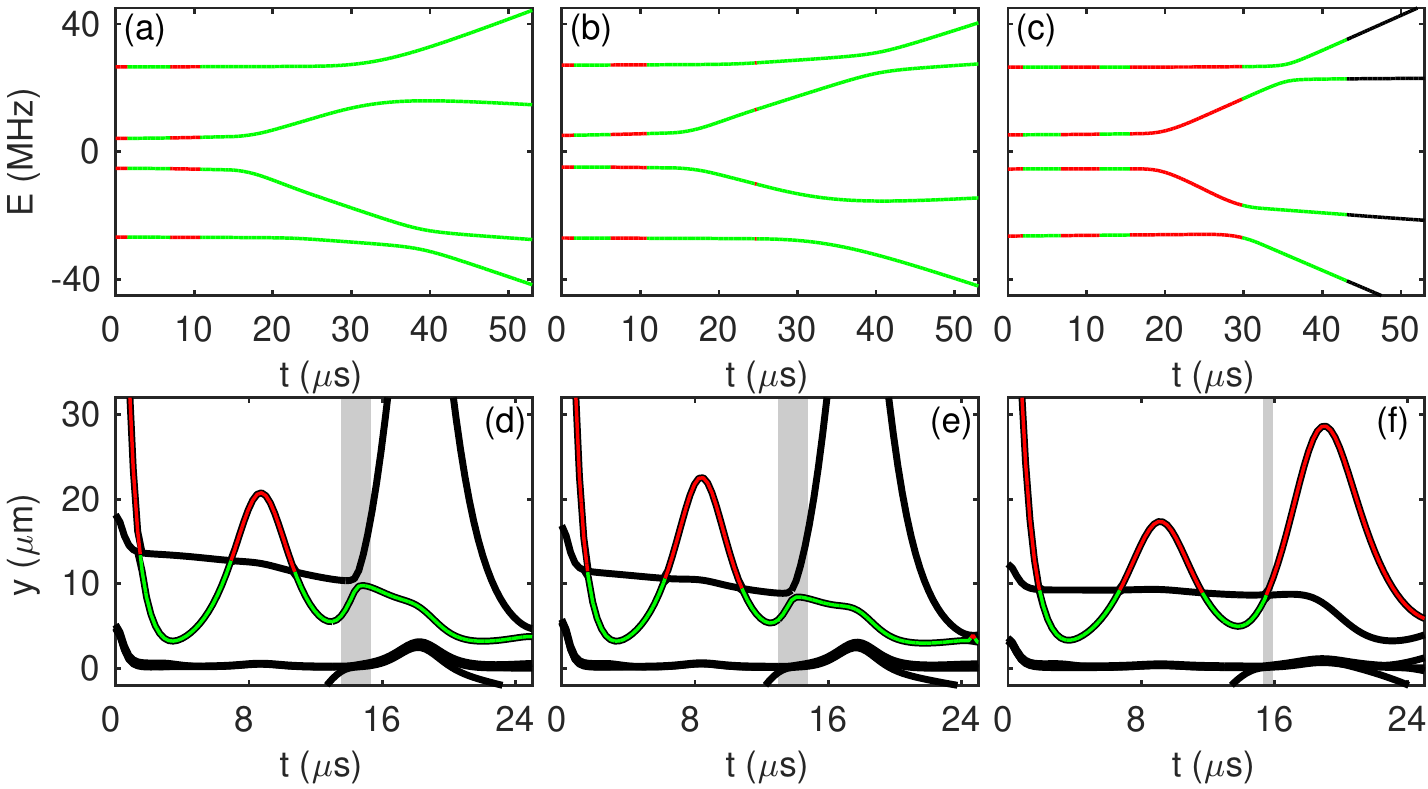}
\caption{\label{fig:Nges7_single_trajectories}
Atomic positions from single trajectories for atoms~(4--6) (a)--(c) with corresponding eigenenergies in (d)--(f). (a) and (b)~Two different trajectories ending up on the adjacent surface. (c)~A trajectory returning to the repulsive surface after traversing the vicinity of the CI. (d)--(f) Time-resolved energy spectra (black lines) and potential energy (colored line), for position trajectories above. For both, positions and potential energy, evolution on the repulsive surface is marked with red and evolution on the adjacent surface with green.
The trajectories shown in (c) evolve in the end on a third surface, shown as black lines. The gray area marks the CI and its vicinity. The earlier transitions between repulsive and adjacent surface are trivial, see the discussion in \sref{sec:doubledimer:typicaltrajs} and \aref{part:app::calcs_planar_aggr__trivial_crossings}
}
\end{figure}
Two examples of trajectories propagating along the adjacent surface after passing the vicinity of the CI are shown in \frefplural{fig:Nges7_single_trajectories}{a}{, }{b}, with corresponding energy spectra in \frefplural{fig:Nges7_single_trajectories}{d}{, }{e}, respectively. As for the double dimer aggregate, the adjacent surface is more likely to be populated by trajectories with larger shifts of the vertical chain away from its symmetric position relative to the horizontal chain. However, the repulsion of atoms~(5,6) is much less asymmetric than in the four atom aggregate, which is due to the addition of vdW~interactions. The off-resonant contribution to the interactions are important here to ensure that atoms without excitation do not ionize when motion decreases the interatomic spacings. We discuss the dynamics for the trajectory in \fref{fig:Nges7_single_trajectories}(a) for which atom~5 has a larger vertical distance to atom~3 than to atom~6.
The propagation along the adjacent surface transfers excitation from atom~3 to atoms~(5,6) very quickly, within the vicinity of the CI. As a feature of the adjacent surface, the vertical atom with the larger distance to the excitation inducing atom of the horizontal chain, shares more excitation than the remaining atom of the trimer subunit [as apparent from the excitation inset in \fref{system_sketch_4}(d)]]. For the trajectory under discussion, this is atom~5 and resonant interactions let the atom be repelled towards atom~4.
Different to the four atom aggregate, we observe also a repulsion of atom~6, which is due to vdW~forces.
In the four atom aggregate we were allowed to neglect vdW~forces and thus observed the repulsion of only one atom.
Atom~5 experiences still a stronger repulsion here and approaches atom~4 faster. Excitation then gets localized for a while on atoms~(5,4) and we expect that momentum and excitation are transferred together in the downwards direction, eventually carried by atom~4. However, this is not the case which we found out by a time-resolved tracking of both atomic motion and excitation transfer from which we subsequently generated a movie\cite{leonhardt:switch}~(not append to this thesis). The movie revealed that due to the occurrence of a second CI the excitation is swapped to the upper dimer pair.
%

A trajectory propagating along the repulsive surface is shown in \fref{fig:Nges7_single_trajectories}(c). When atom~3 reaches the position of the first nontrivial CI crossing, a return transition to the repulsive surface (red line) is performed, apparent from the energy spectrum in \fref{fig:Nges7_single_trajectories}(f). Different to the propagation along the adjacent surface, all the three atoms~3, 5, and 6 simultaneously share excitation after the transition back to the repulsive surface.
Since atoms~(5,6) reach the outer lying atoms~4 and 7 almost at the same time, the excitation gets \emph{delocalized} over all atoms on the vertical chain. However a small asymmetry of the spatial distribution of the vertical atoms localizes excitation on atoms~(5,4).
Atom~3 is reflected off the vertical chain and eventually moves back towards $x<0$. We generated for this trajectory also a movie which revealed that several further CIs occur and finally cause that most excitation reside on atoms~(2,3), hence on the horizontal chain.

To clarify the different distributions of excitation on the two surfaces, we calculate the partial excitation density,
\begin{equation}
 n^{\mathrm{exc}}_{\varphi}(\vc{r},t) := \dfrac{1}{N}\sum_{j=1}^{N}\int \dscalar^{N-1}\vc{R}_{\{j\}}|\braket{\varphi(\vc{R})|\Psi(\vc{R},t)}|^2   
|\braket{\varphi(\vc{R})|\pi_{j}}|^2
\bigr|_{\vc{R}_j = \vc{r}},
\label{eq:planar_aggr_nonad_dyn:def_partial_excitation_density_quant}
\end{equation}
which describes the spatial distribution of the \lstate{p} excitation in state $\ket{\varphi(\vc{R})}$, weighted with the probability density that the aggregate is in the stated exciton. Note that we actually have to adjust \eref{eq:planar_aggr_nonad_dyn:def_partial_excitation_density_quant} for the use with the FSSH method, as it was necessary for the atomic densities.

\begin{figure}[!t]
\centering
\includegraphics{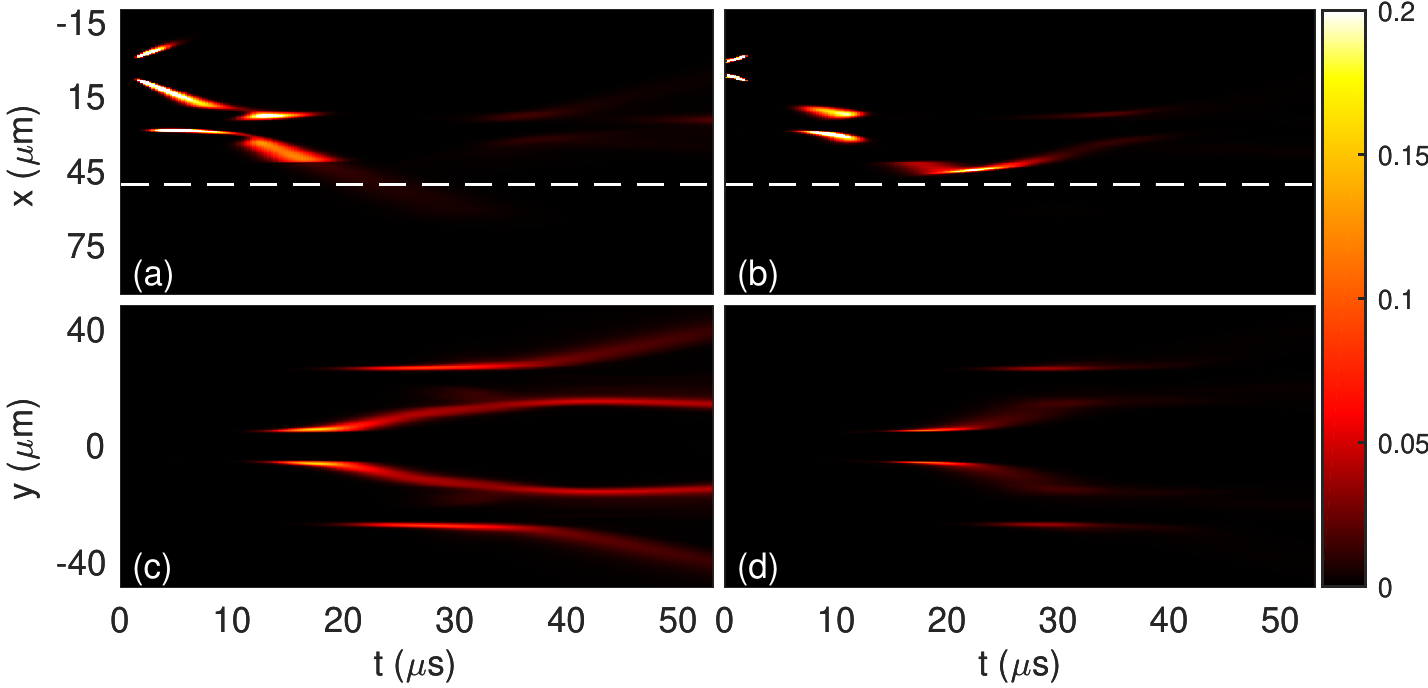}
\caption{\label{fig:N7_part_exc_dens_symmetric_case_both_chains}
Partial excitation densities for atoms~(1--3) in (a), (b) and atoms~(4--7) in (c), (d), according to \eref{eq:planar_aggr_nonad_dyn:def_partial_excitation_density_quant}.
(a) and (c) Spatial distribution of excitation for propagation along the adjacent surface with exciton $\varphi(\vc{R}) =\varphi_{\mathrm{adj}}(\vc{R})$.
(b) and (d) Spatial distribution of excitation for propagation along the repulsive surface with exciton $\varphi(\vc{R}) =\varphi_{\mathrm{rep}}(\vc{R})$. All densities all scaled to have the same global maximum. Initially the excitation is localized on atoms~(1,2) and resides exclusively on the repulsive BO~surface, however, later on, it distributes among two BO~surfaces and more than two atoms, such that the peak values decrease drastically. This makes the implementation of a colormap cutoff necessary, to highlight distribution of excitation at later times. Specifically, we set all density values above $0.2$ to the same color. The dashed white lines in (a), (b) mark the horizontal position of atoms~(4-7).
}
\end{figure}

The excitation distribution of the entire wave paket is shown in \fref{fig:N7_part_exc_dens_symmetric_case_both_chains}, where \linebreak[4]\frefplural{fig:N7_part_exc_dens_symmetric_case_both_chains}{a}{, }{b} 
show results for atoms~(1--3) and \frefplural{fig:N7_part_exc_dens_symmetric_case_both_chains}{c}{, }{d} for atoms~(4--7). The description we gave with single trajectories, of how each surface distributes excitation, is confirmed to hold true for the wave packet as well. 

\begin{figure}[!t]
\centering
\includegraphics{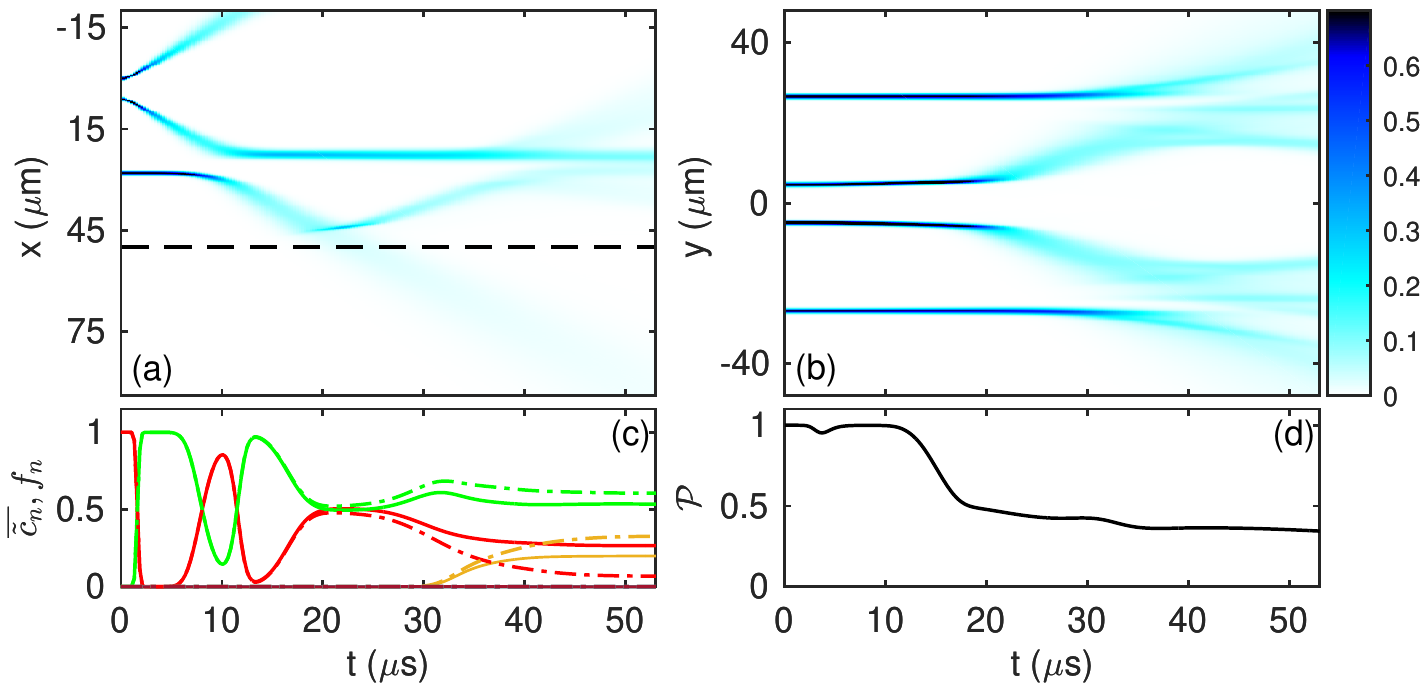}
\caption{\label{fig:Nges7_total_densities_pops}
Dynamics of atomic motion and populations. (a) Total atomic density of atoms~(1--3).
(b) Total atomic density of atoms~(4--7).
(c) Adiabatic populations (solid lines), $\sigma_{n,n}$, according to \eref{eq:planar_aggr_nonad_dyn:traj_average_binary_diab_coeffs}, together with fractions (dashed dotted lines) of repulsive surface (red), adjacent surface (green) and third most energetic surface (yellow).
(d) Purity\index{purity}, $\mathcal{P}(t)$, of the electronic density matrix, according to \eref{eq:planar_aggr_nonad_dyn:def_purity}.
All data is averaged over $10^6$ realizations. The maximum value of the data in (a), (b) is individually set to one. To highlight details at lower densities, all values between 0.7 and 1 are represented with the color of maximum density.
The dashed white line in (a) marks the horizontal position of atoms~(4--7).
}
\end{figure}

The adjacent surface provides a transfer of excitation to the outer atoms of the vertical chain, as apparent from \fref{fig:N7_part_exc_dens_symmetric_case_both_chains}(c).
Moreover, \fref{fig:N7_part_exc_dens_symmetric_case_both_chains}(a) indicates that almost all excitation is removed from atom~3 when it is in the vicinity of the CI. However, the motion of atom~3 can proceed differently. Either it traverses the vertical chain or it is reflected off from it, as evident from the atomic density of the horizontal chain, shown in \fref{fig:Nges7_total_densities_pops}(a). Without vdW~interactions, the atom would always traverse the vertical chain, since resonant interactions can only induce forces on atoms which share excitation~(see \aref{part:app::calcs_planar_aggr__forces_resonant_interactions}).

The repulsive surface transfers a fraction of excitation to atoms~(5,6) on the vertical chain only for a short time, however, before both of them reach atoms~4 and 7, respectively, the excitation is almost completely relocalized, as it can be seen in \fref{fig:N7_part_exc_dens_symmetric_case_both_chains}(d). Most of the excitation stays on atom~3, which is reflected off the vertical chain to move backwards, evident from \fref{fig:N7_part_exc_dens_symmetric_case_both_chains}(b).
The atomic density of atoms~(4--7) in \fref{fig:Nges7_total_densities_pops}(b) shows that almost surely motion is induced on atoms~4 and 7 regardless whether excitation is localized on them or not. The adiabatic populations in \fref{fig:N7_part_exc_dens_symmetric_case_both_chains}(c) confirm that the CI leads to a half-half splitting of population on the repulsive and adjacent surface at around $20\ \mu $s. The population changes before are again due to trivial crossings. The purity\index{purity}, shown in \fref{fig:N7_part_exc_dens_symmetric_case_both_chains}(d), finally confirms the coherent splitting of the wave packet to evolve on two surfaces simultaneously. 

We conclude, that the exciton pulse can not be completely continued on the vertical chain with this symmetric configuration, where $\Delta y = 0$. The transport of excitation is not only undirected on the vertical chain, but also a significant fraction of excitation remains on the horizontal chain.

\newpage
\section{Exciton switch}
\label{sec:exciton_switch}
We found in \sref{sec:nges8} that the continuation of exciton pulse propagation on the vertical chain for a symmetric T-shape aggregate with $\Delta y=0$ is not possible. More precisely, the fraction of excitation which is transferred to the outer two atoms on the vertical chain is very low. This is due to the population of two surfaces after the first nontrivial CI crossing, where only one of them, the adjacent surface, transfers excitation to atoms~4 and 7. However, the repulsive surface transfers the excitation finally back to the horizontal chain and this excitation is lost for the exciton pulse propagation on the vertical chain.

The objective of this section is to engineer the dynamics in the vicinity of the CI to establish exciton pulse propagation on the vertical chain with high fidelity and to control the propagation direction. 
We still use the seven atom T-shape aggregate as sketched in \fref{system_sketch_7}, but take use of the asymmetry parameter $\Delta y$ to guide the exciton pulse.
Moreover we systematically vary all parameters $a_1$, $a_2$ and $\Delta y$ to perform high fidelity exciton pulse propagation after redirection on the vertical chain. In order to quantify exciton pulse propagation, we require a measure of characteristic properties, which is coherently shared excitation interlinked with atomic motion. To obtain this measure, we sample the coherence properties of the pulse on both ends of the vertical chain, a distance $a_{\mathrm{E}}$ away from atoms~4 and 7~[see \fref{system_sketch_7}]. We call this location ``detector'' in the following.
The distance $a_{\mathrm{E}}$ is useful to verify atomic motion. If the atoms are not set into motion, or the motion is interrupted such that the atoms can not reach the detector, exciton pulses are not established. The second feature --- coherently shared excitation--- is quantified through the \emph{bipartite entanglement of formation}\cite{hill:wooters:qbits,wooters:mixed}\index{entanglement of formation}, which contains how much excitation is shared and how it is distributed between two atoms.
To calculate it, we extract from the electronic density matrix information about the exciton pulse restricted on two atoms. This is end we calculate reduced electronic density matrices,
\begin{equation}
\hat{\beta}_{ab}:={\mbox{Tr}}^{\{a,b\}}\big[\hat{\sigma}\big],
\label{beta}
\end{equation}
for two atoms, labeled with $a,b$. Technically, the calculation is a partial trace, \linebreak[4]${\mbox{Tr}}^{\{a,b\}}\big[\cdots\big]$, over electronic states for all atoms other than $a$, $b$. 
Suppose $\ket{\Phi}$ is a pure state of a bipartite system $A+B$, containing the two subsystems $A$ and $B$. With the help of the reduced density matrix in subsystem $X \in \{A,B\}$, $\hat{\rho}_{X}:= {\mbox{Tr}}^{\{X\}}\big[\ket{\Phi}\bra{\Phi}\big]$, the bipartite entanglement is defined as
\begin{equation}
 E(\ket{\Phi}) := {\mbox{Tr}}\big[\hat{\rho}_{A}\log_{2}\hat{\rho}_{A}\big] = {\mbox{Tr}}\big[\hat{\rho}_{B}\log_{2}\hat{\rho}_{B}\big],
\end{equation}
which reveals that the entanglement of formation is basically the entropy associated with the state. A mixed state of the bipartite system is described by a density matrix, denoted with $\hat{\rho}$, and the entanglement of formation is the minimum average entanglement over all pure state decompositions, 
\begin{equation}
 E(\hat{\rho}) := \min_{\{\Phi_i\}}\sum_{i}p_i E(\Phi_i),
\end{equation}
where $p_i$ is the probability weight of the pure state $\Phi_i$. 
The minimization procedure can be performed analytically for arbitrary two qubit states, which are realized by the electron state of any two chosen atoms of the Rydberg aggregate.
The entanglement of formation is then closely related to the \emph{concurrence}\index{concurrence}, which we denote with $\mathcal{C}$. Specifically, the concurrence of atoms~$(a,b)$ is given by $\mathcal{C}_{a,b} = 2|\sigma_{a,b}|$, with $\sigma_{a,b}$ the coherence between electronic states  with excitation residing in atom~$a$ and~$b$, respectively and measures simultaneous excitation on sites $a$ and $b$. The entanglement of formation is finally a nonlinear function of the concurrence,
\begin{align}
 E(\hat{\beta}_{a,b}) &= h\left(1/2 + \sqrt{1-\mathcal{C}_{a,b}^2}/2\right), \mathrm{ with}\\
h(x):&=-x\log_{2} x - (1-x)\log_{2}(1-x).
\end{align}
While using the FSSH method to solve the quantum dynamics, the concurrence simplifies to $\mathcal{C}_{a,b} = 2 \left| \overline{c_{a}c^{*}_{b}} \right|$. Both, the concurrence and the bipartite entanglement take values betweeen $0$ and $1$. Moreover, the bipartite entanglement monotonically increase with the concurrence. When the entire excitation is localized on atoms~(a,b) and the excitation is furthermore shared in equal parts, the bipartite entanglement is maximal.

\begin{figure}[!t]
\centering
\includegraphics{{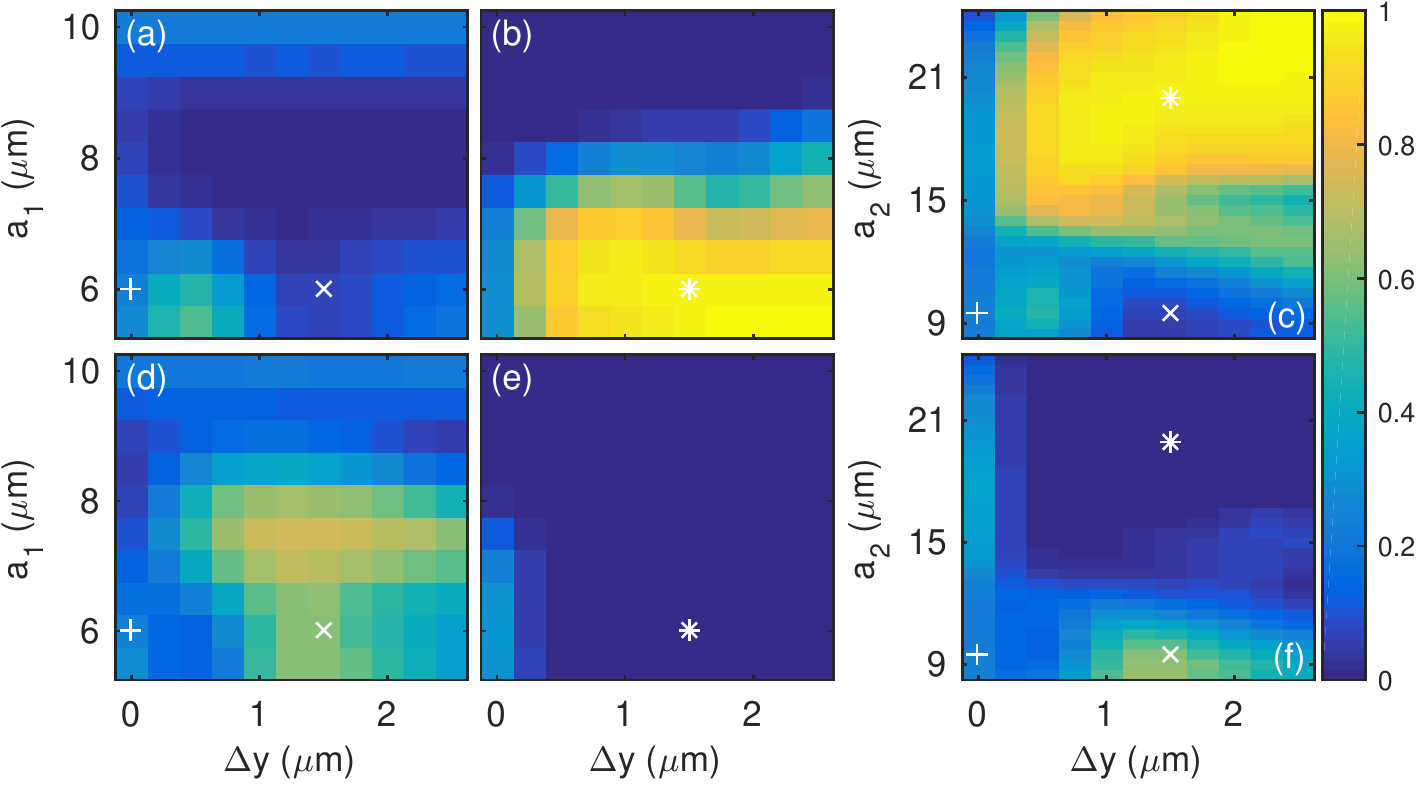}}
\caption{\label{switch_parameterspace}
Response of the exciton switch to control parameters. We show the bipartite entanglement transported (a)--(c) downwards in $y$ direction, $\bar{E}_{34}$, and (d)--(f) upwards in $y$ direction, $\bar{E}_{56}$. Parameters are $\nu=80$ and $d=22~\mu$m. Both entanglement readouts (see \fref{system_sketch_7}) are placed on the vertical chain in a distance of $a_\mathrm{E}=0.3d$ from atoms~4 and~7, respectively. (a)--(b) and (d)--(e) Entanglement as a function of $a_1$ and $\Delta y$, for $a_2=9.5~\mu$m in (a) and (d) and $a_2=20~\mu$m in (b) and (e).
(c) and (f) Entanglement as a function of $a_2$ and $\Delta y$, for fixed $a_1=6~\mu$m.
The markers ($\ast$, $\times$, $+$) highlight extreme cases:
($\ast$) high entanglement transport to atoms~(4,5),
($\times$) high entanglement transport to atoms~(6,7),
($+$) equal entanglement transport towards atoms~(4,5) and atoms~(6,7).
All entanglement measures were obtained by a simulation with FSSH with $10^3$ trajectories.
}
\end{figure}
High values of bipartite entanglement for atoms~(4,5) (atoms~(6,7)) at the detectors, which we denote with $\bar{E}_{45}$ ($\bar{E}_{67}$), indicate exciton pulse propagation with high fidelity along the vertical chain.
In \fref{switch_parameterspace}, the entanglement of formation is shown for varying parameters, \frefplural{switch_parameterspace}{a}{--}{c} are the results for $\bar{E}_{45}$ and \frefplural{switch_parameterspace}{d}{--}{f} show the results for $\bar{E}_{67}$. The figures indicate a complex dependency on the parameters, such that an a priori guess of optimal values for the parameters would be difficult. We can however understand the dependencies qualitatively. In \frefplural{switch_parameterspace}{a}{--}{b}, the bipartite entanglement for atoms~(4,5) is shown for varying interatomic distances of atoms~(1,2), $a_1$, and varying vertical shifts of the horizontal chain, $\Delta y$, whereas the interatomic distance of atoms~(5,6), $a_2$, is fixed, in \fref{switch_parameterspace}(a) to $a_2=9.5 \ \mu$m and in \fref{switch_parameterspace}(b) to $a_2=20 \ \mu$m. Recall that the parameter $\Delta y$ takes positive values for shifts in the downwards $y$ direction. A first comparison indicates larger entanglement in \fref{switch_parameterspace}(b) than in \fref{switch_parameterspace}(a), which shows that for larger values of $a_2$ the entanglement for atoms~(4,5) increases. This is also confirmed in \fref{switch_parameterspace}(c), where the entanglement is shown in dependence on $a_2$.
The reason why the change of the parameter $a_2$ leads to such different entanglement measures is that the ratio $a_2/d$ controls whether atoms~3, 5, and 6 can build an almost isolated trimer subunit or not. For large values of this ratio, as it is realized in \fref{switch_parameterspace}(b), the three atoms are not well isolated in the vicinity of the trimer CI configuration. As a result, at the position where we expect an avoided crossing/CI between repulsive and adjacent surface for an isolated trimer, we observe no crossing at all due to a much stronger coupling between atoms~(2,3) compared to the couplings of atoms~3 with atoms~(5,6). The energy spectra is then more or less given by the dimer states of atoms~(2,3) and atoms~(5,6), respectively. The repulsive and adjacent surface remain far separated during propagation and the transport occurs almost adiabatically, even during the redirection to the vertical chain. This finally initiates an exciton pulse on the vertical chain in the downwards direction, since atom~3 traverses atom~5 more closely, due to a shift of the horizontal chain in the downwards vertical direction ($\Delta y >0$), and the repulsive surface repels the vertical atom with smallest distance [compare with the forces in \fref{system_sketch_4}(c)]. For obtaining directed transport on the vertical chain, a preset asymmetric configuration due to vertically shifting the horizontal chain is thus crucial. This explains why in \fref{switch_parameterspace}(b) the entanglement starts to increase above a certain value of $\Delta y$, around one~micrometer. When the asymmetry is too small, almost equal repulsion of atoms~(5,6) occurs and directed transport is not possible. Moreover, the repulsive surface will more likely reflect atom~3 off the vertical chain together with the excitation remaining on it, as described in \sref{sec:nges8_nondirectional}. 

That directed transport actually can occur at all is confirmed by high entanglement obtain for atoms~(4,5) and low entanglement for atoms~(6,7) apparent by comparing \fref{switch_parameterspace}(b) with \fref{switch_parameterspace}(e) for equal specific choices of $(a_1,\Delta y)$.
Finally we also observe in \fref{switch_parameterspace}(b) a dependency of the entanglement on the parameter $a_1$.
This parameter adjusts the maximum velocity of the atoms and thereby it sets the timescale for the transport. Smaller values lead to faster transport. Since the lifetime of the Rydberg aggregate\footnote{The lifetime of the Rydberg aggregate is roughly given by the lifetime of a single Rydberg atom divided by the total number of atoms in the aggregate.}
is limited, the transport has to be fast enough to reach the entanglement detectors within the lifetime of the aggregate. Technically we set the simulation time to the lifetime of the aggregate. If the transport is too slow to reach the detector, no entanglement is measured. This explains why the entanglement increases almost monotonically in \fref{switch_parameterspace}(b) with decreasing values of $a_1$.

In contrast to the situation realized in \frefplural{switch_parameterspace}{b}{ and }{e} where the dynamics is not affected by the CI, in \frefplural{switch_parameterspace}{a}{ and }{d} the ratio of $a_2/d$ is small enough for atoms~3, 5, and 6 to be well isolated from the remaining atoms in the vicinity of the trimer CI configuration. As a consequence the exciton pulse enters the trimer subunit on the \emph{adjacent} surface, as it is evident from the energy spectra of single trajectories in \frefplural{fig:Nges7_single_trajectories}{d}{--}{f}.
Since we obtain in \fref{switch_parameterspace}(d) high entanglement for configurations with preset asymmetry, as in \fref{switch_parameterspace}(b), we can assume the excitation is dominantly transferred along the adjacent surface after passing the nonadiabatic region, which we explain in the following.
In average the trajectories have an asymmetry of $\bar{b} = 1-2\Delta y /a_2$ and for parameters where we observe high entanglement in  \fref{switch_parameterspace}{d} $\bar{b} \approx 0.7$, which reveals high asymmetry. Importantly the variation of asymmetry among all trajectories is with $\sigma_b /b \sim \sigma_0 /a_2 \approx 0.05$ very low and hence the energy gap in the vicinity of the CI is for all of them approximately given by $\overline{\Delta E /E} \sim 1-\bar{b} = 2\Delta y /a_2$~(see \aref{part:app::calcs_planar_aggr__rel_energy_gap_CI}), which is approximately $30\%$ for parameters of high entanglement transport. Besides this large energy gap we can extract from \fref{switch_parameterspace}(d) that for high entanglement larger values of $a_1$ are required, which corresponds to smaller atomic velocities. Both large energy gap and small velocities ensure that the majority of trajectories after they entered the trimer subunit on the adjacent surface continue propagation on the same surface when they traverse the nonadiabatic region and hence excitation transport remains largely adiabatic.

\begin{table}[!t]
\centering
\begin{tabular}{C{\widthof{XconfigurationX}}C{\widthof{X$(\ast)$X}}C{\widthof{X$(\times)$X}}C{\widthof{X$(+)$X}}}
\hline
\hline
\vspace{0.2cm}configuration\vspace{0.2cm} & $(\ast)$ & $(\times)$ & $(+)$\\
\hline
parameters &&&\\
$a_2\ (\mathrm{\mu m})$ & 20 & 9.5 & 9.5 \\
$\Delta y\ (\mathrm{\mu m})$ & 1.5 & 1.5 & 0 \\
\hline
Entanglement &&&\\
up, $\bar{E}_{67}\ (\%)$  &0&60&24\\
down, $\bar{E}_{45}\ (\%)$  &97&7&24\\
\hline
\hline
\end{tabular}
\caption{
\label{tab:entanglement_transport_values}
Compilation of values for the parameters $a_2$ and $\Delta y$ 
which distinguish the three configurations marked with $(\ast),\ (\times)$ and $(+)$ in \fref{switch_parameterspace} together with the entanglement obtained at both detectors on the vertical chain due to wave packet dynamics. For all three configurations, $a_1 = 6\ \mu$m.
}
\end{table}

Comparing \frefplural{fig:Nges7_single_trajectories}{a}{, }{d} with \frefplural{fig:Nges7_single_trajectories}{b}{, }{e} clearly indicates that large values of $a_2$ lead to entanglement transport in the upwards direction and low valued of $a_2$ to entanglement transport in the downwards direction.
In \frefplural{fig:Nges7_single_trajectories}{c}{, }{f} we show the entanglement transport in downwards and upwards direction, respectively,
in dependence of the vertical shift $\Delta y$ and the parameter $a_2$, to investigate the directionality of entanglement transport at intermediate values of $a_2$. The value of $a_1$ is set to $6\ \mu$m, for which we obtained in both direction high entanglement.
Above $a_2=14\ \mu $m the transport is clearly in downwards direction, almost not dependent on $\Delta y$. In a small region where $a_2<10\ \mu$ m and $1\ \mu \mathrm{m} < \Delta y < 2\ \mu \mathrm{m}$ entanglement is efficiently transported in upwards direction. 

Ultimately, we can maximize the entanglement\index{entanglement of formation} and control the exciton pulse propagation direction by only tuning the parameter $a_2$ and fixing the value for the asymmetry to $\Delta y = 1.5 \mu m$. Choosing $(a_1,a_2,\Delta y) = (6,20,1.5)\ \mu$m we get high entanglement in downwards direction [marked as $(\ast)$ in \fref{switch_parameterspace}(c)], with exciton pulse propagation along the repulsive surface. To achieve high entanglement in the upwards direction, we set $(a_1,a_2,\Delta y) = (6,9.5,1.5)\ \mu$m
[marked as $(\times)$ in \fref{switch_parameterspace}(f)].
The plots confirm also that for symmetric configurations with $\Delta y=0$ the entanglement is low in both directions, almost independently of the values $a_2$. To compare the two configurations, $(\ast), (\times)$ with the entanglement values of a symmetric configuration, we mark the configuration $(a_1,a_2,\Delta y)=(6,9.5,0)\ \mu$m with $(+)$. A comparison of atomic configuration parameters together with obtained entanglements at both detectors on the vertical chain are given in \Tref{tab:entanglement_transport_values}.
\begin{figure}[!t]
\centering
\includegraphics{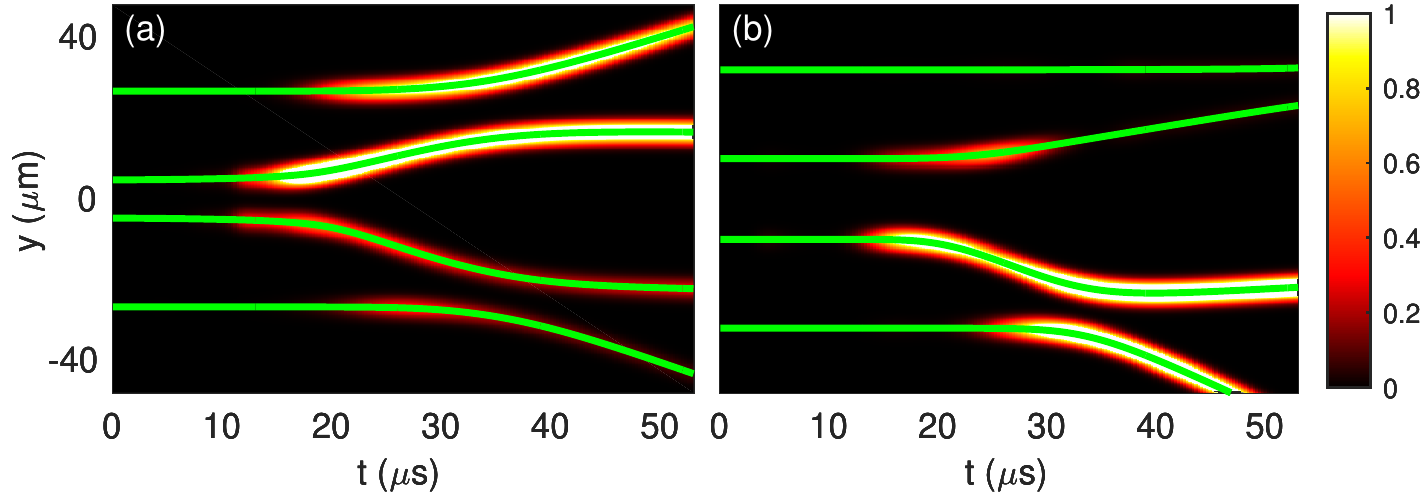}
\caption{\label{fig:Nges7_partial_densities_chain2}
Mean atomic positions (green lines) with \lstate{p} excitation probability of the combined repulsive and adjacent surfaces (shading according to colormap) on each atom of the vertical chain for two different cases: (a)~for the configuration marked with $(\times)$ in \fref{switch_parameterspace}, whose dynamics mainly evolves on the adjacent surface after traversing the nonadiabatic region which leads to exciton pulse propagation in upwards direction, (b)~for the configuration marked with $(\ast)$ in \fref{switch_parameterspace}, for which the wave packet performs a jump back to the repulsive surface in the nonadiabatic region, ultimately leading to exciton pulse propagation in downwards direction. The excitation probability for atom $n$ is represented by Gaussians of a selected fixed width, normalized to $\sum_n |\braket{\pi_n|\sub{\varphi}{n}(\bv{R})}|^2$ (where $\sum_n$ runs over the repulsive- and adjacent surfaces) and centered on the mean position of atom $n$. 
}
\end{figure}
To highlight how the excitation is actually transferred for the cases $(\ast)$ and $(\times)$, we show in \fref{fig:Nges7_partial_densities_chain2} the time-resolved excitation probability~(shading according to colormap) together with the mean positions~(green line) for each atom of the vertical chain. \fref{fig:Nges7_partial_densities_chain2}(a) clearly reveals that almost all excitation is transferred in upwards direction for the configuration  marked with $(\times)$. The repulsion of atoms~(5,4) is ensured due to vdW~interactions, when they approach each other. In \fref{fig:Nges7_partial_densities_chain2}(b), the excitation transfer is shown for the configuration marked with $(\ast)$. It clearly shows localization of excitation and exciton pulse propagation in downwards direction, whereas upwards almost no excitation is transferred. Atoms~(5,6) are repelled regardless of whether they share excitation or not, which is a signature of vdW~interactions.

\begin{figure}[!t]
\centering
\includegraphics{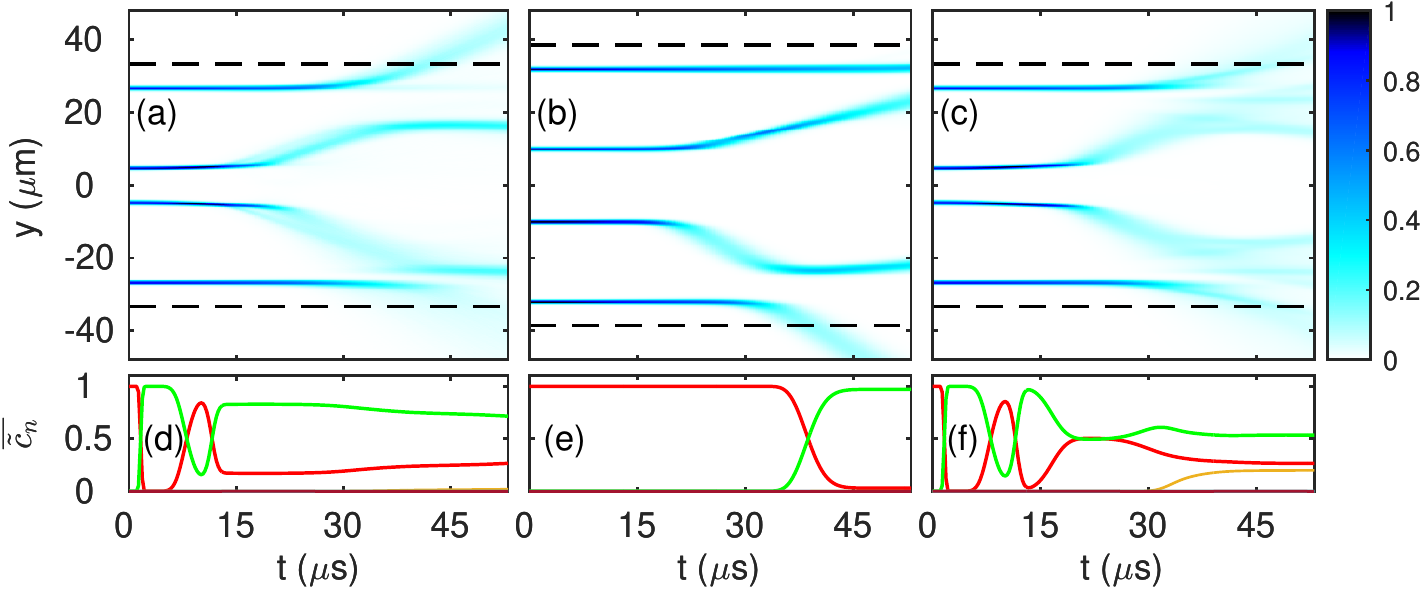}
\caption{\label{fig:exciton_switch_compare_atom_dens_three_cases}
Atomic densities on the vertical chain (a)--(c) with corresponding adiabatic populations in (d)--(f).
(a) and (d) Results for the configuration $(\times)$, which leads to exciton pulse propagation dominantly along the adjacent surface after traversing the nonadiabatic region.
(b) and (e) Results for the configuration $(\ast)$, which leads to exciton pulse propagation along the repulsive surface. Note that the population inversion in (d) is due to a trivial crossing.
(c) and (f) Results for the symmetric configuration $(+)$, which leads to nondirectional transport with low entanglement on both ends of the vertical chain due to a half-half splitting of population on the repulsive and adjacent surface after traversing the nonadiabatic region.
}
\end{figure}

Finally we present for all three cases the atomic densities of atoms~(4--7), together with the corresponding adiabatic populations in \fref{fig:exciton_switch_compare_atom_dens_three_cases}. 
After dynamics was initiated on the vertical chain, the adiabatic populations confirm adiabatic transport on the repulsive surface for the configuration marked with $(\ast)$, as apparent from \fref{fig:exciton_switch_compare_atom_dens_three_cases}(e), and dominant adiabatic transport on the adjacent surface for the configuration marked with $(\times)$, as seen in \fref{fig:exciton_switch_compare_atom_dens_three_cases}(d).
Note that the population inversion around $40\ \mu$s in \fref{fig:exciton_switch_compare_atom_dens_three_cases}(e) is due to a trivial crossing which does not change the dynamics. The atomic density for $(\ast)$ in \fref{fig:exciton_switch_compare_atom_dens_three_cases}(b) shows clearly a strong repulsion of atom~5 which eventually repels atom~4. Atom~6 experiences also a repulsion, which is caused by vdW~interactions. However, excitation follows the atomic motion downwards. 
For the propagation along the adjacent surface, the atomic density of the vertical chain in \fref{fig:exciton_switch_compare_atom_dens_three_cases}(a) shows a strong repulsion of atoms~(5,6), yet the asymmetry is enough to guide the excitation upwards. We observe that the repelling smallest distance between atoms~(5,4) is much smaller than between atoms~(6,7). The reason is that due to excitation being guided upwards, atomic repulsion is ensured by resonant interactions in upwards and by vdW~interactions in downwards directions.
The atomic density for $(+)$ is shown in \fref{fig:exciton_switch_compare_atom_dens_three_cases}(c) with corresponding adiabatic populations in \fref{fig:exciton_switch_compare_atom_dens_three_cases}(f). The dynamics of this configuration we already discussed in \sref{sec:nges8_nondirectional}, leading to nondirectional transport with low entanglement on both ends of the vertical chain.

To conclude, we demonstrated that exciton pulses can be redirected onto an orthogonal direction with high fidelity. The key
is to avoid the formation of a trimer subunit in the vicinity of the trimer CI configuration by increasing the interatomic distance of atoms~(5,6). Creating an asymmetry by shifting the horizontal chain vertically decreases the distance between one of the vertical placed atoms and atom~3, which finally allows unidirectional exciton pulse propagation along the vertical chain. 
On the other hand, fixing the asymmetry but allowing the formation of a trimer subunit allows exciton pulse propagation on the adjacent surface which finally redirects the pulse in \emph{opposite} direction than the propagation direction along the repulsive surface. Remarkably, we can control the BO~surface and direction of exciton pulse propagation by tuning only the interatomic distance between atoms~(5,6).

\chapter{A T-shape aggregate with unconstrained dynamics}\label{part:rs::chap:unconstrained_aggregate}
In \Chref{part:rs::chap:planar_aggregates} we demonstrated the manipulation of exciton pulses on planar T-shaped aggregates. Spatial constraints ensured low-dimensionality allowing to guide exciton pulses and to restrict nonadiabatic dynamics to occur in a small volume. This chapter is dedicated to the investigation of exciton pulse propagation and interaction with a CI in higher dimensions, which complicates directing transport of excitation and controlling the influence of a CI on it.

We will demonstrate that if initiated in a low dimensional space, entangled atomic motion in the continuum will remain confined to this space despite the possibility for all particles (ions and electrons) to move in full space. Together with advances in the newest generation experiments on Rydberg gases beyond the frozen gas regime, involving microwave\index{microwave} spectroscopy\cite{celistrino_teixeira:microwavespec_motion} or position sensitive field ionization\cite{thaicharoen:trajectory_imaging}, our results enable the quantum simulation of chemical processes in flexible Rydberg aggregates as an experimental science.
These recent efforts\cite{celistrino_teixeira:microwavespec_motion,thaicharoen:trajectory_imaging} extend earlier pioneering studies of motional dynamics in Rydberg gases\cite{Fioretti:pillet:longrangeint:prl,li_gallagher:dipdipexcit,mudrich:pillet:backforth,marcassa:collidingdistrib:pra,nascimento:longrangemotion:pra,amthor:mechatt:prl,amthor:mechrepulsive:pra,park:dipdipbroadening,park:dipdipionization} and render now the rich dynamics of Rydberg aggregates fully observable\cite{leonhardt:3dswitch}.

We present the theoretical framework in \sref{unconstr:theoretical_framework}, describing the treatment of an\-isotropic interactions in \sref{part:rs::chap:3D_aggregate_anisotropic_interactions} and how to select an initial exciton state with a microwave\index{microwave} in \sref{part:rs::chap:3D_aggregate_select_exciton}. Subsequently, \sref{nonadiabdyn} discusses the dynamics of the aggregate with a focus on the comparison to the planar aggregates. Time-resolved observables accesible by experiments are reviewed in \sref{experimental_signatures}. In \sref{unconstr:perturb_ground_state_atoms} the influence of perturbing ground state atoms is estimated, for the case that the experimental setup does not employ isolated Rydberg atoms, but instead Rydberg atoms excited out of an ultracold gas of ground-state atoms and subsequently embedded therein.
Finally we conclude in \sref{unconstrained_conclusion}.
We will thus show in this chapter, that two central elements of the Rydberg aggregate, nonadiabatic motional dynamics on several coupled BO~surfaces\cite{wuester:CI,leonhardt:switch} and entanglement transport are now experimentally accessible, as we show here. %

%
\begin{figure}[!t]
\centering
\includegraphics[width=14.5cm]{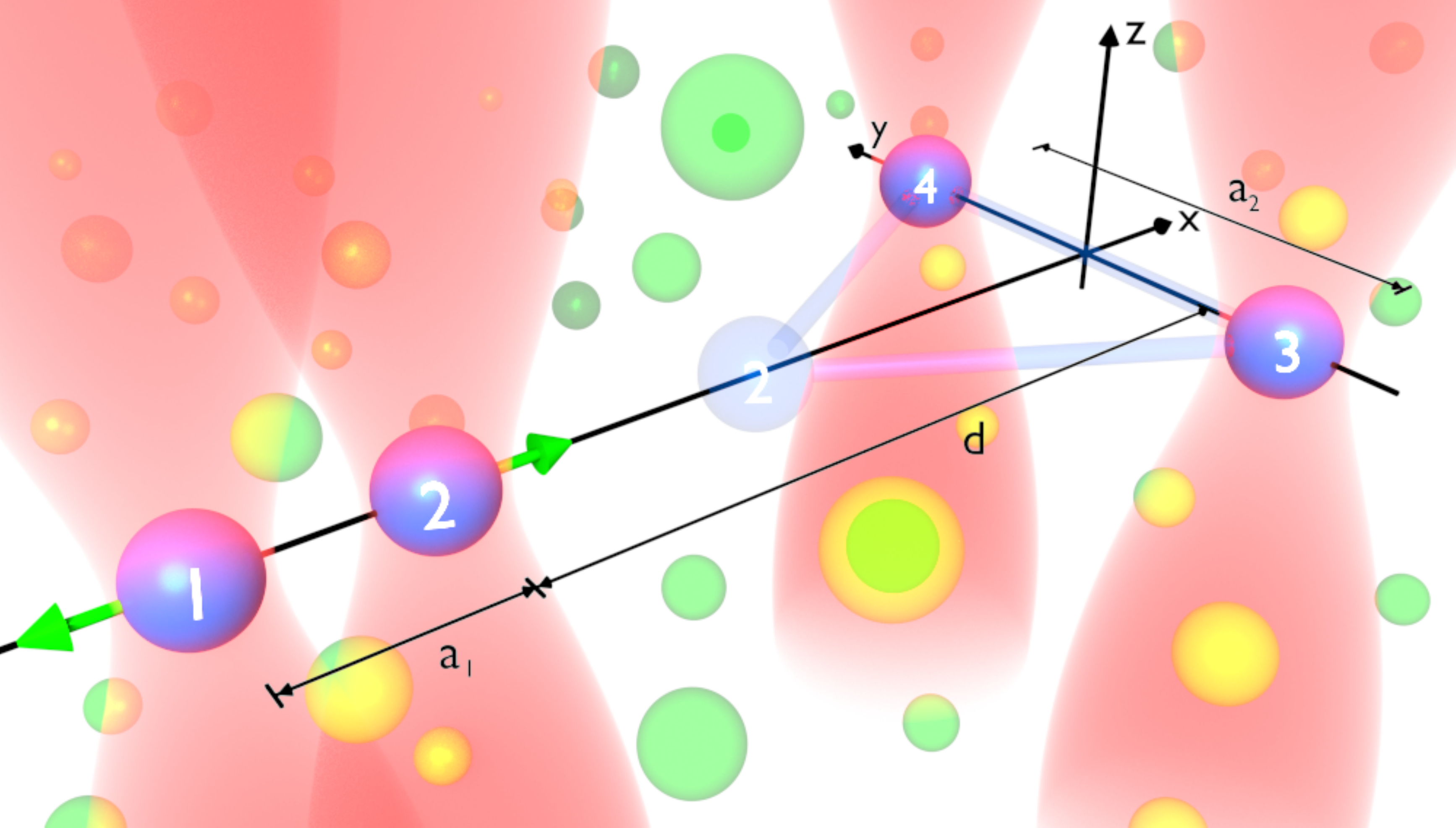} 
\caption{Embedded flexible Rydberg aggregate. Four excitation beams (red shades) define focus volumes in which exactly one atom is excited to a Rydberg state (blue balls, atoms~(1--4)), within a cold gas (green balls).  Our co-ordinate system has its origin at the mean position of atom~1, several geometrical parameters are explained in the text. Subsequent to Rydberg excitation, dipole-dipole interactions will cause acceleration along the green arrows, causing atom~2 to reach the position shown in light blue, where a CI will cause strong nonadiabatic effects.  
\label{geometry}}
\end{figure}

\section{Theoretical Framework}
\label{unconstr:theoretical_framework}
In \sref{sec:nges4}, we already described the dynamics of two perpendicular Rydberg dimers constrained to a plane and moreover restricting motional dynamics to be one-dimensional. Here we study a system with the same initial geometry, but remove all spatial constraints, allowing for motion in the full three-dimensional space.

To be specific, we investigate a flexible Rydberg aggregate consisting of $N=4$ $^{7}$Li Rydberg atoms (mass $M = 11000~$a.u.), excited to principal quantum number $\nu=80$, embedded within a host cold atom cloud of ground state atoms, see \fref{geometry}. The typical experimental situation excites Rydberg atoms from a cold atom cloud. The ground state atoms may have benefits for detecting Rydberg atoms\cite{guenter:EIT,guenter:EITexpt,schoenleber:immag}

This configuration is created with tightly focused Rydberg excitation lasers. We assume that the focus volumes are small enough to deterministically excite just a single atom within each to an angular momentum $l=0$ Rydberg state $\ket{s} \equiv \ket{\nu,s}$, exploiting the dipole-blockade\cite{jaksch:dipoleblockade,lukin2001:dipole_blockade}.
We place the origin of our coordinate system at $\vc{R}_{1}^{(0)}$, the laser focus positioning atom~1, such that other focus positions shown in the figure are
\begin{align}
 \vc{R}_{2}^{(0)} &= (a_1,0,0),\label{eq:3D_aggregate:init_pos_R2}\\
 \vc{R}_{3}^{(0)} &= (a_1+d,-a_2/2,0),\label{eq:3D_aggregate:init_pos_R3}\\
 \vc{R}_{4}^{(0)} &= (a_1+d,a_2/2,0).
\label{eq:3D_aggregate:init_pos_R4}
\end{align}
The vectors of atomic positions, $\vc{R}_{\alpha}= (x_\alpha,y_\alpha,z_\alpha)$, are represented in the cartesian basis $\{\vc{e}_x,\vc{e}_y,\vc{e}_z\}$, as it is depicted in \fref{geometry}. 
In particular we use the following values for the geometrical parameters: $a_1=10~\mu$m, $a_2=37~\mu$m and $d=51~\mu$m. We will refer to atoms~(1,2) as the horizontal Rydberg dimer and atoms~(3,4) as the vertical one, to conform with nomenclature from \Chref{part:rs::chap:planar_aggregates}. As before, the positions of all atoms are collected into the vector $\vc{R} = (\vc{R}_{1},  \dots , \vc{R}_{N})^T$. Co-ordinates of ground-state atoms are not required since these will be merely spectators for the dynamics of Rydberg atoms, as shown in \cite{moebius:bobbels} and found experimentally in \cite{thaicharoen:trajectory_imaging,celistrino_teixeira:microwavespec_motion}. 

Exciting the aggregate to the single \lstate{p} excitation manifold causes resonant dipole-dipole interactions between the atoms and eventually initiates entanglement transport and atomic motion. The interactions scale quadratically with the \lstate{s}-\lstate{p} radial transition matrix element, which specifically for $\nu =80$ is given by $\drad \equiv \drad_{\nu,1;\nu,0} = 8250$~atomic units. For an experimentally treatment we remove the assumption of isotropic interactions, which we used for the planar aggregates in \Chref{part:rs::chap:planar_aggregates}, instead we take the full anisotropy of the dipole-dipole interactions into account.

\subsection{Anisotropic dipole-dipole interactions}
\label{part:rs::chap:3D_aggregate_anisotropic_interactions}
%
The use of isotropic interactions in \Chref{part:rs::chap:planar_aggregates} was a simplification and is achievable either for linear or planar aggregates by choosing a particular alignment of the \lstate{p} orbital and the spatial arrangements.
Another method is to apply an external magnetic field, tuning the interactions from anisotropic to isotropic with increasing magnetic field strength, as we will demonstrate in \Chref{part:app::chap:tun_int_mag_field}.

However, resonant dipole-dipole interactions are in general anisotropic due to the possibility of different orientation of angular momentum orbitals. For a full description of the anisotropy, the magnetic quantum numbers have to be taken into account and for \lstate{s}-\lstate{p} transitions within a single \lstate{p} excitation manifold, the electronic basis needs to be enlarged to $B_{\mathrm{el}}:= \{\ket{\pi_\alpha,m}\}_{\alpha=1,\dots,N}^{m=-1,0,1}$, where
\begin{equation}
 \ket{\pi_\alpha,m} = \ket{\lstate{s}\dots(\lstate{p},m)\dots \lstate{s}}
\label{eq:3D_aggregate:aggregate_states}
\end{equation}
denotes the aggregate state where all but the $\alpha$th atom are in \lstate{s} states and the remaining atom carries the \lstate{p} excitation with magnetic quantum number $m$. The relative orientation of the angular momentum vector $\vc{L}$ of the \lstate{p} state to the quantization axis \quantax\ is described by the magnetic quantum number $m$\footnote{Specifically, the magnetic quantum number $m=0,\ 1,\ -1$ relates to an included angle of $90^{\circ},\ 45^{\circ},\ 135^{\circ}$, respectively.}.

The resonant interactions for \lstate{s}-\lstate{p} transitions are presented in \sref{part:fd::chap:ryd_a::sec:dip_dip_int_subsec_transition_dip_moments} by \eref{eq:alk_ryd_atoms:dip_dip_interaction_qm_matrix_elements_s_p}.
The Hamiltonian containing all resonant interactions can therefore be written as
\begin{equation}
 \hat{H}_{\mathrm{dd}}(\vc{R})= \sum_{\substack{\alpha,\beta=1;\\ \alpha\neq \beta}}^{N} \sum_{m,m' =-1}^{1} V_{m,m'}(\vc{R}_{\alpha, \beta}) \ket{\pi_\alpha,m}\bra{\pi_\beta,m'},
  \label{eq:3D_aggregate:elechamiltonian-dd_anisotrop}
\end{equation}
We retain the simple structure of \eref{eq:planar_aggr:setup_single_p_state__vdw_Hamiltonian} to treat vdW~interactions, \linebreak[4]$\hat{H}_{\mathrm{vdw}}=-\sum_{\alpha,\beta=1}^{N}C_6/2R_{kl}^6\hat{\id}$, which consequently induce an atomic configuration dependent energy shift, ensuring globally repulsive interactions at small interatomic spacings. As in \Chref{part:rs::chap:planar_aggregates}, the vdW~interactions have only the purpose to prevent two excitation-less atoms from collisions.
The electronic Hamiltonian is then again as in \eref{eq:planar_aggr:setup_single_p_state__def_Hel} given by $\hat{H}_{\mathrm{el}}(\vc{R}) := \hat{H}_{\mathrm{dd}}(\vc{R}) + \hat{H}_{\mathrm{vdw}}(\vc{R})$.

Note that for Lithium the spin-orbit coupling is very small and we thus can in good approximation neglect it\cite{haroche:li_finesplitting,leonhardt:orthogonal}.

\subsection{Initial preparation}
\label{part:rs::chap:3D_aggregate_select_exciton}
%
The trapping potentials are approximately harmonic in the vicinity of the laser foci and therefore the spatial wave functions can be assumed as the ground state of a harmonic oscillator, leading to the initial nuclear (atomic) wave function\index{wave function!nuclear}
\begin{equation}
 \phi_{\mathrm{nuc}}(\vc{R},t=0) = (2\pi \sigma_0^2)^{-3N/4}\mathrm{e}^{-(\vc{R} - \vc{R}_{0})^2/(4\sigma_0^2)},
\end{equation}
with $\vc{R}_{0} \equiv (\vc{R}^{(0)}_{1},\dots,\vc{R}^{(0)}_{N})^T$ containing the initial positions and a width of the Gaussians of $\sigma_0= 0.5~\mu\mathrm{m}$, which is challenging but in reach for standard techniques. Besides trapping and arranging the atoms, all four atoms have to be excited to Rydberg \lstate{s} states, such that the electronic state of the aggregate is $\ket{\lstate{S}}\equiv \ket{\lstate{s} \dots \lstate{s}}$. The next step is to excite with a microwave\index{microwave} to a specific exciton, an eigenstate of \eref{eq:3D_aggregate:elechamiltonian-dd_anisotrop}.
We will restrict ourselves to \emph{linearly} polarized light and as a natural choice, we set the quantization axis to the microwave polarization direction. The anisotropy of the dipole-dipole interactions considers the alignment of the \lstate{p} orbital, explicitly, the including angles of the interatomic distance vector with the polarization axis, denoted by $\theta({\quantax},\vc{r}),\phi({\quantax},\vc{r})$ in $V_{m,m'}(\vc{r}) \sim Y_{2,m' - m}(\theta({\quantax},\vc{r}),\phi({\quantax},\vc{r}))/r^3$ describe the orientation of the \lstate{p} orbital to the quantization axis, on which the strength of the interaction is dependent. Since the linearly polarized microwave orients the angular momentum \emph{along} the polarization direction, it is helpful to represent all distance vectors in a microwave fixed frame of reference. Specifically, we use the basis $B_{\quantax} := \{\vc{q}_x,\vc{q}_y,\vc{q}_z=\quantax\}$ such that the microwave populates the $m=0$ states.
So far we specified the representation of spatial directions in the cartesian basis $B:=\{\vc{e}_x,\vc{e}_y,\vc{e}_z\}$.
The basis change $B \to B_{\quantax}$ is technically performed by
$\vc{\tilde{r}} = \vc{Q}_{\quantax}\vc{r}$, where $\vc{r}$ is represented in $B$ and $\vc{\tilde{r}}$ is represented in $B_{\quantax}$ by setting $\vc{Q}_{\quantax} =\left(\braket{\vc{q}_{k},\vc{e}_l}\right)_{k,l =x,y,z}$. The including angles $\theta({\quantax},\vc{r}),\ \phi({\quantax},\vc{r})$ are then the standard expressions for azimuthal and polar angles of $\vc{\tilde{r}}$ in a spherical representation, respectively. For reasons of numerical stability it is best to express the spherical harmonics in cartesian co-ordinates of the interatomic vectors in the basis $B_{\quantax}$\footnote{To obtain atomic forces and nonadiabatic coupling vectors, the gradient of the Hamiltonian and hence the binary interactions has to be determined. A spherical representation can cause problems for the polar angle in the transition region from $2\pi$ to $0$.}, which are then given by
\begin{equation}
 Y_{2,m}(\theta({\quantax},\vc{r}),\phi({\quantax},\vc{r})) =  Y_{2,m}(\tilde{\vc{r}}) = 
\begin{dcases}
\sqrt{15/32\pi}\,\left[\left(\tilde{r}_x\pm \im\tilde{r}_y\right)/\tilde{r}\right]^2, & m=\pm 2 \\
\mp\sqrt{15/8\pi}\,\tilde{r}_z\left(\tilde{r}_x\pm \im\tilde{r}_y\right)/\tilde{r}, & m=\pm 1 \\
\sqrt{5/16\pi}\,\left(-1+3[\tilde{r}_z/\tilde{r}]^2\right),& m=0
\end{dcases}.
\end{equation}

The smallest interatomic distance of the aggregate is between atoms~(1,2) initially. This ensures that excitation is localized on the horizontal dimer when creating the exciton corresponding to the repulsive surface. The microwave\index{microwave} performs the transition $\ket{\lstate{S}} \to \ket{\varphi_\mathrm{rep}}$, with $\ket{\varphi_\mathrm{rep}}$ denoting the exciton, which finally yields repulsive forces. 
%
We restrict the microwave polarization direction to be perpendicular to the interatomic distance of the horizontal dimer, for two: Firstly, only perpendicular- or parallel alignments of the polarization to the interatomic distance lead to a decoupled subspace of the electronic Hamiltonian in \eref{eq:3D_aggregate:elechamiltonian-dd_anisotrop}, for aggregate states given in \eref{eq:3D_aggregate:aggregate_states} with $m=0$, as shown in \rref{moebius:cradle}. Thus, the linearly polarized microwave accesses excitons that populate only the $\lstate{p}_{m=0}$ orbitals, for which the nodal plane is perpendicular to the polarization direction. Due to the necessity to excite repulsive excitons with positive energy, only one of the two alignments is feasible: The microwave is restricted to excite symmetric states\footnote{\label{discuss_microwave_exc} see \aref{part:app::chap:exc_mw}} and only for the perpendicular alignment, the repulsive surface corresponds to the symmetric exciton state of the form
$\ket{\varphi_{\mathrm{ini}}}\equiv \ket{\varphi_\mathrm{rep}} \approx \left(\ket{\pi_1,0} + \ket{\pi_2,0}\right)/\sqrt{2}$. To populate the exciton completely, a Rabi-$\pi$ pulse of the microwave is needed with its frequency being detuned by the exciton energy $U_\mathrm{rep} \approx \drad/3R_{12}^3$ from the \lstate{s}-\lstate{p} transition\textsuperscript{\ref{discuss_microwave_exc}}.

The total initial state of the Rydberg aggregate is then given by
\begin{equation}
  \ket{\Psi_{\mathrm{tot}}(\vc{R},t=0)} = \phi_{\mathrm{nuc}}(\vc{R},t=0)\ket{\sub{\varphi}{ini}},
\end{equation}
and repulsive forces initially lead to a repulsion of the horizontal dimer, such that atom~2 can approach the vertical dimer. Allowing the atoms to move in full space increases the quantum mechanical complexity even more than for the restricted motional dynamics of the planar aggregates in \Chref{part:rs::chap:planar_aggregates}. Therefore, we again rely on quantum-classical methods to solve the dynamics, using Tully's fewest switching algorithm~(FSSH)\cite{tully:hopping2,tully:hopping:veloadjust,barbatti:review_tully}, which we outlined in \sref{part:rs::chap:planar_aggregates_dynamical methods_quantum_classical_method}.

The results in the following section are performed with a microwave\index{microwave} polarization in the $y$ direction, $\quantax = \vc{e}_y$, for which the excited exciton corresponds to the second most energetic BO~surface in the global energy spectrum of the Hamiltonian given in \eref{eq:3D_aggregate:elechamiltonian-dd_anisotrop}.

For the employed parameters of the aggregate, the mean value of the potential energy is initially $\bar{U}_{\mathrm{rep}}(\vc{R}_0) \approx 22.27$~MHz.


\section{Nonadiabatic dynamics}
\label{nonadiabdyn}
Following Rydberg excitation to $\ket{\lstate{s}}$ and $\ket{\lstate{p}}$, the four aggregate atoms will move essentially unperturbed through the background gas\cite{moebius:bobbels}. 

\begin{figure}[!t]
\centering
\includegraphics{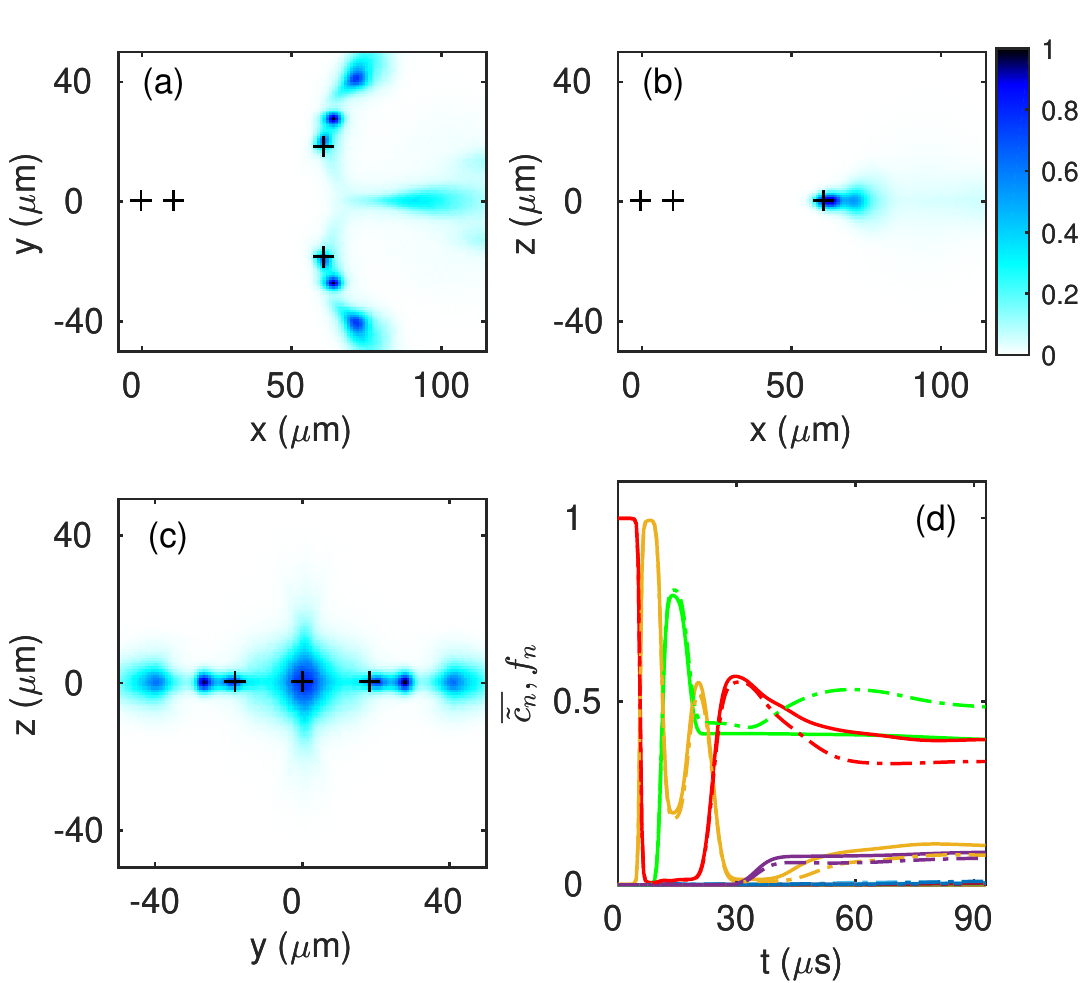} 
\caption{Atomic density of the final state at $t=92.9~\mu$s in~(a)--(c) and adiabatic populations (solid lines) and fractions (dashed dotted lines) in~(d). Shown are column densities in the $x$-$y$ plane (a),  $x$-$z$ plane (b) and  $y$-$z$ plane (c). The black '+' in~(a)--(c) mark the initial atomic positions.
The maximal densities are set to $1$ and the microwave\index{microwave} polarization direction is set to $\vc{q} = \vc{e}_y$.
\label{fig:column_densities}}
\end{figure}

The BO~surface of initial preparation exerts repulsive forces on atoms~(1,2) and although the motion is unconstrained, the large interatomic spacing compared to the very localized nuclear wave function, $\sigma_0/a_1=0.05$, leads to a directed motion of the atomic wave packets 
within~(or very close to) the $x$-$y$~plane, with almost no dynamics off-the-plane.
In \frefplural{fig:column_densities}{a}{--}{c} we show for the final state at $t\approx 93 \ \mu $s the column atomic densities, which are projections of the full three dimensional atomic density onto a plane.
In particular the $x$-$z$ density shown in \fref{fig:column_densities}~(b) confirms almost no dynamics off the $x$-$y$ plane. However the $x$-$y$ density and also the $y$-$z$ density 
in \fref{fig:column_densities}(a) and \fref{fig:column_densities}(c), reveal a branching of each atom's density in the vertical dimer into three parts, which we already observed for the one-dimensional confined case of the planar four atom aggregate in \fref{fig:Nges4_total_densities_pops_purity}, discussed in \sref{sec:nges4}.
It can be assumed that the splitting as previously is due to nonadiabatic dynamics and propagation along two different BO~surfaces. The time-resolved adiabatic populations and fractions shown in \fref{fig:column_densities}(d) in fact reveal a drastic change in the populations of BO~surfaces, such that after $30\ \mu$s mainly two BO~surfaces participate equally in the dynamics. The reason is, as before, that when atoms~(2--4) form an equilateral triangle configuration, the total wave packet hits a CI\index{conical intersection} causing the splitting. The mechanism is thus similar to the dynamics of the planar aggregates with isotropic interactions. 

\begin{figure}[!t]
\centering
\includegraphics{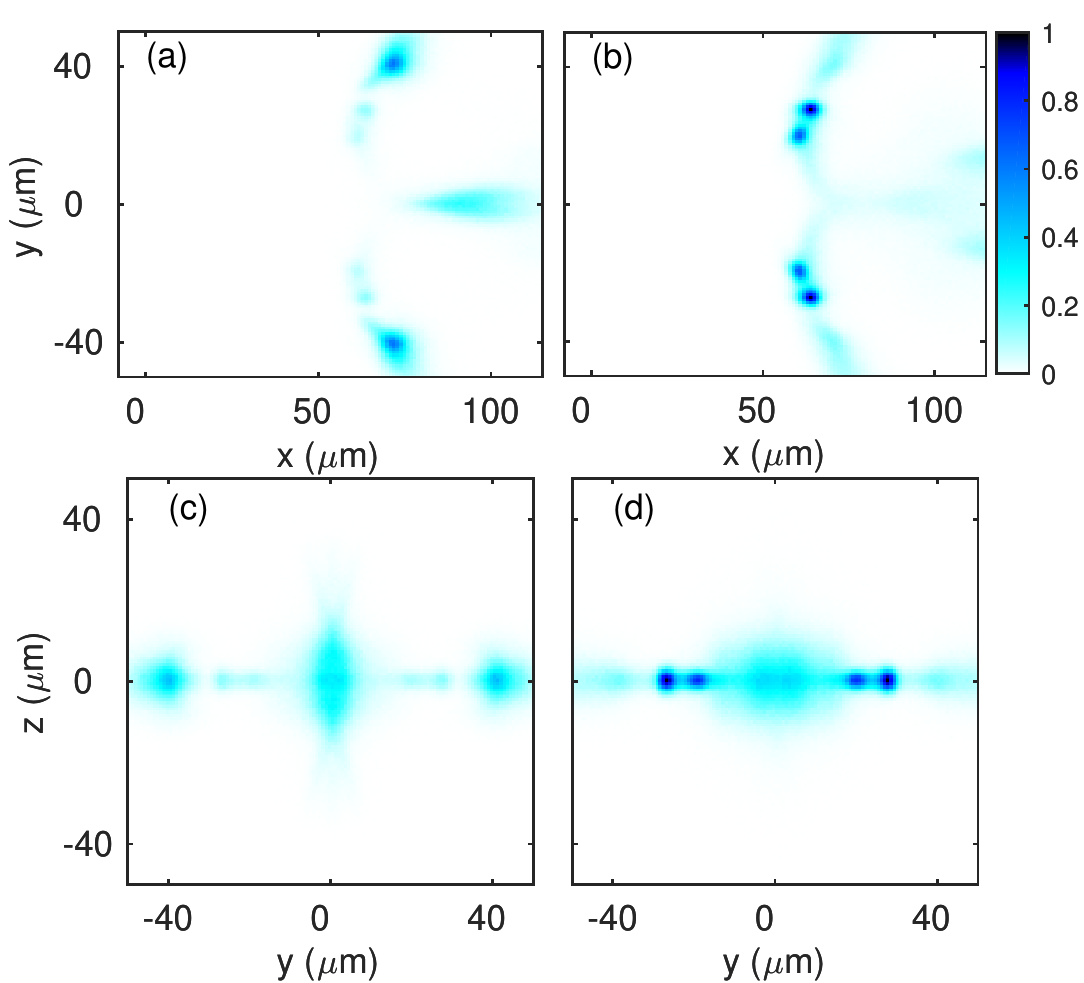} 
\caption{BO~segregated atomic column densities at final time of $t=92.9~\mu$s, for (a) and (c) the second most energetic, and (b) and (d) the fourth most energetic BO~surface. 
(a) and (b) column densities in the $x$-$y$ plane.
(c) and (d) column densities in the $y$-$z$ plane.
The densities are normalized to the global maximum value of all of them.
\label{fig:partial_atomic_densities}}
\end{figure}

In contrast to the aggregate with isotropic interactions, here there are three BO~surfaces involved in the dynamics before the trimer CI configuration is reached, instead of two.
A study of single trajectories and their time resolved energy spectra revealed that the first two transitions are due to trivial (avoided) crossings~(see \aref{part:app::calcs_planar_aggr__trivial_crossings}). The first population transfer around $\approx 5.5 \ \mu$s is a trivial crossing of the repulsive surface~[red line in \fref{fig:column_densities}(d)] with a surface in close proximity~[orange line in \fref{fig:column_densities}(d)]. The latter is another surface of the horizontal dimer, which corresponds to an exciton state with population of $m=\pm 1$ states. For an isolated dimer, these two surfaces would be genuinely degenerate, however, the presence of the vertical dimer introduces a small splitting between them $\sim 0.57$~KHz. During the dynamics the initially populated BO~surface crosses the other and changes the energetic ordering from being the second to the third most energetic one, which is visible as population transfer in \fref{fig:column_densities}(d). The second trivial crossing around $ 11 \ \mu$s eventually populates the adjacent surface~[green line in \fref{fig:column_densities}(d)], the surface which adiabatically is connected to the adjacent surface of the trimer. The wave packet approaches the CI and gets split to approximately propagate on two BO~surfaces afterward. With high fidelity the total wave function gets coherently split such that it is given by
$\ket{\Psi_{\mathrm{fin}}(\vc{R})} \approx \chi_{\mathrm{rep}}(\vc{R})\ket{\varphi_{\mathrm{rep}}(\vc{R})} + \chi_{\mathrm{adj}}(\vc{R})\ket{\varphi_{\mathrm{adj}}(\vc{R})}$. 
To assign components of the atomic density to its surface of propagation we show in \fref{fig:partial_atomic_densities} BO~segregated atomic column densities for the repulsive surface in \frefplural{fig:partial_atomic_densities}{a}{, }{c} for the adjacent surface in \frefplural{fig:partial_atomic_densities}{b}{, }{d}. The repulsive surface corresponds to the components which are the most exterior parts of the density in $y$ direction for atoms~(3,4). Only a small fraction remains at its initial position. Conversely, the adjacent surface corresponds to the components inside the $y$ direction. Moreover, the components have a double peak structure. Overall, the essential characteristics of the atomic density are similar to the aggregates with one-dimensional spatial confinement. The reason is that the BO~surfaces induce the same forces as shown in \frefplural{sec:nges4}{c}{, }{d}, such that each BO~surface leaves its mark in the atomic density.

We can conclude that the CI\index{conical intersection} operates as an exciton splitter, as it was observed for the planar four atom T-shape aggregate. This is an important theoretical result, since it opens the way for \emph{direct} observation of nonadiabatic signatures in \linebreak[4]experiments\cite{nogrette2014:single_atom_trapping_2D,thaicharoen:trajectory_imaging,celistrino_teixeira:microwavespec_motion,guenter:EITexpt} due to much inflated length and time scales compared to the typical quantum chemical systems.
Optical confinement of atoms in one-dimensional traps along with a reduction of the electronic state space assumed for the planar aggregates\cite{wuester:CI,leonhardt:switch} studied in \Chref{part:rs::chap:planar_aggregates} constitute a significant experimental challenge. The present results show that these restrictions are not required. It is simply the symmetry of the initially prepared system which keeps the motion similarly planar and hence accessible. The successful splitting into different motional modes through the CI\index{conical intersection} is a sensitive measure for the extent to which the atomic motion remains in a plane.
Additionally the dynamics leading to \fref{fig:column_densities} entails entanglement transport. At $t=0$, atoms~(1,2) in state $\ket{\varphi_{\mathrm{ini}}}$ are maximally entangled\cite{wuester:cradle}, at the final time this has been transported to atoms~(3,4) for the outermost lobes\cite{leonhardt:switch}.
\section{Experimental signatures}
\label{experimental_signatures}
%
The total atomic density, shown as column densities in \frefplural{fig:column_densities}{a}{--}{c}, is experimentally accessible if the focus positions $\vc{R}^{(0)}_{n}$ are sufficiently reproducible to allow averaging over many realisations. Additionally, one requires near single atom sensitive position detection. A shot-to-shot position uncertainty $\sigma_0$ in 3D \emph{within} each laser focus is already taken into account in our simulation. Recent advances in position sensitive field ionisation enable $\sim1$ $\mu$m resolution, clearly sufficient for an image such as \fref{fig:column_densities}. Panel~(c) in \fref{fig:column_densities} could alternatively also be observed by waiting for atoms~(3,4) to impact on a solid state detector. 

The background gas can also act as a probe for position and state of the embedded moving Rydberg atoms \cite{olmos:amplification,guenter:EIT,guenter:EITexpt,schoenleber:immag,schempp:spintransport}, offering resolution sufficient for \fref{fig:column_densities} as well.

\begin{figure}[!t]
\centering
\includegraphics{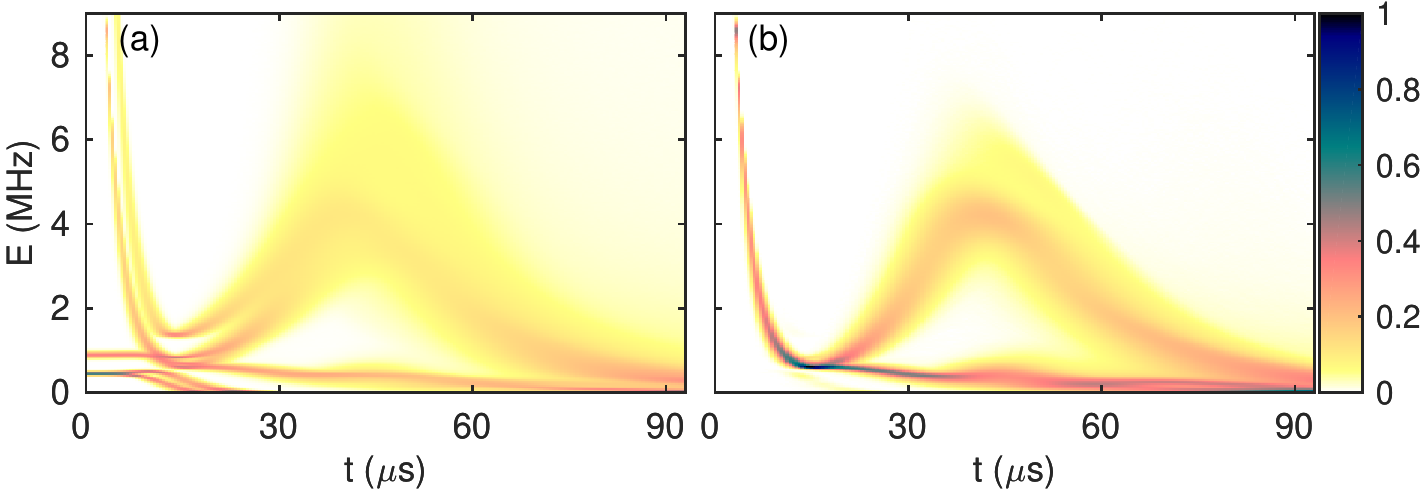}
\caption{Time-resolved exciton density of states $g(E,t)$ (a) and potential energy density $v(E,t)$ (b). The densities are normalized to have a maximum value of one. Furthermore, to emphasize low density features, we plot the square root, $\sqrt{g(E,t)}$ and $\sqrt{v(E,t)}$, respectively.
\label{fig:Epotdens_Espectr_dens_Li4_vdw_zxy}}
\end{figure}

\begin{figure}[!t]
\centering
\includegraphics{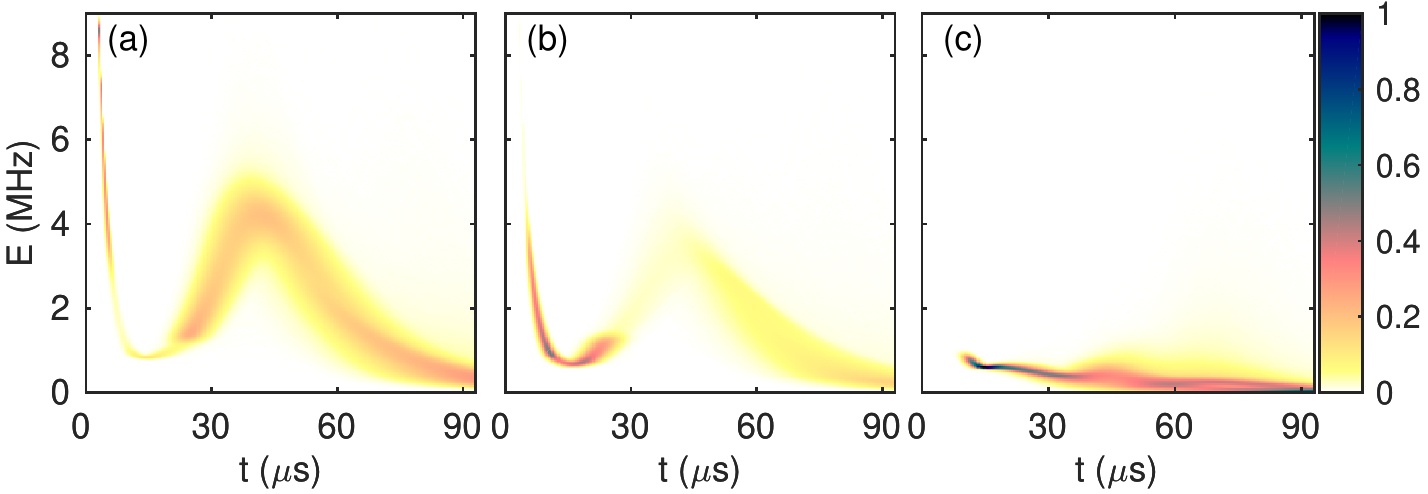} 
\caption{BO~segregated, time-resolved densities of potential energy, for (a) second, (b) third, and (c) fourth most energetic BO~surface.
The normalization is performed with the maximum value of the total potential energy density, which is shown in \fref{fig:dens_partial_Epot_partial_atomic_dens_Li4_vdw_zxy}(c).
Furthermore, to emphasize low density features, we plot the square root densities.
\label{fig:dens_partial_Epot_partial_atomic_dens_Li4_vdw_zxy}}
\end{figure}

Nonadiabatic dynamics discussed here can not only be monitored in position space, but also in the excitation spectrum of the system, similar to \rref{celistrino_teixeira:microwavespec_motion}. The observable is the time-resolved potential energy density $u(E,t)$, shown in \fref{fig:Epotdens_Espectr_dens_Li4_vdw_zxy}(b). 
Observation of $u(E,t)$ could proceed by monitoring the time- and frequency resolved outcome of driving the \lstate{p}-\lstate{d} transition. 
To obtain $u(E,t)$, we bin the potential energy $U_{\zeta(t)}$ of the currently propagated BO~surface $\zeta(t)$ into a discretized energy grid $E$ and average over all trajectories. We see a clear splitting into several features within $u(E,t)$ around $15\ \mu$s, the density branches from there on into two parts which is a clear signatures of the CI.
To understand this in more detail, we show in \fref{fig:dens_partial_Epot_partial_atomic_dens_Li4_vdw_zxy} the potential energy density segregated to the three most strongly participating BO~surfaces. They confirm the propagation exclusively along the repulsive surface at the beginning~[high density values until $5\ \mu$s in \fref{fig:dens_partial_Epot_partial_atomic_dens_Li4_vdw_zxy}(a)], but before the CI is hit around $15\ \mu$s, another surface is populated~[high density values between $5\ \mu$s to $30\ \mu$s in \fref{fig:dens_partial_Epot_partial_atomic_dens_Li4_vdw_zxy}(b) which corresponds to the orange adiabatic population line in \fref{fig:column_densities}(d)] due to a trivial crossing with a surface corresponding to an exciton state populating $m=\pm 1$ \lstate{p} orbitals. However, the transition is not as trivial as for the planar aggregates with isotropic interactions. The initially prepared exciton has a small admixture of \lstate{p} orbitals in $m=\pm 1$ direction due to the anisotropy of the interactions and the dispersion of the wave packet in all three dimensions. Therefore a small fraction of the wave packet does not undergo the transition and remains on the more energetic surface.
A second appearance of such a trivial crossing around $11\ \mu $s populates the fourth most energetic surface, which adiabatically connects to adjacent trimer surface. On the other hand the third most energetic surface connects adiabatically to the repulsive trimer surface in the vicinity of the trimer CI configuration which is reached around $30 \ \mu $s and thus leads to a repopulation of the second most energetic surface.

To visualize distances between energy surfaces, we additionally present in \fref{fig:Epotdens_Espectr_dens_Li4_vdw_zxy}(a) the density of the global energy spectrum, $g(E,t)$. Technically it is obtained through the same procedure as the density of the potential energy. It reveals the close spacing and overlapping of three BO~surfaces
between $10$--$20\ \mu$s, which finally is a source for the nonadiabatic dynamics.

An alternative experiment to observe our nonadiabatic dynamics, would individually trap four atoms at positions $\vc{R}^{(0)}_n$, with trapping width $\sigma_0$, prior to Rydberg excitation, see e.g.~\cite{nogrette2014:single_atom_trapping_2D}. 

\newpage
\section{Switch Born-Oppenheimer surfaces}
%

\begin{figure}[!t]
\centering
\includegraphics{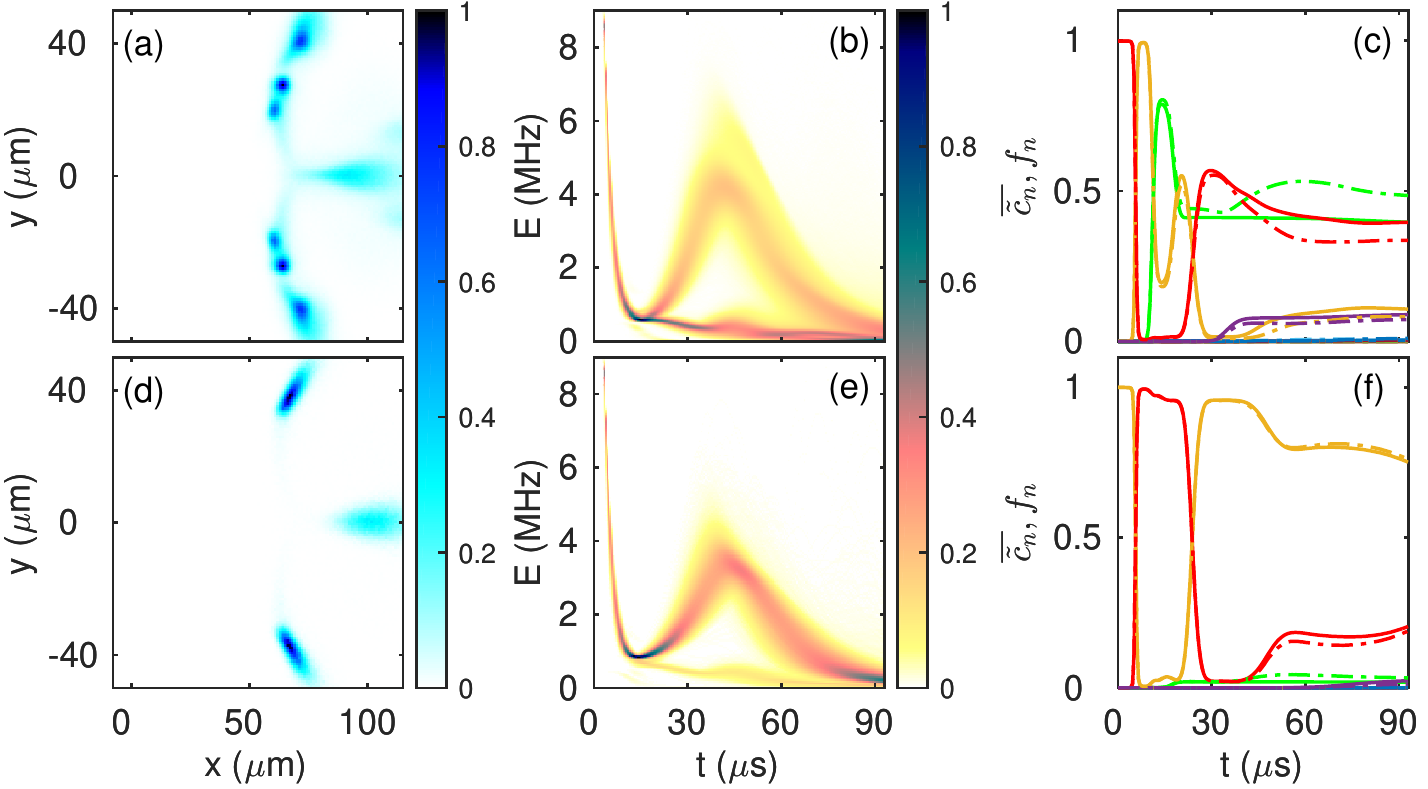} 
\caption{Comparison of $x$-$y$ column density at final time of $t=92.9~\mu$s in (a) and (d), potential energy density in (b) and (e), and adiabatic populations/fractions in (c) and (f), for two different choices of microwave polarization\index{microwave} directions. Results for the polarization direction $\vc{q} \parallel \vc{e}_y$ are shown in (a)--(c), and for $\vc{q} \parallel \vc{e}_y$ in (d)--(f). We plot again the square roots of the normalized potential energy densities.
\label{axis_dependence}}
\end{figure}

So far we presented results for a microwave polarization\index{microwave} 
in $y$ direction. We show in \fref{axis_dependence} a comparison with results for polarization in $z$ direction.
For a purely isolated dimer aligned along the $x$ direction, microwave polarization in both, $y$ and $z$ direction, is perpendicular to the horizontal dimer and thus linear polarization excites the same exciton with the same energy value of the BO~surface. The only difference would be the direction of the excited $p$ orbital, which is aligned along the polarization direction. The degeneracy of these two polarization directions is lifted through the presence of the vertical dimer in the T-shape aggregate. For polarization in $y$ direction the exciton corresponds to the second most energetic BO~surface, whereas for polarization in$z$ direction the third most energetic surface gets populated. Switching between populating both surfaces leads to very different dynamics. The $x$-$y$ atomic column density for polarization in $z$ direction, shown in \fref{axis_dependence}(d), indicates a strong repulsion of atoms~(3,4). Moreover, there is no three-fold branching for each atom, as for polarization in $y$ direction~[for comparison the corresponding $x$-$y$ atomic column density is shown in \fref{axis_dependence}(a)]. The repulsion without branching of the wave packet is due to fact that the biggest part of the wave packet propagates along the same physical BO~surface, as apparent from the adiabatic populations shown in \fref{axis_dependence}(f). The population only changes are due to trivial crossings and actually ensure the following of the \emph{same} excitonic state continuously.
A comparison of the potential energy densities for polarization in $y$ and $z$ direction shown in \fref{axis_dependence}(b) and \fref{axis_dependence}(e), respectively, reveals almost no branching for the $z$ direction polarization and high population density on the repulsive surface, consistent with the adiabatic populations.

The splitting between second and third most largest BO~surface is small, with a mean energetic separation of only $\Delta \bar{E} = 0.57~$KHz.
This provokes an overlap of their individual energy density profiles and in the energy spectrum, shown in \fref{fig:Epotdens_Espectr_dens_Li4_vdw_zxy}(a), where both surfaces are indistinguishable initially.

This close proximity of the two surfaces allows to switch between them by only changing the polarization direction and keeping the microwave\index{microwave} frequency unchanged, assuming a sufficiently broad pulse is used for addressing. Consequently, the polarization direction controls the subsequent evolution, in particular the branching of the wave packet and the nonadiabaticity.

\section{Perturbation by ground state atoms}
\label{unconstr:perturb_ground_state_atoms}
%
We expect the dynamics of the embedded Rydberg aggregate discussed here not to be significantly perturbed by its cold gas environment. Rydberg-Rydberg interactions substantially exceed elastic Rydberg ground-state atom interactions\cite{Greene:LongRangeMols,balewski:elecBEC} for separations $d>200$~nm, and dipole-dipole excitation transport disregards ground state \linebreak[4]atoms\cite{moebius:bobbels}. The kinetic energies of ${\mathcal O}$($10$ MHz) are still low enough to render inelastic $\nu$ or $l$ changing collisions very unlikely\cite{balewski:elecBEC}, leaving molecular ion- or ion pair creation as main Rydberg excitation loss channel arising from collisions with ground state atoms\cite{niederpruem:giantion,balewski:elecBEC}. Even including those and assuming a moderate background gas density, we can extrapolate experimental data from Rb~(\rref{celistrino_teixeira:microwavespec_motion}) and still find a sufficiently large lifetime of the Rydberg aggregate, which we will estimate in the following. Rydberg excited atoms with $\nu=80$ in $l=0,1$ states move through a background gas of ground state atoms, which we assume to be of density $\rho=4\times 10^{18}$ m$^{-3}$, at a maximal velocity of about $v_{\mathrm{ini}} \sim \sqrt{ {U}_{\mathrm{ini}}(R_0)/2 }\approx 0.85$ m/s. We can deduce a maximal cross-section for ionizing collisions between Rydberg atoms and ground state atoms of $\sigma(\nu) = 610$ nm$^2$ at $\nu=60$ from experiment \cite{celistrino_teixeira:microwavespec_motion}.
Assuming scaling with the size of the Rydberg orbit \cite{niederpruem:giantion}, we extrapolate this value to our $\nu=80$, thus $\sigma(80) =\sigma(60) (80/60)^2 10^3$ nm$^2$.
The total decay rate of four atom aggregate under investigation is then $\sub{\Gamma}{tot}=2\sub{\Gamma}{coll} + 4 \sub{\Gamma}{0}$, with spontaneous decay rate $\sub{\Gamma}{0}$ and collisional decay rate $\sub{\Gamma}{coll}$ for single atoms. We have assumed that only two atoms ever move with the fastest velocity. Using $\sub{\Gamma}{coll}=\rho \sub{v}{ini} \sigma(80)$, we finally arrive at a total lifetime $\tau=1/\sub{\Gamma}{tot}=130 \mu$s for the aggregate.

However, detrimentally large cross sections for the same processes were found in \rref{niederpruem:giantion,balewski:elecBEC} for much larger densities $\rho$. Further research on ionization of fast Rydberg atoms within ultra cold gases is thus of interest for the setup assumed here.

\section{Conclusions}
\label{unconstrained_conclusion}
%
In summary, controlled creation of a few Rydberg atoms in a cold gas of ground state atoms will allow to initiate
coherent motion of the Rydberg atoms without external confinement as demonstrated here with the unconstrained motion of four Rydberg atoms, forming coupled excitonic BO~surfaces. This enables 
nonadiabatic motional dynamics and entanglement transport in assemblies of a few Rydberg excited atoms as an experimental platform for studies of 
quantum chemical processes inflated to convenient time (microsconds) and spatial (micrometers) scales, with the perspective to shed new light on relevant processes such as ultra-fast vibrational relaxation or quantum control schemes for embedded systems. Experimental observables are atomic density distributions or exciton spectra.
Different degrees of nonadiabaticity can be accessed from the same initial atomic positions through the choice of the initial exciton state. 

The effects explored will be most prominent with light Alkali species, such as Li discussed here, but also the more common Rb can be used. Here a slightly smaller setup
would sufficiently accelerate the motion to fit our scenario into the Rb system life-time. Rb would, however, pose a greater challenge for the theoretical modelling, making the inclusion of spin-orbit coupling necessary\cite{park:dipdipbroadening,park:dipdipionization}.

Beyond the controlled scenario discussed here, illuminating a 3D gas entirely with a single Rydberg excitation laser, followed by microwave\index{microwave} transitions to the \lstate{p} state, should also quickly result in nonadiabatic effects. They would arise through the abundant number of CIs in random 3D Rydberg assemblies\cite{wuester:CI}.

\chapter{Tuning interactions with magnetic fields}\label{part:app::chap:tun_int_mag_field}
The dipole-dipole interactions between Rydberg atoms are in general anisotropic. For linear\cite{cenap:motion,wuester:cradle,moebius:cradle} (planar) spatial configurations of the aggregates and excitation of the \lstate{p} orbitals parallel or perpendicular (perpendicular) to the configuration directions, the dipole-dipole interactions are isotropic.
However, for planar aggregates resonant interactions are only istropic with positive amplitude. As depicted in \fref{fig:trimer_4atom_aggr_spectra_pl_min_int} of \Chref{part:rs::chap:planar_aggregates}, this prevents access to a CI during dynamics on the repulsive BO~surface. The necessary negative sign of the binary interactions can thus not be achieved by the simple choices alone. We demonstrate in this chapter how this problem can be overcome by applying an external magnetic field.

The effect of the magnetic field on the anisotropic interactions together with a comparison to an isotropic interaction model is presented in \sref{part:app::chap:tun_int_mag_field__isotropic_aniso_model} by simulating the dynamics for a four atom aggregate similar to the one of \sref{sec:nges4} using both interaction models. Subsequently, \sref{part:app::chap:tun_int_mag_field__anal_deriv} quantitatively investigates the effect of the magnetic field on the dipole-dipole interactions. We start in \sref{part:app::chap:tun_int_mag_field__just_neglect_so} with a discussion of spin-orbit coupling, finestructure, and where both can be neglected, and then derive with these prerequisites an analytical interaction model for planar aggregates in \sref{part:app::chap:tun_int_mag_field__planar_strong_fields}. Finally, in \sref{part:app::chap:tun_int_mag_field__compar_iso_eff_compl}, we compare isotropic and effective interactions with the complete interaction model, which also takes spin-orbit coupling into account.

\section{Isotropic versus anisotropic dipole-dipole model}
\label{part:app::chap:tun_int_mag_field__isotropic_aniso_model}
In the electronic Hamiltonian, $\hat{H}_{\mathrm{el}}(\vc{R}) := \hat{H}_{\mathrm{dd}}(\vc{R}) + \hat{H}_{\mathrm{vdw}}(\vc{R})$ in \Chref{part:rs::chap:planar_aggregates}, we employed isotropic binary interactions of the form $V(r) = -\drad^2/6r^3$ , for the investigation of planar aggregates. However naturally, only isotropic dipole-dipole interactions of the form $V(r) = \drad^2/3r^3$, thus $V(r)>0$, can be realized by exciting to the single \lstate{p} excitation manifold with the \lstate{p} orbitals aligned perpendicular to the $x$-$y$ plane of the aggregate. This corresponds to the excitation of $(\lstate{p},m=0)$ states, with a quantization axis \quantax\ perpendicular to the $x$-$y$ plane spanned by the initial configuration of the aggregate, $\quantax \perp \vc{R}_{0}$. To achieve a negative sign the $(\lstate{p},m=\pm 1)$ manifold has to be excited when using the same orientation of the quantization axis. However, the coupling between $m=-1$ and $m=1$ states produces an anisotropy for the interactions. The idea is to use a magnetic field to detune both \lstate{p} orbital orientations energetically, which eventually yields approximately isotropic interactions with negative sign by selectively exciting one of the two \lstate{p} orientations. 

We have to start with using the full anisotropy of the resonant dipole-dipole interactions given by \eref{eq:alk_ryd_atoms:dip_dip_interaction_qm_matrix_elements_s_p}, $V_{m,m'}(\vc{r}) \sim Y_{2,m' - m}(\theta,\phi)/r^3$, and extend the electronic basis to $\{\ket{\pi_\alpha,m}\}$, with $\ket{\pi_\alpha,m} = \ket{\lstate{s}\dots (\lstate{p},m)\dots\lstate{s}}$ the aggregate states which include all \lstate{p} orbital orientations, already defined in \eref{eq:3D_aggregate:aggregate_states}. Applying a magnetic field perpendicular to the $x$-$y$ plane, $\vc{B}=B_z\vc{e}_z$, the $m=\pm 1$ states are energetically shifted by $\Delta E = \mu_{\mathrm{B}}B_z m$, such that the total Hamiltonian can be written as
\begin{equation}
 \hat{H}(\vc{R},B_z) = \hat{H}_{\mathrm{el}}(\vc{R}) + \mu_{\mathrm{B}}B_z\sum_{\alpha=1}^N\sum_{m=-1}^{1} m \ket{\pi_{\alpha},m}\bra{\pi_{\alpha},m}.
\label{eq:tun_int_mag::hamiltonian_aniso_mag}
\end{equation}
\begin{figure}[!t]
\centering
\includegraphics{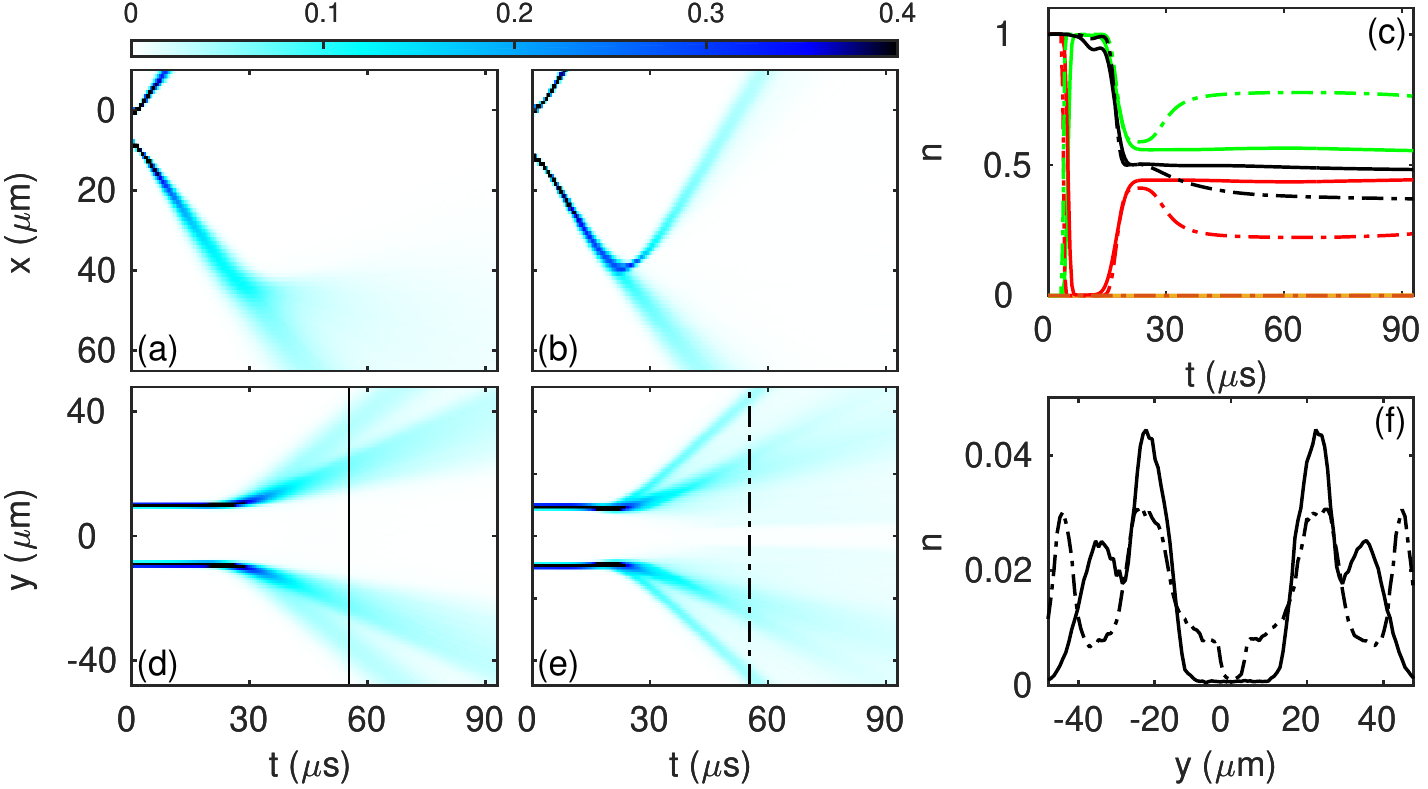}
\caption{\label{Bfield_doubledimer}Comparison of aggregate's dynamics employing two different interaction models, isotropic interactions according to \eref{eq:planar_aggr:theoretical_framework_elec_setup_binary_interactions}, and anisotropic interactions together with a magnetic field shift according to the Hamiltonian in \eref{eq:tun_int_mag::hamiltonian_aniso_mag}. The aggregate consists of two perpendicular dimers, as in \sref{sec:nges4}, with the atomic motion constrained to $1D$. In both models we used the parameters $\nu=80$, $a_2=19~\mu$m and $d=40~\mu$m.
To obtain qualitatively the same dynamics in both, $a_1$ is adjusted individually, compensating quantitative differences in both potentials. The anisotropic model uses $a_1=8~\mu$m and the isotropic one $a_1=11.8~\mu$m. (a) and (b) Atomic density of atoms~(1,2). (d) and (e) Atomic density of atoms~(3,4), where (a) and (d) are for the anisotropic model and (b) and (e) for the isotropic one. To highlight details at lower densities, all value between 0.4 and 1 are represented with the color of the highest density. (c) Adiabatic populations on the  repulsive-~(red line) and adjacent energy surface~(green line) and purity (black lines). We compare the anisotropic (solid line) with the isotropic model (dash-dotted line). (f) Cut through the atomic densities at the time indicated by black lines in (d) and (e) for both models, with line-styles as in (c).
The applied magnetic field strength for the anisotropic model is $B_z = 160$~G.
}
\end{figure}

In \fref{Bfield_doubledimer} we present the dynamics of a double dimer system similar to the double dimer aggregate in \sref{sec:nges4}, employing once the isotropic interaction model and then comparing it with a model that takes the full anisotropy of the interactions into account. Additionally it includes the effect of an external magnetic field, according to the Hamiltonian in \eref{eq:tun_int_mag::hamiltonian_aniso_mag}.
The atomic motion for each dimer is still constrained to $1D$. Different from the aggregate in \sref{sec:nges4}, we use adjusted parameters~(see caption of \fref{Bfield_doubledimer}) to realize a situation which is adapted to the exciton switch of \sref{sec:exciton_switch}/
The results in \fref{Bfield_doubledimer} reveal that qualitatively the same main features are found as for the isotropic model. However, for the greatest resemblance the parameters of both models have to be chosen slightly different due to the quantitative difference of potentials, which affect most importantly the initial acceleration of atoms~(1,2), in turn controlling the relative population of the two energy surfaces after CI crossing, seen in \fref{Bfield_doubledimer}(e).
Thereby, the interatomic distance of the horizontal dimer, $a_1$, is in both models separately adjusted to achieve a rough $50$-$50$ splitting on the two surfaces.
Both variants then qualitatively agree, in particular regarding clear signatures of multiple populated BO~surfaces in the snapshot shown in \fref{Bfield_doubledimer}(f).

\section{Analytical derivation of effective interactions}
\label{part:app::chap:tun_int_mag_field__anal_deriv}
The numerical comparison of the dynamics for an aggregate in \fref{Bfield_doubledimer}, of both interaction models, already showed the qualitative effect of the magnetic field, decreasing the anisotropy. In the following we quantitatively investigate the effect of the magnetic field on the dipole-dipole interactions. We derive a model with effective interactions and demonstrate that in the limit of infinitely strong magnetic fields it results in the isotropic model according to \eref{eq:planar_aggr:theoretical_framework_elec_setup_binary_interactions}.

\subsection{Negligible spin-orbit coupling in the regime of strong magnetic fields}
\label{part:app::chap:tun_int_mag_field__just_neglect_so}
\begin{figure}[!t]
\centering
\includegraphics{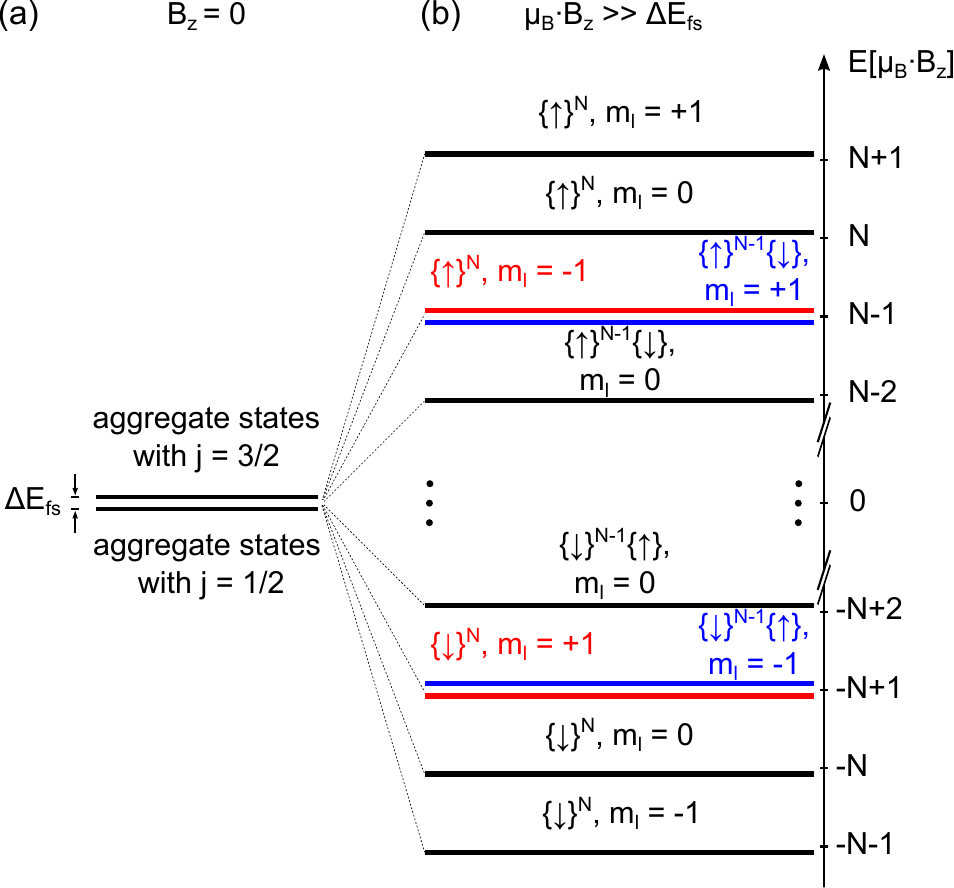} 
\caption{\label{fig:magnetic_field:energy_levels}
Sketch of energy splittings for the aggregate states.
(a) Without magnetic field, the finestructure separates aggregate states with $\lstate{p}_{j=1/2}$-excitation from the ones with $\lstate{p}_{j=1/2}$-excitation. (b) Magnetic field shifts of aggregate states in the strong field regime, where $E_{\mathrm{mf}} = \mu_B B_z\gg \Delta E_{\mathrm{fs}}$. The aggregate states energetically separate with energy gaps $E_{\mathrm{mf}}$ for neighboring states with $(\Delta M_S,\Delta m) = (0, \pm 1)$ and energy gaps $\sim \Delta E_{\mathrm{fs}}$, for neighboring states with $(\Delta M_S,\Delta m) = (\pm 1, \mp2)$ (energy gap between neighboring blue and red lines).
The notation $\{\uparrow\}^{N_\uparrow}\{\downarrow\}^{N_\downarrow}$ is a short form for a spin configuration with $N_\uparrow$ spins oriented upwards ($m_s=1/2$) and $N_\downarrow$ spins oriented downwards ($m_s=-1/2$). 
}
\end{figure}

The spin-orbit coupling leads to a finestructure\index{finestructure} of \lstate{p} states with different azimuthal quantum number, $j$, of the total angular momentum, $\hat{\vc{J}} = \hat{\vc{L}} + \hat{\vc{S}}$. We denote the energy splitting between \lstate{p} states with $j=1/2$ and $j=3/2$ with $\Delta E_{\mathrm{fs}}$. The single \lstate{p} excitation manifold is thus also energetically split by $\Delta E _{\mathrm{fs}}$.
Considering the spins of the atoms, the interaction Hamiltonian describing the coupling of atoms with a magnetic field pointing in $z$ direction can be written as
\begin{equation}
 \hat{H}_{\mathrm{mf}}(B_z) = \mu_\mathrm{B}B_z\sum_{\alpha=1}^{N}\hat{L}_{z}^{(\alpha)} + 2\hat{S}_{z}^{(\alpha)},
\label{eq:H_MF}
\end{equation}
where $\mu_\mathrm{B}$ is the Bohr magneton\index{Bohr magneton}\nomenclature[A]{$\mu_\mathrm{B}$}{Bohr magneton\nomunit{$9.27400968(20)\cdot 10^{24}$ J/T $= 1.4$ MHz/G}}, $B_z$, $\hat{L}_{z}^{(\alpha)}$ and $\hat{S}_{z}^{(\alpha)}$ denote the magnetic field, orbital angular momentum and spin component in $z$ direction, respectively, and $\alpha$ labels the atom.
Evaluation of the energy shifts due to the magnetic field requires an extension to many-body spin states. It is sufficient to label the magnetic quantum number of each spin which we denote with $m_s^{(\alpha)}$ for the $\alpha$th atom. A spin configuration for the aggregate is uniquely defined by the tuple $\vc{M}_S:=[m_{s}^{(1)}\dots m_{s}^{(N)}]^\mathrm{T}$. We denote the corresponding state with $\ket{\vc{M}_S}:=\ket{m_s^{(1)}}_{(1)}\dots\ket{m_s^{(N)}}_{(N)}$, which is the product state of all single atom spin states, labeled for the $\alpha$th atom with $\ket{m_s^{(\alpha)}}_{(\alpha)}$.
Introducing the quantum number for the $z$ component of the 
aggregate spin,
\begin{equation}
 M_S(\vc{M}_S) := \sum_{i=1}^N m_s^{(\alpha)},
\label{eq:M_S_def}
\end{equation}
which is the sum over all individual spin quantum numbers, the energy shift for aggregate states of the form $\ket{\pi_{\alpha},m,\vc{M}_S}$ in the magnetic field is given by 
\begin{equation}
\Delta E_{\mathrm{mf}}(B_z,M_S,m) = \mu_\mathrm{B}B_z (2 M_S + m),
\label{eq:delta_E_MF}
\end{equation}
with $m$ the orbital magnetic quantum number of the \lstate{p} states.
The detuning between the $m=1$ and $m=-1$ states inside a single $M_S$-manifold is $2\mu_\mathrm{B} B_z$. An effective decoupling of both $m$-manifolds is achieved if the detuning between them is significantly larger than the squared coupling elements. If furthermore the magnetic field shifts, $E_{\mathrm{mf}}=\mu_\mathrm{B} B_z$ are much larger than the finestructure splitting\index{finestructure}, than a strong field regime is attained, where spin and angular momentum couple separately to the magnetic field, which effectively removes the finestructure and gives an energy level structure sketched in \fref{fig:magnetic_field:energy_levels}(b). The energy spacing in this strong field regime are $E_{\mathrm{mf}}=\mu_\mathrm{B} B_z$, which is of the order of $\sim100..250$~MHz for adequate magnetic field strengths.

The finestructure\index{finestructure} furthermore yields doublets for neighboring states with $(\Delta M_S, \Delta m)= (\pm 1,\mp 2)$, sketched as red and blue lines in \fref{fig:magnetic_field:energy_levels}(b). The only singlet states with $m\neq 0$ are for completely upwards or downwards oriented spins and the \lstate{p} orbital oriented in the same direction: $(M_S,m) = (N/2,1)$ and $(M_S,m)=(-N/2,-1)$. Hence we concentrate on the $(M_S,m) = (N/2,1)$ manifold, which can be well addressed during the Rydberg excitation process.
The magnetic field yields increasing decoupling of the $(M_S,m) = (N/2,1)$ manifold from the $(M_S,m) = (N/2,-1)$ manifold with increasing magnetic field strength. We study this decoupling in detail, giving the Hamiltonian structure and derive effective interactions in \sref{part:app::chap:tun_int_mag_field__planar_strong_fields}. The only coupling of the $(M_S,m)= (N/2,1)$ state to other manifolds than $(M_S,m) = (N/2,-1)$ is through spin-orbit interactions and can thus be neglected.
%
%
\subsection{Planar Rydberg aggregates in strong magnetic fields}
\label{part:app::chap:tun_int_mag_field__planar_strong_fields}
%
Here we derive the Hamiltonian for a Rydberg aggregate in an external magnetic field pointing in the $z$ direction, where the magnetic field shift is much larger than the finestructure. This strong field regime allows a reduction of the electronic Hilbert space to a single spin manifold. All atoms are assumed to be located within a plane, as in all cases considered here, with quantization axis perpendicular to that plane. Then, the $m=0$ manifold is already completely decoupled\cite{moebius:cradle}. We are interested in the $m=\pm 1$ states and thus neglect the $m=0$ manifold in the following. 
We use the decomposition
$\ket{\pi_{k},m} = \ket{\pi_{k}}\otimes \ket{m}$ for the aggregate states in the following section, where $\ket{-1}, \ket{1}$ are the states for magnetic quantum number $m=\pm 1$ of orbital angular momentum.
We consider two Hilbert spaces: The ``pure'' aggregate space, $\spaces$, spanned by the basis $\basiss := \{\ket{\pi_{k}}\}_{k=1}^N $ and the space including the angular momentum magnetic quantum numbers, $\spacel$, spanned by $\basisl := \{\ket{-1},\ket{1}\}\otimes\basiss$.

The dipole-dipole interaction Hamiltonian with magnetic field shift for \emph{fixed} $m=\pm 1$ is given by
\begin{equation}
\hat{H}_{m}(\vc{R}) :=\hat{H}_{\mathrm{el}}(\vc{R}) + m E_{\mathrm{mf}}\id[{\spaces}],
\end{equation}
with $\hat{H}_{\mathrm{el}}(\vc{R})$ the Hamiltonian defined in \eref{eq:planar_aggr:setup_single_p_state__def_Hel} and $\id[{\spaces}]$ the identity operator, acting on states of $\spaces$. The electronic Hamiltonian $\hat{H}_{\mathrm{el}}(\vc{R})$ is the restriction of the full anisotropic Hamiltonian to aggregate states with fixed $m$, such that $\ket{m}\bra{m} \otimes \hat{H}_{m}(\vc{R}) = \hat{H}(\vc{R},B_z)\bigr|_{m}$, with $\hat{H}(\vc{R},B_z)$ the Hamiltonian given in \eref{eq:tun_int_mag::hamiltonian_aniso_mag}.

The dipole-dipole transitions from $m=1$ to $m=-1$ are described by
\begin{equation}
 \hat{W}(\vc{R}) := \frac{\drad^2}{2}\sum_{\substack{\alpha,\beta=1\\ \alpha \neq \beta}}^N R_{\alpha \beta}^{-3}\mathrm{e}^{-2\im \phi_{\alpha \beta}}\ket{\pi_{\alpha}}\bra{\pi_{\beta}},
\end{equation}
where $R_{\alpha \beta}=|\bv{R}_{\alpha \beta}|$ and $\phi_{\alpha \beta}$ are the modulus and azimuthal angle of the separation $\bv{R}_{\alpha \beta}$ between atoms $\alpha$ and $\beta$, 
within the co-ordinate system defining our quantization axes.

We now treat $\hat{H}_{m}(\vc{R})$ as unperturbed system and $\hat{W}(\vc{R})$ as perturbation and set up operators:
\begin{align}
 \hat{\mathcal{H}}_0(\vc{R})&:=\sum_{m \in \{-1,1\}} \ket{m}\bra{m}\otimes\hat{H}_{m}(\vc{R}),
\label{eq:H0_big}\\
\hat{\mathcal{W}}(\vc{R})&:=\ket{1}\bra{-1}\otimes \hat{W}(\vc{R}) + \ket{-1}\bra{1}\otimes \hat{W}^{\dagger}(\vc{R}),
\label{eq:W_MF_big}\\
\hat{\mathcal{H}}(\vc{R})&:=\hat{\mathcal{H}}_0(\vc{R}) + \hat{\mathcal{W}}(\vc{R}).
\label{eq:H_MF_big}
\end{align}
The Hamiltonian \eref{eq:H_MF_big} describes Rydberg aggregates with magnetic field shifts for the $m=\pm 1$ states, but neglecting finestructure\index{finestructure} shifts. Rewriting it in block-structure with block basis $\ket{1}, \ket{-1}$,
\begin{equation}
 \hat{\mathcal{H}}(\vc{R}) = 
\begin{pmatrix}
\hat{H}_{1}(\vc{R})& \hat{W}(\vc{R})\\
\hat{W}^{\dagger}(\vc{R}) &\hat{H}_{-1}(\vc{R})
\end{pmatrix} = 
\begin{pmatrix}
\hat{H}_{\mathrm{el}}(\vc{R}) + E_{\mathrm{mf}}\id[{\spaces}]& \hat{W}(\vc{R})\\
\hat{W}^{\dagger}(\vc{R}) &\hat{H}_{\mathrm{el}}(\vc{R}) - E_{\mathrm{mf}}\id[{\spaces}]
\end{pmatrix},
\end{equation}
indicates that a block-diagonalization for large magnetic field shifts is possible, such that a perturbation series in orders of $E_{\mathrm{mf}}^{-1}$ can analytically be derived. The structure of the Hamiltonian is already in the form to apply the block-diagonalization scheme derived in \aref{part:app::chap:block_diag__effective_interactions}. Restricting ourselves to the block $\hat{H}_{1}'\equiv \bra{1}\hat{\mathcal{H}}'\ket{1}$, where 
$\hat{\mathcal{H}}'$ denotes the block-diagonalization of $\hat{\mathcal{H}}$, we get up to second inverse order of the magnetic field shifts
\begin{equation}
 \hat{H}_{1}' \approx \hat{H}_{\mathrm{el}} + E_{\mathrm{mf}}\id[{\spaces}] + \hat{W}\hat{W}^\dagger/2E_{\mathrm{mf}} + \left(\hat{W}\hat{H}_{\mathrm{el}}\hat{W}^\dagger - \{\hat{H}_{\mathrm{el}},\hat{W}\hat{W}^\dagger\}/2\right)/4E^2_{\mathrm{mf}},
\label{eq:Hprimeml1}
\end{equation}
according to \eref{eq:app::blockdiagonalization__perturb_solution_with_V_ops}, where $\{\hat{A},\hat{B}\} = \hat{A}\hat{B} + \hat{B}\hat{A}$ denotes the anticommutator for arbitrary operators $\hat{A},\hat{B}$.
It is useful to rescale all operators to get an intuition when the correction terms get small. To do so we define a maximum dipole-dipole interaction element, $\Eddmax:=\drad^2/(2\Rmin^3)$, with $\Rmin:=\min_{\alpha, \beta: \alpha\neq \beta}R_{\alpha \beta}$ and introduce dimensionless operators $\hat{\tilde{X}}$, by defining $\hat{X} := \Eddmax \hat{\tilde{X}}$, with $\hat{X} \in \{ \hat{H}_{1}', \hat{H}_{\mathrm{el}}, \hat{W}\}$. 
This yields $|\bra{\pi_{\alpha}}\hat{\tilde{H}}_{\mathrm{el}}\ket{\pi_{\beta}}| \leq 1$ and $|\bra{\pi_{\alpha}}\hat{\tilde{W}}\ket{\pi_{\beta}}| \leq 1$. Introducing further a decoupling parameter, $\alpha:=\Eddmax/(2E_\mathrm{mf})$ and setting $E_{\mathrm{mf}}$ as our zero of energy, we find up to $\alpha^2$ the following effective interaction Hamiltonian for $m=+1$:
\begin{equation}
 \hat{\tilde{H}}_{1}' \approx \hat{\tilde{H}}_{\mathrm{el}} + \alpha\hat{\tilde{W}}\hat{\tilde{W}}^\dagger
+\alpha^2 \hat{\tilde{W}}\hat{\tilde{H}}_{\mathrm{el}}\hat{\tilde{W}}^\dagger
-\alpha^2 \{\hat{\tilde{H}}_{\mathrm{el}},\hat{\tilde{W}}\hat{\tilde{W}}^\dagger\}/2
\label{eq:Hprimeml1_2_rescaled}
\end{equation}
For $B_z\to \infty$,  which is equivalent to $\alpha \to 0$, we finally find $\lim_{B_z \to \infty}\hat{\tilde{H}}_{1}' = \hat{\tilde{H}}_{\mathrm{el}}$, where $\hat{\tilde{H}}_{\mathrm{el}}$ is the electronic Hamiltonian defined in \eref{eq:planar_aggr:setup_single_p_state__def_Hel}, which uses negative binary resonant interactions of the form $V(r) = -\drad^2/6r^3$. 

To get an idea, which magnetic field strengths are necessary to suppress the remaining anisotropy, we rewrite the decoupling parameter,
\begin{equation}
 \alpha(\nu,R,B_z) = 731.4 \dfrac{(\nu/40)^4}{(R/\mu\mathrm{m})^3 (B_z/\mathrm{G})},
\end{equation}
which is valid for ${}^7$Li. For the double dimer in \sref{sec:nges4} we used $\nu =44 $ and a minimal distance of $a_1 = 2.16\ \mu $m. The decoupling parameter is smaller than $0.27$ for field strengths above $400$~G. The exciton switch in \sref{sec:exciton_switch} worked with $\nu =80$ and a minimal distance of $a_1 = 6\ \mu $m. For a decoupling parameter smaller than $0.27$, a field strength above $200$~G is required. The specified field strengths uniformly decrease the anisotropy. However, the strength of the anisotropic contributions is dependent on the atomic configuration and therefore a fixed field strength suppresses the anisotropic contributions for some atomic configuration better than for others. We will demonstrate this in the next section, where we finally compare the isotropic model with the analytically derived Hamiltonian in this section, given in \eref{eq:Hprimeml1}. We then proceed to compare both models with the complete Hamiltonian, describing anisotropy of the interactions, an additional applied magnetic field and also the spin-orbit coupling.
%
\subsection{Comparison between isotropic, effective and complete Hamiltonian}
\label{part:app::chap:tun_int_mag_field__compar_iso_eff_compl}
%
In this section we assess how good the isotropic or the effective Hamiltonian (both without spin degrees of freedom) approximate the complete Hamiltonian, which includes spin-orbit coupling in addition to the magnetic field.

We denote the space of the electron spin states of the $\alpha$th~atom with $\spacepinsmaller{\alpha}$, the basis of which is spanned by $\basisspinsmaller{\alpha}~=~\{\ket{-1/2}_{(\alpha)},~\ket{1/2}_{(\alpha)}\}$. The states $\ket{-1/2}_{(\alpha)}$ denote downwards oriented electron spin and $\ket{1/2}_{(\alpha)}$ upwards oriented electron spin for the $\alpha$th~atom.
The space of all $N$ electron spins is then given by $\spaceallspins = \otimes_{\alpha=1}^{N}\spacepinsmaller{\alpha}$, with the product basis $\basissallspins =\otimes_{\alpha=1}^{N} \basisspinsmaller{\alpha}$.\newline
The spin-orbit interaction destroys the decoupling of the $m=0$ states, such that we have to redefine some quantities of \sref{part:app::chap:tun_int_mag_field__planar_strong_fields}. The space $\spacel$ is now spanned by $\basisl:=\{\ket{-1},\ket{0},\ket{1}\}\otimes \basiss$, with $\ket{m}$ the states of the quantum number $m \in \{-1,0,1\}$. The Hamiltonian for the $m=0$ states is given by
\begin{equation}
 \hat{H}_{0}(\vc{R}):=-2\hat{H}_{\mathrm{dd}}(\vc{R}) + \hat{H}_{\mathrm{vdw}}(\vc{R}),
\end{equation}
with $\hat{H}_{\mathrm{dd}}(\vc{R})$ the resonant dipole-dipole Hamiltonian in \eref{eq:elechamiltonian-dd} and $\hat{H}_{\mathrm{vdw}}(\vc{R})$ the off-resonant vdW~Hamiltonian in \eref{eq:planar_aggr:setup_single_p_state__vdw_Hamiltonian}. Note that $\hat{H}_{0}(\vc{R})$ experiences no magnetic field shift. We
redefine $\hat{\mathcal{H}}_{0}(\vc{R})$ from \eref{eq:H0_big}, such that it includes the $m=0$ states:
\begin{equation}
 \hat{\mathcal{H}}_0(\vc{R}):=\sum_{m \in \{-1,0,1\}} \ket{m}\bra{m}\otimes\hat{H}_{m}(\vc{R}).
\label{eq:H0_big_redef}
\end{equation}
The Hamiltonian in \eref{eq:H_MF_big}, $\hat{\mathcal{H}}(\vc{R})$, is now calculated with the redefined $\hat{\mathcal{H}}_0(\vc{R})$.

With these definitions, we can span the complete space $\spacefull := \spacel\otimes \spaceallspins$, describing both, the orientation of the \lstate{p} states and the spins of the electrons. The product basis, where spin and orbital angular momentum of the \lstate{p} states are not combined to a total angular momentum, is then given by $B_{ls}[\spacefull]:=\basisl \otimes \basissallspins$.

The Hamiltonian $\mathcal{H}(\vc{R})$ in \eref{eq:H_MF_big} includes the magnetic field shifts for the orbital angular momentum only. The magnetic field shift for the spins is described by
\begin{equation}
 \mathcal{H}_{\mathrm{mf-s}}:=E_{\mathrm{mf}}\sum_{\ket{\vc{M}_S} \in \basissallspins}M_S(\vc{M}_S)\ket{\vc{M}_S}\bra{\vc{M}_S}
\end{equation}
The dipole-dipole interactions together with the total magnetic field shift is then given by
\begin{equation}
 \hat{\stylecs{H}}_{\mathrm{dd+mf}}(\vc{R}) := \mathcal{H}(\vc{R})\otimes \mathcal{H}_{\mathrm{mf-s}}.
\end{equation}

To set up the spin-orbit interaction Hamiltonian in a simple way, it is useful to employ yet another basis. First we define the spin spaces $\spacepinsmaller{\neq \alpha }$, which describe all spins except those of the 
the $\alpha$th~atom, $\spacepinsmaller{\neq \alpha }:=\otimes_{\beta=1;\beta\neq \alpha}^{N}\spacepinsmaller{\beta}$. Their product basis $\basisspinsmaller{\neq \alpha}$ is spanned by $\basisspinsmaller{\neq \alpha}=\otimes_{\beta=1;\beta\neq \alpha}^{N} \basisspinsmaller{\beta}$.
The spin-orbit coupling yields a total angular momentum, $\hat{\vc{J}} := \hat{\vc{L}} + \hat{\vc{S}}$ per atom. The pair $(j,m_j)$ are the quantum numbers of $\hat{\vc{J}}$, with $j \in \{ 1/2, 3/2\}$ and $m_j \in M_j:=\{ - |j|, - |j|+1, \dots, |j|\}$.
The result of the spin-orbit coupling is the finestructure splitting\index{finestructure} $\Delta E_{\mathrm{fs}}$, between \lstate{p} states with $j=3/2$ and \lstate{p} states with $j=1/2$. To write down the spin-orbit Hamiltonian in its eigenbasis, we first introduce aggregate states which include the spin of the \lstate{p} states, $\ket{\pi_k,j,m_j}:=\ket{s\dots (p,j,m_j)...s}$. We now define spaces $\spacefullj{j} \subset \spacefull$, which we span with the basis $\basissfullj{j}:=\otimes_{\alpha=1}^N \{\ket{\pi_k,j,m_j}\}_{m_j \in M_j}\otimes \basisspinsmaller{\neq \alpha}$. The orthogonal sum of both '$j$-spaces' spans the complete space, $\spacefull=\spacefullj{1/2}\oplus \spacefullj{3/2}$. This yields the eigenbasis of the spin-orbit Hamiltonian,
$B_{j,mj}[\spacefull]:=\basissfullj{1/2}\cup\basissfullj{3/2}$. Introducing the unitary transformation $\hat{\stylecs{U}}$, which performs the basis transformation from $B_{j,mj}[\spacefull]$ to $B_{ls}[\spacefull]$, the spin-orbit Hamiltonian in the basis $B_{ls}[\spacefull]$ is given by
\begin{equation}
 \hat{\stylecs{H}}_{\spinorbitsymbol} = \Delta E_{\mathrm{fs}}\hat{\stylecs{U}}\biggl(\nullop[\spacefullj{1/2}] \oplus \id[\spacefullj{3/2}]\biggr)\hat{\stylecs{U}}^\dagger,
\label{eq:H_SO_complete}
\end{equation}
where $\nullop[\spacefullj{1/2}]$ is the null operator acting on elements in $\spacefullj{1/2}$ and transforming them into its zero. Note that we thus shift the origin of energy to the $j=1/2$ manifold. The complete Hamiltonian is then given by
\begin{equation}
 \hat{\stylecs{H}}(\vc{R}):=\hat{\stylecs{H}}_{\mathrm{dd+mf}}(\vc{R}) + \hat{\stylecs{H}}_{\spinorbitsymbol}.
\label{eq:H_complete}
\end{equation}
\begin{figure}[!t]
\centering
\includegraphics{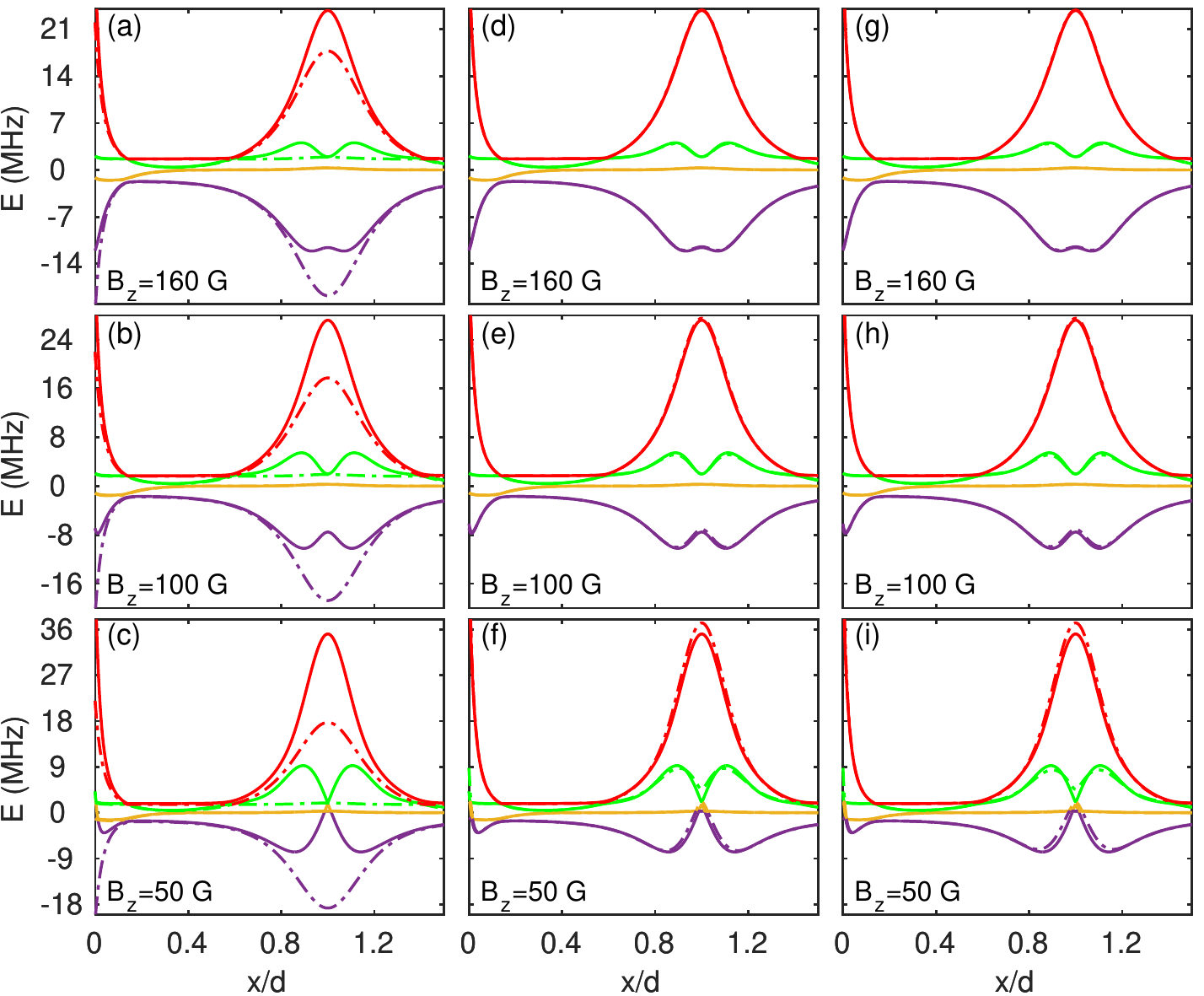} 
\caption{\label{fig:N4_compare_interactions}
 Comparison of energy spectra from different approximations of the complete Hamiltonian, for a double dimer as sketched in \fref{system_sketch_4}(a). We use $\nu=80$, which yields a finestructure splitting\index{finestructure} of $\Delta E_{\mathrm{fs}}=0.15$~MHz~\cite{haroche:li_finesplitting}. Further parameters are $\ b=0, \ d=40~\mu$m, $a_1=8~\mu$m and $a_2=19~\mu$m. The atoms~(3,4) are fixed, whereas the positions of atoms~(1,2) are parameterized, such that $x_1=-x$ for atom~1 and $x_2=a_1+x$ for atom~2. We compare the complete model (solid lines) with different approximate models (dashed dotted lines) in different columns. 
The approximate models are: (a)--(c) purely isotropic model used for the main results in \sref{part:rs::chap:planar_aggregates}, with Hamiltonian given in \eref{eq:planar_aggr:setup_single_p_state__def_Hel}. (d)--(f) Corrected model with effective Hamiltonian according to \eref{eq:Hprimeml1_2_rescaled} up to first order in $\alpha$. (g)--(i) Same as (d)--(f), but using the effective Hamiltonian up to second order in $\alpha$.
We consider three different magnetic field strengths:
$B_z=50$~G in the lower row, $B_z=100$~G in the middle row and $B_z=160$~G in the upper row. The energy of infinitely separated atoms is set to zero.
}
\end{figure}

We compare the three different Hamiltonians in \eref{eq:planar_aggr:setup_single_p_state__def_Hel}, \eref{eq:Hprimeml1_2_rescaled} and \eref{eq:H_complete} by using them to calculate the eigenenergies for a four atom system with a symmetric configuration ($b=0$), as sketched in \fref{system_sketch_4}(a). 
We show for \eref{eq:H_complete} only the $(M_S,m) = (N/2,1)$ manifold, which is the one that we propose to work with.
The positions of atoms~(3,4) are fixed, whereas the positions of atoms~(1,2) are parameterized as $x_1=-x$ and $x_2=a_1+x$. The eigenenergies of all three Hamiltonians are plotted as a function of the co-ordinate $x$ in \fref{fig:N4_compare_interactions}. The isotropic model in \eref{eq:planar_aggr:setup_single_p_state__def_Hel} approximates \eref{eq:H_complete} well for all locations crucial in our simulations, as shown in \frefplural{fig:N4_compare_interactions}{a}{--}{c}. Crucial for the simulations are configurations with small $x$ values, where the atoms are accelerated due to the interactions, and in the neighborhood of the conical intersection. The agreement is not good for the equidistant linear trimer configuration $x = d$. The excitation is there mostly delocalized and the phase of the dipole-dipole interaction plays a role.
The effective Hamiltonian in \eref{eq:Hprimeml1_2_rescaled} approximates the complete one for this configuration very well, as shown in \frefplural{fig:N4_compare_interactions}{d}{--}{g}. It appears that the Hamiltonian of order $\alpha$ in \eref{eq:Hprimeml1_2_rescaled} approximates \eref{eq:H_complete} better than the order $\alpha^2$ version. This may be since \eref{eq:Hprimeml1_2_rescaled} does not take the spin-orbit coupling into account and the finestructure\index{finestructure} is of the order of the $\alpha^2$ corrections. A better description beyond the $\alpha$ correction would then require a block-diagonalization, which explicitly includes spin-orbit coupling.

As expected, increasing the magnetic field strength improves the decoupling of the $(M_S,m) = (N/2,1)$ manifold. This results in a better agreement between the reduced models and the complete model for higher field strengths.

\addchap{Conclusions and Outlook}\label{part:smry} 
The fast progress in cooling, trapping and coherent excitation of Rydberg atoms is opening up the field of quantum simulation with Rydberg atoms. Whereas the focus so far was mainly on quantum information processing or simulators for condensed matter Hamiltonians, this thesis demonstrates suitability of flexible Rydberg aggregates as test bench for quantum transport. To this end we have investigated here flexible Rydberg aggregates in higher-dimensional arrangements.
Quantum transport is typically a feature of molecular aggregates. For plants it is important to absorb photons and ultimately perform with the gained energy photosynthesis. Before the energy can be converted, LHCs\index{light-harvesting complex} transfer photo-induced excitation resonantly to certain reaction centers. Under debate is whether quantum features are essential for this transport. 
The numerous degrees of freedom complicate the identification of the essential components of transport. Nevertheless, efficient models explained already many transport features in molecular aggregates, e.g. for LHCs\index{light-harvesting complex}.
The difficulty is to adequately take into account the strong coupling between electronic and nuclear degrees of freedom for latter. As a result, models are often designed to contain either energy transport or CI-dynamics. With flexible Rydberg aggregates we found a toy model which allows us to investigate both.
Specifically in this thesis, we were able to study the effect of CIs on exciton pulses.
The feasibility to consider both features is in fact a result of using flexible Rydberg aggregates, which are simple enough in their structure to solve combined dynamics of spatial and electronic degrees of freedom with the help of quantum-classical methods, such as FSSH, and to treat the strong interactions non-perturbatively. In fact, the observed exciton pulses\cite{wuester:cradle,moebius:cradle,leonhardt:switch,leonhardt:orthogonal} rely on an adequate description of the coupling between electronic states and nuclear degrees of freedom. The pulses are initiated by preparing a strong diatomic proximity of an atom pair, which also completely localizes the excitation on this dimer.
 Such a large displacement is hard to capture with typical phonon modes. Early studies\cite{weidlich:pulse,davydov:soliton} already revealed combined exciton-phonon pulses. However, in these models the lattice displacements affect only the on-site excitation energies and not the transition matrix elements of the excited state manifold and for this reason they differ in the way of coupling electronic and nuclear degrees of freedom.
Furthermore, a restriction to nearest neighbor interactions excludes the investigation of CIs.
The ability to fully treat the interactions between nuclear and electronic degrees of freedom in flexible Rydberg aggregates allowed us to discover the coherent splitting of an exciton pulse caused by a CI. The coherent superposition is on the scale of several micrometers and therefore in a regime were physics is usually classical.
Recently, an experiment demonstrated coherent superpositions even on the half-metre scale\cite{kovachy2015:quantum_superposition_half_meter}. However, the exciton pulse splitting demonstrated in this thesis differs from other demonstrations of the quantum superposition principle by superimposing both exciton states and distinct spatial states on a mesoscopic scale simultaneously.
\begin{figure}[!t]
\centering
\includegraphics{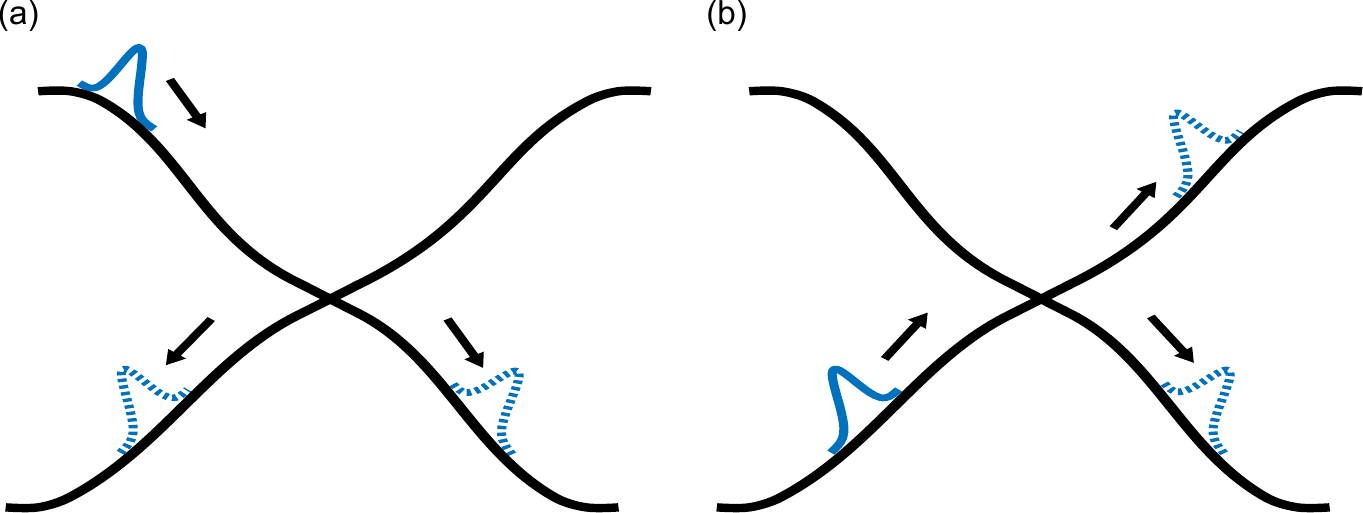}
\caption{\label{branching_CI}
Comparison of different CI splitting mechanisms.
Illustration of a wave packet~(blue, solid line) and its splitting into parts~(blue, dashed line) by hitting a CI of two BO~surfaces(black lines). (a) A wave packet approaches the CI on an excited surface and splits such that two partial wave packets evolve on an energetically lower BO~surface. The CI serves as an ultrafast decay channel as observed in photodissociation processes\cite{polli2010:ci_photoisomerization,domcke2012:review_ci}.
 (b) A wave packet carries kinetic energy and can therefore approach the CI from a lower-lying state. After the splitting, both excited- and lower-lying states are populated and hence the CI is utilized to partly populate the excited state. This type of splitting occurs for instance in quantum reactive scattering processes\cite{wrede1997:reaction_H_D2_CI,kendrick2000:geometric_phase_effects_H_D2,adhikari2002:nonadiabatic_effects_chemical_reactions,kendrick2003:geometric_phase_effects_chem_reac_dynamics} and is also the type of wave packet splitting observed for the studied T-shape aggregates in this thesis.
}
\end{figure}
We were able to demonstrate exciton splitting for both planar and unconstrained aggregates. For the planar aggregates the orientation of the \lstate{p} orbitals was fixed along a chosen direction in the plane of the aggregate. For the unconstrained aggregate, we varied the alignment of the \lstate{p} orbitals and observed a dependency of the exciton splitting on this variation. When the orbital of the excitation is aligned perpendicular to the aggregate's plane of initial configuration, propagation occurs along a decoupled sub-manifold of the electronic Hamiltonian with isotropic interactions. However, in contrast to the interaction model employed for the planar aggregates, the amplitude of the binary interactions is positive, such that the repulsive surface does not feature a CI and therefore no exciton splitting occurs. An alignment of the excitation orbital within the plane genuinely features anisotropy of the interactions which is large in the vicinity of configurations with equidistant diatomic spacings. The consequence is a CI which splits the exciton, similarly as for the planar aggregates. 

In exciton splitting the CI does not act as an ultrafast decay channel, as it is depicted in \fref{branching_CI}(a), and would for instance be the case for the photoisomerization of rhodopsin in vision. Further, the exciton pulse splitting is similar to quantum reactive scattering\cite{wrede1997:reaction_H_D2_CI,kendrick2000:geometric_phase_effects_H_D2,adhikari2002:nonadiabatic_effects_chemical_reactions,kendrick2003:geometric_phase_effects_chem_reac_dynamics} processes where the wave packet approaches the CI from a lower-lying BO~surface with the help of kinetic energy. The splitting partly populates both the excited and the lower neighboring BO~surface, as depicted in \fref{branching_CI}(b). 

This orbital sensitivity of the exciton splitting might be useful to draw conclusions about the initial orbital orientation of excitation and to even measure it without the need of ionizing the atoms. Detecting the atomic density spatially resolved with a CCD camera, perpendicularly oriented to the aggregate's plane of initial configuration and orthogonal to the initial pulse propagation could distinguish between both orbital orientations of the excitation. The experiment has to be repeated several times with the same excitation scheme which subsequently resamples the atomic density. Counting the number of spatially connected domains of atomic hits finally distinguishes both alignments of the excitation orbital.
%
%
%
%

With the extended planar T-shape aggregates with more atoms on the vertical Rydberg chain we investigated the possibility to establish exciton pulse propagation after redirecting it on an orthogonal direction. We observed no pulse propagation for symmetric T-shape aggregates where the pulse exactly hits the CI. The conclusion is that adiabaticity is essential for the pulse propagation. However, the knowledge of how the CI affects the pulse can be used to alter the aggregates' geometry in order to achieve exciton pulse propagation after redirection. The key is to preset an asymmetry of the horizontal chain relative to the vertical chain, such that the pulse avoids exactly hitting the CI. Additionally, we are able to select the direction of the pulse on the vertical chain by varying the interatomic distance of the two atoms closest positioned to the horizontal chain. This distance controls whether a trimer subunit of these two atoms and the approaching excitation carrying atom from the horizontal chain is formed or not. A large spacing prevents the formation of the subunit and ensures that the pulse propagates over the whole time along the repulsive surface, since the energy spacing to other BO~surfaces remains large. On the other hand, for smaller interatomic distances of the inner vertical dimer the trimer subunit is formed but the preset asymmetry ensures that the trimer configurations sufficiently deviates from the equilateral configuration where the CI is located and hence the energy gap to the neighboring BO~surface remains large enough to guide the pulse almost adiabatically on the adjacent surface. Therefore, the diatomic spacing of the inner vertical dimer switches between either populating the repulsive surface or the adjacent surface and consequently controls the direction of exciton pulse propagation on the vertical chain.
We could confirm high fidelity pulse propagation by measuring the bipartite entanglement on both ends of the vertical chain.

\section*{Future perspectives}
The results obtained in this thesis have stimulated further questions. In the following we present some interesting ideas for further investigations.

\subsection*{Exciton pulses with superatoms}
Rydberg aggregates in this thesis are based on single atoms as constituents. Trapping single atoms per lattice site in an optical lattice is already experimentally feasible\cite{schauss2012:observation_spatially_ordered_structures_Rydberg_gas,parsons2015:site_resolved_imaging_Li,greif2015:site_resolved_imaging_fermionic_mott}. However, preparation of specific spatial arrangements, e.g. T-shape aggregates which we studied in this thesis, might still be a difficult experimental task. To overcome this issue, spatially separate atom clouds could be realized as a replacement for single atoms. The dipole blockade ensures that a single Rydberg excitation is coherently shared between all atoms within each cloud such that a superatom emerges\cite{robicheaux2005:superatom,vuletic2006:superatoms,deiglmayr2006:superatoms_evidence,heidemann:strongblockade,johnson2008:rabi_oscilltation_ground_rydberg,reetzlam:rabiblockade,dudin2012:superatom,barredo2014:superatom,zeiher2015:superatoms} with similar properties to a single Rydberg atom.
To ensure that superatoms can move similar to a single atom\cite{moebius:cat,genkin:dressedbobbles,moebius2015:diss}, they are allowed to only contain a small fraction of Rydberg excitation. This is achieved by dressing techniques\cite{nils2010:supersolids,pupillo:strongcorr:dressed,wuester:dressing} where the coupling to Rydberg states is off-resonant. So far a dimer based on Rydberg dressed superatoms was studied\cite{moebius:cat,genkin:dressedbobbles,moebius2015:diss}. An interesting question is whether exciton pulse propagation could be realized with superatoms. Following studies could investigate the possibility for exciton splitting.

\subsection*{Entangled spin transport on mesoscopic scales}
Generation of exciton pulses was theoretically demonstrated with lithium atoms so far. The advantage of using Lithium is that spin-orbit interactions can be neglected. 
However, we expect interesting new physics from studying flexible Rydberg aggregates based on Rubidium which exhibits strong spin-orbit coupling. Dipole-dipole interactions do not directly operate on the spins of the electron as long as spin-orbit coupling plays no role.
In contrast, the transfer of orbital momentum excitation in Rydberg aggregates based on Rubidium might open the way to also transfer entangled spin states. In this way couplings between spin and spatial degrees of freedom are effectively introduced. Flexible Rydberg atoms based on Rubidium are also appropriate from an experimental point of view, simply because most ultracold experimental setups employ them.

\subsection*{Investigation of Rydberg dressed exciton-phonon pulses}
Experiments with ultracold atoms reached the point to trap single atoms in optical lattices and enables the possibility to study exciton-phonon pulses for different spatial structures. Systems with a large number of atoms would 
require Rydberg dressed ground state atoms\cite{nils2010:supersolids,henkel2013:phd_thesis,nils2010:supersolids,wuester:dressing}, since they have a longer lifetime compared to Rydberg atoms. For dressed states, the possibility for exciton pulse propagation was demonstrated for a linear five atom chain\cite{wuester:dressing} so far. Studying the dynamics in an optical lattice would allow for expanding the spatial dynamics around the equilibrium configuration, such that the spatial dynamics can be described as an expansion of phonon modes. Taking into account terms beyond the harmonic expansion\cite{hooton1955:anharmonic_lattice,born1955:stats_dyns_several_periodic_systems,cowley1963:lattice_dynamics_anharmonic_crystal} might be crucial for a complete localization of an angular momentum excitation between an atom and its nearest neighbors, as pointed out earlier. Nonlinear lattice dynamics is responsible for many interesting physical phenomena such as soliton formation and enhanced superconductivity\cite{fleischer2003:2d_sollitons_nonlinear_photonic_lattices,chen2014:nonlinear_lattice_dynamics_pbte,mankowsky2014:nonlinear_lattice_dynamics_superconductivity}. However, combined exciton-phonon pulses in the nonlinear regime were not studied so far. Ultracold Rydberg dressed atoms in an optical lattice could serve as a toy model for such investigations and to derive conclusions about genuine condensed matter systems.

\subsection*{Investigations of three-state CIs}
In this thesis CIs appeared between two BO~surfaces. For three\hyp{}dimensional configurations, three BO~surfaces can conically intersect, which were located for certain\linebreak[4]molecules\cite{matsika2003:three_stateCI,matsika2005:three_stateCI_nucleic_acid_bases,matsika2008:three_stateCI_uracil_cation,kistler2008:three_stateCI_pyrimidinone}. However their role and impact for excited state dynamics is not completely clear. With the help of flexible Rydberg aggregates a study of three-state CIs could be performed with the advantage that the intersection can be studied in a reduced system where influences from an environment can be highly suppressed. A model system would be a four atom aggregate with tetrahedral configuration. An interesting question is whether excitation can be quickly transferred through a sudden transition to a exciton state mediated by a three-state CI.

\subsection*{Comparison of the observed exciton splitting with singlet fission in organic semiconductors}
One of the main result of this thesis is the coherent splitting of an exciton pulse through a CI. In the following we want to compare this process with an exciton splitting process which occurs in organic semiconductors.
%
The mechanism behind solar cells is the creation of excitons by absorbing photons and a subsequent charge separation. The resulting energy difference at the electrodes eventually creates a desired electric potential. High efficiency of solar cells necessitates stable charge separation, implying an exciton lifetime longer than the migration time to the electrodes, which is equivalent to a sufficient mobility of the excitons. The materials used for organic semiconductors lead to a population of molecular orbitals with an overall net spin zero or one, respectively, which consequently creates singlet~(net spin zero) and triplet~(net spin one) excitons\cite{koehler2009:triplet_excitons_review,mikhnenko2012:thesis_singlet_triplet_excitons}.
Although the number of triplet states is three times the number of singlet states due to the quantum mechanical rules of angular momentum addition, triplet excitons usually do not contribute to charge separation, because they are not likely to be excited by absorbed photons and furthermore a direct excitation leads to small mobility of them and the tendency for non-radiative decay\cite{friend1999:electroluminescence_conjugated_polymers}. This limits the efficiency of solar cells already to $25\%$. A more precise theoretical calculation of an upper bound for the efficiency is the \emph{Shockley-Queisser limit}\cite{shockley1961:shockley_queisser_limit}, which also includes the lack of triplet excitons for charge separation. However, for particular materials, a process called \emph{singlet fission}\cite{smith2010:singlet_fission_review,smith2013:singlet_fission_review2} turns singlet excitons into two triplet excitons conserving the total net spin to zero. This process doubles the number of charge carriers and finally overcomes the difficulties of a direct excitation of triplet exciton by creating excitons which are able to move apart such that they do not annihilate. Ultimately, excitons are generated with a long lifetime. Singlet fission can be observed in organic materials in which the molecular singlet excited state has an energy of at least twice the one of the triplet excited state, such that the process is exergonic\cite{chan2012:singlet_fission_exergonic}.  It was theoretically shown that it increases the maximally allowed efficiency to $44\%$, hence it circumvents the Shockley-Queisser limit by far. Based on this predictions, experimental evidence for a drastically increased efficiency was found\cite{hanna2006:solar_conversion_efficiency_photovoltaic,wilson2011:ultrafast_singlet_fission_pentacene,jadhav2012:triplet_exciton_dissociation,ehrler2012:singlet_exciton_fission-sensitized_infrared_quantum_dot,yost2014:transferable_model_singlet_fission_kinetics,congreve2013:quantum_efficiency_above_100_singlet_exciton_fission}. For a long time the detailed mechanism of singlet fission was unclear and several mechanisms were under debate\cite{paci2006:singlet_fission_dye_sensitized_solar_cells,zimmerman2010:singlet_fission_pentacene,greyson2010:maximizing_singlet_fission_organic_dimers,chan2011:observing_multiexciton_state_singlet_fission,zimmerman2011:mechanism_singlet_fission_pentacene,zimmerman2013:correlated_electron_view_singlet_fission,beljonne2013:charge-transfer_excitations_davydov_splitting_single_exciton_splitting,renaud2013:singlet_fission_suggested_CI,walker2013:singlet_exciton_fission_solution}. Recent experiments found evidence for a CI mediating the process\cite{musser2015:singlet_fission_CI} and could verify that strong couplings between nuclear and electronic degrees of freedom are crucial. 
However, the couplings have to be optimal in the sense that on the one hand they need to be strong enough to enable the exciton splitting process but on the other hand they should be sufficiently weak to prevent annihilation of both triplet excitons.

Although our model of exciton dynamics is very simple and the excitons here are of a different kind than in organic semiconductors, we are also able to demonstrate exciton splitting caused by a CI. The difference is that in the process of singlet fission the number of charge carriers is increased, whereas the exciton splitting in T-shape Rydberg aggregates as demonstrated in this thesis does not multiply the number of excitons. However, the comparison of both processes reveals some similarities. For both, high energetic excitons are converted into low energetic ones which therefore suppresses thermalization losses. Furthermore they share a reorganization of the exciton structure.
It is a further open question whether exciton fission can be realized with Rydberg aggregates. To dynamically change the number of excitons, the constituents should feature molecular orbitals such that more than one electron are available to be excited. This could for instance be realized with aggregates of interacting Rydberg molecules\cite{Greene:LongRangeMols,boisseau2002:rydberg_molecules_dimer,liu:prediction_trylobite_ryd_mol,rittenhouse2010:polyatomic_ryd_atoms,bendkowsky2009:rydberg_molecules_observ,bendkowsky2010:rydberg_trimer_quantum_refl,butscher2010:rydberg_molecule,butscher2011:phd_thesis,nipper2012:phd_thesis,li2011:homonuc_ryd_mol_perm_electr_dip}.

\begin{appendices}
\chapter{Calculations for chapter 'Rydberg atoms'}\label{part:app::calcs_ryd_atoms}
\section{Adjusting dipole matrix elements for the treatment of spin-orbit coupling}
\label{part:app::calcs_ryd_atoms__spherical_matrix_element}
In \sref{part:fd::chap:ryd_a::sec:alk_ryd_a:calc_dip_matrix_elements}, we presented the evaluation of the dipole matrix elements without considering finestructure. Here we show the calculation accounting for finestructure.
We only have to use the translation from the orbital angular momentum basis into the total angular momentum basis. Let $\ket{\Gamma_{\spinorbitsymbol}} = \ket{\nu,\ell,j,m_j}$ be a bound state ket of an alkali or hydrogen atom, the finestructure of which is considered. These kets are related to the product basis of spin and the kets of the atom without including finestructure, $\ket{\nu,\ell,m}$, in the following way:
\begin{equation}
 \ket{\nu,\ell,j,m_j} = \sum_{m_s=\pm 1/2}\ClebchGordanSymb_{\ell,m_j-m_s;s,m_s}^{j,m_j}\ket{\nu,\ell,m=m_j-ms}\ket{m_s}.
\label{eq:app::calcs_ryd_atoms__SO_ket_in_LS_basis}
\end{equation}
This is the same transformation as between the spherical harmonics and its generalized versions, given in \eqref{eq:ryd_atoms:def_generalized_spherical_harmonics}. The transformation is the same here, since we neglect a dependency of the bound state radial wave functions on the quantum numbers of the total angular momentum, such that only the spherically dependent part of the wave functions differ. This yields for the dipole matrix elements
\begin{equation}
 \braket{\hat{\vc{d}}}_{\Gamma_{\spinorbitsymbol}}^{\Gamma_{\spinorbitsymbol}'} = \dreduced_{\nu,\ell}^{\nu',\ell'}\tilde{\vc{D}}_{\ell,j,m_j}^{\ell',j',m_j'},
\label{eq:app::calcs_ryd_atoms__dipole_matrix_elements_SO_basis}
\end{equation}
where $\dreduced_{\nu,\ell}^{\nu',\ell'}$ is the radially dependent reduced matrix element, defined in \eref{eq:alk_ryd_atoms:reduced_matrix_element}.
The formula of dipole matrix elements without the treatment of spin-orbit interactions, given in \eref{eq:alk_ryd_atoms:radial_matrix_element} has be adjusted only in the vector, containing the anisotropy, from $\vc{D}_{\ell,m}^{\ell',m'} \to  \tilde{\vc{D}}_{\ell,j,m_j}^{\ell',j',m_j'}$, where $\vc{D}_{\ell,m}^{\ell',m'}$ is defined in \eref{eq:alk_ryd_atoms:anisotropy_dipole_matrix_element} and 
$\tilde{\vc{D}}_{\ell,j,m_j}^{\ell',j',m_j'}$ is given by
\begin{equation}
 \tilde{\vc{D}}_{\ell,j,m_j}^{\ell',j',m_j'} = \sum_{m_s=\pm 1/2} \ClebchGordanSymb_{\ell',m_j'-m_s;s,m_s}^{j',m_j'}\ClebchGordanSymb_{\ell,m_j-m_s;s,m_s}^{j,m_j}\vc{D}_{\ell,m_j-m_s}^{\ell',m_j'-m_s}.
\label{eq:app::calcs_ryd_atoms__anisotropy_dipole_matrix_element_SO_basis}
\end{equation}

\section{Coupling of spherical harmonics}
\label{part:app::calcs_ryd_atoms__coupling_spherical_harmonics}
For the evaluation of matrix elements of the dipole-dipole interactions, it is suitable to recouple the product of two spherical harmonics, which have the same argument, into a sum of single spherical harmonics. Suppose we have the spherical harmonics $Y_{\ell_1,m_1}$ and $Y_{\ell_2,m_2}$, where in general we use the short notation $Y_{\ell,m}:=Y_{\ell,m}(\theta,\phi)$, such that we skip the arguments.
The product of both can be expanded in the following way:
\begin{equation}
 Y_{\ell_1,m_2}\cdot Y_{\ell_2,m_2} = \sum_{\ell = 0}^{\infty}\sum_{m=-\ell}^{\ell} c_{\ell,m}Y_{\ell,m},
\label{eq:app::coupl_spherical_harmonics__general_expansion_scheme}
\end{equation}
with the expansion coefficients
\begin{align}
 c_{\ell,m} &= \int \dscalar \Omega\, Y^{*}_{\ell,m}(\theta,\phi) Y_{\ell_1,m_2}(\theta,\phi)Y_{\ell_2,m_2}(\theta,\phi)
\label{eq:app::coupl_spherical_harmonics__expansion_coeff_general_integral}\\
&=\sqrt{\dfrac{3}{4\pi}}\sqrt{\dfrac{2\ell_2+1}{2\ell+1}} \ClebchGordanSymb_{\ell_1,0;\ell_2,0}^{\ell,0} \ClebchGordanSymb_{\ell_1,m_1;\ell_2,m_2}^{\ell,m}.
\label{eq:app::coupl_spherical_harmonics__expansion_coeff_general_eval}
\end{align}
An important special case is for $\ell_1=\ell_2=1$, where we get
\begin{equation}
 Y_{1,m_1}\cdot Y_{1,m_2} = \dfrac{(-1)^{m1}}{4\pi}\delta_{m_1,-m_2} + \sqrt{\dfrac{3}{10\pi}}\,\ClebchGordanSymb_{1,m_1;1,m_2}^{2,m_1+m_2}\,Y_{2,m_1+m_2}.
\label{eq:app::coupl_spherical_harmonics__special_case_Y1m1_Y1m2}
\end{equation}

\section{Evaluation of matrix elements of dipole-dipole interactions}
\label{part:app::calcs_ryd_atoms__eval_matrix_elements_dip_dip_interaction}
In \sref{part:fd::chap:ryd_a::sec:dip_dip_int}, we presented in \eref{eq:alk_ryd_atoms:dip_dip_interaction_qm} the dipole-dipole Hamiltonian. Its first term is a purely isotropic interaction, whereas the second term makes the interaction anisotropic in general. We can evaluate the second term further by using for both, the dipoles and the distance vector, their spherical representation. Remember that the distance vector has the spherical representation $\vc{R}_{12} = \sqrt{\frac{4\pi}{3}}R_{12}\sum_{\mu=\pm 1,0} Y_{1,\mu}\vc{b}_{\mu}$ and the scalar product of two vectors $\vc{a},\vc{b}$ with spherical representation is given by $\braket{\vc{a},\vc{b}}:=\sum_{\mu=\pm 1,0}a_{\mu}^{*}b_{\mu}$, where $x_\mu = \braket{\vc{b}_\mu,\vc{x}}$ for all $\vc{x} \in \mathbb{C}^3$.
With this, we can write the second term of \eref{eq:alk_ryd_atoms:dip_dip_interaction_qm} as:
\begin{align}
 \dfrac{3}{R_{12}^5}\braket{\vc{R}_{12},\hat{\vc{d}}^{(1)}} \braket{\vc{R}_{12},\hat{\vc{d}}^{(2)}} = \dfrac{3}{R_{12}^5}\dfrac{4\pi}{3}R_{12}^2\sum_{\mu,\mu' = \pm 1,0}Y_{1,\mu}^{*}Y_{1,\mu'}^{*}\hat{d}_{\mu}^{(1)}\hat{d}_{\mu'}^{(2)}.
\end{align}
Using the formula in \eref{eq:app::coupl_spherical_harmonics__special_case_Y1m1_Y1m2} for the product $Y_{1,\mu}^{*}Y_{1,\mu'}^{*}$, we get
\begin{align}
 3\dfrac{\braket{\vc{R}_{12},\vc{F}^{(1)}} \braket{\vc{R}_{12},\vc{F}^{(2)}}}{R_{12}^3} &= \dfrac{4\pi}{R_{12}^3}\sum_{\mu,\mu' = \pm 1,0}\left\{ \dfrac{(-1)^{\mu}}{4\pi}\delta_{\mu,-\mu'} + \sqrt{\dfrac{3}{10\pi}}\,\ClebchGordanSymb_{1,\mu;1,\mu'}^{2,\mu+\mu'}\,Y_{2,\mu+\mu'}^{*}   \right\}   \hat{d}_{\mu}^{(1)}\hat{d}_{\mu'}^{(2)}\\
& = \dfrac{1}{R_{12}^3}\sum_{\mu=\pm 1, 0}(-1)^{\mu}\hat{d}_{\mu}^{(1)}\hat{d}_{-\mu}^{(2)}\nonumber\\
&\phantom{=} + \dfrac{\sqrt{24\pi}}{R_{12}^3}\sum_{\mu,\mu'=\pm 1, 0}
\begin{pmatrix}
 1 & 1 & 2\\
\mu & \mu' & -(\mu + \mu')                                                                                                                                                                                                                                                                                                                                                                                                         \end{pmatrix}
Y_{2,-(\mu + \mu')}\hat{d}_{\mu}^{(1)}\hat{d}_{\mu'}^{(2)},
\label{eq:app::eval_dip_dip_interactions__eval_second_term_dip_dip_hamiltonian}
\end{align}
where we additionally used the relation $Y_{\ell,m}^{*} = (-1)^{-m}Y_{\ell,-m}$ and transformed the Clebsch-Gordan coefficient\index{Clebsch-Gordan coefficient} into a 3j-symbol via
\begin{equation}
 \ClebchGordanSymb_{\ell_1,m_1;\ell_2,m_2}^{\ell_3,m_3} := (-1)^{\ell_1-\ell_2+m_3}\sqrt{2\ell_3+1}\begin{pmatrix}
 \ell_1 & \ell_2 & \ell_3\\
m_1 & m_2 & -m_3                                                                                                                                                                                                                                                                                                                                                                                                         \end{pmatrix}.
\label{eq:app::eval_dip_dip_interactions__relation_clebsch_gordan_wigner_3j}
\end{equation}
We show further that the sum in the first term of \eref{eq:app::eval_dip_dip_interactions__eval_second_term_dip_dip_hamiltonian} is nothing else than the inner product of the two dipoles:
\begin{align}
 \sum_{\mu=\pm 1, 0}(-1)^{\mu}\hat{d}_{\mu}^{(1)}\hat{d}_{-\mu}^{(2)}&= \hat{d}_{0}^{(1)}\hat{d}_{0}^{(2)} - \sum_{\mu=\pm 1}\hat{d}_{\mu}^{(1)}\hat{d}_{-\mu}^{(2)}\\
&= \hat{d}_{0}^{(1)}\hat{d}_{0}^{(2)} - \sum_{\mu=\pm 1} \left(\mu\hat{d}_{x}^{(1)} + \im\hat{d}_{y}^{(1)} \right)\left(-\mu\hat{d}_{x}^{(2)} + \im\hat{d}_{y}^{(2)} \right)\\
&=\hat{d}_{0}^{(1)}\hat{d}_{0}^{(2)} - \sum_{\mu=\pm 1} -\mu^2\hat{d}_{x}^{(1)}\hat{d}_{x}^{(2)} + \im\mu\left(\hat{d}_{x}^{(1)}\hat{d}_{y}^{(2)}-\hat{d}_{y}^{(1)}\hat{d}_{x}^{(2)} \right)-\hat{d}_{y}^{(1)}\hat{d}_{y}^{(2)}\\
&=\sum_{\xi \in\{x,y,z\}}\hat{d}_{\xi}^{(1)}\hat{d}_{\xi}^{(2)} = \hat{\vc{d}}^{(1)}\cdot{\hat{\vc{d}}^{(2)}}
\end{align}
Thus, this term cancels with the purely isotropic part of the dipole-dipole interaction and we finally arrive at
\begin{equation}
 \hat{H}_{\mathrm{dd}}(\vc{R}_{12}):=-\dfrac{1}{4\pi \epsilon_0}\dfrac{\sqrt{24\pi}}{R_{12}^3}\sum_{\mu,\mu'=\pm 1, 0}
\begin{pmatrix}
 1 & 1 & 2\\
\mu & \mu' & -(\mu + \mu')                                                                                                                                                                                                                                                                                                                                                                                                         \end{pmatrix}
Y_{2,-(\mu + \mu')}\hat{d}_{\mu}^{(1)}\hat{d}_{\mu'}^{(2)}
\label{eq:app::eval_dip_dip_interactions__final}
\end{equation}
The matrix elements between the states $\ket{\Gamma_1;\Gamma_2}$ and $\ket{\Gamma_1';\Gamma_2'}$ with $\ket{\Gamma_k} = \ket{\nu_k,\ell_k,m_k}$ and $\ket{\Gamma_k'} = \ket{\nu_k',\ell_k',m_k'}$ are simpler than \eref{eq:app::eval_dip_dip_interactions__final}, since only the dipole matrix elements for the spherical component with $\Delta m_k :=m_k' - m_k, \ k \in \{1,2\}$ are non-vanishing, $\braket{\hat{d}_\mu}_{\Gamma_k}^{\Gamma_k'} = \delta_{\mu,-\Delta m_k}\braket{\hat{d}_{-\Delta m_k}}_{\Gamma_k}^{\Gamma_k'}$. This terminates the sums in \eref{eq:app::eval_dip_dip_interactions__final} and we finally get
\begin{multline}
 \bra{\Gamma_1;\Gamma_2}\hat{H}_{\mathrm{dd}}(\vc{R}_{12})\ket{\Gamma_1';\Gamma_2'} = 
%
%
-\sqrt{24\pi}\dfrac{\braket{\hat{d}_{-\Delta m_1}}_{\Gamma_1}^{\Gamma_1'} \braket{\hat{d}_{-\Delta m_2}}_{\Gamma_2}^{\Gamma_2'}}{R_{12}^3}\\
\times
\begin{pmatrix}
 1 & 1 & 2\\
-\Delta m_1 & -\Delta m_2 & \Delta m_1 + \Delta m_2                                                                                                                                                                                                                                                                                                                                                                                                       \end{pmatrix}Y_{2,\Delta m_1 + \Delta m_2}(\theta_{12},\phi_{12}).
\label{eq:app::eval_dip_dip_interactions__general_dip-dip_elem_eval_final_specified}
\end{multline}
\subsection{Evaluation of matrix elements of dipole-dipole interactions between \lstate{s} and \lstate{p} states with the same principal quantum number}
\label{part:app::calcs_ryd_atoms__eval_matrix_elements_dip_dip_interaction_s_p}
In this work, the focus is on dipole-dipole interactions between $\lstate{s}$ and $\lstate{p}$ states with the same principal quantum number. In particular we are dealing with $^7$Li.
We set $\Gamma_1 = \{\nu,1,m\}, \Gamma_1' = \{\nu,0,0\}$ for atom~1 and $\Gamma_2 = \{\nu,0,0\}, \Gamma_2' = \{\nu,1,m'\}$ for atom~2. This implies $\Delta m_1 = -m$ and $\Delta m_2 = m'$.
The dipole transition matrix elements simplify to
\begin{align}
 \braket{\hat{d}_{-\Delta m_1}}_{\Gamma_1}^{\Gamma_1'} \braket{\hat{d}_{-\Delta m_2}}_{\Gamma_2}^{\Gamma_2'} &=\dreduced_{\nu,1}^{\nu,0}\,\ClebchGordanSymb_{0,0;1,m}^{1,m}\, \dreduced_{\nu,0}^{\nu,1}\,\ClebchGordanSymb_{1,m';1,-m'}^{0,0}\\
&=\drad_{\nu,0;\nu,1}^2\,\dfrac{\ClebchGordanSymb_{0,0;1,0}^{1,0}}{\sqrt{3}}\overbrace{\ClebchGordanSymb_{0,0;1,m}^{1,m}}^{=1}\sqrt{3}\underbrace{\ClebchGordanSymb_{1,0;1,0}^{0,0}}_{=-1/\sqrt{3}}\overbrace{\ClebchGordanSymb_{1,m';1,-m'}^{0,0}}^{=(-1)^{m'+1}/\sqrt{3}}\\
&=\dfrac{(-1)^{m'}}{3}\,\drad_{\nu,0;\nu,1}^2.
\end{align}
We abbreviate $V_{m,m'}(\vc{R}_{12}):=\bra{\Gamma_1;\Gamma_2}\hat{H}_{\mathrm{dd}}(\vc{R}_{12})\ket{\Gamma_1';\Gamma_2'}$ and get as matrix elements
\begin{align}
 V_{m,m'}(\vc{R}_{12})&=
 -\sqrt{\dfrac{8\pi}{3}}\dfrac{\drad_{\nu,0;\nu,1}^2}{R_{12}^3}(-1)^{m'}
\begin{pmatrix}
 1 & 1 & 2\\
m & -m' & m' - m                                                                                                                                                                                                                                                                                                                                                                                                       \end{pmatrix}Y_{2,m' - m}(\theta_{12},\phi_{12}).
\end{align}

\section{Formula for calculating van-der-Waals interactions via block\hyp{}diagonalization}
\label{part:app::calcs_ryd_atoms__vdw_ints_from_block_diag}
We start by introducing the Hamiltonian for two dipole-dipole interacting atoms,
\begin{equation}
\hat{\mathcal{H}}(\vc{R}):=\hat{H}^{(\alpha)}\otimes \hat{\id}^{(\beta)} + \hat{\id}^{(\alpha)}\otimes\hat{H}^{(\beta)} + \hat{V}_{\mathrm{dd}}(\vc{R}),
\label{eq:alk_ryd_atoms:vdW_interactions__total_hamiltonian}
\end{equation}
where $\hat{H}^{(\alpha)}$ is the Hamiltonian of the single alkali atom $\alpha$ and $\hat{\id}^{(k)}$ denotes the identity operator over the bound state space for atom $k=\alpha,\beta$. The Hamiltonian can be expanded in the pair state basis $\{\ket{X}\}$, where each $\ket{X}$ is a tensor state of an individual state from atom $\alpha$ and atom $\beta$.
Since the eigenenergies are degenerate with respect to the magnetic quantum numbers, we rewrite pair states as $\ket{X} = \ket{x,M} \equiv \ket{x}\otimes\ket{M}$, where $\ket{x}$ contains the principal and azimuthal- and $\ket{M}$ the magnetic quantum numbers.
The explicit structure of these pair states is $\ket{x} = \ket{\gamma_\alpha}\otimes\ket{\gamma_\beta}$, with $\ket{\gamma_{k}} = \ket{\nu_k,\ell_k}$ and $\ket{M} = \ket{m_\alpha}\otimes\ket{m_\beta}$ for atom $k= \alpha, \beta$.
Note that the range of magnetic quantum numbers is dependent on the azimuthal quantum numbers, $M = M(x)$.
The expansion of the total Hamiltonian in the pair bound state basis yields
\begin{equation}
 \hat{\mathcal{H}} = \sum_{\ket{x},\ket{x'}}\hat{\mathcal{H}}_{x,x'}\otimes \ket{x}\bra{x'},
\label{eq:alk_ryd_atoms:vdW_interactions__total_hamiltonian_matrix_representation}
\end{equation}
where the sub-Hamiltonians are defined as
\begin{equation}
 \hat{\mathcal{H}}_{x,x'}:=\sum_{M,M'}
%
%
\mathcal{H}_{X,X'}
\,\ket{M}\bra{M'}
\label{eq:alk_ryd_atoms:vdW_interactions__def_block_hamiltonians}
\end{equation}
with the matrix elements $\mathcal{H}_{X,X'} \equiv \bra{x,M}\hat{\mathcal{H}}\ket{x',M'}$.
We omit the dependency on $\vc{R}$ and the atom numbers and abbreviate $\hat{\mathcal{H}}:={}^{(\alpha,\beta)}\hat{\mathcal{H}}(\vc{R})$ for better readability.


For the diagonal sub-Hamiltonians with $x=x'$, no dipole-dipole interactions are involved and the structure is simply given by the pair state energy of $\ket{x}$, which we denote with $E_x$, times the identity of the $M(x)$-subspace, $\hat{\id}_{x} := \sum_{M(x)}\ket{M}\bra{M}$. For transitions, $x \to x'$, dipole-dipole interactions might appear. We can thus write 
\begin{equation}
 \hat{\mathcal{H}}_{x,x'}=
\begin{cases}
 E_{x}\hat{\id}_{x}, & x= x' \\
\hat{V}_{x,x'}, & x \neq x',
\end{cases}
\label{eq:alk_ryd_atoms:vdW_interactions__def_blocks_H}
\end{equation}
with
\begin{equation}
 \hat{V}_{x,x'}:=\sum_{M,M'}
V_{X,X'}\,\ket{M}\bra{M'}
\label{eq:alk_ryd_atoms:vdW_interactions__def_blocks_dip-dip}
\end{equation}
the dipole-dipole operator containing as matrix elements the dipole transitions\linebreak[4] $V_{X,X'} \equiv \bra{x,M}\hat{V}_{\mathrm{dd}}\ket{x',M'}$.
However, since each of these sub-Hamiltonians does not decouple from the others, the total Hamiltonian given in the structure of \eref{eq:alk_ryd_atoms:vdW_interactions__total_hamiltonian_matrix_representation} has to be block-diagonalized via a unitary transformation $\mathcal{U}$. The new Hamiltonian, $\mathcal{H}' = \mathcal{U}^\dagger\mathcal{H}\mathcal{U}$ is the result of a basis change, which decouples the sub-Hamiltonians and modifies them, such that the previous couplings to the other sub-Hamiltonians are included.
Unfortunately, the block-diagonalization can analytically be performed only in a perturbative way, where the effective couplings between the sub-Hamiltonians is weak. However, we found the van-der-Waals interactions from the two state model in \sref{part:fd::chap:ryd_a::sec:resonant_and_vdW_interactions} with their dependency $\sim R^{-6}$ as a perturbation of large detunings compared to the dipole transition strength. This justifies to block-diagonalize
perturbatively, according to the scheme outlined in \Aref{part:app::chap:block_diag__van_der_Waals}, to find van-der-Waals interactions up to terms $\sim R^{-6}$.
We can write 
\begin{equation}
\mathcal{H}'_{x,x'} \approx \mathcal{H}_{x,x'} + W_{x,x'}
\end{equation}
with $\hat{W}_{x,x'}$ the so called van-der-Waals interaction for the corresponding sub-Hamiltonian, given by
\begin{equation}
 \hat{W}_{x,x'}:=
\dfrac{1}{2}\sum\limits_{\substack{\ket{y}: E_y \neq E_x,E_x'}} \left(\Delta_{x,y}^{-1} - \Delta_{y,x'}^{-1}\right)\hat{V}_{x,y}\hat{V}_{y,x'}.
\label{eq:alk_ryd_atoms:vdW_interactions__def_vdW_sub-blocks}
\end{equation}
The binary products of sub-Hamiltonians with dipole-dipole transition matrix elements are weighted with the inverse energy detunings, which are defined as
\begin{equation}
\Delta_{x,y}:=E_{x} - E_{y}.
\label{eq:alk_ryd_atoms:vdW_interactions__def_detuning_vdW}
\end{equation}
The van-der-Waals corrections in \eref{eq:alk_ryd_atoms:vdW_interactions__def_vdW_sub-blocks} are non-vanishing for the following selection rules: $(\ell_1-\ell_1'),(\ell_2-\ell_2') \in \{0,\pm 2\}$.

We can now confirm the dependency $\hat{W}_{x,x'} \sim R^{-6}$, since it consists of binary products of dipole-dipole interaction sub-Hamiltonians, which are $\sim R^{-3}$. The strength of the van-der-Waals interaction is indicated by the so called Dispersion coefficient\index{dispersion coefficient}, which we denote with $C_6$ and can be calculated by
\begin{equation}
 C_6 = - R^6\cdot \hat{W}_{x,x'}
\label{eq:alk_ryd_atoms:vdW_interactions__def_C6},
\end{equation}
for a specific pair $(x,x')$, following the structure of \eref{eq:alk_ryd_atoms:vdW_interactions__vdw_general_formula}. 
%
The most relevant van-der-Waals interaction for our purposes is between \lstate{s}-states. We abbreviate with $ss' \equiv ((\nu,s),(\nu',s))$ the pair of quantum numbers and with $\ket{ss'}$ the pair state. For $x=x'=ss'$, the calculation of the $C_6$-coefficient for this state is, according to \eref{eq:alk_ryd_atoms:vdW_interactions__def_C6}
\begin{equation}
 C_{6}^{ss'}= -R^{6}\sum_{\ket{y}: E_y \neq E_{ss'}}\dfrac{|\bra{ss'}\hat{V}_{\mathrm{dd}}\ket{y}|^2}{E_{ss'}-E_{y}},
\label{eq:alk_ryd_atoms:vdW_interactions__C6_ss_states}
\end{equation}
which is the formula of standard second order Rayleigh-Schr{\"o}dinger perturbation theory. Note that \eref{eq:alk_ryd_atoms:vdW_interactions__C6_ss_states} is also valid for $\nu'=\nu$, such that $s'=s$. We can use the structure of \eref{eq:alk_ryd_atoms:vdW_interactions__C6_ss_states} to calculate dispersion coefficients\index{dispersion coefficient} for other pair states, as long as we do not consider the dependency of the magnetic quantum number. Calculating the dispersion coefficient of the pair state $\ket{x}$, we have to replace $\bra{ss'}\to \bra{x}$ and $E_{ss'} \to E_{x}$ in \eref{eq:alk_ryd_atoms:vdW_interactions__C6_ss_states} to get $C_6^{x}$.

We have to discuss the solution for the van-der-Waals corrections given in \eref{eq:alk_ryd_atoms:vdW_interactions__def_vdW_sub-blocks}, in order to use it for practical calculations. The $C_6$-coefficient is per definition independent of the interatomic distance, by multiplying the van-der-Waals matrices with $R^6$. In the sum, couplings to other states can appear which have very small detunings. If the coupling to a specific state is too strong,
 the coupling to this state has to be removed as a contribution to the $C_6$-coefficient, since the the treatment is limited to the weak coupling regime. We have to successively remove all contributions until the sum converges. All the removed states which have a small energy detuning, couple rather resonantly than off-resonantly to the state of consideration, which lead to interactions $\sim R^{-n}$ with $3<= n <6$. In practice, this means that if resonances to other states appear, all these (quasi)-resonant states have to be collected to one subspace and diagonalized as a whole, which gives then the (quasi-)resonant contribution of the interactions.

\chapter{Calculations for chapter 'Planar aggregates with isotropic interactions'}\label{part:app::calcs_planar_aggr}
\section{Deriving the van-der-Waals interaction formula for localized, singly excited many-body states}
\label{part:app::calcs_planar_aggr__vdw_aggr}
Van-der-Waals interactions can be calculated by second-order Rayleigh-Schrödinger perturbation theory. For the aggregate's basis states, which are localized $N$-body states with a single \lstate{p} excitation, the formula is given in \eref{eq:planar_aggr:formula_vdW_aggr_states_start}, which we repeat here:
\begin{equation}
 h_{\mathrm{vdw}}^{\ket{\pi_\alpha}} = \sum_{\substack{\ket{Y}:\\E_{Y}\neq E_{\mathrm{aggr}}}}\dfrac{\left|\bra{\pi_\alpha}\hat{\mathcal{V}}\ket{Y}\right|^2}{E_{\mathrm{aggr}} - E_{Y}},
\end{equation}
For evaluating the van-der-Waals shift, it is useful to rewrite the $N$-body state as $\ket{Y} = \ket{Y_{1}\dots Y_{N}}$, where $\ket{Y_{k}}$ is the state of atom $k$.
Furthermore we denote with $\nu_{k}$ the principal quantum number and with $\ell_{k}$ the azimuthal quantum number of the state $\ket{Y_{k}}$.
The matrix element can be written as
\begin{align}
 \bra{\pi_\alpha}\hat{\mathcal{V}}\ket{Y} &= \dfrac{1}{2}\sum_{\substack{i,j:\\ i \neq j}}
\Bigl(
\left(
\delta_{\alpha,i}
\bra{ps}\hat{V}_{\mathrm{dd}}^{(i,j)}\ket{Y_i Y_j}+
\delta_{\alpha,j}
\bra{sp}\hat{V}_{\mathrm{dd}}^{(i,j)}\ket{Y_i Y_j}\right)
\prod_{k\neq i,j}\braket{s|Y_{k}}\nonumber\\
&\quad \quad \quad +(1-\delta_{\alpha,i}-\delta_{\alpha,j})\bra{ss}\hat{V}_{\mathrm{dd}}^{(i,j)}\ket{Y_i Y_j}\braket{p|Y_{\alpha}}\prod_{k\neq i,j,\alpha}\braket{s|Y_{k}}
\Bigr)\nonumber\\
&= \dfrac{1}{2}\sum_{i \neq \alpha}
\left(
\bra{ps}\hat{V}_{\mathrm{dd}}^{(\alpha,i)}\ket{Y_\alpha Y_i} + 
\bra{sp}\hat{V}_{\mathrm{dd}}^{(i,\alpha)}\ket{Y_i Y_\alpha}
\right)\prod_{k\neq \alpha,i}\braket{s|Y_{k}}\nonumber\\
&+\dfrac{1}{2}\sum_{i\neq \alpha}\sum_{j \neq i}
\bra{ss}\hat{V}_{\mathrm{dd}}^{(i,j)}\ket{Y_i Y_j}\braket{p|Y_{\alpha}}\prod_{k\neq i,j,\alpha}\braket{s|Y_{k}}
\label{eq:app::calcs_planar_aggr__vdw_aggr_eval_transition_element}
\end{align}
The term $\bra{ps}\hat{V}_{\mathrm{dd}}^{(\alpha,i)}\ket{Y_\alpha Y_i}$ contributes equally to the summation as $\bra{sp}\hat{V}_{\mathrm{dd}}^{(i,\alpha)}\ket{Y_i Y_\alpha}$ and they are only nonvanishing for $\ell_{\alpha}=0,2$, according to dipole-dipole selection rules. However, the transition $\bra{ss}\hat{V}_{\mathrm{dd}}^{(i,j)}\ket{Y_i Y_j}$ contributes only for $\ell_{\alpha}=1$. This gives two cases, depending on how we choose the azimuthal quantum number on atom $\alpha$:
\begin{equation}
 \bra{\pi_\alpha}\hat{\mathcal{V}}\ket{Y}=
\begin{dcases}
 \sum_{i \neq \alpha} \bra{ps}\hat{V}_{\mathrm{dd}}^{(\alpha,i)}\ket{Y_\alpha Y_i}\prod_{k\neq \alpha,i}\braket{s|Y_{k}}, & \ell_\alpha =0,2\\
\dfrac{1}{2}\sum_{i\neq \alpha}\sum_{j \neq i}
\bra{ss}\hat{V}_{\mathrm{dd}}^{(i,j)}\ket{Y_i Y_j}\prod_{k\neq i,j,\alpha}\braket{s|Y_{k}}, & \ell_\alpha =1
\end{dcases}.
\label{eq:app::calcs_planar_aggr__vdw_aggr_eval_transition_element_final}
\end{equation}
Calculating the absolute square values of the transition element doubles the number of summations. 
For the case of $\ell_\alpha=0,2$, this would lead to an additional summation, whose index we denote with $i'$, and terms \[\bra{ps}\hat{V}_{\mathrm{dd}}^{(\alpha,i)}\ket{Y_\alpha Y_i} \bra{Y_\alpha Y_i'}\hat{V}_{\mathrm{dd}}^{(\alpha,i')}\ket{ps} \prod_{k\neq \alpha,i'}\braket{s|Y_{k}} \prod_{k'\neq \alpha,i'}\braket{s|Y_{k'}}\] under the double sum. However, all mixed terms with $i'\neq i$ cancel. The reason is that we need to choose the azimuthal quantum numbers of the states $\ket{Y_{i}}$ and $\ket{Y_{i'}}$ equal to one, $\ell_i, \ell_{i'} = 1$, to have nonvanishing dipole-dipole transitions. On the other hand we need to choose $\ell_{i'}=0$, if $i' \notin \{i,\alpha\}$, such that $\prod_{k\neq \alpha,i}\braket{s|Y_{k}}$ is nonvanishing. This is a contradiction. 
Only for $i'=i$ we do not have to choose $\ell_{i'}=0$, because the state $\ket{Y_{i'}}$ is then not included in the product $\prod_{k\neq \alpha,i}\braket{s|Y_{k}}$. Nonvanishing contributions are thus only from diagonal elements with $i'=i$.
The argumentation is the same for the term with fixed $\ell_{\alpha}=1$. However, here we have two additional sums, whose summation index we denote with $i'$ and $j'$, respectively. As in the previously discussed case, only diagonal elements contribute, $(i',j')=(i,j)$. Since we can swap the indices $i' \leftrightarrow j'$ under the sum, also the contribution with $(i',j') = (j,i)$ is a diagonal element.
After these considerations, we can write for the absolute square of the transition matrix element
\begin{equation}
 \left|\bra{\pi_\alpha}\hat{\mathcal{V}}\ket{Y}\right|^2=
\begin{dcases}
 \sum_{i \neq \alpha} \left|\bra{ps}\hat{V}_{\mathrm{dd}}^{(\alpha,i)}\ket{Y_\alpha Y_i}\right|^2\prod_{k\neq \alpha,i}\braket{s|Y_{k}}, & \ell_\alpha =0,2\\
\dfrac{1}{2}\sum_{i\neq \alpha}\sum_{j \neq i}
\left|\bra{ss}\hat{V}_{\mathrm{dd}}^{(i,j)}\ket{Y_i Y_j}\right|^2\prod_{k\neq i,j,\alpha}\braket{s|Y_{k}}, & \ell_\alpha =1
\end{dcases}.
\label{eq:app::calcs_planar_aggr__vdw_aggr_eval_abs_square_transition_element}
\end{equation}
The initial formula with second-order perturbation theory over $N$-body states decomposes into a sum over second-order perturbation theory with pair states. Due to the products in \eref{eq:app::calcs_planar_aggr__vdw_aggr_eval_abs_square_transition_element}, $N-2$ elements of the state $\ket{Y}$ are fixed, in the case $\ell_\alpha =0,2$ all to $s$ states and for $\ell_\alpha =1$ one to a $p$ state and the rest to $s$ states. Since the aggregate's energy is $E_{\mathrm{aggr}} = (N-1)E_{s} + E_{p}$, we get for the denominator in the case of $\ell_\alpha=0,2$,  $E_{\mathrm{aggr}} - E_{Y} = E_{ps}-E_{Y_{\alpha}Y_i}$ and in the case of  $\ell_\alpha=1$,  $E_{\mathrm{aggr}} - E_{Y} = E_{ss}-E_{Y_{\alpha}Y_i}$, where $E_{ps},E_{ss}$ and $E_{Y_{\alpha}Y_i}$ are pair state energies.
Summing over all different pair states, which we denote with $\ket{y}$, we get the following van-der-Waals formula:
\begin{equation}
 h_{\mathrm{vdw}}^{\ket{\pi_\alpha}}  = \sum_{i \neq \alpha}\sum_{\substack{\ket{y}:\\E_{ps}\neq E_{y}}}\dfrac{|\bra{ps}\hat{V}_{\mathrm{dd}}^{(\alpha,i)}\ket{y}|^2}{E_{ps}-E_{y}} + \dfrac{1}{2}\sum_{i \neq \alpha}\sum_{j\neq i}\sum_{\substack{\ket{y}:\\E_{ss}\neq E_{y}}}\dfrac{|\bra{ss}\hat{V}_{\mathrm{dd}}^{(i,j)}\ket{y}|^2}{E_{ss}-E_{y}},
\end{equation}
which is \eref{eq:planar_aggr:formula_vdW_aggr_states_sum_over_binary_interactions} in \sref{part:rs::chap:planar_aggregates_electronic_Hamiltonian} of \Chref{part:rs::chap:planar_aggregates}.

\section{Tail distribution of the relative energy gap for a trivial avoided crossing}
\label{part:app::calcs_planar_aggr__trivial_crossings}
Here we discuss analytically the appearance of the trivial transitions in the four atom T-shape aggregate.
The Hamiltonian of the four atom aggregate can be written down as the two Hamiltonians of the dimers and the interactions between them.
Defining the dimer Hamiltonian,
\begin{equation}
 \hat{H}^{(k,l)}_{\mathrm{dimer}}(R):= -\mu^2 R^{-3}\left(\ket{\pi_k}\bra{\pi_l} + \mathrm{H.c.}\right)
\label{eq:app::calcs_planar_aggr__trivial_crossings_Hdimer}
\end{equation}
As already pointed out in the main text, the dimer energies are $U_{\pm}(R) = \pm \mu^2/R^3$ with corresponding excitons $\ket{\varphi^{(k,l)}_{\pm}} = \left(\ket{\pi_k}\mp \ket{\pi_l}\right)/\sqrt{2}$.
The atoms on the horizontal dimer are labeled with 1,2 and for the vertical dimer with 3,4, such that $\hat{H}^{(1,2)}_{\mathrm{dimer}}(R_{1,2})$ is the Hamiltonian for the horizontal and $\hat{H}^{(3,4)}_{\mathrm{dimer}}(R_{3,4})$ for the vertical dimer, with $R_{1,2}$ and $R_{3,4}$ the interatomic distance, respectively.
We see that if a total system consists of two completely decoupled dimers, the energies for the repulsive and attractive surfaces are degenerate, when the two interatomic distances are equal, $R_{1,2} = R_{3,4}$. 

\begin{figure}[!t]
\centering
\includegraphics{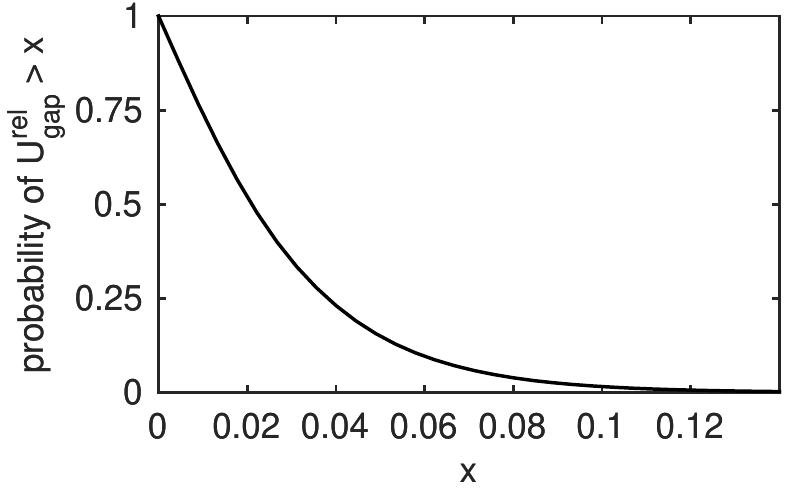} 
\caption{Tail distribution~(exceedance) of the energy gap relative to the unperturbed repulsive dimer energy, defined in \eref{eq:app::calcs_planar_aggr__trivial_crossings_Ugap_rel}. The distribution measures the probability that the relative energy gap is above a varying level. We used for the atomic positions Gaussian probability distributions and calculate for each realization the relative energy gap, with adjusted positions of the horizontal dimer, 
$x_1 = x_1^{(0)} - \Delta_x/2, x_2 = x_2^{(0)} + \Delta_x/2$, such that its interatomic distance equals the one of the other dimer.
The mean positions of the initial configuration are adjusted to $x_1^{(0)} = 0, x_2^{(0)}  = a_1, y_3^{(0)} = -a_2/2, y_4^{(0)}  = a_2/2$ and their standard deviation is $\sigma_0$, where the parameters ($a_1,a_2,d,\sigma_0$) are the ones of the two perpendicular dimer aggregate, discussed in \sref{sec:nges4}. 
\label{fig:prob_distr_Ugap_rel}
}
\end{figure}

If the two dimers are not completely decoupled, the interaction between them can be written as
\begin{align}
\hat{W}(\vc{R}_{1,2}^{3,4})&:=-\mu^2
\sum_{k=1,2}\sum_{l=3,4}R_{k,l}^{-3}\ket{\pi_k}\bra{\pi_l}\\
&=-\mu^2
\begin{pmatrix}
 R_{1,3}^{-3} & R_{1,4}^{-3}\\
R_{2,3}^{-3} & R_{2,4}^{-3},
\end{pmatrix}
\label{eq:app::calcs_planar_aggr__trivial_crossings_interaction_matrix}
\end{align}
where the vector $\vc{R}_{1,2}^{3,4}$ contains all interatomic distances from one to the other dimer.
The Hamiltonian of the total system is then given by
\begin{equation}
 \hat{H}(\vc{R}) = 
\begin{pmatrix}
\hat{H}^{(1,2)}_{\mathrm{dimer}} & \hat{W}\\
\hat{W}^{\dagger} & \hat{H}^{(3,4)}_{\mathrm{dimer}}
\end{pmatrix}.
\label{eq:app::calcs_planar_aggr__trivial_crossings_total_ham}
\end{equation}
To find out what happens with the degeneracy of the two decoupled dimer, we use the following extended dimer states,
\begin{align}
 \ket{\phi^{(1,2)}_{+}} = \ket{\varphi^{(1,2)}_{+}}\otimes 0^{(3,4)}
\label{eq:app::calcs_planar_aggr__trivial_crossings_extended_dimer_state1}\\
 \ket{\phi^{(3,4)}_{+}} = 0^{(1,2)} \otimes\ket{\varphi^{(3,4)}_{+}},
\label{eq:app::calcs_planar_aggr__trivial_crossings_extended_dimer_state2}
\end{align}
which are the repulsive eigenstates of the noninteracting total system. Note that $0^{(k,l)}$ is used as the zero vector of the system $k,l$. We apply quasidegenerate perturbation theory for $R_{1,2} \approx R_{3,4}$ and restrict ourselves to the two repulsive surfaces. We get
\begin{equation}
 \hat{H}_{QDPT} = \left(\bra{\phi^{x}_{+}}\hat{H}\ket{\phi^{y}_{+}}\right)_{x,y} = 
\begin{pmatrix}
 U_{+}(R_{1,2}) & w(\vc{R}_{1,2}^{3,4})\\
w(\vc{R}_{1,2}^{3,4}) & U_{+}(R_{3,4})
\end{pmatrix}, \quad x,y \in \{(1,2),(3,4)\},
\label{eq:app::calcs_planar_aggr__trivial_crossings_H_QDPT}
\end{equation}
with
\begin{equation}
 w(\vc{R}_{1,2}^{3,4}):= -\dfrac{\mu^2}{2}\left( (R_{1,3}^{-3} - R_{1,4}^{-3}) - (R_{2,3}^{-3} - R_{2,4}^{-3})\right).
\label{eq:app::calcs_planar_aggr__trivial_crossings_interaction_scalar}
\end{equation}
The coupling between both dimers vanishes when both are infinitely separated and also for completely symmetric configurations, where $b=1$, which is equivalent to $R_{1,3} = R_{1,4}$ and $R_{2,3} = R_{2,4}$. In this case, the degeneracy of the states is not lifted such that a real crossing of eigenenergies would  appear, when varying one of the dimer's interatomic distance and fixing the other. To discuss the case of nonvanishing interactions, $w \neq 0$, it is useful to introduce the detuning $\Delta:=U_{+}(R_{3,4}) - U_{+}(R_{1,2})$ and the mean energy value, $\bar{U}_{+} :=\left( U_{+}(R_{3,4}) + U_{+}(R_{1,2})\right)/2$. The eigenenergies and -states of $\hat{H}_{QDPT}$ are then given by
\begin{align}
 U^{\pm}_{QDPT}  &= \bar{U}_{+} \pm \sqrt{w^2 + \left(\frac{\Delta}{2}\right)^2},\label{eq:app::calcs_planar_aggr__trivial_crossings_U_QDPT}\\
\ket{\phi^{\pm}_{QDPT}} &= \dfrac{1}{\sqrt{\alpha_{\pm}^2+1}}\left(\alpha_{\pm}\ket{\phi_{+}^{(1,2)}} + \ket{\phi_{+}^{(3,4)}}\right), \mathrm{\ with }\label{eq:app::calcs_planar_aggr__trivial_crossings_phi_QDPT}\\
\alpha_{\pm} &:= -\frac{\Delta}{2w}\pm \frac{1}{w}\sqrt{w^2 + \left(\frac{\Delta}{2}\right)^2}
\label{eq:app::calcs_planar_aggr__trivial_crossings_alphapm_QDPT}
\end{align}
We are specifying now the formulas for the case $R_{1,2} = R_{3,4}$, where the degeneracy is located for the noninteracting total system. This is equivalent to $\Delta = 0$ and $\bar{U}_{+} = U_{+}(R_{1,2}) = U_{+}(R_{3,4})$.
The solutions are then given by
\begin{align}
 U^{\pm}_{QDPT}  &= \bar{U}_{+} \pm |w|\label{eq:app::calcs_planar_aggr__trivial_crossings_U_QDPT_special}\\
\ket{\phi^{\pm}_{QDPT}} &= \left(\pm \mathrm{sgn}(w)\ket{\phi_{+}^{(1,2)}} + \ket{\phi_{+}^{(3,4)}}\right)/\sqrt{2},
\label{eq:app::calcs_planar_aggr__trivial_crossings_phi_QDPT_special}
\end{align}
and we see that the degeneracy is lifted and the energy gap between both states is $U_{\mathrm{gap}} = 2 |w|$. The gap relative to the unperturbed energy is then given by
\begin{equation}
 U^{\mathrm{rel}}_{\mathrm{gap}} := \dfrac{U_{\mathrm{gap}}}{U_{+}(R_{1,2})} = \dfrac{|(R_{1,3}^{-3} - R_{1,4}^{-3}) - (R_{2,3}^{-3} - R_{2,4}^{-3})|}{R_{1,2}^{-3}}.
\label{eq:app::calcs_planar_aggr__trivial_crossings_Ugap_rel}
\end{equation}
In \fref{fig:prob_distr_Ugap_rel} we show the tail distribution of the relative energy gap, measuring the probability that the relative gap is above a varying level. The distribution reflects the situation of the two perpendicular dimer aggregate in \sref{sec:nges4}. We find the relative gap is smaller than $0.08$ with about $96\%$ probability and smaller than $0.022$ with a probability of $52\%$, confirming that the avoided crossing has a small energy gap.

\section{Trimer conical intersection: energy gap considerations for single trajectories and wave packets}
\label{part:app::calcs_planar_aggr__rel_energy_gap_CI}
\begin{figure}[!t]
\centering
\includegraphics{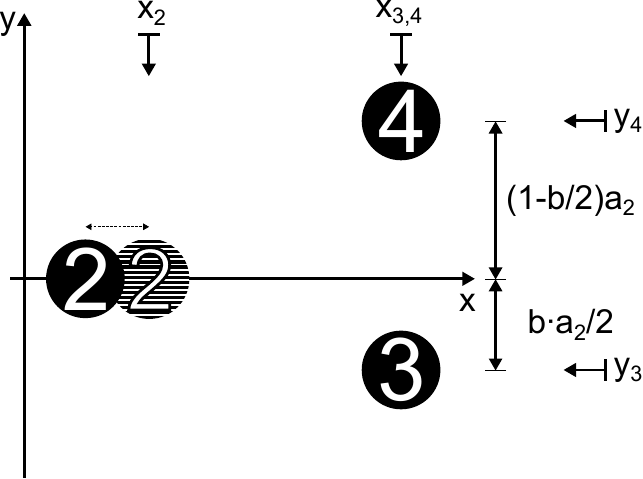} 
\caption{Sketch of a trimer with atoms~2, 3 and 4, near a configuration with CI, as it could be realized as a subunit of the four atom aggregate, discussed in \sref{sec:nges4}. The parameter $b$, defined in \eref{eq:planar_aggr_nonad_dyn:def_asymmetry_param}, measures the asymmetry of the configuration by comparing with the mirror configuration, obtained by mirroring along the horizontal axis. Configurations with $b \approx 1$ are symmetric and with $b\lessgtr 1$ asymmetric configurations. The mean distance between atom~3 and 4 is $a_2$, as in \sref{sec:nges4}. The CI is located at the equilateral triangle configuration, which implies $b=1$. 
\label{fig:sketch_asymmetric_trimer}
}
\end{figure}

\begin{figure}[!t]
\centering
\includegraphics{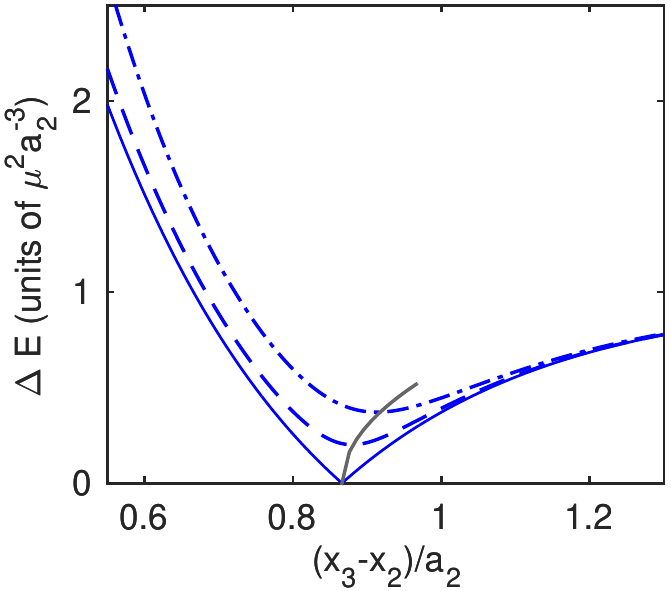} 
\caption{
Energy spacing between repulsive and adjacent eigenenergy for different asymmetry parameters, $b =1$ (solid), $b=0.88$ (dashed), $b=0.76$ (dashed dotted). The blue lines are numerical results.
The minimal energy spacing~(black dots) is shifted to bigger horizontal distances, $x_3-x_2$ for higher asymmetry, which is well described by the analytical result~\eref{eq:analyt_delta_E_min}~(grey line)
\label{fig:energy_gap_trimer_single_traj}
}
\end{figure}

The trimer is the minimal aggregate system allowing for a CI of two BO surfaces. We already discussed an isosceles trimer in \sref{part:fd::chap:theoretical_framework_spatially_unfrozen system} with the corresponding energy spectrum shown in \fref{fig:isosceles_trimer}. 
However, the intersection appears only for the equilateral triangle configuration which requires a completely symmetric positioning of the vertical dimer relative to the horizontal axis. This is only one out of infinitely many configurations which form the wave packet and thus, the intersection is almost never exactly present. For all asymmetric configurations an energy gap appears which is dependent on the order of asymmetry. We first want to derive how the size of the energy gap scales with the asymmetry, which is quantified by the parameter $b$ defined in \eref{eq:planar_aggr_nonad_dyn:def_asymmetry_param}. We set the atomic positions of atoms~2, 3, and 4 to 
\begin{align}
\vc{R}_2 &= ({x}_2,{y}_2) = (0,0),\\
\vc{R}_3 &= ({x}_3,{y}_3) = (\sqrt{3}a_2/2 + \Delta x, -b a_2/2),\\
\vc{R}_4 &= ({x}_4,{y}_4) = (\sqrt{3}a_2/2+ \Delta x,(1-b/2) a_2),
\end{align}
to explicitly account for asymmetric configurations, as sketched in \fref{fig:sketch_asymmetric_trimer}. 
The configurations can be parameterized with $\vc{K}~:=~(\Delta x,b)^{\mathrm{T}}$, where $\vc{K}_0~:=~(0,1)^{\mathrm{T}}$ sets the CI~configuration.
The electronic Hamiltonian of the trimer is given by 
\begin{align}
 \hat{H}_{\mathrm{trimer}}(\vc{K}) &:=
-\mu^2 \sum_{\substack{k,l=2\\ k \neq l}}^{4}
R_{k,l}^{-3}(\vc{K})\ket{\pi_{k}}\bra{\pi_{l}}.
\end{align}
For the CI~configuration the value of the degenerate eigenenergy is $E^{(0)}_{\mathrm{CI}}~=~\mu^2/\mathrm{a_2}^3$. The corresponding eigenstates can be set to 
\begin{align}
 \ket{\varphi^{(0)}_{\mathrm{CI},1}} &= \frac{1}{\sqrt{2}}\begin{pmatrix}-1 & 0 &1 \end{pmatrix}^{\mathrm{T}},\\
\ket{\varphi^{(0)}_{\mathrm{CI},2}} &=\frac{1}{\sqrt{6}}\begin{pmatrix}-1 & 2 &-1 \end{pmatrix}^{\mathrm{T}}.
\end{align}
We Taylor expand the electronic Hamiltonian around the CI configuration,
\begin{equation}
\hat{H}^{\mathrm{as}}_{\mathrm{trimer}}(\vc{K})\approx \hat{H}_{\mathrm{trimer}}(\vc{K}_0) + \hat{H}^{\mathrm{PT}}_{\mathrm{trimer}}(\vc{K}),
\end{equation}
and restrict the further calculations to the the term  $\hat{H}^{\mathrm{PT}}_{\mathrm{trimer}}(\vc{K})$ which is responsible for lifting the degeneracy.
Building matrix elements with the CI eigenstates,
\begin{equation}
S_{\mathrm{trimer}}(\vc{K})_{\alpha\beta}:=\bra{\varphi^{(0)}_{\mathrm{CI},\alpha}}\op{H}^{\mathrm{PT}}_{\mathrm{trimer}}(\vc{K})\ket{\varphi^{(0)}_{\mathrm{CI},\beta}},
\end{equation}
the eigenenergies $E^{\mathrm(1)}_{1}(\vc{K}), E^{\mathrm(1)}_{2}(\vc{K})$ of $S_{\mathrm{el}}(\vc{K})$ are the first order corrections to the energy and lift the degeneracy. Thus the energy gap is given by $\Delta E^{\mathrm{as}}(\vc{K})~:=~| E^{\mathrm(1)}_{1}(\vc{K}) - E^{\mathrm(1)}_{2}(\vc{K})|$ up to first order. Consistently expanding this expression to second order around $\vc{K}_0$, we get
\begin{equation}
{\Delta E^{\mathrm{as}}(\vc{K})}\approx \dfrac{\mu^2}{a_2^3}\left[
\sqrt{3} (1 - b) - 
 31(1 - b)\Delta x/4a_2   + 
\dfrac{2705-2321(2 - b)b}{64 \sqrt{3} (1 - b)}(\Delta x/a_2)^2\right],
\end{equation}
for $b\lesssim 1$. For every small given asymmetry, there is a $\Delta x_{\mathrm{min}}$ where the energy gap becomes minimal:
\begin{align}
\begin{split}
 \Delta x_{\mathrm{min}}/a_2 &\approx 1.12\cdot(1-b)^2,\\
\Delta E^{\mathrm{as}}_{\mathrm{min}}(b)& \approx \mu^2 a_2^{-3}\sqrt{3}\cdot(1-b)
\label{eq:analyt_delta_E_min}.
\end{split}
\end{align}
Thus, the horizontal distance between atom~2 and the two others has to be larger compared to the CI configuration to achieve a minimal energy gap, as evident in \fref{fig:energy_gap_trimer_single_traj}.
For adjusting the horizontal distance with $\Delta x$ such that for every asymmetry it adjusts the minimal energy gap, we find a linear scaling of the energy gap with the asymmetry. This is still valid for higher asymmetry values, as apparent from \fref{fig:energy_gap_trimer_single_traj}.

Since the nonadiabatic dynamics due to the CI is not controlled by a single trajectory but by the complete wave packet, we calculate in the following the energy gap distribution for positioning the wave packet at the CI configuration with mean positions $(\bar{x}_2,\bar{y}_2) = (0,0)$, $(\bar{x}_3,\bar{y}_3) = (\sqrt{3}a_2/2, -a_2/2)$, $(\bar{x}_4,\bar{y}_4) = (\sqrt{3}a_2/2, a_2/2)$ of atoms~2, 3, and 4. The ratio between the length scale $a_2$ and the width of each atomic Gaussian, $\sigma_0$ is set to $\sigma_0/a_2 =0.0952$ to reflect the situation of the double dimer system  studied in \sref{sec:nges4}
Note that atom~2 has no distribution in vertical and atoms~3 and 4 no distribution in horizontal direction, such that $(y_2,x_3,x_4) = (\bar{y}_2,\bar{x}_3,\bar{x}_4)$.
We diagonalize the trimer Hamiltonian for $10^6$ realizations, where the atomic positions are randomly distributed according to the atomic Gaussians. Afterward, for each configuration the relative energy gap is calculated, $U^{\mathrm{rel}}_{\mathrm{gap}} = U_{\mathrm{gap}}/\bar{U} = (U_{\mathrm{rep}} - U_{\mathrm{adj}})/\bar{U}$, with $\bar{U} := \left(U_{\mathrm{rep}} - U_{\mathrm{adj}}\right)/2$. We finally get the tail distribution of the relative energy gap determined through statistical sampling. The tail distribution gives the probability that the relative energy gap is above a varying level.
\begin{figure}[!t]
\centering
\includegraphics{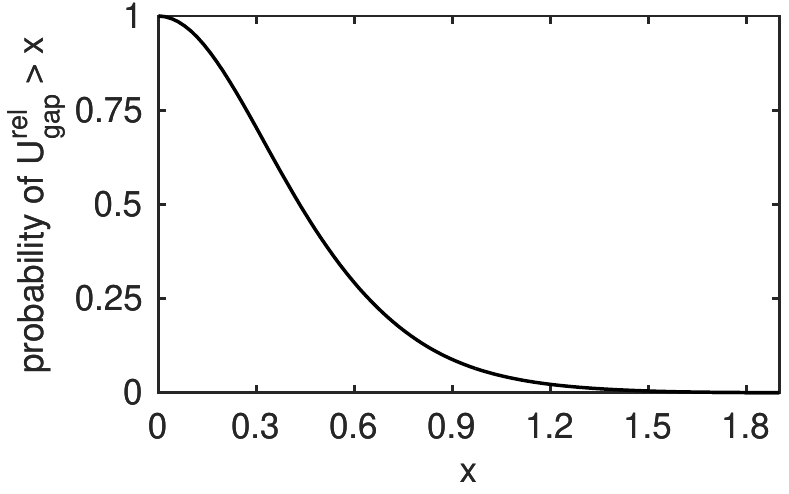}
\caption{Tail distribution~(exceedance) of the energy gap relative to the mean energy between repulsive and adjacent BO~surface for a trimer whose atomic mean positions are in the CI configuration.
 The distribution measures the probability that the relative energy gap is above a varying level. We used for the atomic positions Gaussian probability distributions and calculate for each realization the relative energy gap. The varying atomic positions are $x_2,\ y_3, \ y_4$. The distribution is only dependent on the ratio between the standard deviation of the atomic spatial distributions, $\sigma_0$, and the mean interatomic distance of the vertical dimer, $a_2$. This ratio is adjusted to $\sigma_0/a_2 =0.0952$ to reflect the situation for the two perpendicular dimer aggregate, discussed in \sref{sec:nges4}.
\label{fig:asymmetric_trimer_tail_distr_energy_gap}
}
\end{figure}
The result is shown in \fref{fig:asymmetric_trimer_tail_distr_energy_gap} and we see that the distribution is much broader than the one for the trivial (avoided) crossing, shown in \fref{fig:prob_distr_Ugap_rel}, finding large energy gaps with significant probability. This is one of the reasons, why almost surely the transition between repulsive and adjacent BO~surface occurs for the trivial (avoided) crossing, but in the vicinity of the CI the transition can be avoided with significant probability.

\section{Forces on atoms from resonant interactions}
\label{part:app::calcs_planar_aggr__forces_resonant_interactions}
When the aggregate is in the exciton state $\ket{\varphi(\vc{R})}$, the force from resonant interactions on atom $k$ can be calculated with $\vc{F}^{\ket{\varphi}}_{k} =\bra{\varphi(\vc{R})}\nabla_{\vc{R}_k} \hat{H}_{\mathrm{dd}}(\vc{R}) \ket{\varphi(\vc{R})}$. Evaluating this formula for the use with isotropic binary interactions, $V(\vc{R}_{k,l}) = V(R_{k,l})$, and the basis $\{\ket{\pi_k}\}$ yields 
\begin{equation}
 \vc{F}^{\ket{\varphi}}_{k} = 2\Re\left\{\braket{\varphi(\vc{R})|\pi_{k}}\sum_{l \neq k}\left(\nabla_{\vc{R}_k}V({R_{k,l}})\right)\braket{\pi_{l}|\varphi(\vc{R})}\right\}.
\end{equation}
For anisotropic binary interactions, $V_{m,m'}(\vc{R}_{k,l})$, and the basis $\{\ket{\pi_k,m}\}$, we get 
\begin{equation}
 \vc{F}^{\ket{\varphi}}_{k} = 2\Re\left\{\sum_{m}\braket{\varphi(\vc{R})|\pi_{k},m}\sum_{l \neq k}\sum_{m'}\left(\nabla_{\vc{R}_k}V_{m,m'}(\vc{R}_{k,l})\right)\braket{\pi_{l},m'|\varphi(\vc{R})} \right\}.
\end{equation}
For both cases we find, that a force from resonant interactions is only acting on an atom when it shares excitation.

\chapter{Microwave excitation of excitons}
\label{part:app::chap:exc_mw}
We demonstrate here how to get access to excitons with a microwave. The microwave wavelength has to be large to ensure homogeneous field strengths for more than one atom. This also allows to use the dipole approximation such that the electric field directly couples to the atomic dipoles without spatial dependence. We furthermore restrict ourselves to the treatment of linearly polarized and monochromatic light in $\vc{q}$ direction, such that we can write $\vc{E}(t) = \mathcal{E}\cos\left( \omega t\right)\vc{q}$. Setting the co-ordinate system in which the dipoles are defined such that the $z$ axis is set to the microwave polarization direction, the microwave couples only to the $d_{0}$ components of the dipole. 
The aggregate-microwave interaction can thus approximately be written as
\begin{align}
 \hat{H}_{\mathrm{mw}}(t) = \mathcal{E}\cos\left( \omega t\right) \sum_{\alpha=1}^{N} \hat{d}_{0}^{(\alpha)},
\label{eq:app::exc_microwave:Hmw}
\end{align}
with $\alpha$ labeling the aggregate's atoms. 
The excitation is to be performed from the Rydberg ground state $\ket{\lstate{S}} = \ket{\lstate{s}\dots\lstate{s}}$ to an exciton of the electronic Hamiltonian, which makes it necessary to enlarge the electronic basis and also the electronic Hamiltonian by adding the state $\ket{\lstate{S}}$. 
The aggregate-microwave interaction represented in the extended electronic basis makes it necessary to first determine the matrix elements. Since the microwave couples exclusively to the $d_{0}$ components of the dipoles, matrix elements of the coupling operator in \eref{eq:app::exc_microwave:Hmw} are non-vanishing only for aggregate states $\ket{\pi_{\alpha},m}$ with $m=0$. We therefore abbreviate $\ket{\pi_{\alpha}} \equiv \ket{\pi_{\alpha},m}$ and can evaluate the matrix elements:
\begin{align}
\bra{\lstate{S}}\hat{H}_{\mathrm{mw}}(t)\ket{\lstate{S}} &=0,\\
\bra{\pi_k}\hat{H}_{\mathrm{mw}}(t)\ket{\pi_l} &= 0,\\
 \bra{\pi_k}\hat{H}_{\mathrm{mw}}(t)\ket{\lstate{S}} &= \hbar \Omega \cos\left( \omega t\right),
\end{align}
with the \emph{Rabi frequency}\index{Rabi frequency} defined as $\Omega:=\mathcal{E}\vc{q}\bra{\lstate{s}}\hat{\vc{d}}\ket{\lstate{p}}/\hbar = E\,\drad_{\nu,1;\nu,0}/\sqrt{3}\hbar$.
Most importantly, we need to get access to Rydberg \emph{dimer} excitons. The dimer states are necessary for initiating the exciton pulses. Therefore it is enough to restrict the treatment to two atoms. The resonant dipole-dipole transition matrix elements between aggregate states with $m=0$ are given by $V_{0,0}(\vc{r}) = \left(1-3\cos^2\theta\right)\drad_{\nu,1;\nu,0}^2/3r^3$, where $\theta$ is the angle between microwave polarization direction and direction of the interatomic distance vector $\vc{r}$. Altough the microwave couples only to $m=0$ states, the resonant dipole-dipole interactions can still couple to states with $m=\pm1$. A complete decoupling of the $m=0$ subspace is only ensured for perpendicular or parallel adjustment of the microwave polarization direction relative to the interatomic distance vector, $\theta =0,\pi/2$. We restrict the description of the excitation scheme to these two orientations, which are the ones used in the main text.
We furthermore neglect the van-der-Waals interactions in the excitation scheme description, which gives only a small shift to the eigenenergy values, but does not change the excitons.
\paragraph{perpendicular alignment} When the polarization direction is perpendicular to the interatomic distance vector such that $\theta =\pi/2$, the eigenenergies are given by $E_{\pm}(r) := \pm \drad_{\nu,1;\nu,0}^2/3r^3$ and the corresponding excitons $\ket{\varphi_{\pm}} = \left(\ket{\pi_{1}} \pm \ket{\pi_{2}}\right)/\sqrt{2}$.
\paragraph{parallel alignment} For polarization direction and the interatomic distance being parallel, $\theta =0$, the eigenenergies are given by $E_{\pm}(r) := \pm \drad_{\nu,1;\nu,0}^2 2/3r^3$, whereas the excitons are the same, however, the assignment to the energies changes, \linebreak[4]$\ket{\varphi_{\pm}} = \left(\ket{\pi_{1}} \mp \ket{\pi_{2}}\right)/\sqrt{2}$.

We specifically need to access the surface/energy $E_{+}$ which exerts repulsive forces on the atoms. In the following we restrict the treatment to the perpendicular alignment. Changing the representation to the eigenbasis $\{\ket{\varphi_{+}},\ket{\varphi_{-}},\ket{S}\}$, the atomic levels and interactions are captured by the Hamiltonian
\begin{equation}
 \hat{H}_0 =
\begin{pmatrix}
 E_{+} & 0 & 0\\
0 & E_{-} & 0\\
0 & 0 & -\Delta_{\lstate{S}}
\end{pmatrix},
\end{equation}
assuming that $\ket{\lstate{S}}$ is detuned by $-\Delta_{\lstate{S}}$ from the single excitation manifold. In the same basis, the atom-microwave interaction is given by
\begin{equation}
 \hat{H}_{I}(t) =\hbar \Omega \cos\left( \omega t\right)
\begin{pmatrix}
 0 & 0 & \sqrt{2}\\
0 & 0 & 0\\
\sqrt{2} & 0 & 0
\end{pmatrix}.
\label{eq:microwave_coupling_exciton_basis}
\end{equation}
We realize, that the antisymmetric exciton, $\ket{\varphi_{-}}$, is not accessible with a spatially homogeneous field since it decouples here in the description. This is the reason why we chose to treat the case of perpendicular alignment of the microwave, since for parallel alignment the repulsive surface is not accessible due to its antisymmetric exciton.
The decoupling of the antisymmetric state from the microwave allows us to remove it in the further description.

To eliminate the trivial oscillations of state populations which arise due to the Hamiltonian $\hat{H}_0$, we transform the Schrödinger equation $\im \hbar\, \partial \ket{\psi(t)}/ \partial t  =(\hat{H}_0+\hat{H}_{I}(t))\ket{\psi(t)}$ to the \emph{interaction picture}\index{interaction picture}, which is then given by $\im \hbar\, \partial \ket{{\psi}(t)}/ \partial t  =\hat{\tilde{H}}_{I}(t)\ket{{\psi}(t)}$, with $\hat{\tilde{H}}_{I}(t) = \mathrm{e}^{\im \hat{H}_0 t /\hbar}\hat{H}_{I}(t))\mathrm{e}^{-\im \hat{H}_0 t /\hbar}$. In the basis $\{\ket{1}\equiv\ket{\varphi_{+}},\ket{0}\equiv\ket{\lstate{S}}\}$, evaluation yields the dynamics governing Hamiltonian
\begin{align}
 \hat{\tilde{H}}_{I}(t) &= 
\sqrt{2}\hbar \Omega \cos\left( \omega t\right)
\begin{pmatrix}
 0 &  \mathrm{e}^{\im \omega_{\mathrm{A}} t }\\
\mathrm{e}^{-\im \omega_{\mathrm{A}} t } & 0 
\end{pmatrix}\\
&=
\sqrt{2}\hbar \Omega 
\begin{pmatrix}
0 &  \mathrm{e}^{\im \left(\omega_{\mathrm{A}} + \omega \right)t }/2 + \mathrm{e}^{-\im \Delta t }/2\\
 \mathrm{e}^{-\im \left(\omega_{\mathrm{A}} + \omega \right)t }/2 + \mathrm{e}^{\im \Delta t }/2 & 0 
\end{pmatrix}
\end{align}
where $\omega_{\mathrm{A}} = \left(E_{+}+\Delta_{\lstate{S}}\right)/\hbar$ is the frequency corresponding to the transition $\ket{0}\to\ket{1}$. In a second step we rewrote the cosine in its frequency representation and defined the detuning between microwave frequency and the transition frequency corresponding to $\ket{0}\to\ket{1}$, $\Delta:=\omega - \omega_{\mathrm{A}}$. Performing an additional unitary transformation, we arrive at the Schrödinger equation $\im \hbar\, \partial \ket{\psi(t)}/ \partial t  =\hat{H}(t)\ket{{\psi}(t)}$, with 
\begin{align}
  \hat{H}(t) &=   -\hbar \Delta \hat{\sigma}_{1,1} + \mathrm{e}^{\im \Delta \hat{\sigma}_{1,1} t}\hat{\tilde{H}}_{I}(t))\mathrm{e}^{-\im \Delta \hat{\sigma}_{1,1}t}\\
&= \hbar
\begin{pmatrix}
 -\Delta  & \Omega/\sqrt{2} + \mathrm{e}^{2\im \omega t}\Omega/\sqrt{2}\\
\Omega/\sqrt{2} + \mathrm{e}^{-2\im \omega t}\Omega/\sqrt{2} & 0
\end{pmatrix}
\end{align} where $\hat{\sigma}_{1,1} = \ket{1}\bra{1}$ is the projector onto the subspace of $\ket{1}$. We thus were able to reduce the three state excitation problem to the standard Rabi two level system. In a last step, we perform a \emph{rotating wave approximation}\index{rotating wave approximation} and neglect all terms rotating with $2\omega$. They average out on a much shorter time scale and are therefore negligible. Finally, we arrive at a time independent Hamiltonian,
\begin{equation}
 \hat{H}(t) \approx \hbar
\begin{pmatrix}
 -\Delta  & \Omega/\sqrt{2}\\
\Omega/\sqrt{2} & 0
\end{pmatrix},
\end{equation}
and the population in the exciton state, $P_{1}(t):=|\braket{1|\psi(t)}|^2$, is given by\cite{metcalf1999:laser_cooling_book}
\begin{equation}
 P_{1}(t) = \dfrac{2\Omega^2}{\Delta^2 + 2\Omega^2}\sin^2\left(\sqrt{\Delta^2 + 2\Omega^2}t/2\right),
\end{equation}
for the initial condition where all population is in the state $\ket{\lstate{S}}$, $\ket{\psi(0)} = \ket{\lstate{S}}$. The exciton state $\ket{\varphi_{+}}$ can only be populated completely for the microwave frequency being resonant with the $\ket{S} \to \ket{\varphi_{+}}$~transition frequency. 

With this excitation scheme we get access to the required initial exciton state for the aggregate with unconstrained dynamics, discussed in \Chref{part:rs::chap:unconstrained_aggregate}. There, the initial symmetric exciton state gave access to the CI during the dynamics. However, using isotropic interactions, as in \Chref{part:rs::chap:planar_aggregates} for the planar aggregates, requires to prepare the aggregate initially in the antisymmetric exciton state, which is not accessible by a spatially homogeneous microwave, as it is apparent from \eref{eq:microwave_coupling_exciton_basis}. Nevertheless, excitation schemes into asymmetric states are available. One way is to change the relative phase of the exciton state by driving a resonant transition between the \lstate{p} state and the absolute ground state locally, which means only on \emph{one} of the two atoms\cite{kruse2010:site_selective_manipulation_atomic_quantum_system}.

\chapter{Conical intersections}\label{part:app::chap:con_int}
The theoretical description of dynamics in atomic and molecular systems is a difficult task due to the necessity of using quantum mechanics and the large number of constituents. Apart from Hydrogen, already single atoms are at least three body systems and the feasibility of a full description saturates quickly with increasing number of constituents. Nevertheless, in many situations approximations can be performed to get a solution close to the exact one. In atomic and molecular systems, the dynamics of the nuclei occurs often on a much slower time scale than for electrons due to the much bigger masses of nuclei. The contribution to the forces on the nuclei from the electrons is therefore well described by averaging over the electron configurations. The nuclei move according to forces from potential energy surfaces~(PES), in this context they are also called Born-Oppenheimer~surfaces, which we already introduced in \Chref{part:fd::chap:theoretical_framework}. The restriction that the nuclei propagate along exactly a single PES is the \emph{Born-Oppenheimer approximation}. However, due to the quantum nature several coupled BO~surfaces exist and the Born-Oppenheimer approximation is only valid under two conditions: First, the motion of the nuclei remains small during the dynamics and second, the PESs are energetically sufficiently separated. Although the first condition is usually met, the second is in higher dimensional systems often not fulfilled. \emph{Conical intersections}\index{conical intersection}\cite{neumann1929:ci_first_paper,teller1937:ci_crossing_of_pes,yarkony1996diabolical,yarkony2001conical,domcke2004:book_cis,spiridoula2007:ci_molec_systems_review} are crossings of BO~surfaces in configuration space and their appearance invalidates the Born-Oppenheimer approximation. 
They provide ultra-fast relaxation from excited to lower lying states for chemical and biological processes\cite{hahn2002:ultrafast_cis_trans_photoswitching,ben-nun2002:cis_trans_photoisomerization,levine2007:isomerization_through_CI,polli2010:ci_photoisomerization}. In general it is today widely accepted that in organic photochemistry, conical intersections serve as decay channel from excited to ground states\cite{robb1995:CI_as_mechanistic_feature_organic_photochemistry}.

Owing to its significance, we want to state the criteria when two BO~surfaces conically intersect. Denoting the electronic Hamiltonian with $\hat{H}$ which is dependent on configurations of the nuclear degrees of freedom, which we denote with $\vc{R}$, the eigen value problem can be written as
\begin{equation}
 \hat{H}(\vc{R})\ket{\varphi_{k}(\vc{R})} = U_k(\vc{R})\ket{\varphi_{k}(\vc{R})}, \quad k=1,2,
\end{equation}
with $\ket{\varphi_{k}(\vc{R})}$ electronic eigenstates and $U_k(\vc{R})$ BO~surfaces. Although the BO~surfaces can be closely spaced and can even intersect, the Hamiltonian itself is in most situations continuously differentiable due to containing binary interactions between constituents, which often scale with powers of inverse spacings e.g. the Coulomb potential or the potential of dipole-dipole interactions. The differentiability of the Hamiltonian implies that there is a basis in which all matrix elements and the basis states are differentiable. The so called \emph{diabatic basis} is constructed to fulfill this requirement, which feature the even stronger condition that all derivative couplings vanish. Technically speaking, for $\{\ket{\pi_k}\}$ being a diabatic basis, $\bra{\pi_k}\nabla_{\vc{R}}\ket{\pi_l} = 0$ for all states. The single excitation states, defined in \eref{eq:planar_aggr:def_pi_states} and in \eref{eq:3D_aggregate:aggregate_states} are not dependent on the configurations $\vc{R}$ and thus form a diabatic basis.

Suppose the electronic Hamiltonian can be written as
\begin{equation}
 \hat{H} =
\begin{pmatrix}
H_{11} & H_{12}\\
H_{12}^{*} & H_{22} 
\end{pmatrix},
\label{eq:app_ci:el_H}
\end{equation}
with $H_{kl} = \bra{\pi_k}\hat{H}\ket{\pi_l}$ the matrix elements in a diabatic basis $\{\ket{\pi_k}\}$. The Hamiltonian and the diabatic matrix elements dependent parametrically on $\vc{R}$. We omit to write this dependency for better readability. The BO~surfaces are given by
\begin{equation}
 U_{\pm} = \bar{H} \pm \sqrt{(\Delta H)^2 + |H_{12}|^2},
\label{eq:app_ci:BOsurf_sol}
\end{equation}
with the mean value $\bar{H} := (H_{11} + H_{22})/2$ and the half distance between the diagonal elements $\Delta H := (H_{11} - H_{22})/2$. The electronic eigenstates can be expressed as
\begin{align}
 \ket{\varphi_{+}} &= \cos(\alpha/2)\ket{\pi_{1}} + \sin(\alpha/2)\ket{\pi_{2}}\\
 \ket{\varphi_{-}} &= - \sin(\alpha/2)\ket{\pi_{1}} + \cos(\alpha/2)\ket{\pi_{2}},
\end{align}
where the trigonometric coefficients are given by
\begin{align}
 \sin\alpha &= \dfrac{H_{12}}{\sqrt{(\Delta H)^2 + |H_{12}|^2}},\\
\cos\alpha &= \dfrac{\Delta H}{\sqrt{(\Delta H)^2 + |H_{12}|^2}}.
\end{align}
The BO~surfaces are degenerate, when\cite{spiridoula2007:ci_molec_systems_review,yarkony1998:ci_often_misunderstood} 
\begin{align}
 \Delta H &= 0,\label{eq:app_ci:cond1_degeneracy}\\
H_{12} &= 0.\label{eq:app_ci:cond2_degeneracy}
\end{align}
The subspace of the full configuration space containing configurations for which the BO~surfaces are degenerate is called the \emph{seam space}\index{conical intersection!seam space}, whereas all configurations lifting the degeneracy are collected in the \emph{branching plane}\index{conical intersection!branching plane}. If the configuration space is $N$-dimensional, the seam space is $(N-2)$-dimensional and the branching plane two-dimensional. Thus, the entire configuration space has to be at least two-dimensional in order to show degeneracies.

The change of BO~surfaces in the vicinity of a degeneracy configuration, which we denote with $\vc{R}^{\star}$, can be calculated by taylor expanding
the Hamiltonian in \eqref{eq:app_ci:el_H} around $\vc{R}^{\star}$ and using \eref{eq:app_ci:cond1_degeneracy} and \eref{eq:app_ci:cond2_degeneracy}. It is useful to abbreviate the gradients of the three quantities which determine the BO~surface,
\begin{align}
 \vc{s} &:= \nabla_{\vc{R}} \bar{H},\\
\vc{g} &:= \nabla_{\vc{R}}(\Delta {H}),\\
\vc{h} &:= \nabla_{\vc{R}}{H}_{12},
\end{align}
to write down the BO surfaces in the vicinity of a CI,
\begin{equation}
U_{\pm}(\vc{R}^{\star} + \delta \vc{R}) \approx \bar{H}^{\star} +  \vc{s}^{\star}\delta \vc{R} \pm \sqrt{(\vc{g}^{\star}\delta \vc{R})^2 + |\vc{h}^{\star}\delta \vc{R}|^2},
\end{equation}
where quantities with ${}^\star$ denote evaluation at $\vc{R} = \vc{R}^{\star}$. The degeneracy is a conical intersection when $\vc{g}^{\star}, \vc{h}^{\star}$ are nonvanishing and linearly independent.
The two vectors then span the branching plane, which is therefore also known as $g$--$h$ space\cite{yarkony1996diabolical,yarkony2001conical}.
For displacements from the intersecting point within the branching plane, we can set $\delta \vc{R} = x \vc{e}_{g}^{\star} + y \vc{e}_{h}^{\star}$, where $\vc{e}_{g}^{\star} := 
\vc{g}^{\star}/\|\vc{g}^{\star}\|_{2},\ \vc{e}_{h}^{\star} := 
\vc{h}^{\star}/\|\vc{h}^{\star}\|_{2}$ are unit vectors in $\vc{g}^{\star},\ \vc{h}^{\star}$ direction and with $s^{\star}_{g}:=\vc{e}_{g}^{\star}\vc{s}^{\star},\  s^{\star}_{h}:=\vc{e}_{h}^{\star}\vc{s}^{\star}$ we find
\begin{equation}
U_{\pm}(\vc{R}^{\star} + x \vc{e}_{g}^{\star} + y \vc{e}_{h}^{\star}) \approx \bar{H}^{\star} +  xs^{\star}_{g} + ys^{\star}_{h} \pm \sqrt{x^2\|\vc{g}^{\star}\|_{2}^{2} + y^2\|\vc{h}^{\star}\|_{2}^{2}},
\end{equation}
Therefore, the degeneracy is lifted linearly for displacements along the branching.


\chapter{Effective interactions from block-diagonalization}
\label{part:app::chap:block_diag}
When a system is coupled to other systems, the treatment of the total system is often not tractable, since it is just too large.
If the interactions between the systems are small, a derivation of a new basis is possible, in which the systems (almost) completely decouple. This procedure is an iterative block-diagonalization of the total system, known as van Vleck perturbation theory\cite{vanVleck:PT}. Here we demonstrate the scheme in a canonical way, outlined by Shavitt et al\cite{shavitt:vanVleck}.

Suppose we have a Hamiltonian $\hat{\mathcal{H}} = \hat{\mathcal{H}}_{\mathrm D} + \hat{\mathcal{W}}$ of the total system,
which consists of a block-diagonal part, $\hat{\mathcal{H}}_{\mathrm D}$, and an interaction $\hat{\mathcal{W}}$. 
We denote for every linear operator $\hat{\mathcal{A}}$ in the space of the total system with $\hat{\mathcal{A}}_{\mathrm X} := \hat{\mathcal{A}} - \hat{\mathcal{A}}_{\mathrm D}$ the corresponding operator with vanishing block-diagonal entries.

In the following we assume the block-diagonal part of the interactions vanishes, such that $\hat{\mathcal{W}} = \hat{\mathcal{W}}_{\mathrm X}$. The objective is to find a unitary transformation $\hat{\mathcal{U}}$, such that $\hat{\mathcal{H}}' =\hat{\mathcal{U}}^\dagger \hat{\mathcal{H}} \hat{\mathcal{U}} $ is block-diagonal, i.e. $\hat{\mathcal{H}}'_{\mathrm X}=0$. In canonical van Vleck perturbation theory, the transformation is rewritten as $\hat{\mathcal{U}} = \exp(\hat{\mathcal{G}})$ with the property $\hat{\mathcal{G}} = -\hat{\mathcal{G}}^\dagger$ and $\hat{\mathcal{G}}_{\mathrm D} = 0$. The generator fullfills the commutation relation
\begin{equation}
 \left[\hat{\mathcal{H}}_{\mathrm D},\hat{\mathcal{G}}\right] = -\sum_{n=0}^{\infty} c_n \left[\hat{\mathcal{W}},\hat{\mathcal{G}}\right]_{2n}
\label{eq:app::blockdiagonalization__komm_G},
\end{equation}
where $\left[\hat{A},\hat{B}\right]_{k} = \left[\left[\hat{A},\hat{B}\right]_{k-1},\hat{B}\right]$ denotes the recursive commutator, with $\left[\hat{A},\hat{B}\right]_{0} = \hat{A}$ and $c_n$ are related to the Bernoulli numbers $\mathfrak{b}_{2n}$\cite{abramowitz2006:handbook_mathematical_functions}:
\begin{equation}
 c_n = \dfrac{2^{2n}}{(2n)!}\mathfrak{b}_{2n}.
\label{eq:app::blockdiagonalization__expansion_coeffs_cn}
\end{equation}
Note that \eref{eq:app::blockdiagonalization__komm_G} already exploits that in our case $\hat{\mathcal{W}}_{D} = 0$. An expansion of \eref{eq:app::blockdiagonalization__komm_G} order by order leads to the equations
\begin{align}
 [\hat{\mathcal{H}}_{\mathrm D},\hat{\mathcal{G}}^{(1)}] &= -\hat{\mathcal{W}}
\label{eq:app::blockdiagonalization__det_Gl1}\\
[\hat{\mathcal{H}}_{\mathrm D},\hat{\mathcal{G}}^{(2)}] &= 0
\label{eq:app::blockdiagonalization__det_Gl2}\\
[\hat{\mathcal{H}}_{\mathrm D},\hat{\mathcal{G}}^{(3)}] &= -1/3[[\hat{\mathcal{W}},\hat{\mathcal{G}}^{(1)}],\hat{\mathcal{G}}^{(1)}],
\label{eq:app::blockdiagonalization__det_Gl3}\\
&\ \ \vdots\nonumber
\end{align}
which can be used to determine the $\hat{\mathcal{G}}^{(\xi)}, \xi \in \{1,2, \dots\}$. The block-diagonalized Hamiltonian is order by order determined by
\begin{align}
\hat{\mathcal{H}}'_{(0)} &= \hat{\mathcal{H}}_0,
\label{eq:app::blockdiagonalization__Hprime0}\\
\hat{\mathcal{H}}'_{(1)} &= 0,
\label{eq:app::blockdiagonalization__Hprime1}\\
\hat{\mathcal{H}}'_{(2)} &= 1/2[\hat{\mathcal{W}},\hat{\mathcal{G}}^{(1)}],
\label{eq:app::blockdiagonalization__Hprime2}\\
\hat{\mathcal{H}}'_{(3)} &= 1/2[\hat{\mathcal{W}},\hat{\mathcal{G}}^{(2)}],
\label{eq:app::blockdiagonalization__Hprime3}\\
\begin{split}
\hat{\mathcal{H}}'_{(4)} &= 1/2[\hat{\mathcal{W}},\hat{\mathcal{G}}^{(3)}]\\
&-1/24[[[\hat{\mathcal{W}},\hat{\mathcal{G}}^{(1)}],\hat{\mathcal{G}}^{(1)}],\hat{\mathcal{G}}^{(1)}],
\end{split}
\label{eq:app::blockdiagonalization__Hprime4}
\end{align}
such that $\hat{\mathcal{H}}' = \sum_{\xi =0}^{\infty}\hat{\mathcal{H}}'_{(\xi)}$.

\section{Effective interactions of off-resonant coupled systems}
\label{part:app::chap:block_diag__effective_interactions}
We consider now $N\geq2$ coupled systems, $\hat{H}_{k}, \ k=1,...,N$. The interaction between system $k$ and $l$ is given by $\hat{W}_{k,l}$, such that we can write
\begin{align}
 \hat{\mathcal{H}}_{\mathrm D} &= \sum_{k=1}^{N}\hat{H}_{k}\otimes \ket{k}\bra{k},
\label{eq:app::blockdiagonalization__Hdiag_decomposition}
\\
\hat{\mathcal{W}} &=\sum_{k=1}^{N}\sum_{l=k+1}^N \hat{W}_{k,l}\otimes \ket{k}\bra{l} + \hat{W}_{k,l}^\dagger\otimes \ket{l}\bra{k},
\label{eq:app::blockdiagonalization__W_decomposition}
\\
\hat{\mathcal{G}}^{(\xi)} &=\sum_{k=1}^{N}\sum_{l=k+1}^N \hat{G}^{(\xi)}_{k,l}\otimes \ket{k}\bra{l} - \hat{G}^{\dagger(\xi)}_{k,l}\otimes \ket{l}\bra{k},
\label{eq:app::blockdiagonalization__G_decomposition}
\end{align}
where $\{\ket{k}\}_{k=1}^N$ is an arbitrary orthonormal basis, where state $\ket{k}$ adresses the system labeled with k.
With this composition, we can rewrite the following commutators
\begin{align}
 \left[\hat{\mathcal{H}}_{\mathrm D},\hat{\mathcal{G}}^{(\xi)}\right] &= \hat{H}_{k}\hat{G}^{(\xi)}_{k,l} - \hat{G}^{(\xi)}_{k,l}\hat{H}_{l}
\label{eq:app::blockdiagonalization__rewrite_komm_Hdiag_G}
\\
\bra{k}\left[\hat{\mathcal{W}},\hat{\mathcal{G}}^{(\xi)}\right]\ket{k}&=\sum_{\substack{l=1,\\l\neq k}}^{N}\left(\hat{W}_{k,l}\,\hat{G}^{(\xi)}_{l,k} + \mathrm{H.c.}\right),
\label{eq:app::blockdiagonalization__rewrite_komm_W_G}
\end{align}
where $\mathrm{H.c.}$ is the abbreviation for Hermitian adjoint.
Up to third order, the block-diagonalization of each system $k$, is given by
\begin{equation}
 \hat{H}_k' \approx \hat{H}_k + \dfrac{1}{2}\sum_{\xi =1}^{2}\sum_{k=1}^{2}\sum_{\substack{l=1,\\l\neq k}}^{N}\left(\hat{W}_{k,l}\,\hat{G}^{(\xi)}_{l,k} + \mathrm{H.c.}\right)
\end{equation}

In many applications of van Vleck perturbation, the systems energy is detuned from each other. To account for this, we first rewrite the systems Hamiltonians
\begin{equation}
 \hat{H}_{k}:=E_k \cdot\hat{\id} + \hat{V}_{k},
\label{eq:app::blockdiagonalization__hamiltonian_detuning}
\end{equation}
where $\hat{V}_{k}$ is Hermitian. This structure of the Hamiltonians in \eref{eq:app::blockdiagonalization__hamiltonian_detuning} turns \eref{eq:app::blockdiagonalization__det_Gl1} and \eref{eq:app::blockdiagonalization__det_Gl3} into fix point equations of the contained generator. Furthermore it gives $\hat{G}^{(2)} = 0$.
 We consider now only \eref{eq:app::blockdiagonalization__det_Gl1}, which yields
\begin{equation}
 \hat{G}^{(1)}_{k,l}=\Delta_{k,l}^{-1}\left( \hat{G}^{(1)}_{k,l}\hat{V}_{l} -\hat{V}_{k}\hat{G}^{(1)}_{k,l}-\hat{W}_{k,l}\right),
\label{eq:app::blockdiagonalization__fix_point_equation_G1kl}
\end{equation}
where $\Delta_{k,l}:=E_k - E_l$ is the energy detuning between system $k$ and $l$. This equation can iteratively be solved in powers of $\Delta_{k,l}^{-1}$. For infinite detuning, $\Delta_{k,l} \to \infty$, the generator has to vanish. So we chose as an iteration start $\hat{G}^{(1),(0)}_{k,l}=0$. Inserting this in the rhs of \eref{eq:app::blockdiagonalization__fix_point_equation_G1kl}, we get the generator up to first order in $\Delta_{k,l}^{-1}$. Doing this recursion over and over reveals the generator in higher inverse powers of the detuning. We stop with the second order and take this as an approximation of the generator itself, which gives
\begin{equation}
 \hat{G}^{(1)}_{k,l} \approx -\Delta_{k,l}^{-1}\hat{W}_{k,l} + \Delta_{k,l}^{-2}\left(\hat{V}_{k}\hat{W}_{k,l} - \hat{W}_{k,l}\hat{V}_{l}\right).
\label{eq:app::blockdiagonalization__solution_G1kl_approximate}
\end{equation}
With this solution at hand, we see from \eref{eq:app::blockdiagonalization__det_Gl3} and the evaluation of the lhs commutator with \eref{eq:app::blockdiagonalization__rewrite_komm_Hdiag_G}, that the leading order of $\hat{G}^{(3)}_{k,l}$ is proportional to $\Delta_{k,l}^{-3}$. This implies that the block-diagonal solutions, $\hat{H}'_k$, up to the second inverse orders of the detunings require only an evaluation of \eref{eq:app::blockdiagonalization__Hprime2}, which finally yields
\begin{equation}
 \hat{H}_{k}' \approx \hat{H}_{k} + \sum_{\substack{l=1,\\l\neq k}}^{N}\left(\Delta_{k,l}^{-1}\hat{W}_{k,l}\hat{W}_{k,l}^{\dagger}+\Delta_{k,l}^{-2}\left[\hat{W}_{k,l}\hat{V}_{l}\hat{W}_{k,l}^{\dagger}-\dfrac{1}{2}\{\hat{V}_{k},\hat{W}_{k,l}\hat{W}_{k,l}^{\dagger}\}\right]\right).
\label{eq:app::blockdiagonalization__perturb_solution_with_V_ops}
\end{equation}
If instead of the operators $\hat{V}_l,\ l\neq k $, we use the operators $\hat{H}_l$ on the rhs of \eref{eq:app::blockdiagonalization__perturb_solution_with_V_ops}, the formula changes to
\begin{equation}
 \hat{H}_{k}' \approx \hat{H}_{k} + \sum_{\substack{l=1,\\l\neq k}}^{N}\left(2\Delta_{k,l}^{-1}\hat{W}_{k,l}\hat{W}_{k,l}^{\dagger}+\Delta_{k,l}^{-2}\left[\hat{W}_{k,l}\hat{H}_{l}\hat{W}_{k,l}^{\dagger}-\dfrac{1}{2}\{\hat{H}_{k},\hat{W}_{k,l}\hat{W}_{k,l}^{\dagger}\}\right]\right).
\label{eq:app::blockdiagonalization__perturb_solution_with_H_ops}
\end{equation}
Note that if the perturbative series is truncated already with the first inverse order of the detunings, then the factor $2$ in front of $\Delta_{k,l}^{-1}$ in \eref{eq:app::blockdiagonalization__perturb_solution_with_H_ops} has to be neglected.

\section{A block-diagonalization scheme for treating off-resonant interactions between an atom\hyp{}pair}
\label{part:app::chap:block_diag__van_der_Waals}
Considering two dipole-dipole interacting atoms of the same species, we have as starting point a Hamiltonian of the form given in \eref{eq:alk_ryd_atoms:vdW_interactions__total_hamiltonian}. 
The block-diagonalization scheme is a bit more complicated to evaluate due to the pair state basis. We use as basis $\{\ket{x}\}$, where $\ket{x}$ is a pair state without magnetic quantum number dependency, explicitly, a pair state for atoms~($\alpha, \beta$) is given by $\ket{x} = \ket{\gamma_\alpha}\otimes\ket{\gamma_\beta}$, with $\ket{\gamma_{k}} = \ket{\nu_k,\ell_k}$, $k= \alpha, \beta$.
We repeat the structure of the sub-Hamiltonians, given in \eref{eq:alk_ryd_atoms:vdW_interactions__def_blocks_H}, which is
\begin{equation}
 \hat{\mathcal{H}}_{x,x'}=
\begin{cases}
 E_{x}\hat{\id}_{x}, & x= x' \\
\hat{V}_{x,x'}, & x \neq x',
\end{cases}
\label{eq:alk_ryd_atoms:vdW_interactions__def_blocks_H_repeat}
\end{equation}
The difference now to the previous scheme of block-diagonalization is, that to each pair state $\ket{x}$ there exists another pair state, which is energetically resonant to it. This is the pair state with swapped quantum numbers on the atoms between the atoms. To account for this, we introduce for each $\ket{x} = \ket{\gamma_{\alpha}}\otimes\ket{\gamma_{\beta}}$ the swapped state, $\ket{T(x)} := \ket{\gamma_{\beta}}\otimes\ket{\gamma_{\alpha}}$, with $T(\gamma_\alpha,\gamma_\beta) := (\gamma_\beta,\gamma_\alpha)$ the transposition of the atomic quantum numbers.
The sub-Hamiltonians projecting to both states are identical, $\hat{\mathcal{H}}_{x,x} = \hat{\mathcal{H}}_{T(x),T(x)} = E_{x}\hat{\id}_{x}$, with $E_{x}$ the energy corresponding to the pair states $\ket{x}$, $\ket{T(x)}$ and $\hat{\id}_{x}$ the identity of the magnetic quantum number subspace.
To avoid issues in the block-diagonalization procedure, the pair state with swapped quantum numbers for each pair state has to be included in the diagonal part of the total block Hamiltonian.
this leads to the following decomposition of the total Hamiltonian,
\begin{align}
 \hat{\mathcal{H}}_{\mathrm D} &= \sum_{\ket{x}} \hat{\mathcal{H}}_{x,x}\otimes \ket{x}\bra{x} + \hat{\mathcal{H}}_{x,T(x)}\otimes \ket{x}\bra{T(x)},
\label{eq:app::blockdiagonalization__Hdiag_decomposition_vdW}
\\
\hat{\mathcal{V}} &=\sum_{\substack{\ket{x},\ket{x'}:\\x'\neq x,T(x)}}\hat{V}_{x,x'}\otimes \ket{x}\bra{x'},
\label{eq:app::blockdiagonalization__W_decomposition_vdW}
\end{align}
such that $\mathcal{H} = \mathcal{H}_{\mathrm D} + \mathcal{V}$.
The expansion of the generators in the pair state basis yields
\begin{equation}
 \hat{\mathcal{G}}^{(\xi)} =\sum_{\substack{\ket{x},\ket{x'}:\\x'\neq x,T(x)}}\hat{G}^{(\xi)}_{x,x'}\otimes \ket{x}\bra{x'}.
\label{eq:app::blockdiagonalization__G_decomposition_vdW}
\end{equation}
Note that the generators, $\hat{\mathcal{G}}^{(\xi)}$, are anti-Hermitian operators. The perturbation order is given by the number $\xi$. The generators have vanishing entries for the diagonal part, which is $\hat{G}^{(\xi)}_{x,x} = \hat{G}^{(\xi)}_{x,T(x)}=0$.
The commutator for two general operators, $\hat{\mathcal{A}}$ and $\hat{\mathcal{B}}$ with matrix elements $\hat{A}_{x,x'}$, $\hat{B}_{x,x'}$ in the pair state basis $\{\ket{x}\}$ is given by
\begin{equation}
 \left[\hat{\mathcal{A}},\hat{\mathcal{B}}\right] = \sum_{\ket{x},\ket{x'}}\left(\sum_{y:\ket{y}}\hat{A}_{x,y}\,\hat{B}_{y,x'} - \hat{B}_{x,y}\,\hat{A}_{y,x'} \right)\ket{x}\bra{x'},
\label{eq:app::blockdiagonalization__general_commutator_eval}
\end{equation}
where in the braket on the rhs we find the matrix element $\left[\hat{\mathcal{A}},\hat{\mathcal{B}}\right]_{x,x'}$. With this we can evaluate the lhs of \eref{eq:app::blockdiagonalization__det_Gl1} - \eref{eq:app::blockdiagonalization__det_Gl3}, which gives
\begin{align}
 \left[\hat{\mathcal{H}}_{\mathrm D},\hat{\mathcal{G}}^{(\xi)}\right]_{x,x'} &=(E_x-E_{x'})\hat{G}^{(\xi)}_{x,x'} + \hat{\mathcal{H}}_{x,T(x)}\hat{G}^{(\xi)}_{T(x),x'} - \hat{G}^{(\xi)}_{x,T(x')}\hat{\mathcal{H}}_{T(x'),x'}\\
&=\Delta_{x,x'}\hat{G}^{(\xi)}_{x,x'} + \hat{V}_{x,T(x)}\hat{G}^{(\xi)}_{T(x),x'} - \hat{G}^{(\xi)}_{x,T(x')}\hat{V}_{T(x'),x'},
\label{eq:app::blockdiagonalization__rewrite_komm_Hdiag_G_vdW}
\end{align}
where we introduced the notation for the energy detunings with
\begin{equation}
\Delta_{x,x'}:=E_{x}-E_{x'}.
\label{eq:app::blockdiagonalization__def_detuning_vdW}
\end{equation}
We restrict ourselves to the evaluation of equation \eref{eq:app::blockdiagonalization__det_Gl1} and abbreviate $\hat{\mathcal{G}}\equiv\hat{\mathcal{G}}^{(1)}$, $\hat{G}_{x,x'}\equiv \hat{G}{}^{(1)}_{x,x'}$. For a fixed transition from $x \to x'$
we get four equations from \eref{eq:app::blockdiagonalization__rewrite_komm_Hdiag_G_vdW} and \eref{eq:app::blockdiagonalization__det_Gl1}, which can be presented in the following compact form,
\begin{equation}
 \Delta_{x,x'}\hat{\vc{G}}_{x,x'} +  \uuline{\hat{\vc{A}}}{}_{x} \hat{\vc{G}}_{x,x'} - \left(\hat{\vc{G}}^T_{x,x'}\uuline{\hat{\vc{B}}}{}_{x}^{T}\right)^T = -\hat{\vc{Z}}_{x,x'},
\label{eq:app::blockdiagonalization__fix_point_equation_G1_matrix_elements_compact_vdW}
\end{equation}
with 
\begin{align}
\hat{\vc{Y}}_{x,x'} &:=\begin{pmatrix}\hat{Y}_{x,x'} &\hat{Y}_{x,T(x')} & \hat{Y}_{T(x),x'} &\hat{Y}_{T(x),T(x')}\end{pmatrix}^T, \quad Y \in\{G,Z\}
\label{eq:app::blockdiagonalization__def_compact_vector_Gxy_vdW}
\end{align}
and
\begin{align}
 \uuline{\hat{\vc{A}}}{}_{x}&:= \begin{pmatrix}
                                  0 & 0 & \hat{V}_{x,T(x)}& 0\\
0 & 0 & 0 & \hat{V}_{x,T(x)}\\
\hat{V}_{x,T(x)}^\dagger & 0 & 0 & 0\\
0 & \hat{V}_{x,T(x)}^\dagger & 0 & 0
                                 \end{pmatrix}
\label{eq:app::blockdiagonalization__def_A_matrix_vdW}
\\
 \uuline{\hat{\vc{B}}}{}_{x'}&:= \begin{pmatrix}
                                  0 & \hat{V}_{x',T(x')}^{\dagger} & 0& 0\\
\hat{V}_{x',T(x')} & 0 & 0 & 0\\
0 & 0 & 0 & \hat{V}_{x',T(x')}^{\dagger}\\
0 & 0 & \hat{V}_{x',T(x')} &0
                                 \end{pmatrix}.
\label{eq:app::blockdiagonalization__def_B_matrix_vdW}
\end{align}
Solving \eref{eq:app::blockdiagonalization__fix_point_equation_G1_matrix_elements_compact_vdW} perturbatively by starting with infinite energy detuning and iterate then the fix point equation with the start $\hat{\vc{G}}_{x,T(x')} = 0$, yields up to second order
\begin{equation}
 \hat{\vc{G}}_{x,x'} \approx \Delta_{x,x'}^{-2}\left(
 \uuline{\hat{\vc{A}}}{}_{x} \hat{\vc{Z}}_{x,x'} - \left(\hat{\vc{Z}}_{x,x'}^T\uuline{\hat{\vc{B}}}{}_{x'}^{T}\right)^T
\right)-\Delta_{x,x'}^{-1} \hat{\vc{Z}}_{x,x'},
\label{eq:app::blockdiagonalization__sol_G_vector}
\end{equation}
which is equivalent to write for each component
\begin{equation}
 \hat{G}_{x,x'} \approx \Delta_{x,x'}^{-2}\left(\hat{V}_{x,T(x)} \hat{V}_{T(x),x'} - \hat{V}_{x,T(x')} \hat{V}_{T(x'),x'}\right) - \Delta_{x,x'}^{-1}\hat{V}_{x,x'}
\label{eq:app::blockdiagonalization__sol_G_elements}
\end{equation}
Evaluating the first correction given in \eref{eq:app::blockdiagonalization__Hprime2} with this expression we get up to first order in the inverse detunings the following block-diagonalized interactions:
\begin{equation}
 \hat{\mathcal{H}}'_{x,x'} \approx \begin{cases}
                             E_{x}\hat{\id}_{x} +\sum\limits_{\ket{y}: E_{y} \neq E_{x}} \Delta_{x,y}^{-1}\hat{V}_{x,y}\hat{V}^{\dagger}_{x,y}, & x' = x\\
\\
\hat{V}_{x,T(x)} +\sum\limits_{\ket{y}: E_{y} \neq E_{x}} \Delta_{x,y}^{-1}\hat{V}_{x,y}\hat{V}_{y,T(x)}, & x' = T(x)\\
                            \end{cases}
\label{eq:app::blockdiagonalization__sol_Hxx_HxTx}
\end{equation}
This result has no assumption about the interaction between the atoms so far. Let us in the following assume that the atoms are dipole-dipole coupled. We specify the pair state $\ket{x} = \ket{\gamma_\alpha,\gamma_\beta}$, where $\gamma_k = (\nu_k,\ell_k)$ are the principal and azimuthal quantum number of atom $k=\alpha, \beta$.
The resonant interactions can be evaluated by \eref{eq:alk_ryd_atoms:dip_dip_interaction_qm_matrix_elements_general} and   the dipole selection rules apply for these interactions, such that for each atom the azimuthal quantum number has to change its value by one for nonvanishing transitions.
 If we set $\ket{y} = \ket{\tilde{\gamma}_\alpha,\tilde{\gamma}_\beta}$, with $\tilde{\gamma}_k = (\tilde{\nu}_k,\tilde{\ell}_k)$, $k=\alpha, \beta$, the product $\hat{V}_{x,y}\hat{V}^{\dagger}_{x,y}$ 
is only nonvanishing, when the following selection rules are fulfilled:
\begin{equation}
 \tilde{\ell}_{k} = \ell_{k} \pm 1, \quad k= \alpha, \beta.
\label{eq:app::blockdiagonalization__selection_rules_diag}
\end{equation}
The product $\hat{V}_{x,y}\hat{V}_{y,T(x)}$ is nonvanishing for the selection rules
\begin{align}
 \tilde{\ell}_{p} = \ell_p + s_p,
\label{eq:app::blockdiagonalization__selection_rules_offdiag1}\\
\tilde{\ell}_{p} = \ell_q + s_q,
\label{eq:app::blockdiagonalization__selection_rules_offdiag2}
\end{align}
with $s_p, s_q \in \{-1,1\}$ and $ q \neq p = \alpha, \beta$. Substracting 
\eref{eq:app::blockdiagonalization__selection_rules_offdiag2} from \eref{eq:app::blockdiagonalization__selection_rules_offdiag1}, we get
\begin{equation}
 \ell_p - \ell_q \overset{!}{=} s_q - s_p = \begin{cases}
                                            0, & s_p = s_q\\
					    \pm 2, & s_q = -s_p = \pm 1.
                                           \end{cases}
\end{equation}
For pair states $\ket{x}$ with a difference in the azimuthal quantum number of one, the product $\hat{V}_{x,y}\hat{V}_{y,T(x)}$ vanishes and we get no corrections for the transition $\ket{x} \to \ket{T(x)}$ in first order, such that $\hat{\mathcal{H}}'_{x,T(x)} \approx \hat{\mathcal{H}}_{x,T(x)}$.


\end{appendices}

\backmatter


\renewcommand*{\bibname}{References}
 \bibliography{dissertation_karsten_leonhardt_v2}


\cleardoublepage
\addcontentsline{toc}{chapter}{Index}
\printindex

\addcontentsline{toc}{chapter}{Nomenclature}
\printnomenclature

\addchap*{Acknowledgements/Danksagung}
Ich möchte hier die Gelegenheit nutzen um mich bei allen zu bedanken, die das Entstehen dieser Arbeit überhaupt erst ermöglicht haben und mit denen ich eine wunderbare Zeit in Dresden verbringen konnte.
Zuallererst möchte ich meinem Doktorvater Prof.~Dr.~Jan-Michael Rost danken, der mir in seiner Abteilung die Möglichkeit gegeben hat an einem interessanten und herausfordernden Thema zu arbeiten. Vor allem ist es aber meinem Betreuer Dr.~Sebastian Wüster zu verdanken, dass diese Arbeit realisiert werden konnte. Über die letzten Jahre hast Du mir ein Profil gegeben, warst geduldig mit mir und hattest immer das Ziel im Auge, mir etwas beizubringen. Durch Dich habe ich einen anderen Zugang zu physikalischen Problemstellungen kennen und schätzen gelernt. Ich danke Dir weiterhin für Dein sorgfältiges Korrekturlesen dieser Arbeit und den vielen Hinweisen und Verbesserungsvorschlägen bzgl. der englischen Sprache.

Mit vielen Wegbegleitern habe ich auf verschiedenste Arten und Weisen die Zeit in Dresden intensiv erlebt.
Von der \glqq Alten Garde\grqq\ danke ich insbesondere Nils Henkel, nicht nur weil er ein Vorbild als Physiker war, sondern vor allem das er mir einen Zugang zu liberalem Gedankengut vermitteln konnte, mit dem ich vorher nicht vertraut war. Ich möchte das hier so explizit erwähnen, da im Osten der Republik das kollektive Erbe der DDR schwer wiegt und ein wertschätzender Blick auf Individualität nicht vermittelt, ja oft sogar verachtet wird. Ich durfte durch die vielen Gespräche und Diskussionen feststellen, wie oberflächlich das Anti-Establishment Denken hier ist und das rationalen Gedanken oft einer emotionalen Opferkultur weicht.
Fabian Maucher danke ich für seine offene Art und dass er mich schnell bekannt machte mit vielen coolen Leuten in Dresden als ich neu war in der Stadt. Unvergessen sind vor allem die vielen genialen WG Partys und Politdiskussionen. Christian Köhler war mir ein toller Zimmerkollege, und, seit dem ich im Kletterbusiness bin, einer der immer Ratschläge parat hat wie diese verdammte Route doch zu meistern ist.

Mehrdad Baghery, Mozhdeh Massah, Younes Javanmard danke ich für die Einführung in die persische Kultur, von der ich insbesondere das leckere Essen und die grandiose Musik schätzen gelernt habe. Mehrdad, danke das Du mich auf diesen wunderbaren Trip in den Iran mitgenommen hast. Younes, über dich gibt es nichts weiter zu sagen, als  das Du great bist :)

Abraham Camacho, Talía Lezama Mergold Love, Omar Adame Arana, Pablo Carlos López danke ich für Verkostungen mit superscharfen mexikanischen Sachen und ihren Salsa Einlagen, die immer wieder eine Augenweide sind. 
Nicht zu vergessen ist Alan Celestino, der mir ein sehr guter Freund geworden ist, der mich auch in seinen ``circle'' aufgenommen hat. Izaak Neri danke ich für die geilen Abende in der Neustadt, in der Kletterarena und den \glqq intellektuellen Gesprächen\grqq. Aus der Klettercommunity möchte ich mich bei Sabine, Janina Kratzert, Fabian Pitzer, James Taylor, Callum Murray, Gary Klindt, Thomas Baberowski für das gemeinsame \glqq Leistenziehen\grqq\ bedanken, bei der auch Izaak, Younes, Talía, Abraham, Omar, Mehrdad und Christian mit dabei waren :D

Ich verneige mich vor dem großen Thüringer Thomas Pohl alias \glqq TPohl\grqq, der ja in seiner Heimat eigtl. für seine Klöße, Bratwürste und als Kickboxer bekannt ist, aber außerhalb Thüringens leider nur als Physiker rezipiert wird. Danke dass Du mir während der Zeit in der ich die Diss. aufgeschrieben habe sehr viel Motivation gegeben und Beistand geleistet hast. Am Institut hast Du Dich ja auch als Ballkünstler mit unendlicher Leidenschaft und Siegeswillen ausgezeichnet.
Die Kicker Sessions mit Sebastian alias \glqq Herrn Wüschter\grqq,  Ulf Saalmann alias \glqq Uffe\grqq\ bzw. \glqq Professor Ulf\grqq, Callum alias \glqq Gollum\grqq\ und Mehrdad alias \glqq der Perser\grqq\ waren regelrechte Schlachten und bleiben unvergessen. Ich alias \glqq WetHand\grqq\ bzw. \glqq The man on cocaine\grqq\ musste im Tor ganz schön zittern, konnte aber das Leben eines TPohls auch ganz schön schwer machen ;)

Armen Hayrapetyan möchte ich danken für nette Abende in der Neustadt bei Bier und gemeinsamen Schwitzen :D

Ich möchte mich weiterhin bedanken bei Alexander Eisfeld für gute Gespräche über die Unterschiede zwischen Ost- und Westdeutschland und die Einführung in die schwäbische Kultur, vor allem aber als kompetenten Gesprächspartner über alles was Physik angeht.

Dank gilt auch Sebastiaan Vlaming, Colm Mulhern und Marcus Morgenstern für schöne Abende in der Neustadt, meinen neuen wunderbaren Zimmerkollegen David Schönleber und Andreas Rubisch für schöne Ge\-spräche und unkompliziertes Zusammenarbeiten, Martin Winter für Gespräche über den 1. und 2. Weltkrieg und Geschichte im Allgemeinen, Valentin Walther für Wanderausflüge in die Sächsische Schweiz, Laura Gil und Adrián Sanz Mora für nette Koch- und Spielabende, Anna Deluca für schöne spanische ``get togethers'', Pierfrancesco Di Cintio für das Bekannt\-machen mit dem Voynich-Manuskripts, Michael Genkin für eine tolle und relaxte Koordination der IMPRS, und Stefan Skupin, Fabio Cinti, Tommaso Macrì, Zachary Walters, Jörg Götte, welche einfach gute Zeitgenossen waren.

Ich danke meinen WG (Ex-)Mitbewohnern Friederike Dietrich, Ann-Christin Damm und Veronika Sunko für eine wunderbare Zeit zusammen. Sorry, dass ich während Diss. aufschreiben kaum in der WG war. Ihr ward bzw. seid großartige Mitbewohner!

Zum Schluß möchte ich insbesondere meiner gesamten Familie danken, meinen Eltern Christine und Dietmar, meinen Großeltern Christa und Rohland, meiner Schwester Daniela und meinem Schwager Uwe und natürlich meinen Lieblingen, meinen Nichten Alysha und Marleen, dass sie mich immer unterstützt und motiviert haben. Leider hast Du, Opa, es nicht mehr ganz geschafft die Fertigstellung dieser Arbeit zu erleben.

{\centering \large\textit{\glqq's is Feieromd 's is Feieromd 's Tochwark is vullbracht\grqq}\par}
{\raggedleft Anton Günther~(1876-1937), Volksdichter und Sänger des Erzgebirges\par}

\cleardoubleemptypage
\addchap*{Versicherung}\thispagestyle{empty}
\noindent
Hiermit versichere ich, dass ich die vorliegende Arbeit ohne unzulässige Hilfe Dritter und ohne Benutzung anderer als der angegebenen Hilfsmittel angefertigt habe; die aus fremden Quellen direkt oder indirekt übernommenen Gedanken sind als solche kenntlich gemacht. Die Arbeit wurde bisher weder im Inland noch im Ausland in gleicher oder ähnlicher Form einer anderen Prüfungsbehörde vorgelegt.\\
Die Arbeit wurde am Max-Planck-Institut f{\"u}r Physik komplexer Systeme in der
Abteilung \glqq Endliche Systeme\grqq \ angefertigt und von Prof. Dr. Jan Michael Rost betreut.\\
Ich erkenne die Promotionsordnung der Fakult\"at Mathematik und
Naturwissenschaften der Technischen Universit\"at Dresden vom 23.02.2011~an.

\vspace*{1.5cm}

   \noindent{------------------------\hfill------------------------}\\
   \makeatletter
   Datum\hfill{}Unterschrift
   \makeatother

\end{document}